\documentstyle[12pt,epsf,here]{report}
\newcounter{nfig}
\newcommand{\figcap}{\par\refstepcounter{nfig}{Fig.\arabic{nfig}. }}
\newcommand{\cd}{\makebox[0.08cm]{$\cdot$}} 
\title
{\bf {Explicitly Covariant Light-Front Dynamics and 
Relativistic Few-Body Systems}}
\author{J. Carbonell$^1$,
B. Desplanques$^2$,
\\{\small \em Institut des  Sciences Nucl\'{e}aires}$^a$,
\\{\small \em  53 avenue des Martyrs, F-38026 Grenoble Cedex, France}
\and V.A. Karmanov$^3$,
\\{\small \em  Lebedev Physical Institute, Leninsky Prospekt 53, 117924
Moscow, Russia}
\and J.-F. Mathiot$^4$
\\{\small \em Laboratoire de Physique Corpusculaire}$^b$,
\\{\small \em 24 avenue des Landais, F-63177 Aubi\`ere Cedex,
France}}

\begin{figure}[b]


----------------------------------------\\
$^1${\small e-mail: carbonel@isnhp4.in2p3.fr}
\\$^2${\small e-mail: desplanq@isn.in2p3.fr}
\\$^3${\small e-mail: karmanov@sci.lpi.msk.su}
\\$^4${\small e-mail: mathiot@in2p3.fr}
\\$^a${\small Unit\'e mixte de recherche CNRS-Universit\'e Joseph
Fourier}
\\$^b${\small Unit\'e mixte de recherche CNRS-Universit\'e Blaise Pascal}
\end{figure}
\begin{document}
\maketitle
\bibliographystyle{unsrt} 

\begin{abstract}

The wave function of a composite system is defined in relativity on a space-time 
surface. In the
explicitly covariant light-front dynamics, reviewed in the present article, the
wave functions are defined on the plane  $\omega \cd x=0$, where $\omega$  is
an arbitrary four-vector with $\omega^2=0$. The standard non-covariant approach
is  recovered as a particular case for $\omega = (1,0,0,-1)$. Using the
light-front plane is of crucial importance, while the explicit covariance gives
strong advantages emphasized through all the review.

The properties of the relativistic few-body wave functions are discussed in
detail and are illustrated by examples in a solvable model.  The
three-dimensional graph technique for the calculation of amplitudes in the
covariant light-front perturbation theory is presented.

The structure of the electromagnetic amplitudes is studied. We  investigate the
ambiguities which arise in any approximate light-front calculations, and which
lead to a non-physical dependence of the electromagnetic amplitude on the
orientation of the light-front plane. The elastic and transition form factors
free from these ambiguities are found for  spin 0, 1/2 and 1 systems.

The formalism is applied to the calculation of the relativistic  wave functions
of  two-nucleon  systems (deuteron, scattering state), with  particular 
attention to the role of their new components in the deuteron elastic and
electrodisintegration form factors and to their connection with  meson exchange
currents. Straigthforward applications to the pion and nucleon form factors and
the $\rho-\pi$ transition are also made.
\end{abstract}

\newpage
\tableofcontents


\chapter{Introduction }\label{intro}
\section{The relevance of relativity in few-body systems}\label{rel}

The relevance of a coherent relativistic description of few-body  systems, both
for bound and scattering states, is now well recognized  in nuclear as well as
in particle physics.

This is already clear in particle physics for the understanding of the wave
functions of the valence  quarks in the nucleon or in the pion, as revealed for 
instance in exclusive reactions at very high momentum transfer.  The typical
example of such a reaction is elastic  scattering, and the extraction of the
electromagnetic form factors,  with their well known behavior at very high
momentum transfer, the so-called  scaling laws expected from QCD.

The need for a coherent relativistic description of few-body systems has  also
become clear in nuclear physics in order for instance to check the  validity of
the  standard description of the microscopic structure of nuclei in terms of 
mesons exchanged between nucleons. In this case, electromagnetic  interactions
play also a central role in ``seeing'' meson exchanges in  nuclei. The
forthcoming experiments at Thomas Jefferson  National Accelerator Facility
(former CEBAF) at momentum transfer of  a few (GeV/c)$^2$ are here of
particular importance.

In both domains, it is obligatory to have, first, a relativistic  description 
of the bound and scattering states. It is also necessary to have a  consistent
description of the electromagnetic current operator needed to  probe the
system. This is mandatory in order to have meaningful  predictions for the
various cross-sections. 

A few relativistic approaches have been developed in the past ten years  in
order to meet these goals. Among them, two have received particular  attention
in the last few years. 

The first one is based on the Bethe-Salpeter formalism \cite{bs} or its
three-dimensional reductions. The Bethe-Salpeter formalism is four-dimensional
and  explicitly Lorentz covariant. The calculational technique to evaluate 
electromagnetic amplitudes is based on Feynman diagrams and associated rules.
Three-dimensional reductions result in  equations of the quasi-potential
type. 

The second one is Light-Front Dynamics (LFD) \cite{dirac}. In this case, the  
state vector
describing the system is expanded in Fock components with  increasing number of
particles. The state vector is defined on a surface in four-dimensional
space-time which should be indicated explicitly. The Fock components -- the
relativistic wave functions in this formalism -- are the direct generalization
of the non-relativistic wave functions.

In the non-relativistic limit, ($c\to\infty$), the wave function is  defined 
at time $t=0$, and the time evolution is governed by the Schr\"odinger 
equation, once the Hamiltonian of the system is known. This is the  ``instant''
form of dynamics. Physical processes are thus calculated  according to old
fashioned (time ordered) perturbation theory.  This form of dynamics is however
not very well suited for  relativistic systems,  since the interaction of a
probe (the electron, for instance) with the constituents of the system is not
separated from its interaction with the vacuum fluctuations. Moreover, the
plane $t=cte$ is not  conserved by a Lorentz boost.

In the standard version of LFD, the wave function is defined on the 
plane $t+\frac{\displaystyle{z}}{\displaystyle{c}}=cte$  \cite{dirac}.
It is also equivalent to the usual equal time formalism  in the infinite
momentum frame. From a qualitative point of view, all the physical processes
become as slow as possible because of time dilation in  this system of
reference. This greatly simplifies  the description of the system. The
investigation of  the  wave function is equivalent  to make a snapshot of a
system  not spoiled by vacuum fluctuations. It is thus very natural in the
description of  high energy experiments like deep inelastic scattering.  The
calculational technique is here based on the Weinberg rules  \cite{weinberg}.

This formulation has however a serious drawback since the equation of the 
plane $t+z=cte$ (we take here and in the following $c=1$) is still not 
invariant by an arbitrary rotation in space and time (Lorentz boost). 
This plane breaks rotational invariance. As we shall see later on, 
this fact has many important consequences as far as the construction of 
bound states (or scattering states) of definite angular momentum is 
concerned, or in the calculation of electromagnetic amplitudes.

\section{Why light-front dynamics?}\label{why} 

We shall present in this review a covariant formulation of LFD which provides
 a simple, practical and very powerfull tool  to
describe relativistic few-body systems, their bound and scattering states, as
well as  their physical electromagnetic amplitudes. In this formulation, the
state  vector is defined on the plane characterized by  the invariant equation
$\omega \cd x=0$, where $\omega$ is an arbitrary  light-like four vector
$\omega = (\omega_0, \vec \omega)$, with  $\omega ^2=0$ \cite{karm76,karm88}.
With the particular choice  $\omega=(1,0,0,-1)$, we recover the standard LFD
defined on the plane $t+z=0$. The covariance of our approach is realized by
the invariance of the  light-front plane $\omega \cd x=0$ under any Lorentz
transformation of both  $\omega$ and $x$. This implies in particular that
$\omega$ cannot be kept the  same  in any system of reference, as it is the
case in the usual formulation of LFD  with $\omega=(1,0,0,-1)$.

There is of course equivalence, in principle, between these  approaches. Any
exact calculation of a given  electromagnetic process should give the same
physical cross-section,  regardless of the relativistic formalism which is
used. In practical  calculations however, one can hardly hope to carry out an
exact  calculation of physical observables starting from ``first principles'', 
although some simple theoretical systems can  of course be used as test  cases
(the Wick-Cutkosky model for instance).

From a practical point of view, one may thus be led to choose a 
particular formalism depending on the system one is interested in. This 
choice is of course dependent on various criterions, as well as function 
of personal taste. Among these criterions, let us mention those we 
think are the most important:

{\it i)} The formalism should enable us to have physical insights into 
the various processes under consideration, at each step of the 
calculation.

{\it ii)} It should have a direct and transparent non-relativistic limit  in 
order to gain more physical intuition from our present knowledge in the 
non-relativistic domain. This is particularly important in nuclear  physics 
where a lot is already known from the non-relativistic description of  the 
microscopic structure of nuclei and their electromagnetic interactions  at the
scale of 1 GeV or less \cite{jfm}.

{\it iii)} It should provide a as simple as possible calculational 
procedure to evaluate physical processes and compare them with 
experimental results.

We would like to show in this review that the covariant formulation of  LFD is,
according to these criterions, of  particular interest. As we shall see
extensively in the following, it  has  definite advantages as compared to both
the Bethe-Salpeter formalism,  and the usual formulation of LFD. Let us recall
the most important ones  below.

{\it i}) The calculational formalism - time ordered graph technique 
(in the light-front time) in contrast to old fashioned perturbation theory -
does not involve  vacuum fluctuation contributions. This strongly
simplifies a priori the physical picture and calculations.

{\it ii)} In this approach, the wave functions -- the Fock components of  the
state vector -- satisfy a three-dimensional equation, and have the same 
physical meaning (probability amplitudes) as the non-relativistic wave
functions.  Their non-relativistic limit is thus explicit from the general
structure  of  the wave function. This enables a transparent link with
non-relativistic  approaches in first $1/m$ order and in particular with the
contribution of the dominant meson exchange current (the so-called  pair term)
in the two-nucleon systems.

{\it iii)} The explicit covariance of LFD is very important in  practical
applications: it allows to construct states with definite total angular momentum 
to separate
contributions of relativistic origin from non-relativistic ones, and simplifies
very much the calculations in the framework of a special graph technique. In
these respects, the advantages of the covariant formulation in comparison to
ordinary LFD are the same as the advantages of the Feynman graph technique in
comparison to old fashioned perturbation theory.

{\it iv)} A very important property of relativistic wave functions and 
off-shell  amplitudes is their dependence on the orientation of the
light-front  plane. It takes place both in non-covariant and covariant
approaches. In the covariant approach this dependence is parametrized 
explicitly in terms of the four-vector $\omega$. On the contrary, exact 
on-shell physical amplitudes should not depend on  the orientation of
light-front plane. However, in practice,  this dependence survives due to
approximations. The covariant representation of the electromagnetic amplitudes
allows to separate the physical form  factors from the unphysical
contributions.

As already mentioned, the  present approach differs from the standard ones by
the parametrization of the current  choice of the $z$ direction by an arbitrary
one $\vec n$. A second major  difference is the reference to a particular field
theory inspired dynamics,  which implies that our approach describes states
with an unfixed number of  particles. This is a source of many problems,
especially related to the mass  dependence of the interaction. These problems,
which will receive a particular  attention, have a strong relationship with
those encountered in nuclear physics when using an energy dependent
nucleon-nucleon interaction (Bonn-E potential  \cite{bonn}), or with the
contribution of recoil and norm corrections to meson-exchange currents
\cite{chemtob79}.

Our aim in this review is to present the various facets of the covariant 
formulation of LFD, and compared them to the standard formulation. For 
completeness, we shall also make contacts with the BS formalism. The relevance 
of Light-Front Quantization in quantum field theory is the subject of an intense 
present activity. Among others, the non-perturbative problems of how the 
condensates present in 
many theories appear on the light-front, or how renormalization should be 
implemented, are not yet completely solved \cite{zakopane}. However, directions 
of research are clearly identified (zero modes in LFD in particular). These 
problems are outside the scope of this review, although the explicit covariance 
of our formalism may be of particular interest in solving these problems.

We believe that the real advantages of the  covariant formulation of
LFD can be apprehended in practice, i.e., in 
applications to relativistic nuclear and particle physics and 
field theoretical problems. The first applications to relativistic nuclear
and particle physics  are reviewed in the 
present paper, with particular attention to two-nucleon systems for 
which many experimental data exist at present or are expected in the
near future.

Many extensive or review papers have been devoted to the description of
few-body  systems in relativistic approaches. Some of them are based on
different versions of the quasi-potential equations \cite{lt63,blsug66}. Many
applications were made using the three-dimensional Gross equation
\cite{gross6982,gross,acg80,goh92,ig92}. Two- and three-nucleon systems were 
investigated in the framework of the Bethe-Salpeter equation in refs. 
\cite{ztjonall,kt82,htjon89,rtjon92} (see for a review  \cite{tjonconf}).  The
solution of the Bethe-Salpeter equation was also found and applied to deep
inelastic scattering on deuteron and to the deuteron electrodisintegration in
refs. \cite{uk94,ukkk94,kukk95}.

The standard version of the LFD and its applications to  few-body systems was
reviewed in refs.  \cite{kogsus73,ls78,lev83,kp91,coester92,gars93,keist94fb}.
The first results on the covariant formulation of LFD are given in ref.
\cite{karm88}. This approach has been investigated and developed also 
in refs. \cite{fudaall}.

The content of this review is the following.  We develop in chapter 
\ref{cov-graph} the  general properties of the covariant formulation of LFD, 
and  derive the graph technique associated to it. Chapter \ref{wf} is devoted
to the  properties of the two-body wave function: spin structure, equation for 
the  wave function, as well as its connection to the Bethe-Salpeter amplitude.
We pay particular attention to the construction of the angular momentum
operator. We
apply  our formalism to the two-nucleon system in chapter \ref{nnp} by first 
deriving the  nucleon-nucleon potential in this formalism. The two-nucleon
wave  function (deuteron and $^1S_0$ wave function) is then constructed in 
chapter \ref{nns}. The  general structure of the electromagnetic amplitude is
detailed in  chapter \ref{ema},  where we show how to extract the physical form
factors for spin 0, 1/2  and 1  systems, as well as transition form factors. We
discuss  in chapter \ref{emob} the  electromagnetic form factor of the simplest
states with $J=0$ and 1 of the Wick-Cutkosky model, as well as a few hadronic
systems (pion, nucleon, ...). We  apply our  formalism to the two-nucleon
system (deuteron form factors and  electrodisintegration
cross-section). We also discuss, in a $1/m$  expansion,  the relationship
between our formalism and first order  relativistic corrections taken into
account as meson exchange currents  (the so-called pair term) in
non-relativistic approaches. 

The chapters 2, 3 and 6 are quite general and apply to any 
system. Chapters 4, 5 and 7 are direct applications to the two-
nucleon systems, as well as  three quark and quark-antiquark systems.

\chapter{Covariant formulation of light-front 
dynamics}\label{cov-graph}                       

We detail in this chapter the general properties of the covariant  formulation
of LFD. In contrast to the standard approach, the transformations of the
coordinate system and of the light-front plane can be done independently from
each other. These two types of transformations entail the corresponding
transformation properties of the state vector and its Fock components.
Particular attention is paid to the angular momentum operator. This formal
field-theoretical introduction (sect. \ref{trpr}) can be omitted by a reader
not interested in these details.

We  then present  the graph technique associated to this formulation,  for
particles of spin 0, 1/2  and 1.  We illustrate peculiarities of this graph
technique by a few simple examples.

\section{Transformation properties of the state vector}\label{trpr}             
                                                                                
The state vector is defined in general on a hypersurface in
space-time and therefore depends dynamically on its position. For example, the 
non-relativistic wave function $\psi(t)$ depends dynamically on  translations
of the plane $t=cte$, i.e. on time $t$ (by the  trivial phase factor
$\exp(-iEt)$ for a bound state, where the binding  energy $E$ is determined by
dynamics). In ordinary LFD, the wave function is defined on the plane 
$t+z=t^+=cte$.
It  depends dynamically on any translation of this plane.
However, some Lorentz transformations and rotations change  the orientation of 
this
plane \cite{dirac}. These transformations are thus also  dynamical  ones. That 
means lack of {\em explicit} covariance (i.e., the impossibility to  transform
the state vector from one reference system to another one  without knowledge of
the dynamics). This does not mean the absence of  Lorentz covariance at all,
since there exist anyhow a closed system  of  generators of the Poincar\'e
group, and the observable amplitudes  calculated exactly would be covariant. In
practical approximate  calculations, however, the covariance is lost.

In the  covariant formulation of LFD, the wave function  is  defined on the
general plane $\omega\cd x=\sigma$, where $\omega$  is  an arbitrary four
vector restricted to the condition  $\omega^2=0$,and $\sigma$ is the 
''light-front time``. When there is no need to refer to the $\sigma$ evolution, 
we shall take $\sigma=0$. The kinematical
transformations of the system of  reference are thus separated from the
dynamical transformations of  the  plane $\omega\cd x =\sigma$.  The dynamical
dependence of the wave function
 on the light-front plane  results in that case in their dependence on
$\omega$.

This separation of kinematical and dynamical transformations has a definite
advantage in the sense that it  provides a definite prescription for
constructing bound and  scattering  states of definite angular momentum. Since
the total angular momentum  of a  composite system is determined by the
transformation properties of  its wave  function under rotation of the
coordinate system, the construction of  systems  with definite $J$ is now
purely kinematical, in contrast to the  formulation on  the plane $t+z=t^+$. The
dynamical part of this problem, resulting from  the presence of the interaction
in the  generators of the Poincar\'e group which change the position of the
light-front  plane, is separated out and replaced by the so-called angular
condition.

\subsection{Kinematical transformations}\label{kt} 
Let us first specify the transformation properties of the state 
vector with respect to transformations of the coordinate system.
We will use for this purpose a field-theoretical language.
The operators associated to the four-momentum and four-dimensional 
angular momentum are expressed in terms of integrals of the 
energy-momentum $T_{\mu\nu}$ and the angular momentum 
$M^{\rho}_{\mu\nu}$ tensors over the light-front plane $\omega\cd x 
= 
\sigma$, according to:                         
\begin{equation}\label{kt1}                                                     
\hat{P}_{\mu}=\int T_{\mu\nu}\omega^{\nu}\delta(\omega\cd 
x-\sigma)d^4x=          
\hat{P}^0_{\mu} +\hat{P}^{int}_{\mu}\ ,                                         
\end{equation}                                                                  
\begin{equation}\label{kt2}                                                     
\hat{J}_{\mu\nu}=\int M^{\rho}_{\mu\nu}\omega_{\rho} \delta(\omega\cd              
 x-\sigma)d^4x = \hat{J}^0_{\mu\nu} +\hat{J}^{int}_{\mu\nu}\ ,                 
\end{equation}                                                                  
where the $0$ and $int$ superscripts indicate the free and 
interacting 
parts of the operators respectively. For generality, we consider here 
the light-front time $\sigma \neq 0$. 

The description of the evolution along the light-front time $\sigma$ implies  a
fixed value of the length of $\vec{\omega}$, or, equivalently, of  $\omega_0$.
This is necessary in order to have a scale of $\sigma$.  However, the most 
important properties of the physical amplitudes following from covariance do
not require to fix the scale of $\omega$ and will be invariant relative to its
change.

We work in the interaction 
representation in which the operators are expressed in terms of free 
fields. For example, for a scalar field $\varphi(x)$, the free 
operators $\hat{P}^0_{\mu}$ have the form: 
\begin{eqnarray}\label{kt3}
 \hat{P}^0_{\mu} &=&\int 
a^\dagger (\vec{k})a(\vec{k})k_{\mu}\ d^3k\ , \\ 
\hat{J}^0_{\mu\nu}&=&\int                                                       
a^\dagger (\vec{k})a(\vec{k})i\left(k_{\mu}\frac{\partial}{\partial 
k^{\nu}}-          
k_{\nu}\frac{\partial}{\partial k^{\mu}}\right)d^3k\ ,\label{kt4}
\end{eqnarray}                                                                  
where $a^\dagger$ and $a$ are the usual creation and destruction operators, 
 with 
$[a(\vec{k}),a^\dagger(\vec{k'})]=\delta^3(\vec{k}-\vec{k}')$. The operators 
$\hat{P}^{int}$ and $\hat{J}^{int}$ contain the interaction                 
Hamiltonian $H^{int}(x)$:                                                       
\begin{eqnarray}\label{kt5}                                                                
\hat{P}^{int}_{\mu}&=&\omega_{\mu}\int H^{int}(x)\delta(\omega\cd                  
x-\sigma)\ d^4x\ ,\\                                                
\hat{J}^{int}_{\mu\nu}&=&\int H^{int}(x)(x_{\mu}\omega_{\nu} -x_{\nu}           
\omega_{\mu}) \delta(\omega\cd x-\sigma)\ d^4x\ .
\label{kt6}                      
\end{eqnarray}                                                                  
For the particular applications considered in this review,
we do not need to develop the field theory in its full form. We
therefore do not pay attention here to the fact that the field-theoretical 
Hamiltonian  $H^{int}(x)$ is usually singular and requires a regularization.
The  regularization of  amplitudes in our formulation will be illustrated by
the example  of a typical self-energy contribution at the end of this chapter.

\begin{figure}[hbt]
\centerline{\epsfbox{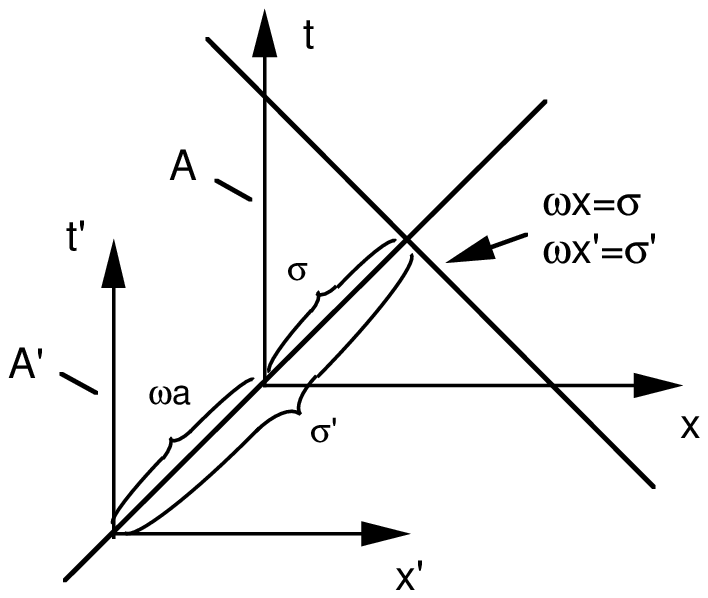}}
\figcap{Translation of the reference system along the 
light-front time.}
\label{refsys} 
\end{figure}

Eq.(\ref{kt5})  is consistent with the expectation that only the component  
$P_\mu$ along the direction $\omega$ of the light-front ''time``  has a 
dynamical character in the light-front 
formalism \cite{dirac}. Under translation $x \rightarrow x'=x+a$ of the
coordinate system  $A\rightarrow A'$, the equation $\omega\cd x = \sigma$ takes
the form   $\omega\cd x'=\sigma'$, where $\sigma'=\sigma+\omega\cd a$. The
state  vector is transformed in accordance with the law:  
\begin{equation}\label{kt7}                                                     
\phi_\omega(\sigma)\rightarrow \phi'_\omega(\sigma ')                                         
  =U_{P^0}(a)\phi_\omega(\sigma)\ ,                                                    
\end{equation}                                                                  
where $U_{P^0}(a)$ contains only the operator of
the four-momentum (\ref{kt3}) of the free field:  
\begin{equation}\label{kt8}
 U_{P^0}(a)=\exp(i\hat{P}^0\cd a)\ .  
\end{equation}                                                                  
The ``prime" at $\phi'(\sigma)$ indicates that $\phi'(\sigma)$ is defined in
the system $A'$ on the plane $\omega \cd x'=\sigma$ in  contrast to
$\phi(\sigma)$ defined in the system $A$ on the plane  $\omega\cd x=\sigma$
(the value of $\sigma$ being the same). The state vector $\phi'(\sigma')$ is
defined in $A'$ on the plane $\omega\cd  x'=\sigma'$, which coincides with
$\omega\cd x = \sigma$. Therefore   no  dynamics  enters into the
transformation (\ref{kt7}). This is  rather  natural, since under translation
of the coordinate system the  plane $\omega\cd x = \sigma$ occupies the same
position in space while it occupies a new position with respect to the axes of
the new  coordinate system, as indicated in fig.~\ref{refsys}. The formal proof
of (\ref{kt7}), (\ref{kt8}) can be  found  in~\cite{karm82}.  

In the case of infinitesimal four-dimensional rotations $x_{\mu} 
\rightarrow x'_{\mu}=gx_{\mu}=x_{\mu}+\epsilon _{\nu\mu}x^{\nu}$, the 
result is similar~\cite{karm82}:                                                
\begin{equation}\label{kt9}                                                     
\phi_\omega(\sigma) \rightarrow 
\phi'_{\omega'}(\sigma)=U_{J^0}(g)\phi_\omega(\sigma)\ ,               
\end{equation}                                                                  
where $\omega'_{\mu}=\omega_{\mu} +\epsilon_{\nu\mu} 
\omega^{\nu}$ and
\begin{equation}\label{kt9a}
U_{J^0}(g)=1+\frac{1}{2}\hat{J}^0_{\mu\nu}\epsilon^{\mu\nu}. 
\end{equation}
The operator
$\hat{J}^0_{\mu\nu}$ is given by (\ref{kt4}).

\subsection{Dynamical transformations}\label{dt} 
The properties of the state vector under transformations of the 
hypersurface are determined by the dynamics and follow from the 
Tomonaga-Schwinger equation \cite{qed}:  
\begin{equation}\label{kt10}                                                    
i\delta\phi/\delta\sigma(x)=H^{int}(x)\ \phi\ .                                 
\end{equation}                                                                  
>From the definition of the variational derivative in (\ref{kt10}) we 
obtain:  
$$ 
i\delta\phi =H^{int}(x)\ \phi\ \delta V(x)\ ,                                   
$$                                                                              
where $\delta V(x)$ is the volume between the initial surface and the 
surface obtained from the original one by the variation 
$\delta\sigma(x)$ around the point $x$.  

Under the translation $\sigma\rightarrow \sigma +\delta\sigma$ of the           
plane, the total increment of the state vector is obtained through            
the increment at each point of the surface:                                     
\begin{equation}\label{kt11}                                                    
i\delta\phi =\int H^{int}(x)\delta(\omega\cd x -\sigma) d^4x\ \phi \ 
\delta       
\sigma\ .                                                                       
\end{equation}                                                                  
This relation gives the Schr\"odinger equation.  In the interaction             
representation in the light-front time, we have:                                 
\begin{equation}\label{kt11p}                                                                              
i\partial\phi/\partial\sigma = H(\sigma)\phi(\sigma)\ ,                         
\end{equation}                                                                               
where:                                                                          
\begin{equation}                                                                              
H(\sigma) =\int H^{int}_{\omega}(x) \delta(\omega\cd x-\sigma)d^4x,            
\end{equation}  
and $H^{int}_{\omega}(x)$ may differ from $H^{int}(x)$ because of 
singularities of the field commutators on the light cone. This point 
is
explained below in sect. \ref{gendev}.

Similarly, in the case of a rotation of the light-front plane, 
$\omega_{\mu}         
\rightarrow \omega'_{\mu}=\omega_{\mu}+ \delta\omega_{\mu},$                    
$\delta\omega_{\mu}=\epsilon_{\nu\mu}\omega^{\nu}$, we find:                    
\begin{equation}                                                                                
\phi_\omega(\sigma) \rightarrow \phi_{\omega+\delta\omega}(\sigma) =\phi_\omega                       
+\delta\phi_\omega,\; \delta\phi_\omega=\frac{1}{2}\epsilon_{\mu\nu}                          
\left( \omega^{\mu}\frac{\partial}{\partial\omega_{\nu}} 
-\omega^{\nu}          
\frac{\partial}{\partial\omega_{\mu}}\right)\phi_\omega(\sigma)\ .                     
\end{equation}                                                                               
The increment of the volume over the point $x$ is:                              
\begin{equation}                                                                               
\delta V =\epsilon_{\mu\nu}\ x^{\mu}\ \omega^{\nu}\ \delta(\omega\cd               
x-\sigma)\ d^4x\ ,                                                             
\end{equation}                                                                               
and it follows from (\ref{kt11}) that~\cite{karm82}:                            
\begin{equation}\label{kt12}                                                    
\hat{J}^{int}_{\mu\nu}            
\ \phi_\omega(\sigma)= \hat{L}_{\mu\nu}(\omega)\phi_\omega(\sigma)\ ,                                                               
\end{equation}                                                                  
where:                                                                          
\begin{equation}\label{kt13}                                                    
\hat{L}_{\mu\nu}(\omega) =i\left(\omega_{\mu}                                   
\frac{\partial}{\partial\omega^{\nu}} -\omega_{\nu}                             
\frac{\partial}{\partial\omega^{\mu}}\right)\ ,                                 
\end{equation}                                                                  
and $\hat{J}^{int}_{\mu\nu}$ is given by (\ref{kt6}).                           
                                                                                
Equation (\ref{kt12}) is called the {\it angular condition}. It plays 
an important role in the construction of relativistic bound states, 
as we shall explain below.  

The transformation of the coordinate system and the simultaneous 
transformation of the light-front plane, which is rigidly related to the 
coordinate axes, correspond to the successive application of the two types of 
transformations considered above (kinematical and dynamical). Thus,
under the infinitesimal  translation
$x\rightarrow x'=x+a$ of the coordinate system, $A\rightarrow A'$, and of the
plane, we have:                                
\begin{equation}\label{kt13a}                                                   
\phi_\omega(\sigma)\rightarrow \phi'_\omega(\sigma)=(1+i\hat{P}\cd 
a)\phi_\omega(\sigma)\ .              
\end{equation}                                                                  
Note that for the state with definite total four-momentum $p$ (i.e.,            
for an eigenstate of the four-momentum operator), the            
equations (\ref{kt7}) and (\ref{kt13a}) give:  
\begin{equation}\label{kt18a}                                                   
\exp(i\hat{P}^0\cd a)\phi(\sigma) =\exp(ip\cd a)\phi(\sigma 
+\omega\cd a)\ .             
\end{equation}
This equation determines the conservation law for the four-momenta of the
constituents, given in chapter \ref{wf}.                                                               
                  
%

\subsection{Role of the angular condition}\label{ac}                            
                                                              
We are interested here in the state vector of a bound system. It 
corresponds to a definite mass $M$, four-momentum $p$, total angular
momentum $J$ with projection $\lambda$ onto the $z$ axis in the rest 
frame, i.e., the 
state vector forms a representation of the Poincar\'e group. This 
means 
that it satisfies the following eigenvalue equations:                                             
\begin{eqnarray}\label{kt14}                                                    
\hat{P}_{\mu}\ \phi^{J\lambda}(p)&=&p_{\mu}\ \phi^{J\lambda}(p)\ ,\\                          
\label{kt15}                                                                    
\hat{P}^2\ \phi^{J\lambda}(p)&=&M^2\ \phi^{J\lambda}(p)\ ,\\                            
\label{kt16}                                                                    
\hat{S}^2\ \phi^{J\lambda}(p)&=&-M^2\ J(J+1)\ \phi^{J\lambda}(p)\ ,\\                   
\label{kt17}                                                                    
\hat{S}_{3}\ \phi^{J\lambda}(p)&=&M\ \lambda\phi^{J\lambda}(p)\ ,                                 
\end{eqnarray}                                                                  
where $\hat{S}_{\mu}$ is the Pauli-Lubanski vector:                             
\begin{equation}\label{kt18}                                                    
\hat{S}_{\mu}= \frac{1}{2}\epsilon_{\mu\nu\rho\gamma}                           
\ \hat{P}^{\nu}\ \hat{J}^{\rho\gamma}\ .                                        
\end{equation}  
The state vector $\vert p,\lambda\rangle_{\omega}$  for a given $J$ is 
normalized as follows:
\begin{equation}\label{orth1}
_{\omega} \langle \lambda',p' \vert p,\lambda\rangle_{\omega}                         
=2p_0\ \delta^{(3)}(\vec{p}- \vec{p}\,')\ \delta^{\lambda'\lambda}\ .  
\end{equation}
For convenience we introduce here another notation for the same state 
vector:
$$
\vert p,\lambda\rangle_{\omega} \equiv\phi_{\omega}^{J\lambda}(p).
$$
We omit in $\phi_{\omega}^{J\lambda}(p)$ the argument $\sigma$, but 
show explicitly the argument $p$. 
For simplicity, we have left out the $\omega$ underscript 
when not absolutely necessary.

We can now use the angular condition (\ref{kt12}) and replace the               
operator $\hat{J}^{int}_{\mu\nu}$, which is contained in                        
$\hat{J}_{\mu\nu}$ in (\ref{kt18}), by $\hat{L}_{\mu\nu}(\omega)$.              
Introducing the notations:                                                       
\begin{eqnarray}\label{kt19}                                                    
\hat{M}_{\mu\nu} &=&\hat{J}^0_{\mu\nu} +\hat{L}_{\mu\nu}(\omega)\ ,\\           
\label{kt20}                                                                    
\hat{W}_{\mu}&=& \frac{1}{2}\epsilon_{\mu\nu\rho\gamma}                         
\ \hat{P}^{\nu}\ \hat{M}^{\rho\gamma}\ ,                                        
\end{eqnarray}                                                                  
we obtain instead of eqs.(\ref{kt16}) and (\ref{kt17}):                         
\begin{eqnarray}\label{kt21}                                                    
\hat{W}^2\phi^{J\lambda}(p)&=&-M^2J(J+1)\ \phi^{J\lambda}(p)\ ,\\                       
\label{kt22}                                                                    
\hat{W}_{3}\ \phi^{J\lambda}(p)&=&M\ \lambda\ \phi^{J\lambda}(p)\ .                                  
\end{eqnarray}                                                                  
{\it These equations do not contain the interaction Hamiltonian, once $\phi$ 
satisfies (\ref{kt14}) and (\ref{kt15})}. The construction of states with 
definite 
angular momentum becomes therefore {\it 
a 
purely kinematical problem}. 

At the same time, the 
state 
vector must  satisfy the dynamical equation (\ref{kt12}). In 
terms 
of the operators $\hat{J}_{\mu\nu}$ and $\hat{M}_{\mu\nu}$, the 
angular 
condition (\ref{kt12}) can be rewritten as:                                     
\begin{equation}\label{kt23}                                                    
\hat{M}_{\mu\nu}(\omega)\phi_{\omega}(\sigma) =\hat{J}_{\mu\nu}                          
\phi_{\omega}(\sigma)\ .                                                                 
\end{equation}                                                                  
The commutation relations between the operators $\hat{P}_{\mu}$,                
$\hat{J}_{\mu\nu}$, $\hat{M}_{\mu\nu}$ have the form:                           
\begin{eqnarray}\label{kt24}                                                    
[\hat{P}_{\mu},\hat{P}_{\nu}]&=&0\ ,\\                                          
\label{kt25}                                                                    
\frac{1}{i}[\hat{P}_{\mu},\hat{J}_{\kappa\rho}]                                 
&=&g_{\mu\rho}\hat{P}_{\kappa} -g_{\mu\kappa}\hat{P}_{\rho}\ ,\\                
\label{kt26}                                                                    
\frac{1}{i}[\hat{J}_{\mu\nu},\hat{J}_{\rho\gamma}] &=&g_{\mu\rho}               
\hat{J}_{\nu\gamma} -g_{\nu\rho}\hat{J}_{\mu\gamma}                             
+g_{\nu\gamma}\hat{J}_{\mu\rho} -g_{\mu\gamma}\hat{J}_{\nu\rho}\ ,\\            
\label{kt27}                                                                    
\frac{1}{i}[\hat{P}_{\mu},\hat{M}_{\kappa\rho}]                                 
&=&g_{\mu\rho}\hat{P}_{\kappa} -g_{\mu\kappa}\hat{P}_{\rho}\ ,\\                
\label{kt28}                                                                    
\frac{1}{i}[\hat{M}_{\mu\nu},\hat{M}_{\rho\gamma}] &=&g_{\mu\rho}               
\hat{M}_{\nu\gamma} -g_{\nu\rho}\hat{M}_{\mu\gamma}                             
+g_{\nu\gamma}\hat{M}_{\mu\rho} -g_{\mu\gamma}\hat{M}_{\nu\rho}\ ,\\            
\label{kt29}                                                                    
\frac{1}{i}[\hat{J}_{\mu\nu},\hat{M}_{\rho\gamma}] &=&g_{\mu\rho}               
\hat{J}_{\nu\gamma} -g_{\nu\rho}\hat{J}_{\mu\gamma}                             
+g_{\nu\gamma}\hat{J}_{\mu\rho} -g_{\mu\gamma}\hat{J}_{\nu\rho}\ .
\end{eqnarray}                                                                  

Equations (\ref{kt24}), (\ref{kt25}) and (\ref{kt26}) are the 
standard 
commutation relations of the Poincar\'e group.  The equations  
(\ref{kt27}), (\ref{kt28}) and (\ref{kt29}) simply reflect the tensor 
nature of the operators. This set also corresponds to the Poincar\'e 
group transformations. The derivation of these equations  is 
explained 
in ref.~\cite{karm88}.
                                                                                
We have thus shown that under the condition (\ref{kt12}) the problem  of 
constructing states with definite angular momentum can be formulated in terms of 
 the
kinematical operators $\hat{M}_{\mu\nu}$ and $\hat{W}_{\mu}$ in  exactly the
same way as in terms of the operators $\hat{J}_{\mu\nu}$  and $S_{\mu}$ which
depend on the interaction.  This naturally reflects the fact that the angular
momentum of a
system determines the kinematical properties of the wave function relative to
transformations of the coordinate system. The dynamics is involved in the
composition of the angular momentum of the system from the spins of its 
constituents.

As already mentioned, the  angular momentum generators in ordinary LFD contain 
the interaction. To avoid any misunderstanding, note that the interaction 
Hamiltonian 
does not disappear in our approach. It is moved into the angular condition 
(\ref{kt23}). However, it is very convenient to first construct on purely 
kinematical grounds the general form of the light-front wave function for a 
given angular momentum. One can then find from  dynamics the coefficients in 
front of all the spin structures. This way of separating dynamical from 
kinematical transformation properties is of particular interest since the 
interaction Hamiltonian is often approximate.

We emphasize that by introducing the operator  $\hat{L}_{\mu\nu}(\omega)$, 
eq.(\ref{kt13}), containing the derivatives over $\omega$, we enlarge the
Hilbert space where the state vector is defined. It is now the  direct sum of
the Hilbert space where  the "normal" field-theoretical operators act and of
the Hilbert  space, where the operator $\hat{L}_{\mu\nu}(\omega)$ acts. Hence,
in this  enlarged space the scalar  product and, correspondingly, the
normalization condition contains integration over $\omega$ with the appropriate
measure  $d\mu_{\omega}$:
\begin{equation}\label{orth2}
\int {_{\omega}\langle \lambda',p' \vert p,\lambda\rangle_{\omega}}
d\mu_{\omega}                        
=2p_0\ \delta^{(3)}(\vec{p}- \vec{p}\,')\ \delta^{\lambda'\lambda}\ .  
\end{equation}
The particular form of the measure $d\mu_{\omega}$  corresponds to integration
over the directions of $\vec{\omega}$ in a particular system of reference. The
angular condition (\ref{kt12}), (\ref{kt23}) just ensures the  equivalence of
the approach developed in the enlarged Hilbert space  to the ordinary approach.
In particular, the orthogonality condition  (\ref{orth2}), which contains
integration over $\omega$, has to be equal  to the orthogonality condition
(\ref{orth1}), where $\omega$ is a  fixed parameter. This means, of course,
that  the product ${_{\omega}\langle \lambda',p'  \vert
p,\lambda\rangle_{\omega}}$ does not depend on $\omega$ at all,(though for an 
arbitrary vector  $\vert
\cdots\rangle_{\omega}$  from the enlarged Hilbert space, the scalar product 
${_{\omega}\langle \cdots \vert \cdots\rangle_{\omega}}$ depends on $\omega$).
However, any separate contribution of the Fock component to  ${_{\omega}\langle
\lambda',p' \vert p,\lambda\rangle_{\omega}}$ depends on $\omega$, while the
$\omega$-dependence disappears in the sum  over all  Fock components.

In the enlarged Hilbert space, the angular momentum operator is given in terms
of derivatives on the momenta and $\omega$ and do not contain the interaction.
The  construction of states with definite angular momentum becomes therefore 
very simple.
In practice, it is convenient first to solve the kinematical part of the
problem -- the construction of  states with definite angular momentum -- and
then satisfy the angular  condition (\ref{kt12}).

The transformation properties of a state  with definite angular momentum are  
discussed in
chapter \ref{wf}. We will see that they completely  determine the structure of
the wave function, and, in particular, the  number of its spin components (six
in the case of the deuteron).  These  components  will be found below by
solving the light-front generalization of the  Schr\"odinger equation. At first
glance, this procedure is  unambiguous  and nothing remains to be determined by
the angular condition. What  is  then the role of this condition?  

Without solving the angular condition, there is an ambiguity in 
finding 
states with definite angular momentum. To show that, let us construct the 
operator:  
\begin{equation}\label{kt30} \hat{A} 
=(\hat{W}\cd\omega)^2,  
\end{equation} 
where $\hat{W}_{\mu}$ is the kinematical Pauli-Lubansky vector 
(\ref{kt20}). It is readily verified that this operator commutes with 
$\hat{P}_{\mu}$, $\hat{M}_{\mu\nu}$, $\hat{W}_{\mu\nu}$ and also with 
the parity operator. It seems therefore that the 
state vector must be characterized not only by its mass, momentum and 
angular momentum, but also by the eigenvalue $\alpha$ of the operator 
$A$:  
\begin{equation}\label{kt31} 
\hat{A}\ \phi_{\alpha} =\alpha\ \phi_{\alpha}\ .  
\end{equation} 
One can show that, for instance, for the total angular momentum  $J=1$, there
are only two states with $\alpha=0$ and $\alpha=1$. However,  these states
$\phi_{\alpha}$ are degenerate.  Since the commutators  of  $\hat{M}_{\mu\nu}$
and of $\hat{J}_{\mu\nu}$ with $\hat{P}_{\mu}$,  $\hat{J}_{\mu\nu}$ are equal
to each other (see  eqs.(\ref{kt24})-(\ref{kt29})), the operator 
\begin{equation}\label{kt32} 
\Delta\hat{J}_{\mu\nu} =\hat{M}_{\mu\nu} -\hat{J}_{\mu\nu} = 
\hat{L}_{\mu\nu}(\omega) -\hat{J}^{int}_{\mu\nu} 
\end{equation} 
commutes with $\hat{P}_{\mu}$ and with $\hat{J}_{\mu\nu}$, but  $[\Delta 
\hat{J}_{\mu\nu},\hat{A}] \neq 0$.   The state $\phi'  =\Delta 
\hat{J}_{\mu\nu}\ \phi_{\alpha}$ is therefore not an eigenvector of the 
operator  $\hat{A}$, i.e., it can be represented in the form $\phi' 
=\sum_{\alpha} \ \beta_{\alpha} \ \phi_{\alpha}$. But it corresponds  to  the
same mass as $\phi_{\alpha}$.  We would thus always get for the  deuteron and
for any state with $J=1$ two degenerate states with  $\alpha =0$ and 1, in
evident contradiction with reality.  The  angular  condition (\ref{kt12}) is
just distinguishing a definite  superposition  of states
$\phi_{\alpha}$,              
\begin{equation}\label{kt33}                                                    
\phi =\sum_{\alpha}c_{\alpha}\phi_{\alpha}\ ,                                   
\end{equation}                                                                  
which is such that $\Delta\hat{J}_{\mu\nu}\phi =0$. This equation                    
eliminates the problem of the ``spurious'' states of               
relativistic composite systems. This procedure will be illustrated in           
section \ref{am-ac} by a simple example.                                        
                                                                                

\section{Covariant light-front graph technique} \label{lfgt} 
The light-front graph technique is a method for calculating  the  $S$-matrix.
In the   framework of perturbation theory, the on-shell amplitude  given by
this graph technique coincides with the one given by the Feynman graph
technique. However, the methods to calculate them  drastically differ from
each other, as we shall see in this section.  The most important difference
lies in the fact that in the  LFD all four-momenta are always on the mass
shell. This  three-dimensional form of the theory has enormous advantages in 
solving several problems. In  applications to relativistic few-body systems,
it provides a direct and  close connection between the relativistic  wave
functions and the non-relativistic ones, as we shall detail in  the next
chapters.  Moreover, the diagrams corresponding to vacuum fluctuations are
absent. This simplifies very much the theory  as compared to other 
three-dimensional  approaches (like the old fashioned time ordered
perturbation theory). Another important advantage results from the
explicitly covariant formulation of the light-front plane defined by $\omega
\cd x=0$ as compared  to the standard formulation on the plane $t+z=0$. 

We shall derive in this section the rules pertinent to the covariant
light-front  graph  technique by transforming the standard expression for the
$S$-matrix.  We explicitly show in appendix \ref{other} the relations between 
the 
covariant light-front amplitudes and the amplitudes given by the  Weinberg and
Feynman rules.  

\subsection{General derivation} \label{gendev}                                  
The graph technique described below was developed by 
Kadyshevsky~\cite{kadysh64} (see ref.~\cite{kms72} for a review) and 
applied to LFD in ref.~\cite{karm76}. It is 
manifestly 
covariant, like the Feynman graph technique, and retains all the 
positive features of the old fashioned perturbation theory developed 
by Weinberg~\cite{weinberg}. Following ref. \cite{kadysh64}, we start 
from the standard expression for the $S$-matrix:  
\begin{eqnarray}\label{rul1}                                                    
S&=&T\exp\left[-i\int H^{int}(x)d^4x\right]=1+  \nonumber
\\  &&\sum_n\int (-i)^n H^{int}(x_1)\theta(t_1-t_2)
H^{int}(x_2) \ldots
\theta(t_{n-1}-t_n)H^{int}(x_n)d^4x_1\ldots d^4x_n\ ,                    
\nonumber\\ 
&&                                                                              
\end{eqnarray}                                                                  
where $H^{int}(x)$ is the interaction Hamiltonian. The sign of the 
$T$-product (and $1/n!$) are omitted, since the time ordering is made 
explicit by means of the $\theta$-functions. The expression  (\ref{rul1}) is
then represented in terms of the light-front time  $\sigma=\omega\cd
x$:                                                             
\begin{eqnarray} 
\label{rul11}                                                              
S&=& 1+ \nonumber
\\  &&\sum_n\int
(-i)^n H^{int}_{\omega}(x_1)                                       
\theta\left(\omega\cd (x_1-x_2)\right) H^{int}_{\omega}(x_2) \ldots 
\theta\left(\omega\cd (x_{n-1}-x_n)\right) H^{int}_{\omega}(x_n) 
\nonumber\\                                                                     
&&\times d^4x_1\ldots d^4x_n \ . \label{rul2}                            
\end{eqnarray}                                                                  
The index $\omega$ at $H^{int}_{\omega}$ indicates that $H^{int}$ and 
$H^{int}_{\omega}$ may differ from each other in order to provide the
equivalence between (\ref{rul1}) and (\ref{rul11}). The region in which this
can happen is a line on  the light cone. Indeed, if $(x_1  -x_2)^2 >0$, the
signs of $\omega \cd (x_1 -x_2)$ and $t_1 -t_2$  are the same and hence
$H^{int}_{\omega}=H^{int}$.  If $(x_1 -x_2)^2<0$,  the operators
commute:                                                              
\begin{equation}\label{rul3}                                                    
[H^{int}(x_1),H^{int}(x_2)]=0,                                                  
\end{equation}                                                                  
and their relative order has no significance. On the light cone, i.e.  if $(x_1
-x_2)^2=0$,   $\omega\cd (x_1 -x_2)$ can be equal to zero while $t_1 -t_2$ may
be  different from zero.  If the integrand has no singularity at $(x_1
-x_2)^2=0$, this line does not contribute to the integral over the volume
$d^4x$. However, if the integrand is singular, some care is needed. To
eliminate the influence of this region on the $S$-matrix, we have introduced in
(\ref{rul2}) a new Hamiltonian  $H^{int}_{\omega}$, such that expressions
(\ref{rul1}) and  (\ref{rul2})  be equal to each other.  The form of
$H^{int}_{\omega}$, which provides  the  equivalence between (\ref{rul1}) and
(\ref{rul2}), depends on the  singularity of the commutator (\ref{rul3}) at
$(x_1 -x_2)^2=0$. For  the  scalar fields, the singularity is weak enough, and
the expressions  (\ref{rul1}) and (\ref{rul2}) are the same, so that 
$H^{int}_{\omega}=H^{int}$. For fields with spins 1/2 and 1 or with  derivative
couplings, the equivalence is obtained with  $H^{int}_{\omega}$ differing from
$H^{int}$ by an additional contribution (counter term) leading  to contact
terms in the propagators (or so called instantaneous  interaction). We shall
come back to this point later on in this
section.                                                             

Introducing the Fourier transform of the Hamiltonian:        
\begin{equation}\label{rul4}                                                    
\tilde{H}_{\omega}(p)=\int H^{int}_{\omega}(x)\exp(-ip\cd x)d^4x\ ,                         
\end{equation}                                                                  
and using the integral representation for the $\theta$ function:      
\begin{equation}\label{rul5}                                                    
\theta\left(\omega\cd (x_1-x_2)\right)=  \frac{1}{2\pi i} 
\int_{-\infty}          
^{+\infty}\frac{\exp\left(i\tau\omega\cd (x_1                                      
-x_2)\right)}{\tau-i\epsilon}\ d\tau\ , \end{equation}                          
we can transform the expression (\ref{rul2}) to the form:                       
\begin{eqnarray}\label{rul6}                                                    
S&=&1+R(0)\nonumber\\                                                           
&=&1-i\tilde{H}_{\omega}(0)\nonumber\\
&+&\sum_{n\geq 2} (-i)^n \int \tilde{H}_{\omega}(-\omega\tau_1)
 \frac{d\tau_1} 
{2\pi i           
(\tau_1-i\epsilon)} \tilde{H}_{\omega}(\omega\tau_1 -\omega\tau_2) \ldots                
\frac{d\tau_{n-1}}{2\pi i(\tau_{n-1}-i\epsilon)} \tilde{H}_{\omega}(\omega               
\tau_{n-1})\ . \nonumber\\                                                      
&&                                                                              
\end{eqnarray}                                                                  
The $\tau$ variable appears here as an auxiliary variable, as defined 
in                    
eq.(\ref{rul5}); $\omega\tau$ has the dimension of a momentum.

The $S$-matrix (\ref{rul6}) gives the state vector
$\phi(\sigma)=                S(\sigma)\ \phi_0$ for asymptotic states,
i.e. for an infinite value  of the light-front time                
$\sigma=\infty$. It determines the on-energy-shell amplitude. The 
off-energy shell amplitude is determined by $S(\sigma)$ at finite
$\sigma$.  Similarly, introducing another $\theta$-function 
$\theta(\sigma-\omega\cd x_1)$ in (\ref{rul2}), one can find easily  that
the $S$-matrix on a finite light-front plane $\sigma$ is  represented in
the
form:                                                                       
\begin{equation}\label{rul6a}                                                   
S(\sigma)=1+\int_{-\infty}^{\infty}\frac{\exp(i\tau\sigma)}                     
{2\pi i(\tau-i\epsilon)}R(\omega\tau)d\tau\
,                                   
\end{equation}                                                                  
where                                                                           
\begin{equation}\label{rul6b}                                                   
R(\omega\tau)=                                                                  
\sum_n (-i)^n \int \tilde{H}_{\omega}(\omega\tau-\omega\tau_1)
\frac{d\tau_1}             {2\pi i (\tau_1-i\epsilon)}
\tilde{H}_{\omega}(\omega\tau_1 -\omega\tau_2)  \ldots        
\frac{d\tau_{n-1}}{2\pi i(\tau_{n-1}-i\epsilon)} \tilde{H}_{\omega}(\omega 
\tau_{n-1})\ .
\end{equation}                                                                 
These formulae give the iterative solution of the equation
(\ref{kt11p}). The matrix elements of
the operator $R(\omega\tau)$ for $\tau\neq 0$  correspond to the
off-energy shell amplitudes, whereas  at  $\tau=0$ they give the on-energy
shell amplitude. We shall precise the difference  between off-energy shell
and off-mass-shell amplitudes below in sect.  \ref{tchan}. At
$\sigma\rightarrow \infty$, the only residue in (\ref{rul6a}) for
$\tau=i\epsilon\rightarrow 0$ survives, and we recover eq.(\ref{rul6}).

We emphasize that despite the presence of the four-vector $\omega$ in 
eq.(\ref{rul6}), the $S$-matrix and any physical amplitude do not  depend on
$\omega$, since eq.(\ref{rul6}) gives the same $S$-matrix,  as  the initial
one given by eq.(\ref{rul1}). Similarly, the  off-shell matrix
$R(\omega\tau)$ in eq.(\ref{rul6b}), depends on  $\omega$ and off-shell
light-front amplitude does not coincide with the Feynman one. This is natural
since $R(\omega\tau)$ determines the  $S$-matrix at a given light-front plane
$\omega \cd x=\sigma$ in the interaction region.  As we  shall see in chapter
\ref{wf}, the same is also true for light-front  wave functions. The latters
depend on $\omega$ since they are always  off-shell objects and are defined
not as asymptotic states, but at  any  finite time (at any given light-front
plane in our case).  

\subsection{Spin 0 system}                                                      
\label{sl-graph}                                                                
The covariant light-front graph technique arises when, as usual, one 
represents the expression (\ref{rul6}) in normal form. Let us consider 
for example the simple case of an interaction Hamiltonian of the form 
$H=-g\varphi^3(x)$, where $\varphi$ is a scalar field.  We introduce 
the Fourier transform $\tilde{\varphi}$ of the field $\varphi$ given by:
\begin{eqnarray}\label{ft1}
\varphi(x)& \equiv &\frac{1}{(2\pi)^{3/2}} \int 
\tilde{\varphi}(k)\exp(ik\cd 
x)d^4k\nonumber\\
&=&\frac{1}{(2\pi)^{3/2}} \int\left[a(\vec{k})\exp(-ik\cd x) +
a^{\dagger}(\vec{k})\exp(ik\cd 
x)\right]\frac{d^3k}{\sqrt{2\varepsilon_k}}\ .
\end{eqnarray}
We thus have:
\begin{equation}\label{ft2}
\tilde{\varphi}(k)= [a(-\vec{k})\theta(-k_0) 
+a^{\dagger}(\vec{k})\theta(k_0)]\sqrt{2\varepsilon_k}\delta(k^2-m^2)\ .
\end{equation}
When the $S$-matrix (\ref{rul6}) is reduced to normal form, 
we obtain the contractions:  
\begin{equation}\label{rul7} 
\underbrace{\tilde{\varphi}(k)\tilde{\varphi}}(p) = 
\tilde{\varphi}(k)\tilde{\varphi}(p) -                                          
:\!\tilde{\varphi}(k)\tilde{\varphi}(p)\!:\; =                                         
\theta(p_0)\delta(p^2 -m^2)\delta^{(4)}(p+k)\ .                                 
\end{equation}                                                                  
It is convenient to replace in the following                                    
$\theta(p_0)$ in the propagator (\ref{rul7}) by $\theta(\omega\cd p)$.
This is always possible, since $p^2=m^2 > 0$.             
                                                                                
We would like to emphasize at this point that the propagator 
(\ref{rul7})       
contains the delta-function $\delta(p^2-m^2)$, and therefore {\em all           
particles are always on their mass shells}.  The reason of this property, which 
drastically differs from the Feynman approach, is the absence of the T-product 
operator in (\ref{rul1}).

In this graph technique, the four-vectors $\omega \tau_j$ in  (\ref{rul6})
are associated with a fictitious particle -- called  spurion -- and the
factors $1/(\tau_j-i\epsilon)$ are interpreted as  the  propagator of the
spurions responsible for taking  the intermediate states off the energy
shell.  This spurion  should be  interpreted as a convenient tool
in order to take into account  off-energy shell  effects in  the covariant
formulation of LFD (in the absence of off-mass shell effects), and not
as a physical particle. It  is absent, by  definition,  in all asymptotic,
on-energy shell, states. We shall show below on  simple  examples how the
spurion should be used in practical calculations.                       

The general invariant amplitude $M_{nm}$ of a transition 
$m\rightarrow n$ is related to the $S$-matrix by:  
\begin{equation}\label{rul8} 
S_{nm}=1+i(2\pi)^4\delta^{(4)}\left(\sum_{i=1}^m 
 k_i -\sum_{i=1}^n k'_i\right)
\frac{M_{nm}}{\left((2\pi)^3 2\varepsilon_{k'_1}\ldots 
(2\pi)^3 2\varepsilon_{k'_n}\ (2\pi)^3 2\varepsilon_{k_1} \dots 
(2\pi)^3 2\varepsilon_{k_m}\right)^{1/2}}, 
\end{equation} 
where, e.g., 
$\varepsilon_{k_1}=\sqrt{m_1^2+\vec{k}_1^2}$.  The cross-section of 
the 
process $1+2\rightarrow 3+\ldots +n$ is thus expressed as:  
\begin{equation}\label{rul9}                                                    
d\sigma= \frac{(2\pi)^4}{4j\varepsilon_{k_1}\varepsilon_{k_2}} 
|M|^2\frac{d^3k_3}{(2\pi)^3 2\varepsilon_{k_3}}\cdots 
\frac{d^3k_n}{(2\pi)^3 2\varepsilon_{k_n}} \delta^{(4)}(k_1+k_2 
-k_3-\ldots-k_n)\ , 
\end{equation} 
where $j$ is the flux density of the 
incident particles:  
$$ 
j\varepsilon_{k_1}\varepsilon_{k_2}=\frac{1}{2}[s-(m_1+m_2)^2]^{1/2}            
[s-(m_1-m_2)^2]^{1/2},\quad s=(k_1+k_2)^2\ .                                    
$$                                                                              

\begin{figure}[htbp]
\centerline{\epsfbox{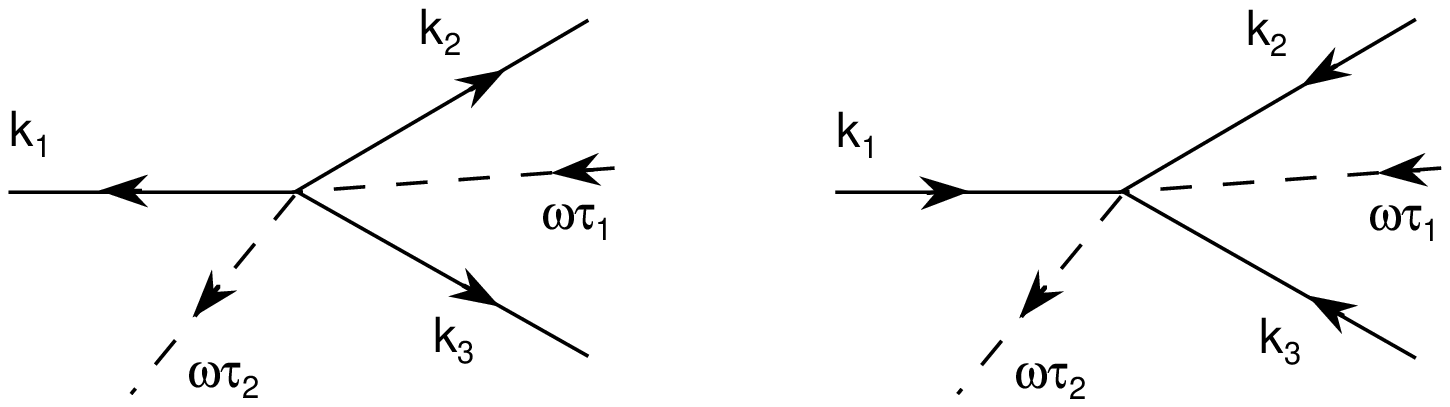}}
\figcap{The vacuum vertices.} 
\label{rulevac} 
\end{figure}

To find the matrix element $M$ of order $n$ one  must proceed as follows 
~\cite{kadysh64,kms72,kadysh68,karm76,karm88}\footnote{In  order to stick
to conventional notations,  the normalization of the  amplitude  given by
the  standard formulae (\ref{rul8}), (\ref{rul9}) and some factors in  the
rules of the graph technique (mainly, the degrees of $2\pi$)   differ from
the ones previously used in  ref.~\cite{karm88,kadysh64,kms72,kadysh68}.
Besides, the numbering of vertices here and in 
~\cite{karm88,kadysh64,kms72,kadysh68} has opposite order.}: 

\begin{enumerate}                                                               
\item \label{r1}                                                                
Arbitrary label by a number the vertices in the Feynman graph of  order 
$n$.  Orientate continuous lines (the lines of physical particles) in  the
direction  from the smaller to the larger number.  Initial particles are 
oriented  as incoming into a graph,  and final particles as  outgoing. 
Connect  by  a directed  dashed line (the spurion line) the vertices in the
order of  decreasing numbers.  Diagrams in which there are vertices with
all  incoming or outgoing particle lines (vacuum vertices, as indicated in 
fig.~\ref{rulevac}) can be omitted. Associate with each continuous  line  a
corresponding four-momentum, and with each $j$-th spurion line a 
four-momentum $\omega \tau_j$.

\item \label{r2} 
To each internal continuous line with four-momentum  $k$, associate the
propagator $\theta(\omega\cd k)\delta(k^2-m^2)$,  and  to each internal
dashed line with four-momentum $\omega\tau_j$  the  factor
$1/(\tau_j-i\epsilon)$.  

\item \label{r3} 
Associate with each vertex the coupling constant $g$. All the four-momenta
at  the  vertex, {\it including the spurion momenta}, satisfy the
conservation  law, i.e., the sum of incoming momenta is equal to the sum of
outgoing  momenta. 

\item \label{r4} 
Integrate (with $d^4k/(2\pi)^3$) over those four-momenta of the  internal
particles which remain unfixed after taking into account the  conservation
laws, and over all $\tau_j$  for the spurion lines from  $-\infty$  to
$\infty$.  

\item \label{r5}  
Repeat the procedure described in \ref{r1}-\ref{r4} for all  $n!$ possible
numberings of the vertices.   \end{enumerate}  We omit here the factorial
factors that arise from the identity of  the particles and depend on the
particular theory.  

The vacuum vertices indicated in fig.~\ref{rulevac} disappear for a 
trivial reason: it is impossible to satisfy the four-momentum  conservation
law for them. Indeed, the conservation law for the vertex  of  
fig. \ref{rulevac} has the form $k_1 +k_2+k_3 =  \omega(\tau_1-\tau_2)$. 
Since the four-momenta are on the mass shell: $k_{1-3}^2=m^2>0$, so that  the
left-hand side is always strictly positive: $(k_1+k_2+k_3)^2\geq  8m^2$,
whereas the right-hand side is zero since $\omega ^2=0$.  However, it will
be seen that the vacuum  contributions that vanish in the light-front
approach leave their track in a different way, making in the cases
discussed below the  light-front interaction  $H_\omega (x)$ in
eq.(\ref{rul11}) different from the usual  interaction $H(x)$ in
(\ref{rul1}).

The light-front diagrams can be interpreted as time-ordered
graphs. As soon as the vertices are labelled by numbers, any deformation of
a diagram changing the relative position of the vertex projections on the
``time direction" from left to right does not change the topology of the
diagram and the corresponding amplitude. Therefore it is often convenient
to deform the diagram so that the vertices with successively increasing
numbers are disposed from left to right. This just corresponds to 
time ordered graphs.

The light-front amplitudes can be also obtained by direct  transformation
of a given Feynman amplitude.  This transformation is  given in  section
\ref{f-lf} of appendix \ref{other}.

\subsection{Spin 1/2 system}\label{rule-spin}                                   
                                                                                
\begin{figure}[htbp]
\begin{center}
\centerline{\epsfbox{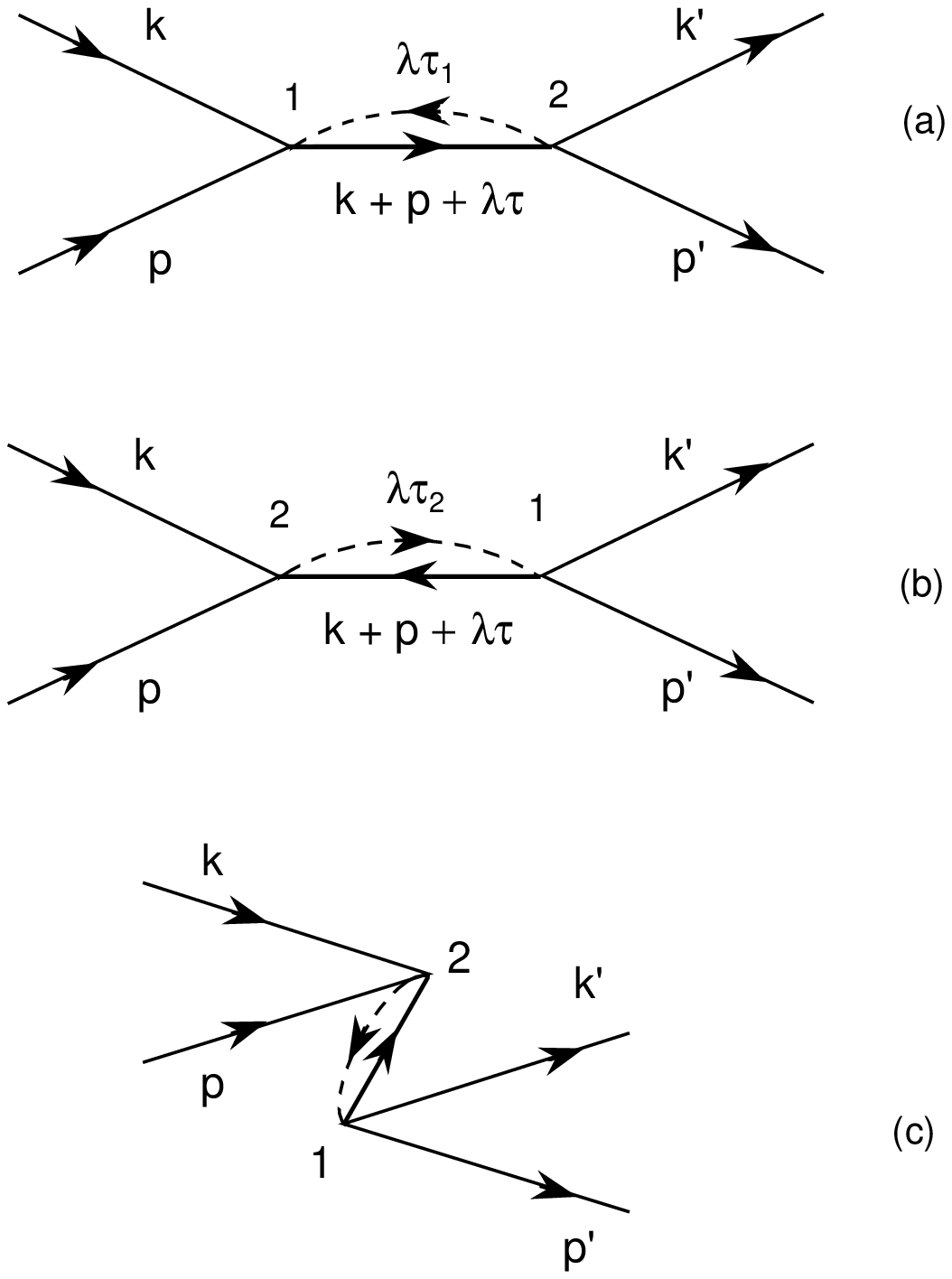}}
\figcap{Exchange of a particle in the $s$ channel; (a) contribution from
intermediate state containing one particle; (b) contribution from antiparticle
intermediate state; (c) same as (b) but rewritten as a time ordered diagram}
\label{pole} 
\end{center}
\end{figure}

The rules of the graph technique for spin 1/2 particles are similar  to 
those given above except for the fact that one has to worry about  contact
interactions we already mentioned in section (\ref{gendev}).  To  see this
from a more practical point of view, let us first consider  the  diagrams
of figs.~\ref{pole}(a) and (b)  for scalar particles and  consider  for a
moment the space-like plane $\lambda\cd x =\sigma$ with  $\lambda^2=1$,
as developed by Kadyshevsky~\cite{kadysh64}.                                                 

In accordance with the rules given above, the amplitude for the                 
diagram of fig.~\ref{pole}(a) gets the form:                                       
\begin{equation}\label{rul10p}                                                  
M_a=g^2\int \delta \left[(k+p+\lambda\tau_1 )^2-m^2 
\right] \theta \left[\lambda\cd (k+p)+\tau_1 \right] 
\frac{d\tau_1}{\tau_1-i\epsilon}\ . \end{equation}                        
Integrating over $d\tau_1$ by means of the $\delta$-function, we get:           
\begin{eqnarray}\label{rul11p}                                                  
M_a=\frac{g^2}{2\tau_1\sqrt{[\lambda\cd (k+p)]^2 
+m^2-(k+p)^2}}\ , \nonumber \\ \tau_1=-\lambda\cd 
(k+p)+\sqrt{[\lambda\cd (k+p)]^2 +m^2-(k+p)^2}\ .  
\end{eqnarray} 
Similarly, we obtain the following expression for the amplitude 
of fig.~\ref{pole}(b):                                                           
\begin{eqnarray}\label{rul12p}                                                  
M_b&=&\frac{g^2}{2\tau_2\sqrt{[\lambda\cd (k+p)]^2 
+m^2-(k+p)^2}}\ , \nonumber \\ \tau_2&=&\lambda\cd 
(k+p)+\sqrt{[\lambda\cd (k+p)]^2 +m^2-(k+p)^2}\ .  
\end{eqnarray} 
The sum of (\ref{rul11p}) and (\ref{rul12p}) gives the usual Feynman 
amplitude in the $s$ channel:  
$$ M_a+M_b = \frac{g^2} {m^2-(k+p)^2}\ .  $$ 
In the case where $\lambda=(1,\vec{0})$, we recover the contribution of
the usual time  ordered graphs in the old fashioned perturbation theory:  
\begin{eqnarray}\label{rul13p} 
M_a&=&\frac{g^2}{2\varepsilon_{\vec{k} +\vec{p}} 
\left[\varepsilon_{\vec{k}+\vec{p}} -\varepsilon_{\vec{k}}- 
\varepsilon_{\vec{p}}\right]}\ , 
\nonumber\\ 
M_b&=&\frac{g^2}{2\varepsilon_{\vec{k}+\vec{p}}                        
\left[\varepsilon_{\vec{k}+\vec{p}} +\varepsilon_{\vec{k}}+                     
\varepsilon_{\vec{p}}\right]}\ .                                                
\end{eqnarray}                                                                  
The diagram of fig.~\ref{pole}(b) and the amplitude $M_b$ correspond to
a vacuum contribution. This is clear if we draw it as a time ordered  diagram,
as it is done in
fig.~\ref{pole}(c).                                                

The light-front case can be obtained by introducing first the  four vector
$\omega$  with  $\omega^2=\delta^2$, replacing in the above formulae
$\lambda$  by  $\omega/\delta$ and then taking the limit $\delta 
\rightarrow 0$, i.e. $\omega^2  \rightarrow 0$. This limit is  however a 
delicate issue. For the scalar case under consideration above, it can  be
checked that the amplitude $M_b$ disappears.  In this limit, the system of
reference where $\lambda=(1,\vec{0})$ moves with a velocity close to $c$.
The graph technique for  $\omega^2=0$  is  therefore naturally related to
the old fashioned time  ordered perturbation theory in the infinite
momentum  frame, corresponding to the particular case $\omega
=(1,0,0,-1)$,  that defines the ``standard" light front
$t+z=0$.                                                

The full consideration of spin 1/2 particles shows that the above  result
has  to be completed. In such a case, indeed,  one should associate to 
each propagator the factor $(\hat{k}+m)$ for a fermion and $(m-\hat{k})$ 
for an antifermion, where $\hat{k}=k_\mu\gamma^\mu$ (see for instance section
\ref{nonzs} in appendix \ref{other}).  

The amplitude $M_a$, in eq.(\ref{rul11p}), is thus multiplied by 
$(\hat{p}_1+m)=(\hat{p}+\hat{k}+\hat{\lambda}\tau_1+m)$ and $M_b$ in 
eq.(\ref{rul12p}), is multiplied by 
$(m-\hat{p}_2)=-(\hat{\lambda}\tau_2-\hat{p}-\hat{k}-m)$ (for 
shortness 
we omit the initial and final spinors). Replacing again $\lambda$ by 
$\omega/\delta$ with $\omega^2=\delta^2$, we get in the limit 
$\delta\rightarrow 0$:                                                          
\begin{eqnarray}                                                                
M_a&=& g^2\frac{\hat{k}+\hat{p}+m} {m^2-(k+p)^2}                                          
+ g^2\frac{\hat{\omega}}{2\omega\cd (k+p)}\ ,         
\label{rul14}\\ 
M_b&=&-                                                         
g^2\frac{\hat{\omega}}{2\omega\cd (k+p)}\ .           
\label{rul15} \end{eqnarray}                                                    
The amplitude (\ref{rul15}) is the contact term we already mentioned  in 
section \ref{gendev}. It is just the track of the disappeared vacuum  diagrams.
The sum of the amplitudes $F_a$ and $F_b$ gives the Feynman  amplitude for spin
1/2 particles.  As we shall see later on, contact  terms are indeed essential
in getting fully covariant results (i.e., independent of  $\omega$). Other 
examples will be considered in the following
sections.                                              

\begin{figure}[htbp]
\centerline{\epsfbox{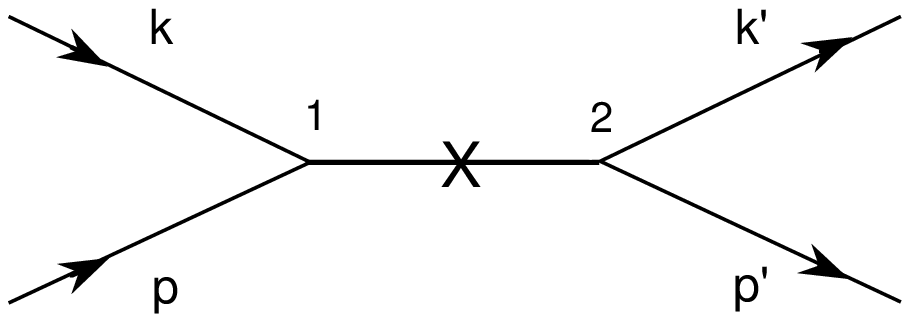}}
\figcap{Contact interaction contributing to the exchange of a spin 1/2 particle
in the $s$ channel}
\label{cont}
\end{figure}

To incorporate the contact term from the very beginning, one should  not 
consider the diagram of fig.~\ref{pole}(b), but add to the diagram of 
fig.~\ref{pole}(a) the diagram indicated in fig.~\ref{cont}, which is 
obtained from fig.~\ref{pole} by deleting the spurion line and  marking the
internal line by a cross. This contribution can be also derived from a
counter term added to  $H^{int}$ in order to get $H^{int}_{\omega}$, as it
was done for spin 1/2 in ref. \cite{amn2}. Associating the crossed line to
a fermion  carrying the momentum $l$, we assign to this line the factor 
$-\hat{\omega}\theta(\omega\cd l)/(2\omega \cd l)$. For the diagram  of
fig.~\ref{cont} this rule gives eq.(\ref{rul15}) back. 

We can thus formulate the rules of the covariant light-front graph
technique for the case of spin 1/2 particles.

\begin{enumerate} 
\item \label{rs1} 
Transform the Feynman graph of a given order to the set of the  light-front
graphs in the same way as described in point 1 of the  rules for  scalar
particles.  Without making  change in the orientation, replace the ordinary
lines corresponding  to antifermions by double lines, so that 
 the number  of fermions minus the number of  antifermions is conserved. Both
initial fermions and antifermions are shown by lines incoming to a
diagram, and final  fermions and antifermions  by outgoing lines.

\item \label{rs1a} 
Consider the diagrams with internal fermion and antifermion lines  labelled at
their extremities by two successive numbers (i.e., the  ends of the lines  are
directly connected by the spurion line, see, e.g., fig.~\ref{pole}).  In
addition to all the diagrams we already have, we create from the latter
diagrams another set of diagrams, by deleting those spurion lines, which 
connect the ends of the fermion (and antifermion) lines. Put on the
corresponding fermion and antifermion lines a cross (see, e.g.,
fig.~\ref{cont}). Draw any diagram with incoming lines at  the left and
outgoing lines at the right. 

\item
The analytical  expression for the amplitude is
written from the left to the right. The factors in this expression are written
in the order where they are met when one goes through a diagram, starting at
the right, from an outgoing fermion line, and  continuing in the direction
opposite to the orientation of the  fermion lines.  For an incoming 
antifermion line, one starts at the left,  and continues in the direction  of
orientation of the (double) antifermion lines.  

\item \label{rs2} 
To each internal continuous line with four-momentum $k$, associate   the 
propagator $(\hat{k}+m)\theta (\omega \cd k)\delta(k^2-m^2) $ for a 
fermion, and the propagator $(m-\hat{k})\theta (\omega\cd k)\delta(k^2-m^2) $ 
for an 
antifermion, the factor $-\hat{\omega}\theta(\omega\cd k)/(2\omega\cd k)$ 
for each crossed fermion line, the factor $\hat{\omega}\theta(\omega\cd
k)/(2\omega\cd k)$  for each crossed antifermion line, and the factor
$1/(\tau_j-i\epsilon)$  for each internal dashed line with four-momentum
$\omega  \tau_j$.  

\item \label{rs3} 
Associate with each vertex the factor $gV$, where $V$ depends on the  type
of coupling (1, $i\gamma_5$, $\gamma_{\mu}$, etc.) in the  Hamiltonian 
$H=-g\bar{\psi}V\psi\cdots$. All the four-momenta at each vertex,  {\it
including the spurion momenta}, satisfy the conservation law,  i.e., the
sum of incoming momenta is equal to the sum of outgoing  momenta (incoming
and outgoing momenta always correspond to the incoming and outgoing
lines). 

\item \label{rs4}
With any outgoing fermion [antifermion] line with momentum $p$, 
associate the spinor $\overline{u}(p)$ [$v(p)$] and with any incoming 
fermion [antifermion] line associate the spinor $u(p)$ $\left[ 
\overline{v}(p) \right]$.

\item \label{rs5} 
Integrate (with $d^4k/(2\pi)^3$) over those four-momenta of the  internal
particles which remain unfixed after taking into account  the 
conservation laws, and over all $\tau_j$  for the spurion lines from 
$-\infty$ to $\infty$.  

\item \label{rs6} 
Repeat the procedure described in \ref{rs1}-\ref{rs5} for all $n!$ 
possible numberings of the vertices. One should also take 
into account the standard sign factors appearing from permutation of 
fermion fields.  
\end{enumerate} 

In the case of a diagram containing boson lines connected to fermion
loops, one should go through the outgoing boson line, and then, having  reached
the  fermion line, pass it as indicated in the rules, i.e., in the
direction opposite to the orientation of the fermion line and along the
orientation of the antifermion line. Both
lines in the loop, single and double, are oriented in the same direction,
but one makes a cycle when passing through them.

We emphasize that the fermion and antifermion are distinguished by the type
of the  continuous lines (single or double). Both the fermion and
antifermion lines are directed  from the smaller to the larger number (in
the time ordered graph they both  propagate from left to right, i.e., in
the direction of time increase in  time ordered graph). In contrast
to the standard versions of the Feynman  rules, any incoming line, {\em
both} for antifermion and fermion,  corresponds to the initial state, and
outgoing lines -- to the final state.  The fermion and antifermion lines
are followed in opposite directions  (opposite to the orientation for a
fermion line and along the orientation for  an antifermion line). The
fermion and antifermion propagators differ from each other. 

In many practical applications,
the contact term (or so called instantaneous interaction) can be 
conveniently incorporated for spin 1/2  particles by replacing the spin
part of the propagator  $(m\pm\hat{k})$  (for those lines where the contact
term contributes at all) by  $[m\pm(\hat{k}-\hat{\omega}\tau)]$ (however,
the  delta-function in  the  propagator $\delta(k^2-m^2)$ still depends on
the argument  $(k^2-m^2)$).  Together with the fact that the contact term 
originates  from the lines which ends are labelled by two successive
numbers,  this rule coincides with the rule given in 
refs.~\cite{brodsky73,brodsky}: the contact terms modify only the propagators 
corresponding to the ``lines extending over a single time interval''. Two
successive numbers just determine a single time interval.

The above replacement to incorporate systematically the contact term  can
always be made in perturbation theory.  However,  it cannot be made in more 
complicated cases. In particular, the  graph  for the wave function,  for
instance, indicated in fig.~\ref{fwf} below, does not contain  any  external
crossed lines. Hence, the contribution of the contact terms  to  the deuteron
electromagnetic form factors or electrodisintegration  amplitude, in the
impulse approximation for example, cannot be  incorporated by the above 
substitution in the propagators, since it would create an object with  an
external crossed line.  The contact terms in the deuteron  electromagnetic form
factors and electrodisintegration will be  properly taken into account in
section~\ref{cted}.

If a diagram contains two fermions (but not fermion and antifermion) turning 
into a boson, like in the case  of the deuteron form factor (the vertex
$NNd$),  one should pass through one fermion line from the right to the left,
and then start again from the same final vertex and pass through the second 
line. 
Both lines are followed in the direction opposite to their orientation.  Such
example is given  below in section \ref{ff}.

\subsection{Spin 1 and coupling with derivatives}\label{sp1-der}                               
For particles with spin 1 the rules are similar to the case of 
particles with spin 1/2. Like in Feynman rules, external lines for
spin 1 particles are associated with polarization
vectors. However, in the case of 
spin 1 particles, there is no difference between propagators of 
particle and antiparticle. The propagator has the form                                   

\begin{equation}\label{bos1}                                                    
D_{\mu\nu}(k,m)=\left(-g_{\mu\nu}                                                  
+\frac{k_{\mu}k_{\nu}}{m^2}\right)\theta(\omega \cd 
k)\delta(k^2-m^2)\ .           
\end{equation}                                                                  
The contact term is given by:                                                   
\begin{equation}\label{bos2}                                                    
\Delta_{\mu\nu}= -\frac{k_{\mu}\omega_{\nu}+k_{\nu}\omega_{\mu}}                 
{2(\omega\cd k)m^2} - \frac{\omega_{\mu}\omega_{\nu}}{4(\omega \cd 
k)^2m^2}          
(k^2-m^2)\ .                                                                    
\end{equation}                                                                  
Like in the case of spin 1/2, the contact term contributes only in 
the          
lines extending over a single time interval. It can be incorporated 
simply      
by  replacing the spin part of the corresponding propagators 
$(g_{\mu\nu}                      
-k_{\mu}k_{\nu}/m^2)$ by $\left[g_{\mu\nu}                                      
-(k-\omega\tau)_{\mu}(k-\omega\tau)_{\nu}/m^2\right]$.                          

In the  case of massless vector boson (the photon for instance)
the form of the propagator, as usual, depends on the
gauge. For a general gauge, the momentum dependence of the gauge term in
the propagator induces also a corresponding contact term. We do
not investigate here the general case and give the photon propagator for
the Feynman gauge:
\begin{equation}\label{bos3}
D_{\mu\nu}(k,m=0)=-g_{\mu\nu}\theta(\omega \cd k)\delta(k^2)\ .           
\end{equation}
There is no contact term in this gauge.

It is sometimes more convenient to use in this formulation of LFD the
light-cone gauge defined by $\omega\cd A=0$. The spin-1 propagator for a
massless particle in this gauge is \cite{brodsky}:
\begin{equation}\label{bos3p}                                                    
D_{\mu\nu}(k,m)=\left(-g_{\mu\nu}                                                  
+\frac{\omega_{\mu}k_{\nu}+\omega_{\nu} k_{\mu}}{\omega\cd k}\right)
\theta(\omega \cd k)\delta(k^2-m^2)           
\end{equation} 
and the corresponding contact term has the form:
\begin{equation}\label{bos3pp}                                                    
\Delta_{\mu\nu}= -\frac{\omega_{\mu}\omega_{\nu}}{(\omega\cd k)^2}\ .                                                                   
\end{equation} 
The contact term (\ref{bos3pp}) can be again incorporated by the replacement
$k\rightarrow k-\omega\tau$ in the spin part of eq.(\ref{bos3p}).

Consider finally the coupling with a derivative, e.g., when the 
interaction Lagrangian contains a term like $\partial_\mu  \varphi(x)$.  
Since all the lines are oriented, any line is either incoming in the 
vertex, or outgoing from it. Let this line be associated with the 
four-momentum $k$.  The derivative results in the multiplication of a 
vertex by the extra factor $ik_{\mu}$ for an outgoing  line, 
and by the factor $-ik_{\mu}$ for an incoming line. The contact term 
contributes to the lines involving the derivative of a field  and having
the same  topology, as in the case of spins 1/2 and 1. That means that the
ends  of  these lines should be connected by the spurion line.  Like in
the case of spin 1 propagator, the contact term can be
incorporated by  the replacement $k_{\mu}\rightarrow
k_{\mu}-\omega_{\mu}\tau$, where  $\omega\tau$ is the spurion momentum, and
this spurion line should  connect the ends of the particle line,
corresponding to the field  with  derivative
coupling. 

\subsection{Ultraviolet and infrared behavior}\label{behav}

A peculiarity of the covariant light-front amplitudes is that they have no any
ultraviolet divergences for  finite values of all the spurion four-momenta. 
All the ultraviolet divergences in light-front diagrams appear after
integrations over $\tau_j$ with infinite limits \cite{kadysh64}. Indeed, the
energy-momentum conservation (including the spurion  four-momentum) is valid at
any vertex.  Since all the four-momenta are on the corresponding mass  shells, 
we have at each vertex a real physical process  as far as  the kinematics is
concerned. For finite initial particle  energies and for finite incoming
spurion energy, the  energies of the particles in the intermediate states are
thus  also finite. Hence, the integrations over the particle momenta for fixed
spurion momenta are constrained by a kinematically allowed finite domain. It is
the same reason that provides  finite imaginary part of a Feynman diagram 
found by replacing the Feynman propagators
$\frac{\displaystyle{1}}{\displaystyle{(k^2-m^2+i\epsilon)}}$ by the
delta-functions $-i\pi\delta(k^2-m^2)$. In both cases the internal particle
lines are associated with the delta-functions.

The only source of the ultraviolet divergences in the light-front amplitudes is
the infinite intermediate spurion energies, i.e., infinite $\tau_j$.  This is
the  reason why divergences may appear at the upper limit of integration over
$\tau_j$. Since  $\tau_j$ are scalar quantities, one can introduce an invariant
cutoff in terms of these variables.  This way of regularizing the divergent
diagrams is another advantage of the covariant  formulation of LFD. We
illustrate this property of the light-front amplitudes in sect. \ref{selfen}
for the example of the self-energy.

For  massless particles,  light-front amplitudes may have infrared
divergences, like in the case of Feynman diagrams.

Another peculiarity of LFD is the appearance of ``zero modes''. For
constituents of zero  mass, for instance, the state vector may contain
components with  $\omega \cd k=0$ for non-zero four-momentum $k$. In the
standard approach, this  corresponds to the finite light-front energy
$k^-=\vec{k}_{\perp}^2/k_+$ for  both $k_+=0$ and $\vec{k}_{\perp}^2=0$.  Zero
modes can also appear  in theories with spontaneously broken symmetry. They
make the equivalence between LFD and the instant form of  quantization in which
nontrivial vacuum structures (condensates) appear
\cite{zakopane,werner,wilson94}. 

The detailed discussion of the above problems is beyond the scope of the
present review.

\section{Simple examples}                                                       
                                                                                
\subsection{Exchange in $t$-channel}\label{tchan}

\begin{figure}[hbtp]
\centerline{\epsfbox{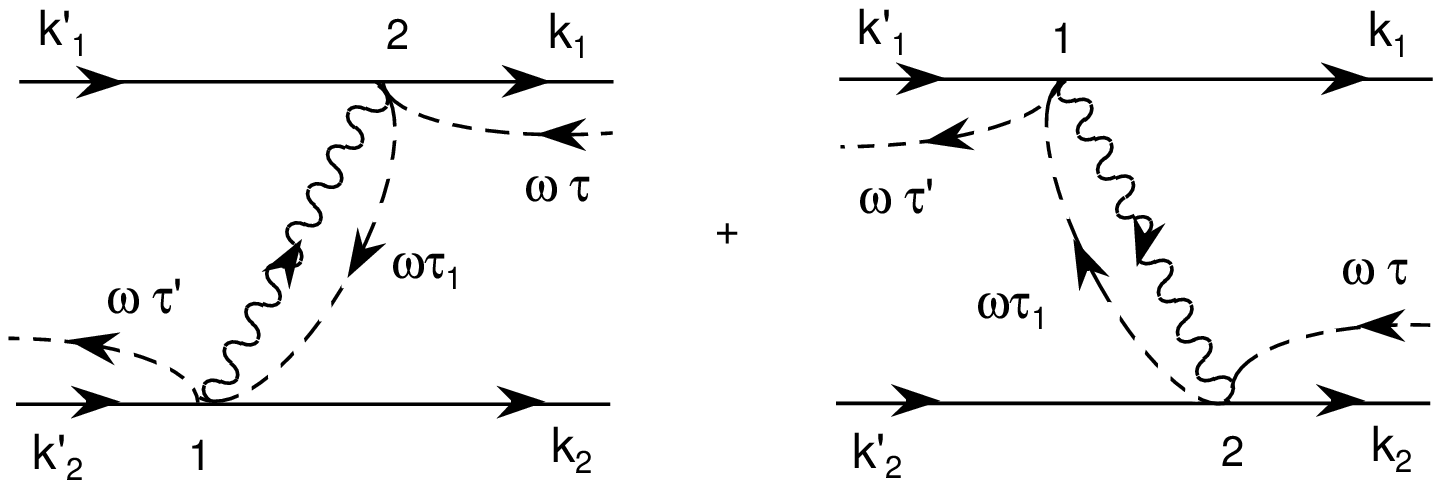}}
\figcap{Exchange by a particle in $t$-channel.}
\label{fkern} 
\end{figure}

Consider the diagrams shown in fig.~\ref{fkern}. It corresponds to the 
exchange of a scalar particle of mass $\mu$ between two scalar particles, in 
the $t$ channel. These diagrams  determine, in the  ladder approximation,
the kernel of the equation for the calculation of the  light-front  wave 
function. The
external  spurion lines indicate that the amplitude is off-energy shell.
According to the light-front graph technique for spinless particles,  the
amplitude has the
form:                                                                           
\begin{eqnarray}\label{k1}                                                      
{\cal K}&=&\quad 
g^2\int \theta\left(\omega\cd (k_1-k'_1)\right) \delta             
\left((k_1-k'_1 +\omega\tau_1- \omega\tau)^2-\mu^2\right)                             
\frac{d\tau_1}{\tau_1-i\epsilon}
\nonumber \\                             
&&+g^2\int \theta\left(\omega\cd (k'_1-k_1)\right) 
\delta\left((k'_1-k_1 +\omega\tau_1- \omega\tau')^2-\mu^2\right)                      
\frac{d\tau_1}{\tau_1-i\epsilon}
\nonumber \\                             
&=&\quad
\frac{g^2\theta\left(\omega\cd 
(k_1-k'_1)\right)}{\mu^2-(k_1-k'_1)^2+2\tau 
\omega\cd(k_1-k'_1) -i\epsilon}
\nonumber\\ 
&&+\frac{g^2\theta\left(\omega\cd (k'_1-k_1)\right)}{\mu^2 
-(k'_1-k_1)^2 +2\tau' 
\omega\cd (k'_1-k_1)-i\epsilon}\ .  
\end{eqnarray} 
The two items in (\ref{k1}) correspond to the two diagrams of 
fig.~\ref{fkern}. They cannot be non-zero simultaneously.
On the energy shell, i.e. for both $\tau=\tau '=0$, 
the expression for the kernel is identical to the Feynman amplitude:                
\begin{equation}\label{k2}                                                      
{\cal K}(\tau=\tau'=0) = \frac{g^2}{\mu^2-(k_1-k'_1)^2-i\epsilon}\ .  
\end{equation} 
Note that the off-shell amplitude (\ref{k1}) depends on $\omega$. This
agrees with the fact that it is related by eqs.(\ref{rul6a}), (\ref{rul6b})
to the $S$-matrix on a finite light-front plane in the interaction
region. It  depends therefore on this plane. 
In the case of a scalar amplitude, this dependence is given in
terms of extra scalar variables (in addition to Mandelstam variables $s$ and
$t$). These variables are the scalar products of $\omega$ with the
four-momenta, as seen from eq. (\ref{k1}). Hence, the amplitude has extra
singularities as a function of this new variables, which can be found similarly
to singularities of the Feynman amplitudes~\cite{karm78}.

On the energy shell,  corresponding to $\tau=\tau'=0$ and to the light-front
plane shifted to infinity, out of the interaction region, the dependence of the
amplitude on $\omega$  disappears. In more complicated cases, when a Feynman
diagram  corresponds  to the sum a few light-front diagrams (like in the case
of the  box  diagrams considered in appendix~\ref{other}), the amplitude for a 
particular light-front diagram may depend on $\omega$ even on the  energy
shell.  This dependence disappears in the sum of all  amplitudes at a given
order.  In this case the singularities of different amplitudes cancel each
other in the sum.                     

The off-energy shell amplitude corresponds to a diagram with external spurion
lines. This term --  off-energy shell -- is borrowed from the old fashioned
perturbation theory. As mentioned above, the latter is  obtained, if one
introduces in eq.(\ref{rul11}), instead of $\omega$, the four-vector $\lambda$
with $\lambda =(1,0,0,0)$. The difference of initial and final four-momenta
differs from zero by $\lambda\tau$.  Due to $\vec{\lambda}=0$ the sums of
initial and final spatial components of momenta is still equal to each other,
but for $\tau\neq 0$  initial and final energies are not equal to each  other
(their difference is equal to $\tau$).  This just characterizes the
``off-energy shell" amplitude, which can be a part of a  bigger diagram. In the
light-front amplitude for arbitrary  $\omega$, all the four-vector components
do not satisfy the conservation law.  Their difference is proportional to
$\omega\tau$. Hence the term "off-energy shell"  may be inappropriate for the 
light-front amplitude, since not only energies, but also three-momenta are not 
conserved. We
emphasize again that the four-momenta are always  on the mass shells, whereas
the Feynman amplitude depends  on the four-momenta which may be off-mass shell,
but the conservation law is always valid.

\subsection{Compton scattering}\label{comptel}

\begin{figure}[htbp]
\epsfxsize = 14.cm
\epsfysize = 4.cm
\centerline{\epsfbox{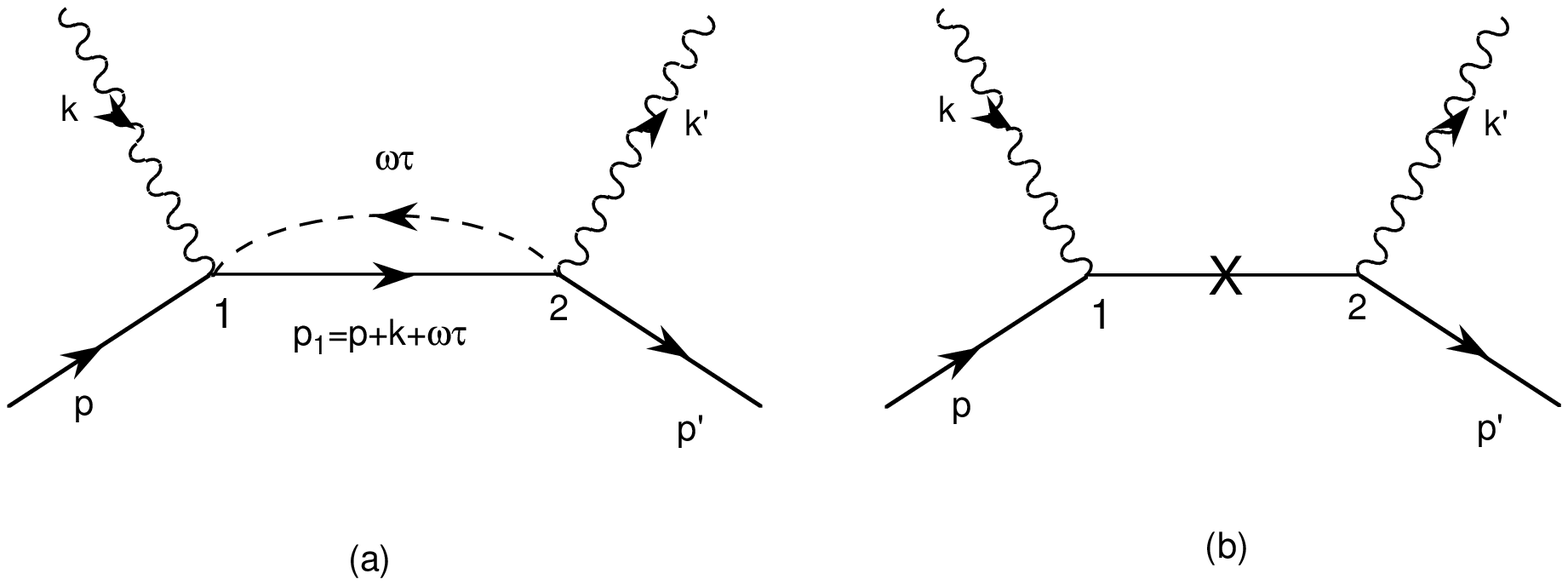}}
\figcap{(a) The direct diagram for electron Compton scattering  with 
an electron in the intermediate state.
(b) The contact term corresponding to the diagram (a).}
\label{compton}
\end{figure}

The  Compton scattering amplitude from an electron (through $s$-channel)
is indicated in  fig. \ref{compton}(a),
and, for the contact term, in fig.\ref{compton}b.
According to the above rules of graph technique,
we associate with fig. \ref{compton}(a) the following expression:
\begin{eqnarray}\label{comp1}
M_a&=&g^2\int \bar{u'}(p')\hat{e}'^*(m+\hat{p}_1)\hat{e}u(p)
\theta(\omega\cd p_1)\delta(p_1^2-m^2)\frac{d\tau}{\tau-i\epsilon}
\delta^{(4)}(p+k+\omega\tau-p_1)d^4p_1
\nonumber\\
&=&g^2\bar{u'}(p')\hat{e}'^*\frac{m+\hat{p}+\hat{k}}
{m^2-(p+k)^2}\hat{e}u(p)
+g^2\bar{u'}(p')\hat{e}'^*\frac{\hat{\omega}}{2\omega\cd (p+k)}\hat{e}u(p),
\end{eqnarray}
where $\hat{e}=\gamma_{\mu}e^{\lambda}_{\mu}(k)$ and similarly for
$\hat{e}'^*$.
The polarization vector of the 
photon is denoted by  $e^{\lambda}_{\mu}(k)$.
The order of factors in (\ref{comp1}) corresponds to
follow the fermion line from the right to the left, in the direction opposite 
to its orientation. 

The amplitude for the contact term contribution, indicated in 
fig. \ref{compton}(b), is:
\begin{equation}\label{comp2}
M_b=-g^2\bar{u'}(p')\hat{e}'^*\frac{\hat{\omega}}{2\omega\cd (p+k)}\hat{e}u(p)
\ .
\end{equation}
We omit here the theta-function $\theta(\omega\cd (p+k))$,
which is always 1.

The expressions (\ref{comp1}), (\ref{comp2}) reproduce the amplitudes
(\ref{rul14}), (\ref{rul15}). In the sum, $M_a+M_b$, the omega-dependent 
items cancel each other.

\begin{figure}[hbtp]
\epsfxsize = 14.cm
\epsfysize = 4.cm
\centerline{\epsfbox{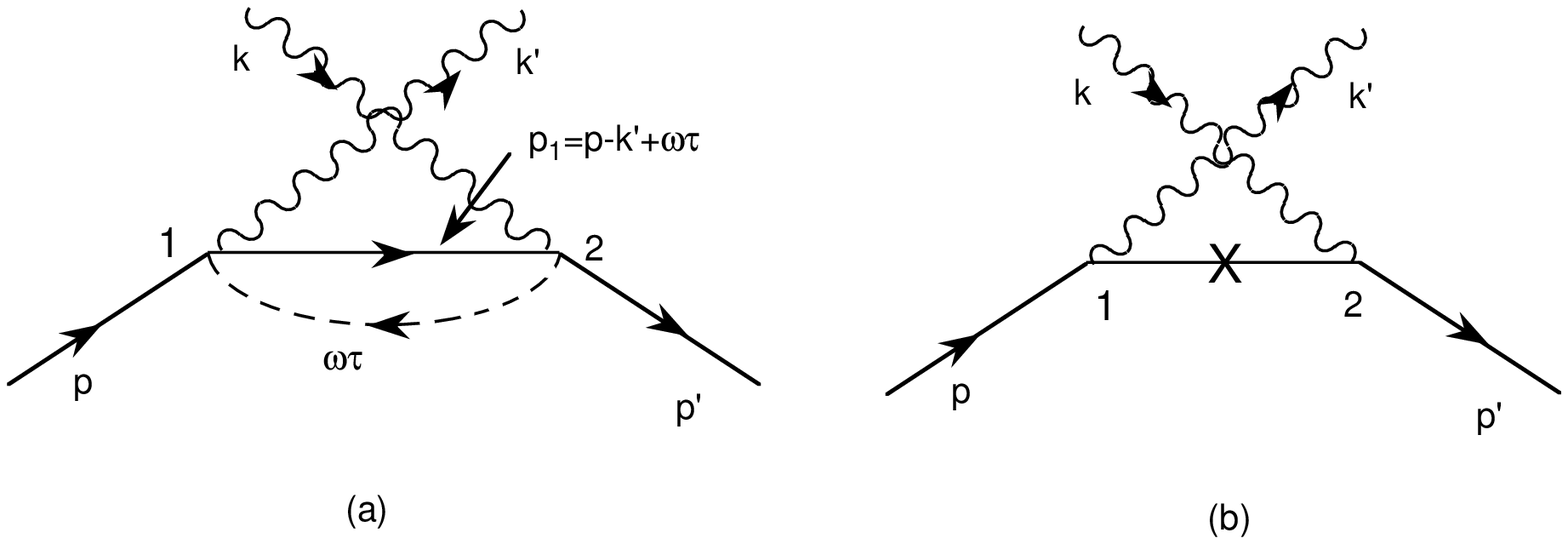}}
\figcap{(a) The cross diagram for electron Compton scattering with 
an electron in the intermediate state.
(b) The contact term corresponding to the diagram (a).}
\label{cross1}
\end{figure}

\begin{figure}[htbp]
\epsfxsize = 14.cm
\epsfysize = 12.cm
\centerline{\epsfbox{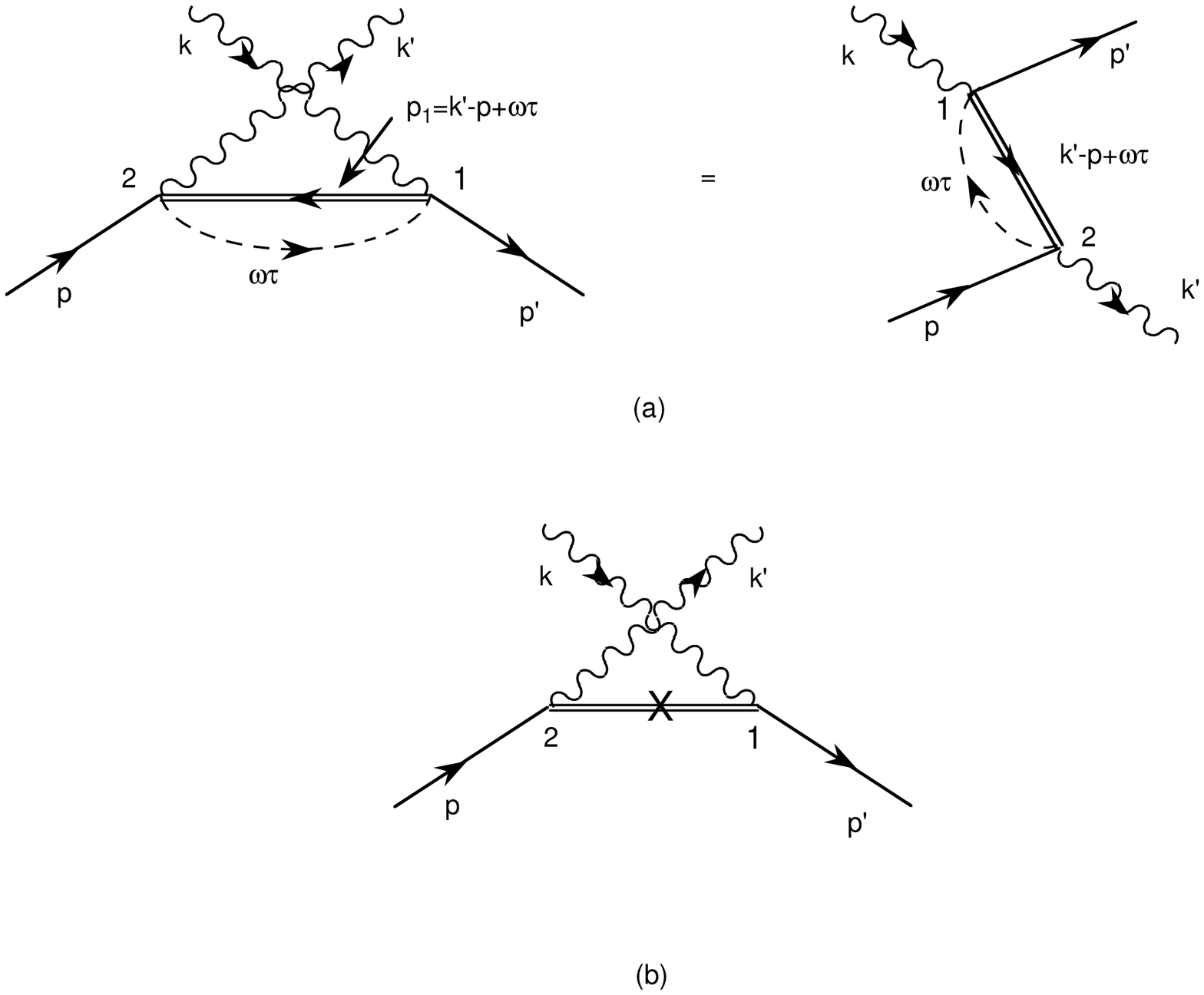}}
\figcap{(a) The cross diagram for electron Compton scattering with
a positron in the intermediate state.
(b) The contact term corresponding to the diagram (a).}
\label{cross2}
\end{figure}

Now consider the crossed amplitude, indicated in figs. \ref{cross1} and 
\ref{cross2}.
The amplitude corresponding to fig. \ref{cross1}(a) has the form:
\begin{eqnarray}\label{comp3}
M_a&=&g^2\int \bar{u'}(p')\hat{e}(m+\hat{p}_1)\hat{e}'^*u(p)
\theta(\omega\cd p_1)\delta(p_1^2-m^2)\frac{d\tau}{\tau-i\epsilon}
\delta^{(4)}(p-k'+\omega\tau-p_1)d^4p_1
\nonumber\\
&=&g^2\bar{u'}(p')\hat{e}\frac{m+\hat{p}-\hat{k}'}
{m^2-(p-k')^2}\hat{e}'^*u(p)\theta(\omega\cd (p-k'))
\nonumber\\
&&+g^2\bar{u'}(p')\hat{e}\frac{\hat{\omega}}{2\omega\cd (p-k')}
\hat{e}'^*u(p)\theta(\omega\cd (p-k')).
\end{eqnarray}
The contact term contribution in fig. \ref{cross1}(b) writes:
\begin{equation}\label{comp4}
M_b=-g^2\bar{u'}(p')\hat{e}\frac{\hat{\omega}}{2\omega\cd (p-k')}
\hat{e}'^*u(p)
\theta(\omega\cd(p-k')).
\end{equation}
It has the form of the second term in (\ref{comp3}) and cancel it in the sum 
$M_a+M_b$. 

The amplitude of fig. \ref{cross2}(a) differs from fig. \ref{cross1}(a) by
the time ordering of the vertices. If we draw the vertices in the order of
increasing numbers from the left to the right, this amplitude appears as a 
time ordered diagram. It contains the creation of a $e^+e^-$-pair by the
photon, i.e., includes a positron in the intermediate state, shown by the
double line, and its subsequent annihilation. The corresponding amplitude
has the form:
\begin{eqnarray}\label{comp5}
M_a&=&g^2\int \bar{u'}(p')\hat{e}(m-\hat{p}_1)\hat{e}'^*u(p)
\theta(\omega\cd p_1)\delta(p_1^2-m^2)\frac{d\tau}{\tau-i\epsilon}
\delta^{(4)}(k'-p+\omega\tau-p_1)d^4p_1
\nonumber\\
&=&g^2\bar{u'}(p')\hat{e}\frac{m+\hat{p}-\hat{k}'}
{m^2-(p-k')^2}\hat{e}'^*u(p)\theta(\omega\cd (k'-p))
\nonumber\\
&&-g^2\bar{u'}(p')\hat{e}\frac{\hat{\omega}}{2\omega\cd (k'-p)}\hat{e}'^*u(p)
\theta(\omega\cd (k'-p)),
\end{eqnarray}
and for the contact term indicated in fig. \ref{cross2}(b):
\begin{equation}\label{comp6}
M_b=g^2\bar{u'}(p')\hat{e}\frac{\hat{\omega}}{2\omega\cd (k'-p)}\hat{e}'^*u(p)
\theta(\omega\cd(k'-p)).
\end{equation}
The first item in eq.(\ref{comp3}) differs from zero in the region 
$\omega\cd (p-k')>0$, whereas in eq.(\ref{comp5}) this occurs in the
region  $\omega\cd (p-k')<0$. The sum of these two contributions
 reproduce the Feynman amplitude
in the $u$-channel, in the whole kinematical range.

In order to show how to deal with antifermion states,  consider now the
Compton scattering amplitude from an antifermion. The $s$-channel  diagram
is similar to fig. \ref{compton} with  the single lines replaced by the
double ones.  The corresponding amplitude has  the form:
\begin{equation}\label{comp7}
M_a=g^2\bar{v}(p)\hat{e}\frac{m-(\hat{p}+\hat{k})}
{m^2-(p+k)^2}\hat{e}'^*v'(p')
-g^2\bar{v}(p)\hat{e}\frac{\hat{\omega}}{2\omega\cd (p+k)}\hat{e}'^*v'(p'),
\end{equation}
while the contact term gives:
\begin{equation}\label{comp8}
M_b=g^2\bar{v}(p)\hat{e}\frac{\hat{\omega}}{2\omega\cd (p+k)}\hat{e}'^*v'(p').
\end{equation}
The order of factors in (\ref{comp7}) and (\ref{comp8}) corresponds to follow
the antifermion line from the left to the right, in the direction along its
orientation. Note that in eq.(\ref{comp8}) the sign of the contact term 
originating from the antifermion line is opposite to the fermion contact
term, eq.(\ref{comp2}). The corresponding  $\omega$-dependent contribution
in (\ref{comp7}) has also an opposite sign  and thus disappears in the sum
$M_a+M_b$.

The amplitude for the cross diagram with an antifermion intermediate state
is also similar to fig. \ref{cross1} and is given by:
\begin{eqnarray}\label{comp9}
M_a&=&g^2\int \bar{v}(p)\hat{e}'^*(m-\hat{p}_1)\hat{e}v'(p')
\theta(\omega\cd p_1)\delta(p_1^2-m^2)\frac{d\tau}{\tau-i\epsilon}
\delta^{(4)}(p-k'+\omega\tau-p_1)d^4p_1
\nonumber\\
&=&g^2\bar{v}(p)\hat{e}'^*\frac{m-(\hat{p}-\hat{k}')}
{m^2-(p-k')^2}\hat{e}v'(p')\theta(\omega\cd(p-k'))
\nonumber\\
&&-g^2\bar{v}(p)\hat{e}'^*\frac{\hat{\omega}}{2\omega\cd (p-k')}\hat{e}v'(p')
\theta(\omega\cd(p-k')).
\end{eqnarray}
The corresponding contact term is:
\begin{equation}\label{comp10}
M_b=g^2\bar{v}(p)\hat{e}'^*\frac{\hat{\omega}}{2\omega\cd (p-k')}\hat{e}v'(p')
\theta(\omega\cd(p-k')).
\end{equation}
The kinematics of the Compton scattering from fermion and antifermion, 
i.e., the expression of the intermediate momentum $p_1$ through the external
momenta, as given in  eq.(\ref{comp9}), is here identical. Besides the
spinors $u$ and $v$, the difference is in the signs of the spin parts of the
propagators and of the contacts terms.

Finally we give the amplitude corresponding to the scattering on an
antifermion with a fermion intermediate state (same as fig. \ref{cross1},
but with single lines replaced by the double ones and vice versa):
\begin{eqnarray}\label{comp11}
M_a&=&g^2\int \bar{v}(p)\hat{e}'^*(m+\hat{p}_1)\hat{e}v'(p')
\theta(\omega\cd p_1)\delta(p_1^2-m^2)\frac{d\tau}{\tau-i\epsilon}
\delta^{(4)}(k'-p+\omega\tau-p_1)d^4p_1
\nonumber\\
&=&g^2\bar{v}(p)\hat{e}'^*\frac{m-(\hat{p}-\hat{k}')}
{m^2-(p-k')^2}\hat{e}v'(p')\theta(\omega\cd(k'-p))
\nonumber\\
&&+g^2\bar{v}(p)\hat{e}'^*\frac{\hat{\omega}}{2\omega\cd (k'-p)}\hat{e}v'(p')
\theta(\omega\cd (k'-p))
\end{eqnarray}
and the contact term:
\begin{equation}\label{comp12}
M_b=-g^2\bar{v}(p)\hat{e}'^*\frac{\hat{\omega}}{2\omega\cd (k'-p)}\hat{e}v'(p')
\theta(\omega\cd (k'-p)).
\end{equation}
The difference between (\ref{comp9}), (\ref{comp10}) and  (\ref{comp11}),
(\ref{comp12}) is in the sign of the argument of the  corresponding
theta-functions. The sum of both contributions gives the Feynman amplitude
for  Compton scattering on an antifermion.

As mentioned above, the contact term can be incorporated 
everywhere by the following replacement in the spin matrix of the
propagator:  $\hat{p}_1\rightarrow \hat{p}_1-\hat{\omega}\tau$. 

%
\subsection{Self-energy contributions}\label{selfen}                            

\begin{figure}[hbtp]
\centerline{\epsfbox{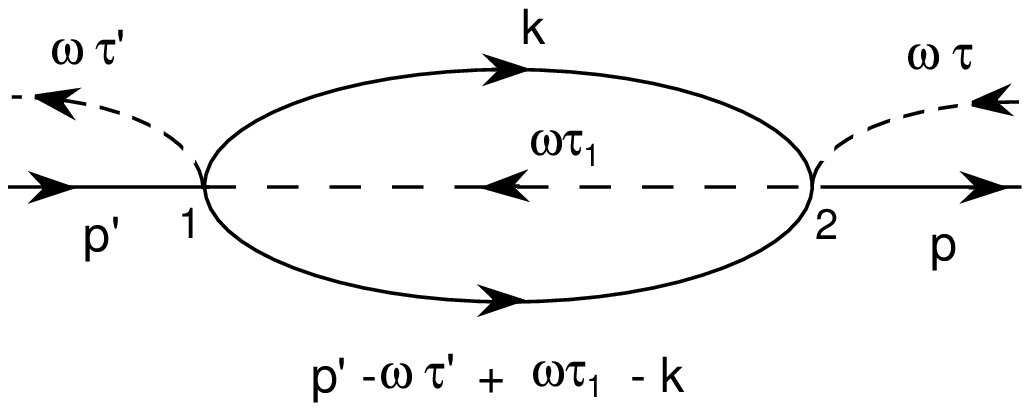}}
\figcap{Self-energy loop.}
\label{selffig} 
\end{figure}

Another simple example is the self-energy diagram shown in 
fig. \ref{selffig}. The corresponding amplitude (equal to the 
self-energy up to a factor) has the form:  
\begin{equation}\label{sel1}                                                    
\Sigma(p')=g^2\int \theta(\omega\cd k)\delta(k^2-m^2)            
\theta\left(\omega\cd (p_1+\omega\tau_1-k)\right)                                    
\delta\left((p_1+\omega\tau_1-k)^2 -m^2\right) \frac{d^4k}{(2\pi)^3} 
\frac{d\tau_1}{\tau_1-i\epsilon}\ ,                                                 
\end{equation}                                                                  
with $p_1=p'-\omega\tau'$.                                                        
                                                                                
Let $q=p_1+\omega\tau_1$.  The integral over $d^4k$ is thus reduced to  the
well known calculation of the imaginary part of the Feynman amplitude, when
all  the propagators are replaced by the
delta-functions:                                                
\begin{equation}\label{sel2}                                                    
\int\delta(k^2-m^2)\delta\left((q-k)^2-m^2\right)d^4k=                         
\frac{\pi}{2\sqrt{q^2}} \sqrt{q^2-4m^2}.                                        
\end{equation}
Inserted in (\ref{sel1}), it gives:                                                      
\begin{equation}\label{sel3}                                                    
\Sigma(p_1) = \frac{g^2}{16\pi^2}\int\limits_{4m^2-p_1^2}^{\infty} 
\frac{\sqrt{p_1^2-4m^2+\tau_1}}{\sqrt{p_1^2+\tau_1}}                                  
\frac{d\tau_1}{\tau_1-i\epsilon}.                                                   
\end{equation}                                                                  
The logarithmic divergence  is at the upper limit of the integration  over
$\tau_1$. This is a particular manifestation of  a general property of the
light-front amplitudes, discussed above in sect. \ref{behav}. One can
introduce the invariant cutoff  in terms of $\tau_1$. In this way, after
renormalization, the standard expression for the self-energy  amplitude is
obtained.


\chapter{The two-body wave function}\label{wf}                                  
Since the wave functions of a composite system are the Fock components of the 
state vector, their
transformation properties under four dimensional rotations are determined  by
the corresponding properties of the state vector investigated in the previous
chapter. This allows one to construct their   angular momentum explicitly.  
The angular condition, necessary to construct the  angular  momentum operator,
is here of particular relevance. We pay also  particular  attention to the
normalization of the wave function. The comparison  with the BS  wave function
is made, and a simple example (Wick-Cutkosky model) is  developed.


\section{General properties of the wave function}\label{wfp}
The wave functions we consider are the Fock components of the state vector
defined on the light-front plane $\omega\cd x=\sigma$.  This means that
they are coefficients in an expansion of the state vector
${\phi}^{J\lambda}(p)$ with respect to the basis of free fields:  
\begin{eqnarray}\label{wfp1} 
|p,\lambda\rangle_{\omega}&\equiv&
{\phi}^{J\lambda}(p) = (2\pi)^{3/2}\int{\mit \Phi}^{J\lambda}_{j_1 \sigma_1 j_2
\sigma_2}(k_1,k_2,p,\omega\tau)
a^\dagger_{\sigma_1}(\vec{k}_1)a^\dagger_{\sigma_2}(\vec{k}_2)|0\rangle 
\nonumber \\
&\times& \delta^{(4)}(k_1+k_2-p-\omega\tau) \exp(i\tau\sigma)2(\omega\cd p)d\tau
\frac{d^3k_1}{(2\pi)^{3/2}\sqrt{2\varepsilon_{k_1}}}
\frac{d^3k_2}{(2\pi)^{3/2}\sqrt{2\varepsilon_{k_2}}} \nonumber \\
&+&(2\pi)^{3/2}\int{\mit \Phi}^{J\lambda}_{j_1 \sigma_1 j_2 \sigma_2 j_3
\sigma_3}(k_1,k_2,k_3,p,\omega\tau)
a^\dagger_{\sigma_1}(\vec{k}_1)a^\dagger_{\sigma_2}(\vec{k}_2)
a^\dagger_{\sigma_3}(\vec{k}_3)|0\rangle \nonumber \\ 
&\times&
\delta^{(4)}(k_1+k_2+k_2-p-\omega\tau) \exp(i\tau\sigma)2(\omega\cd p)d\tau
\nonumber \\ 
&\times& \frac{d^3k_1}{(2\pi)^{3/2}\sqrt{2\varepsilon_{k_1}}}
\frac{d^3k_2}{(2\pi)^{3/2}\sqrt{2\varepsilon_{k_2}}}
\frac{d^3k_3}{(2\pi)^{3/2}\sqrt{2\varepsilon_{k_3}}} + \cdots\ .  
\end{eqnarray}

Here $\lambda$ in ${\phi}^{J\lambda}(p)$ is the projection of the total angular
momentum of the system on the $z$-axis in the rest frame, where $\vec{p}=0$ and
$\sigma_1,\sigma_2,\sigma_3$ are the spin projections of the particles 1, 2 and
3 in the corresponding rest systems.

We emphasize in (\ref{wfp1}) the presence of the delta-function 
$\delta^{(4)}(k_1+ \cdots+k_n -p -\omega\tau)2 (\omega \cd p)d\tau$. Formally it 
can be 
obtained from the 
relation (\ref{kt18a}):                    
\begin{equation}\label{wfp2}
\exp(i\hat{P}^0\cd a)\phi_\omega(\sigma) =\exp(ip\cd a)\phi_\omega(\sigma 
+\omega\cd a)
\end{equation}
with
$$
\hat{P}^0_{\mu} =\int \sum_{\sigma}                                             
a^\dagger_{\sigma}(\vec{k})a_{\sigma}(\vec{k}) k_{\mu}d^3k\ .                         
$$                                                                              
Indeed, after substituting (\ref{wfp1}) in (\ref{wfp2}), the action  of the
operator $\exp(i\hat{P}^0\cd a)$  on the $n$-body  sector gives the factor
$\exp\left[i(k_1+\cdots +k_n)\cd a\right]$ in the  integrand of the l.h.s. of
(\ref{wfp2}), whereas the factor   $\exp\left[i(p +\omega\tau)\cd a \right]$
appears in the r.h.s. The  delta function in (\ref{wfp1}) ensures the equality
of these factors  and, hence, the relation (\ref{wfp2}).  

In the particular case where $\omega =(1,0,0,-1)$, the delta-function 
$\delta^{(4)}(k_1+k_2 -p -\omega\tau)$ gives, after integration over $\tau$,  
the standard  conservation  laws
for the $(\perp,+)$-components of the momenta, but does not  constrain the
minus-components.  

From (\ref{wfp1}) one can see that the wave function depends on the 
orientation of the light front, from its argument $\omega\tau$.  This 
important property of any Fock component is very natural. As explained in the
previous chapter, any off-energy shell amplitude  is related to the $S$-matrix
defined on the finite light-front plane in  the interaction region and
therefore depends on its orientation (see eq.(\ref{rul6a})). The bound state
wave function is always an off-shell object ($\tau\neq 0$ due to binding
energy). Therefore it also depends on the orientation of the light-front plane.
This property is not a peculiarity of the covariant approach. It allows
however to parametrize this dependence explicitly. We will investigate below
this dependence for a few systems.

\subsection{Transformation properties}                                          
We derive in this section the transformation properties of the 
relativistic wave functions under four dimensional rotations. To do 
this, we 
write                               
eq.(\ref{kt9}) in a more explicit form: 
\begin{equation}\label{wfp3}          
\phi^{J\lambda}_\omega(p)\rightarrow \phi'^{J\lambda}_{g\omega}(gp)=              
U_{J^0}(g) \phi^{J\lambda}_\omega(p)\ ,                                         
\end{equation}                                                                  
where $U_{J^0}(g)$ is given for infinitesimal transformations by
eq.(\ref{kt9a}) and $g$ is a Lorentz  transformation and/or rotation.  Here
$\lambda$ is the projection of the angular momentum operator on the $z$-axis in 
the
system at rest, $\vec{p}=0$. After transformation, the new state $\phi'$ does
not correspond to a definite  projection of $J$.  We then expand $\phi'$ with
respect  to the states
$\phi$:                                                               
\begin{equation}\label{wfp4}                                                    
\phi'^{J\lambda}_{g\omega}(gp)                                                    
=\sum_{\lambda'}D^{(J)}_{\lambda'\lambda}\{R(g,p)\} 
\phi^{J\lambda'}_{g\omega}(gp)\ .                                                 
\end{equation}                                                                  
Here $D^{(J)}_{\lambda'\lambda}\{R(g,p)\}$ is the matrix of the 
rotational group and $R(g,p)$ is the rotation operator:  
\begin{equation}\label{rot}                                                     
R(g,p)=L^{-1}(gp)gL(p)\ ,                                                       
\end{equation}                                                                  
where $L(p)$ is the Lorentz transformation, corresponding to the  velocity
$\vec{v}=\vec{p}/p_0$, i.e., for example, $$L(K)(m,0,0,0)  =(mK_0/\sqrt{K^2},
m\vec{K}/\sqrt{K^2}).$$ The Euler angles that  determine the rotation $R(g,p)$
can be expressed in terms of the momentum $p$  and the parameters of the
transformation $g$.    The explicit expression of the Euler angles in terms of
$g$ and $p$ will  not be needed.  

To obtain the transformation properties of the state vector, we first 
substitute $\phi'$ from (\ref{wfp4}) in (\ref{wfp3}), and represent 
$\phi$ in the form of the expansion (\ref{wfp1}). Since 
$a^\dagger_{\sigma}|0\rangle$ in (\ref{wfp1}) is the state vector of a free 
particle with spin $j$, the operator $a^\dagger_{\sigma}$ transforms in 
accordance with the law                                           
\begin{equation}\label{wfp5}                                                    
U_{J^0}(g)a^\dagger_{\sigma}(\vec{k}) U^{-1}_{J^0}(g)                                 
=\sum_{\sigma'}D^{(j)}_{\sigma'\sigma}\{R(g,k)\}a^\dagger_{\sigma'}(g\vec{k
})
\ .  
\end{equation} 
Comparing the left- and right-hand sides of the resulting equation, 
we 
thus obtain for the two-body wave function:  
\begin{eqnarray}\label{wfp6} 
{\mit\Phi}^{J \lambda}_{j_1\sigma_1 j_2\sigma_2} 
(gk_1,gk_2,gp,g\omega\tau)                                                      
=\sum_{\lambda'\sigma'_1\sigma'_2} 
&&D^{(J)*}_{\lambda\lambda'}\{R(g,p)\} 
D^{(j_1)}_{\sigma_1\sigma'_1}\{R(g,k_1)\} 
D^{(j_2)}_{\sigma_2\sigma'_2}\{R(g,k_2)\} 
\nonumber \\ 
&&\times{\mit\Phi}^{J\lambda'}_{j_1\sigma'_1 
j_2\sigma'_2}(k_1,k_2,p,\omega\tau)  
\end{eqnarray} 
and similarly for any $n$-body Fock component. Here 
$\lambda,\sigma_1,\sigma_2$ are the projections of the spins on the  $z$  axis
in the rest frame of each of the particles. We emphasize that  the 
transformation (\ref{wfp6}) is a purely kinematical one. As mentioned  above,
in the usual formulation of  LFD, the  transformation of the system of
reference changes the position of the  light-front plane.  Indeed, under the
transformation $x\rightarrow  x'=gx$, the state vector is transformed from the
plane $t+z=0$ in a  system A to the plane $t'+z'=0$ in a system A$'$.
Introducing  $\omega^{(0)}=(1,0,0,-1)$ we indeed obtain different planes 
$\omega^{(0)}\cd x=0$ and $\omega^{(0)}\cd x' =(g^{-1}\omega^{(0)})\cd  x=0$.  
In the  covariant formulation of LFD, the state vector  remains to be defined on 
one
and the same plane $\omega \cd  x=\omega'\cd x'=0$ in both systems  A and
A$'$.  This property ensures the  kinematical transformation law (\ref{wfp6}).
The dependence on the  surface is then given by the dynamical dependence of the
wave  function  on $\omega$.        

\subsection{Parametrization in the spinless case}\label{sc}                     
We will mainly concentrate on the two-body wave function. 
Generalization to the $n$-body case is straightforward.

\begin{figure}[htbp]
\centerline{\epsfbox{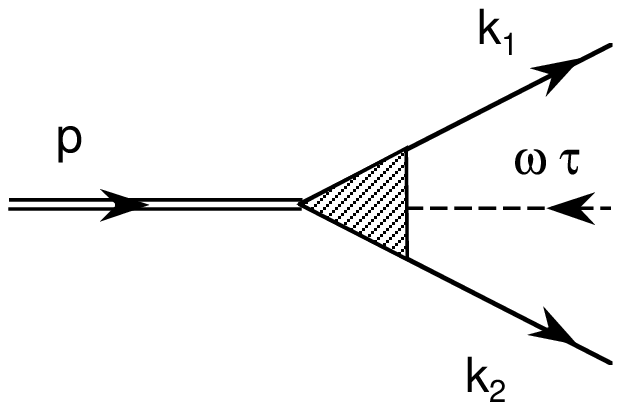}}
\figcap{Graphical representation of the two-body wave function on the
light front. The broken line corresponds to the spurion (see text).}
\label{fwf}
\end{figure}

Due to the conservation law                                                     
\begin{equation}\label{sc1}                                                     
k_1+k_2=p+\omega\tau\ ,                                                         
\end{equation}                                                                  
the light-front wave function can be shown graphically like a  two-body 
scattering amplitude as indicated in fig.~\ref{fwf}. The broken line 
corresponds  to  the fictitious spurion. We emphasize again that although we
assign a  momentum  $\omega\tau$ to the spurion, there is no any  fictitious
particle in the physical state vector. The basis in  eq.(\ref{wfp1}) contains
the particle states only. Due to this  analogy,  the decomposition of the wave
function in independent spin structures  and their parametrization is analogous
to the expansion of a two-body  amplitude in terms of invariant amplitudes. We
will use this  analogy below.  

Let us consider a system of two spinless particles in a state with              
zero angular momentum. According to (\ref{wfp6}), its wave function 
$\psi={\mit \Phi}_{j_1=j_2=J=0}$ is scalar and, hence, depends on 
invariant variables. We can first use the Mandelstam variables:                                                           
\begin{equation}\label{sc2}                                                     
s=(k_1+k_2)^2=(p+\omega\tau)^2,\quad t=(p-k_1)^2,\quad                          
u=(p-k_2)^2\ ,                                                                  
\end{equation}                                                                  
with                                                                            
\begin{equation}                                                                
s+t+u =M^2+2m^2\ .                                                              
\end{equation}                                                                  
The wave function depends therefore on two independent scalar 
variables:        
$\psi=\psi(s,t)$.                                                               
It will be  convenient in the following to introduce another pair of 
variables,  the  relativistic momentum $\vec{k}$ which corresponds, in the
c.m.-system  where   $\vec{k}_1+ \vec{k}_2=0$, to the usual relative momentum
between the  two  particles. Note that this choice of variable does not assume, 
however,
that we  restrict  ourselves to this particular reference frame. We denote by  
$\vec{n}$  the unit  vector in the  direction of $\vec{\omega}$ in this system.
Note that {\em due to the  conservation law (\ref{sc1}), the total momentum
$\vec{p} \neq 0$  of  the system  in this reference frame is not zero}. In
terms of these variables,  the wave  function  takes a form close to the
non-relativistic case. Making  the  appropriate Lorentz transformation, we
get:                                     
\begin{equation}\label{sc4}                                                     
\vec{k}= L^{-1}({\cal P})\vec{k}_1 = \vec{k}_1 -                                
\frac{\vec{\cal P}}{\sqrt{{\cal P}^2}}[k_{10}                                   
- \frac{\vec{k}_1\cd\vec{{\cal P}}}{\sqrt{{\cal P}^2}+{\cal P}_0}]\ ,              
\end{equation}                                                                  
\begin{equation}\label{sc5}                                                     
\vec{n} = L^{-1}({\cal P})\vec{\omega}/|L^{-1}({\cal P})                        
\vec{\omega}| = \sqrt{{\cal P}^2} L^{-1}({\cal P})                              
\vec{\omega}/\omega\cd p\ ,                                                     
\end{equation}                                                                  
where                                                                           
\begin{equation}\label{calp}                                                    
{\cal P} = p + \omega\tau\ ,                                         
\end{equation}                                                                  
and $L^{-1}({\cal P})$ is the Lorentz boost. The unit vector $\vec n$  is 
reminiscent of the unit vector ${\vec p}/\vert {\vec p} \vert$ which  appears
in  the infinite momentum frame.

From these definitions, it follows that under a rotation and a  Lorentz 
transformation $g$ of the four-vectors from which $\vec{k}$ and $\vec{n}$ are
formed, the vectors $\vec{k}$ and $\vec{n}$ undergo  only          
rotations:                                                                      
\begin{equation}\label{sc6}                                                     
\vec{k}\,'=R(g,{\cal P})\ \vec{k}\ ,\quad                                       
\vec{n}'=R(g,{\cal P})\ \vec{n}\ ,                                              
\end{equation}                                                                  
where $R$ is the rotation operator (\ref{rot}). Therefore  $\vec{k}\,^2$ and
$\vec{n}\cd\vec{k}$ are invariants and can be expressed in terms of $s$ and
$t$. For the wave function with zero angular momentum we thus
obtain~\cite{karm76}:                                                                         
\begin{equation}\label{sc7}                                                     
\psi=\psi(\vec{k},\vec{n}) \equiv 
\psi(\vec{k}\,^2,\vec{n}\cd\vec{k})\ .           
\end{equation}                                                                  
It is seen from (\ref{sc7}) that the relativistic light-front wave              
function depends not only on the relative momentum $\vec{k}$ but on             
another variable -- the unit vector $\vec{n}$.                                  
                                                                                
Finally, we intro\-duce the third set of vari\-ables in which the  wave 
func\-tion can be pa\-ra\-met\-rized, in analogy to the equal-time wave
func\-tion in the in\-fi\-ni\-te mo\-mentum frame. We define the           
vari\-ables:                                                                    
\begin{equation}\label{sc8}
 x=\omega\cd k_1/\omega\cd p\ , \quad R_1=k_1-xp\ ,                                                                     
\end{equation}                                                                  
and represent the spatial part of $R$ as $\vec{R}=\vec{R}_{\|} 
+\vec{R_{\perp}}$, where $\vec{R}_{\|}$ is parallel to $\vec{\omega}$  and 
$\vec{R}_{\perp}$ is orthogonal to $\vec{\omega}$.  Since 
$R\cd\omega=R_0\omega_0-\vec{R}_{\|}\cd\vec{\omega}=0$ by definition  of  $R$,
it follows that $R_0=|\vec{R}_{\|}|$, and, hence,  $\vec{R}^2_{\perp} =-R^2$ is
invariant. Therefore, the two scalars on  which the wave function should depend
on can be $\vec{R}^2_{\perp}$  and $x$:                                                    
\begin{equation}\label{sc9}                                                     
\psi=\psi(\vec{R}^2_{\perp},x) \ .                                         
\end{equation}                                                                  
Using the definitions of the variables $\vec{R}^2_{\perp}$ and $x$, we          
can readily relate them to $\vec{k}^2$ and $\vec{n}\cd\vec{k}$:                    
\begin{equation}\label{sc9a}                                                    
\vec{R}^2_{\perp}=\vec{k}\,^2-(\vec{n}\cd\vec{k})^2,\quad                          
x=\frac{1}{2}\left(1-\frac{\vec{n}\cd\vec{k}}{\varepsilon_k}\right).               
\end{equation}                                                                  
The inverse relations are                                                       
\begin{equation}\label{sc10}                                                    
\vec{k}\,^2=                                                                    
\frac{\vec{R}^2_{\perp}+m^2}{4x(1-x)}-m^2,\quad                                 
\vec{n}\cd\vec{k}=                                                                 
\left[\frac{\vec{R}^2_{\perp}+m^2}{x(1-x)}\right]^{1/2}                         
\left( \frac{1}{2}-x\right).                                                    
\end{equation}                                                                  
The variables introduced above can  be easily generalized to the case 
of different masses and an arbitrary number of 
particles~\cite{karm88}. The corresponding variables 
$\vec{q}_i,\vec{n}$ are still constructed according to 
eqs.(\ref{sc4}), 
(\ref{sc5}) and the variables $\vec{R}_{i\perp},x_i$ according to 
(\ref{sc8}) i.e., $x_i=\omega\cd k_i/\omega\cd p$ and $R_i=k_i-x_i 
p$.

\subsection{Normalization}\label{norms}
The state vector is normalized by eq.(\ref{orth1}):
\begin{equation}\label{nor1}
_{\omega}\langle p',\lambda' \vert p,\lambda\rangle_{\omega}                         
=2p_0\ \delta^{(3)}(\vec{p}- \vec{p}\,')\ \delta^{\lambda'\lambda}\ .  
\end{equation}
The Fock components are normalized so as to provide the condition 
(\ref{nor1}). Substituting the state vector (\ref{wfp1}) in the 
left-hand side of eq.(\ref{nor1}), we get:
\begin{equation}\label{nor2}
_{\omega}\langle p', \lambda' \vert p,\lambda\rangle_{\omega}                         
=2p_0\ \delta^{(3)}(\vec{p}- \vec{p}\,')\ N^{\lambda'\lambda}\ ,
\end{equation}
with
\begin{equation}\label{nor2a}
N^{\lambda'\lambda}=\sum_n N_n^{\lambda'\lambda} \equiv \delta^{\lambda'\lambda} 
\ , 
\end{equation}
where $N_n^{\lambda'\lambda}$ is the contribution to the normalization 
integral from the $n$-body Fock component. We represent it in terms 
of 
all three sets of variables introduced above:
\begin{eqnarray}\label{nor3}
N_n^{\lambda'\lambda}&=&(2\pi)^3\int\sum_{\sigma_1\cdots\sigma_n}
{\mit \Phi}^{J\lambda'*}_{j_1 \sigma_1\cdots j_n          
\sigma_n} {\mit \Phi}^{J\lambda}_{j_1 \sigma_1\cdots j_n \sigma_n}
\delta^{(4)}(k_1+\cdots+k_n-p-\omega\tau) 
\nonumber\\                                                                     
&&\hfill \times\prod_{i=1}^n{d^3k_i\over 
(2\pi)^3 2\varepsilon_{k_i}}2(\omega\cd p) d\tau                   
\nonumber\\                                                                     
&=&(2\pi)^3\int \sum_{\sigma_1\cdots\sigma_n}
{\mit \Phi}^{J\lambda'*}_{j_1 \sigma_1\cdots j_n          
\sigma_n} {\mit \Phi}^{J\lambda}_{j_1 \sigma_1\cdots j_n \sigma_n}
\delta^{(3)}(\sum_{i=1}^n\vec{q}_i)2                             
(\sum_{i=1}^n\varepsilon_{q_i}) \prod_{i=1}^n{d^3q_i\over                       
(2\pi)^3 2\varepsilon_{q_i}} 
\nonumber\\                                                 
&=&(2\pi)^3\int\sum_{\sigma_1\cdots\sigma_n}
{\mit \Phi}^{J\lambda'*}_{j_1 \sigma_1\cdots j_n          
\sigma_n} {\mit \Phi}^{J\lambda}_{j_1 \sigma_1\cdots j_n \sigma_n}
\delta^{(2)}(\sum_{i=1}^n\vec{R}_{\perp i})                      
\delta(\sum_{i=1}^n x_i-1)2\prod_{i=1}^n{d^2R_{\perp 
i}dx_i\over(2\pi)^3 2x_i}. 
\nonumber\\
& &
\end{eqnarray}
For the state with zero total angular momentum the normalization 
condition has the form:
\begin{equation}\label{nor4} 
\sum_n N_n=1.  
\end{equation}
In this case, the two-body contribution to the normalization integral 
reads, for constituents of equal masses:  
\begin{eqnarray}\label{nor5} 
N_2&=&{1\over (2\pi)^3} 
\int\psi^2(k_1,k_2,p,\omega\tau)\delta^{(4)}(k_1+k_2-p-\omega\tau)              
{d^3k_1\over 2\varepsilon_{k_1}}{d^3k_2\over 2\varepsilon_{k_2}}                
2(\omega\cd p)d\tau 
\nonumber\\ 
&=& {1\over (2\pi)^3}\int \psi^2(\vec{k},\vec{n}) 
{d^3k\over\varepsilon_k} ={1\over (2\pi)^3}\int 
\psi^2(\vec{R}_{\perp},x) \frac{d^2R_{\perp}dx}{2x(1-x)},
\end{eqnarray}                                                                  
where $\psi={\mit \Phi}^{J\lambda}_{j_1 \sigma_1 j_2 
\sigma_2}(k_1,k_2,p,\omega\tau)\vert_{J,\lambda=0}$ and 
we imply in 
(\ref{nor5}) summation over the spin indices of the constituents. To obtain 
eqs.(\ref{nor5}) we used the fact that the first integral is the 
two-body phase volume:  
$$\int(\cdots)\delta^{(4)}(k_1+k_2-p-\omega\tau) {d^3k_1\over 
2\varepsilon_{k_1}} {d^3k_2\over 2\varepsilon_{k_2}}2(\omega \cd p) 
d\tau= \int(\cdots)\frac{kd\Omega_k}{8\varepsilon_k}2(\omega\cd 
p)d\tau\ ,$$ and then use the equality $M^2+2(\omega\cd 
p)\tau=(k_1+k_2)^2=4\varepsilon_k^2$, which gives:  $2(\omega\cd p) 
d\tau=8kdk$. The change of variables in (\ref{nor3}) is made 
similarly. For the state with 
$J=0$ the integral (\ref{nor5}) and any $N_n$ do not depend on 
$\omega$. 

For the wave function of a system with total angular momentum $J=  1/2$,  the
integral (\ref{nor3}) does not depend on $\omega$ as well, since  it  is
impossible to construct any $\omega$-dependent terms.  At first  glance, one
could construct the following $\omega$ dependent  structure:  
$\bar{u}'(p)\hat{\omega}u(p)/\omega\cd p$ ($\bar{u}'$ and $u$ may  correspond
to different spin projections).  However, since                      
\begin{equation}\label{nor6}
\bar{u}'(p)\gamma_{\mu}u(p)=\frac{p_{\mu}}{m}\bar{u}'(p)u(p)\ ,
\end{equation}
we get:  $\bar{u}^{\lambda'}\hat{\omega}u^{\lambda}/\omega\cd 
p=\bar{u}^{\lambda'}u^{\lambda}/m=2\delta^{\lambda\lambda'}$, i.e., 
this structure does not depend on $\omega$.  Hence, any 
$N_n^{\lambda'\lambda}$ is proportional to 
$\delta^{\lambda'\lambda}$.  
For the sum of them we get $N^{\lambda'\lambda}= 
A\delta^{\lambda'\lambda}$, and the wave function for a system with 
$J=1/2$ is normalized by the condition $A=1$.  As an example, we give 
for $J=1/2$ the normalization integral for a $n$-body system :  
\begin{equation}\label{nor7}  
N_n= (2\pi)^3\int\frac{1}{2}\sum_{\sigma,\sigma_1\cdots \sigma_n} 
\vert 
{\mit\Phi}^{\frac{1}{2}\sigma}_{\sigma_n\cdots \sigma_1}
\vert^2                             
\delta^{(4)}(k_1+\cdots+k_n-p-\omega\tau) \prod_{i=1}^n{d^3k_i\over             
(2\pi)^3 2\varepsilon_{k_i}}2(\omega\cd p) d\tau\ ,                                 
\end{equation}                                                                  
which equals to 1 in the case where the system consists of 
$n$ 
particles only.  

We emphasize that the condition (\ref{nor1}), independent of  the light-front
plane, is a dynamical property of the state vector  provided by the 
Poincar\'e group. When $J$ is different from $0,1/2$, the  integral 
$N_n^{\lambda'\lambda}$ depends in general on $\omega$, though this  dependence
disappears in the sum (\ref{nor2a}) calculated with all  Fock  components. For
an exact state vector one should still write:  
$N^{\lambda'\lambda}=A\delta^{\lambda'\lambda}$, and the  normalization 
condition is reduced to $A=1$, like in the spin 1/2 case. However,  for spins
higher than 1/2 the $\omega$-dependence of  $N^{\lambda'\lambda}$ does not
disappear automatically if the state vector is approximated by a limited
number of Fock components i.e. if a finite sum is retained in  eq.(\ref{nor2a}). 

Consider, for example, a system with $J=1$ (the deuteron for 
instance).  
Extracting from the Fock components the polarization vector 
$e^{(\lambda)}_{\mu}(p)$ of spin 1 system, we represent 
$N_n^{\lambda'\lambda}$ in the form:  
\begin{equation}\label{nor8a}
N_n^{\lambda'\lambda}=e^{*(\lambda')}_{\mu}(p)I_n^{\mu\nu} 
e^{(\lambda)}_{\nu}(p). 
\end{equation}
The general structure of the tensor 
$I_n^{\mu\nu}$ is the following:  
\begin{equation}\label{nor8} 
I_n^{\mu\nu}=-A_ng^{\mu\nu}+B_np^{\mu}p^{\nu}+C_n(p^{\mu}\omega^{\nu}+                 
p^{\nu}\omega^{\mu})+ D_n(p^{\mu}\omega^{\nu}-p^{\nu}\omega^{\mu})               
+E_n\left(\frac{\omega^{\mu}\omega^{\nu}}{(\omega\cd p)^2}+
\frac{g_{\mu\nu}}{3M^2}\right)\ , 
\end{equation}                                   
where $A_n,B_n,C_n,D_n,E_n$ are constants, $M$ is the mass of the composite 
system. 
For example, eq.(\ref{nor8a}) has the form in the rest system:
\begin{equation}\label{nor8b} 
N_n^{\lambda'\lambda}=A_n\delta^{\lambda'\lambda}+\frac{E_n}{M^2}
\left(n^{\lambda'}n^{\lambda}-\frac{1}{3}\delta^{\lambda'\lambda}\right),
\end{equation}
where $n^{\lambda}$ is a unit vector in the direction of 
$\vec{\omega}$. 
After integration of eq.(\ref{nor8b}) over $\vec{n}$, 
according to (\ref{orth2}),
 the irreducible 
structure $(n^{\lambda'}n^{\lambda}-\delta^{\lambda'\lambda}/3)$ 
gives 
zero, and we get the relation $A_n=N_n$. Using eq.(\ref{nor4}), we get $\sum 
A_n=\sum N_n =1$. As it was mentioned 
above
in sect. \ref{ac}, the same condition  has to be valid without 
integration 
over the directions of $\omega$. This means that the exact Poincar\'e 
covariant 
state vector should provide zero values for the sum of the constants
$C_n,D_n,E_n$ ($\sum C_n=\sum D_n=\sum E_n=0$), although each of them, or a 
partial sum of them may not be. 
While it is not required to do so, the $\omega$-dependent structure in 
eq.(\ref{nor8b}) disappears after averaging over $\vec{n}$ (or $\vec{\omega}$).

The general normalization 
condition $A=1$ thus writes, with (\ref{nor8}):  
\begin{equation}\label{nor9}                          
A=\frac{1}{3}\delta_{\lambda'\lambda} N^{\lambda'\lambda}=
-I^{\mu\nu}\frac{1}{3}\left(g_{\mu\nu}-
\frac{p_{\mu}p_{\nu}}{M^2}\right) =1,                
\end{equation} 
where
\begin{equation}
I^{\mu \nu}=\sum_n I_n^{\mu \nu}\ .
\end{equation}
We emphasize that the normalization (\ref{nor9}), according to  (\ref{nor2a})
and (\ref{nor3}), contains the sum over the $n$-body Fock components for all 
$n$. It cannot be generally fulfilled for a two-body component, if the
contribution of other components is not negligible.  The explicit expression
for $A_2$ in terms of the deuteron  wave function is found in
sect.\ref{norm}.

\subsection{New representation}\label{sp} 
One can see from (\ref{wfp6}) that the relativistic wave function, in  contrast
to the non-relativistic one, is transformed in each index  by  different
rotation matrices. It is therefore convenient to use a  representation  in
which the wave function is transformed in each index by one and  the  same
rotation operator $R(g,{\cal P})$, rotating, according to  (\ref{sc6}), the
variables $\vec{k}$ and $\vec{n}$. We define the  wave  function in this new
representation as follows:  
\begin{eqnarray}\label{sp1} 
{\mit\Psi}^{J\lambda}_{j_1\sigma_1 j_2\sigma_2}(k_1,k_2,p,\omega\tau) 
& \equiv & 
\sum_{\lambda',\sigma_1',\sigma_2'} 
D^{(J)*}_{\lambda\lambda'}\{R(L^{-1}({\cal P}), p)\} 
D^{(j_1)}_{\sigma_1\sigma_1'}\{R(L^{-1}({\cal P}),k_1)\} \nonumber \\               
& \times &                                                                      
D^{(j_2)}_{\sigma_2\sigma_2'}\{R(L^{-1}({\cal P}),k_2)\}                        
{\mit\Phi}^{J\lambda'}_{j_1\sigma_1' 
j_2\sigma_2'}(k_1,k_2,p,\omega\tau)\ , 
\end{eqnarray} 
where, e.g.,  $R(L^{-1}({\cal P}),p)$ is given by                               
(\ref{rot}) with $g=L^{-1}({\cal P})$.                                          
                                                                                
Under the transformation $g$, the operator $R\left(L^{-1}({\cal 
P}),k_1         
\right)$ is factorized as follows~\cite{karm79}:                    
\begin{equation}\label{sp2}                                                     
R\left(L^{-1}(g{\cal P}),gk_1\right) =R(g,{\cal P})R(L^{-1}({\cal          
P}),k_1) R^{-1}(g,k_1)\ .                                                   
\end{equation}                                                                  
Using also the property $D\{R_1R_2\}=D\{R_1\}D\{R_2\}$, we find that            
the wave function (\ref{sp1}) is indeed transformed like                        
eq.(\ref{wfp6}), in which the arguments of all the $D$-functions                
are replaced by:                                                                
\begin{equation}                                                                
R(g,{\cal P})=L^{-1}(g{\cal P})gL({\cal P})\ .                                  
\label{sp3}                                                                     
\end{equation}                                                                  
We thus have:                                                                   
\begin{eqnarray}\label{sp4}                                                     
{\mit\Psi}^{\lambda}_{\sigma_1\sigma_2}(gk_1,gk_2,gp,g\omega\tau) & = 
& 
 \sum_{\lambda',\sigma'_1\sigma'_2} 
 D^{(J)*}_{\lambda\lambda'}\{R(g,{\cal P})\}                                    
D^{(j_1)}_{\sigma_1\sigma'_1}\{R(g,{\cal P})\}                                  
\nonumber \\ &                                                                  
\times & D^{(j_2)}_{\sigma_2\sigma'_2}\{R(g,{\cal P})\}                         
{\mit\Psi}^{\lambda'}_{\sigma'_1\sigma'_2}(k_1,k_2,p,\omega\tau)\ .  
\end{eqnarray}                                                                  
The equation (\ref{sp4}) together with (\ref{sc6}) shows that in this  new
representation and in the variables $\vec{k},\vec{n}$ the relativistic wave
function transforms exactly as  a non-relativistic wave function under a
rotation $R$. This strongly  simplifies the spin structure of the relativistic
wave function,  making  it as close as possible  to the non-relativistic one.
The only  difference is the dependence of the wave function on  the extra
variable $\vec{n}$. We shall illustrate this dependence in  the case of the
Wick-Cutkosky model in this chapter, and in the case  of  the deuteron in
chapter \ref{nns}.  

In the particular case of spin 1/2, the matrix
$D^{(\frac{1}{2})}_{\sigma\sigma'}$ coming in the 
transformation (\ref{sp1}), takes the form \cite{shir54}:  
\begin{equation}\label{nz4} 
D^{\frac{1}{2}}\{R(L^{-1}({\cal P}),k_1)\} = \frac{(k_{10}+m)({\cal 
P}_0 + \sqrt{{\cal P}^2}) - \vec{\sigma}\cd\vec{{\cal 
P}}\;\vec{\sigma}\cd\vec{k_1}} {[2(k_{10}+m)({\cal P}_0 + \sqrt{{\cal 
P}^2})(k_{10}{\cal P}_0 - \vec{k_1}\cd\vec{{\cal P}} + m\sqrt{{\cal 
P}^2})]^{1/2}}. 
\end{equation} 
The transformation to the representation (\ref{sp1}) is similar to the
Melosh transformation \cite{melosh}. We will come back to this point in 
sect. \ref{am-ac}.

\section{Equation for the wave function} \label{toto}                                       
                                                                                
\begin{figure}[hbtp]
\centerline{\epsfbox{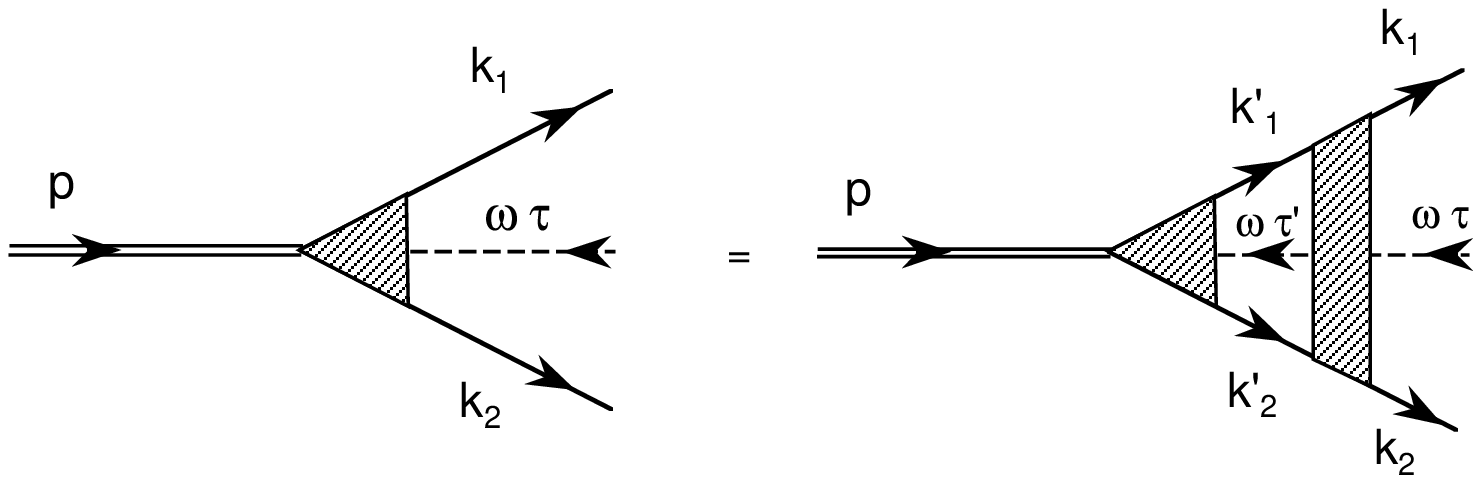}}
\figcap{Equation for the two-body wave function.}
\label{feq}
\end{figure}

The equation for the wave function is obtained 
from the equation for the vertex part shown graphically in 
fig.~\ref{feq}.

It is the analogue, for a bound state, of the 
Lippmann-Schwinger equation for a scattering state. Let us first 
explain its derivation 
for the case of 
spinless particles. In accordance with the rules given in 
sect.~\ref{sl-graph}, we associate with the diagram of fig.~\ref{feq} 
the following analytical expression:                                                
\begin{eqnarray}\label{eqwf1}                                                   
\Gamma(k_1,k_2,p,\omega\tau) =\int \Gamma(k_1',k_2',p,\omega\tau')              
\theta(\omega \cd k_1')\delta(k_1'^2-m^2)\theta(\omega\cd k_2')                      
\delta(k_2'^2-m^2)\nonumber\\                                                   
\times\delta^{(4)}(k_1'+k_2'-p-\omega\tau')d^4k_1' 
{\cal K}(k_1',k_2',\omega\tau';k_1,k_2,\omega\tau)
\frac{d\tau'}{\tau'-i\epsilon}\frac{d^4k_2'}{(2\pi)^3}\ .
\end{eqnarray}                                                                  
The delta-function $\delta^{(4)}(\cdots )d^4k_1'$ is kept for convenience.
The kernel ${\cal K}$ is an irreducible block which is calculated  directly by
the graph technique once the underlying dynamics is  known.   We should then
express the vertex $\Gamma$ through the two-body wave  function.  This can be
done by comparing, for example, two ways of  calculating the amplitude for the
breakup of a bound state by some  perturbation: 1) by means of the graph
technique (the result contains  $\Gamma$); 2) by calculating the matrix element
of the perturbation  operator between the bound state and the free states of
$n$ particles  (the result contains ${\mit \Phi}$).  We thus get:\footnote{The
coefficient  in  the relation (\ref{eqwf2}) differs from previous references
(see  \cite{karm88}) due to the different normalization coefficient in 
(\ref{wfp1}) and the different normalization of the amplitudes given by  the  
rules of the graph technique.} 
\begin{equation}\label{eqwf2} 
{\mit \Phi}(k_1,k_2,p,\omega\tau)= 
\frac{\Gamma(k_1,k_2,p,\omega\tau)}{s-M^2}\ , 
\end{equation} 
where $s=(k_1+k_2)^2=(p+\omega\tau)^2$. The corresponding relation 
for 
the $n$-body case has the same form. In any practical calculation of 
the amplitude, we associate $\Gamma$ with the vertex shown in 
fig.~\ref{fwf} and then express $\Gamma$ in terms of $\phi$ by 
eq.(\ref{eqwf2}). Taking into account that $s-M^2=2(\omega\cd 
p)\tau$, and keeping all the 
spin indices, the equation for the wave function 
${\mit\Phi}^{\lambda}_{\sigma_1\sigma_2}$ has the form~\cite{karm81}:  
\begin{eqnarray}\label{2.16} 
\left[(k_1 + 
k_2)^2-M^2\right]{\mit\Phi}^{\lambda}_{\sigma_1\sigma_2}(k_1,k_2,p, 
\omega\tau) = -\frac{m^2}{2\pi^3} \sum_{\sigma'_1\sigma'_2}\int 
{\mit\Phi}^{\lambda}_{\sigma'_1\sigma'_2}(k'_1,k'_2,p,\omega\tau') 
\nonumber \\ \times V^{\sigma'_1\sigma'_2}_{\sigma_1\sigma_2} 
(k'_1,k'_2,\omega\tau';k_1,k_2,\omega\tau) 
\delta^{(4)}(k'_1+k'_2-p-\omega\tau')\frac{d^3k'_1} 
{2\varepsilon_{k'_1}} \frac{d^3k'_2}{2\varepsilon_{k'_2}}2(\omega\cd 
p)d\tau' \ ,  
\end{eqnarray} 
where we have defined $V=-{\cal K}/4m^2$. This equation can be extended in a 
similar way 
for the many-body component of the state vector.

For many practical applications, it may be usefull to express this equation in 
terms of the variables $\vec{k}$ and
$\vec{n}$, i.e. express the wave function in terms of 
${\mit\Psi}^{\lambda}_{\sigma_1\sigma_2}(k_1,k_2,p,\omega\tau) \equiv 
{\mit\Psi}^{\lambda}_{\sigma_1\sigma_2}(\vec{k},\vec{n})$. With the 
transformation (\ref{sp1}), one thus gets:  
\begin{eqnarray}\label{2.16bb} 
\left[4(\vec{k}\,^2 + m^2)-M^2\right] {\mit\Psi}_{\sigma_1\sigma_2}^\lambda 
(\vec{k},\vec{n}) &=& \\
&-&\frac{m^2}{2\pi^3}\sum_{\sigma'_1\sigma'_2} \int 
{\mit\Psi}_{\sigma'_1\sigma'_2} '^{\lambda}(\vec{k}\,',\vec{n})
V^{\sigma'_1\sigma'_2}_{\sigma_1\sigma_2} 
(\vec{k}\,',\vec{k},\vec{n},M^2) 
\frac{d^3k'}{\varepsilon_{k'}}\ . 
\nonumber 
\end{eqnarray} 
The integration variable $\vec{k}\,'$ in eq.(\ref{2.16bb}) is defined 
analogously to $\vec{k}$ in eq.(\ref{sc4}):  
\begin{equation}\label{4.3a} \vec{k}\,' = L^{-1}({\cal 
P}')\vec{k}\,'_1, \end{equation} 
where ${\cal P}' = k'_1 + k'_2$.  The 
equation (\ref{2.16bb}) can also be obtained from (\ref{2.16}) simply by 
transcribing it in the system of reference where 
$\vec{k}_1+\vec{k}_2=0$.  In this system, 
the rotation operators in (\ref{sp1}) are the unit operators, 
the wave function 
$\mit\Phi^{\lambda}_{\sigma_1\sigma_2}$ turns into 
$\mit\Psi^{\lambda}_{\sigma_1\sigma_2}$.  
However, $\mit\Psi_{\sigma'_1\sigma'_2} '^{\lambda}$ in the integrand 
differs from $\mit\Psi_{\sigma'_1\sigma'_2}^{\lambda}$. 
The wave function $\mit\Psi'(\vec{k}\,',\vec{n})$ obtains the same
form as $\mit\Psi(\vec{k},\vec{n})$ (however, expressed through the variable
$\vec{k}\,'$) in the
system where $\vec{k}'_1 +\vec{k}'_2=0$, but not 
in the system where $\vec{k}_1 
+\vec{k}_2=0$.  This fact is marked by  a ``prime".  The 
function $\mit\Psi_{\sigma'_1\sigma'_2} '^{\lambda}$ is obtained from 
$\mit\Phi^{\lambda}_{\sigma'_1\sigma'_2}(k'_1,k'_2,p,\omega\tau')$ 
by simply expressing the arguments $k'_1,k'_2,p,\omega\tau'$ through 
$\vec{k},\vec{k}',\vec{n}$.  The relation between these variables has 
the form~\cite{ck-deut}:  
\begin{equation}\label{4.10} 
\vec{k}\,'_1 = \vec{k}\,' + \frac{\varepsilon_{k'}^2 - 
\varepsilon_k^2}{2\varepsilon_k}\vec{n} + 
\frac{(\varepsilon_{k'}-\varepsilon_k)^2}{2\varepsilon_k 
\varepsilon_{k'}} (\vec{n}\cd\vec{k}\,')\vec{n},\quad 
k'_{10}=\frac{\varepsilon_{k'}^2+\varepsilon_k^2}{2\varepsilon_k} 
+\frac{\varepsilon_{k'}^2-\varepsilon_k^2}{2\varepsilon_k\varepsilon_{k'}}      
\vec{n}\cd\vec{k}\,',  
\end{equation} 
\begin{equation}\label{4.12}               
\vec{k}_2\,'=-\vec{k}\,' 
+\frac{\varepsilon_{k'}^2-\varepsilon_k^2}{2\varepsilon_k}\vec{n} - 
\frac{(\varepsilon_{k'}-\varepsilon_k)^2}{2\varepsilon_k\varepsilon_{k'}}       
(\vec{n}\cd\vec{k}\,')\vec{n},\quad 
k'_{20}=                                                                       
\frac{\varepsilon_{k'}^2+\varepsilon_k^2}{2\varepsilon_k}                       
-\frac{\varepsilon_{k'}^2 
-\varepsilon_k^2}{2\varepsilon_k\varepsilon_{k'}} \vec{n}\cd\vec{k}\,',  
\end{equation} 
\begin{equation}                           
\vec{p}=-\vec{\omega}\tau = 
-\frac{4\varepsilon_k^2-M^2}{4\varepsilon_k}\vec{n},\quad 
p_0 = \frac{4\varepsilon_k^2+M^2}{4\varepsilon_k}, 
\label{4.7a} 
\end{equation}        
\begin{equation} 
\vec{\omega}\tau =\frac{4\varepsilon_k^2 
-M^2}{4\varepsilon_k}\vec{n},\quad \omega_0\tau= \frac{4\varepsilon_k^2 
-M^2}{4\varepsilon_k},\quad \vec{\omega}\tau' 
=\frac{4\varepsilon_{k'}^2 -M^2}{4\varepsilon_k}\vec{n},\quad 
\omega_0\tau'= \frac{4\varepsilon_{k'}^2 -M^2}{4\varepsilon_k}\ . 
\label{4.7b}                   
\end{equation}                                                                  
Since the formulae (\ref{4.10}-\ref{4.7b}) are the Lorentz 
transformations between the systems with $\vec{k}_1 +\vec{k}_2=0$ and 
$\vec{k}'_1+\vec{k}'_2=0$, the function 
$\mit\Psi_{\sigma'_1\sigma'_2} '^{\lambda}(\vec{k}\,',\vec{n})$ can  be  
also obtained from 
$\mit\Psi_{\sigma'_1\sigma'_2}^{\lambda}(\vec{k}\,',\vec{n})$ by a 
rotation of spins by means of the 
$D$-functions:                                                                  
\begin{equation}\label{3.11aa} 
\mit\Psi'^{\lambda}= D^{(j_2)\dagger}\{R(L^{-1}({\cal 
P}'),k_{2}')\}\mit\Psi^{\lambda} D^{(j_1)}\{R(L^{-1}({\cal 
P}'),k_{1}')\}.                               
\end{equation}                                                                  
For  spin 1/2 particles, the matrix $D^{(\frac{1}{2})}\{R(L^{-1}({\cal P}),k)\}$ 
is given explicitly in eq.(\ref{nz4}).  
                                                   
In the simple case of a scalar particle, the equation for the wave-function in 
terms of the variables $\vec{k}, \vec{n}$ has the following form:  
\begin{equation}\label{eqwf3} 
\left(4(\vec{k}\,^2 + m^2)-M^2\right)\psi(\vec{k},\vec{n}) = 
-\frac{m^2}{2\pi^3} \int \psi(\vec{k}\,',\vec{n}) 
V(\vec{k}\,',\vec{k},\vec{n},M^2) \frac{d^3k'}{\varepsilon_{k'}} \ .  
\end{equation}                                                                  
An equation of such a type was also considered in                                  
refs.~\cite{fudaall,nam78,dn79,fn80,nw80,saw85}.                                             

In the non-relativistic limit,  equation  (\ref{eqwf3}) turns into  the
Schr\"odinger equation in momentum space, the kernel  $V$ being the
non-relativistic potential in momentum space, and the  wave  function no longer
depends on $\vec{n}$.  

We emphasize that the wave function, which is an equal-time wave function on 
the  light front, turns into the ordinary wave function in the 
non-relativistic limit where $c \to \infty$. This reflects the fact  that in
the  non-relativistic limit two simultaneous events in one frame are 
simultaneous in all other frames.  

In the variables $\vec{R}_{\perp}$ and $x$, eq.(\ref{eqwf3})  can be            
rewritten in the form:                                                          
\begin{equation}\label{eq1}                                                     
\left(\frac{\vec{R}^2_{\perp}+m^2}{x(1-x)}-M^2\right)                           
\psi(\vec{R}_{\perp},x)=                                                        
-\frac{m^2}{2\pi^3}\int\psi(\vec{R}'_{\perp},x')                                
V(\vec{R}'_{\perp},x';\vec{R}_{\perp},x,M^2)                                    
\frac{d^3R'_{\perp}dx'}{2x'(1-x')}\ .                                           
\end{equation}                                                                  
In this form, this equation is nothing else than the Weinberg                   
equation~\cite{weinberg}.

The advantages of the equation for the wave function in the form  (\ref{eqwf3})
compared with (\ref{eq1}) are its similarity to the  non-relativistic
Schr\"odinger equation in momentum space, and its  simplicity in the case of
particles with spin. These properties make  eq.(\ref{eqwf3}) very convenient
for practical calculations.

The kernel of eq.(\ref{eqwf3}) depends on the vector variable  $\vec{n}$. We
shall see that this dependence, especially the part which depends on $M^2$,
as  given by eqs.(\ref{4.7a}), (\ref{4.7b}) for instance,  is associated with
the  retardation of the interaction. From this point of view, the  dependence 
of the wave function $\psi(\vec{k},\vec{n})$ on $\vec{n}$ is a  consequence of
retardation. 

In absence of explicit calculations, a very useful information on the global
contribution  of many-body Fock components can be obtained by considering the
dependence of  the interaction on the total mass $M$ of the system. The current
normalization  procedure is then to insert in the norm operator the derivative
of the potential  with respect to $M^2$. In the case of a $J=0$ state made of
two spinless  particles, the normalization is defined by:
\begin{equation} \label{AAA}
\frac{1}{(2\pi)^3}\int\frac{d^3k}{\varepsilon_k}\frac{d^3k'}{\varepsilon_{k'}}
\psi^*(\vec{k}', \vec{n})\left[\varepsilon_k 
\delta(\vec{k}-\vec{k'})-\frac{4m^2}{(2\pi)^3}\frac{\partial 
V(\vec{k}\,',\vec{k},\vec{n},M^2)}{\partial M^2}\right] 
\psi(\vec{k}, \vec{n}) =1\ ,
\end{equation}
where the second term accounts for the many-body contribution to the norm, 
$\sum_{n>2} N_n$. The extension to states with $J\ne 0$ is straightforward. 
There are many justifications for the introduction of the derivative of the 
potential $V$ in eq.(\ref{AAA}). A general derivation can be done similarly to 
the normalization condition of the BS function \cite{nak69}. Apart from the fact 
that it appears in the 
description of a system of two coupled components in which one component
is explicitly retained, it is suggested by the requirement that 
two-state vectors describing two infinitesimally close states should be 
orthogonal. Examination of eq.(\ref{eqwf3}) then shows that such a property is 
fulfilled by introducing in the scalar product of the two state vectors the 
derivative of $V$ with respect to $M^2$ as it is done in (\ref{AAA}). The 
extension to the normalization immediately follows from the fact that the 
orthogonality and the renormalization are defined from the same operator, namely 
the time component of a conserved current. 

The consistency of (\ref{AAA}) with the general normalization condition
(\ref{nor2a}) supposes that the potential $V$ is determined  to all  orders in
the coupling constant entering the interaction. In practice, this is  limited
to the exchange of one or two bosons, which may be already enough for  the
applications considered in the following. Let us just mention here  that the
examination of a few examples (Wick-Cutkosky model, model-perturbative 
calculation) shows that the $\vec{n}$ dependence of the integrand in
(\ref{AAA})  due to the derivative of the potential tends to cancel that of the
first term.  This is in accordance with the general discussion given above in
section \ref{norms}.

When the state is described in relativistic quantum mechanics  with  a fixed
number (two) of particles with $\partial V/\partial M^2=0$, the normalization
for $J=0$ is reduced to  $N_2=1$.

%
\section{Relation with the Bethe-Salpeter function}                       
                                                                                
To find the relation between the light-front wave function and the 
Bethe-Salpeter function we should start from the integral that 
restricts the variation of the arguments of the Bethe-Salpeter 
function 
to the light-front plane:                                                     
\begin{equation} \label{bs1}I=\int                                              
d^4x_1\ d^4x_2\ \delta (\omega \cd x_1)\ \delta (\omega\cd x_2)\                     
\Phi(x_1,x_2,p)\exp (ik_1\cd x_1+ik_2\cd x_2)\ ,                                      
\end{equation} where $k_1,k_2$ are the on-shell momenta:
$k_1^2=k_2^2=m^2$,  and $\Phi (x_1,x_2,p)$ is the                                  
Bethe-Salpeter function \cite{bs},                                              
\begin{equation} \label{bs2}\Phi                                                
(x_1,x_2,p)=\langle 0 \left| T(\varphi (x_1)\varphi (x_2))\right|               
p\rangle\ .                                                                     
\end{equation}                                                                  
Representing the $\delta $       
-functions in (\ref{bs1}) by the integral form
$$\delta(\omega\cd x_1)=
\frac{1}{2\pi}\int \exp(-i\omega\cd x_1\alpha_1)d\alpha_1,\quad
\delta(\omega\cd x_2)=
\frac{1}{2\pi}\int \exp(-i\omega\cd x_2\alpha_2)d\alpha_2\ ,$$
introducing the Fourier       
transform of the Bethe-Salpeter function $\Phi (k,p)$, $$ \Phi                  
(x_1,x_2,p)=(2\pi )^{-3/2}\exp \left[-ip\cd 
(x_1+x_2)/2\right]\tilde{\Phi          
}(x,p)\ ,\quad x=x_1-x_2\ ,$$                                                   
\begin{equation}                                                                
\label{bs3}\Phi (l,p)=\int \tilde \Phi (x,p)\exp (il\cd x)d^4x\ ,                  
\end{equation}                                                                  
where $l=(l_1-l_2)/2$, $p=l_1+l_2$, $l_1$ and $l_2$ are off-mass 
shell 
four-vectors, and making the change of variables 
$\alpha_1+\alpha_2=\tau$, $(\alpha_2-\alpha_1)/2=\beta$ we obtain:    
\begin{equation}                                                                
\label{bs4}                                                                     
I=\sqrt{2\pi}\int_{-\infty }^{+\infty }\delta ^{(4)}(k_1+k_2-p-\omega 
\tau )d\tau \int_{-\infty }^{+\infty }\Phi 
(l_1=k_1-\omega\tau/2+\omega\beta, 
l_2=k_2-\omega\tau/2-\omega\beta)d\beta\ .                             
\end{equation}                                                                  
                                                                                
On the other hand, the integral (\ref{bs1}) can be expressed in terms  of the
two-body light-front wave function. We assume that the  light-front plane is
the limit of a space-like plane, therefore  the  operators $\varphi (x_1)$ and
$\varphi (x_2)$ commute, and, hence,  the  symbol of the $T$ product in
(\ref{bs2}) can be omitted. In the  considered representation, the Heisenberg
operators $\varphi (x)$ in  (\ref{bs2}) are identical on the light front
$\omega\cd x=0$ to the  Schr\"odinger operators (just as in the ordinary
formulation of field  theory the Heisenberg and Schr\"odinger operators are
identical for  $t=0$).  The Schr\"odinger operator $\varphi (x)$ (for the
spinless  case for simplicity), which for $\omega\cd x=0$ is the free field 
operator, is given by (\ref{ft1}):   
\begin{equation} \label{bs5} 
\varphi(x)=\frac 1{(2\pi )^{3/2}}\int \left[a(\vec k)
\exp (-ik\cd  x)+a^{\dagger}(\vec k)\exp (ik\cd  x)\right]
\frac{d^3k}{\sqrt{2\varepsilon_k}}\ .   
\end{equation} 
We represent the state vector $|p\rangle \equiv \phi (p)$ in  (\ref{bs2}) in
the form of the expansion (\ref{wfp1}). Since the vacuum  state on the light
front is always ``bare'', the creation operator,  applied to the vacuum state
$\langle 0|$ gives zero, and in the  operators $\varphi (x)$  the part
containing the annihilation  operators only survives. This cuts out the
two-body Fock component in  the  state vector.  We thus obtain:  
\begin{equation} \label{bs6} 
I=\frac{(2\pi )^{3/2}(\omega\cd p)}{2(\omega\cd k_1)(\omega\cd 
k_2) }\int_{-\infty }^{+\infty }\psi (k_1,k_2,p,\omega \tau )\delta 
^{(4)}(k_1+k_2-p-\omega \tau )\ d\tau\ .  
\end{equation} 
Comparing (\ref{bs4}) and (\ref{bs6}), we find:  
\begin{equation} \label{bs7}
\psi (k_1,k_2,p,\omega \tau )=\frac{(\omega\cd k_1)(\omega\cd 
k_2)}{\pi 
(\omega \cd p)}\int_{-\infty }^{+\infty }\Phi (k+\beta \omega 
,p)d\beta\ , 
\end{equation}                                                                  
where the argument $p$ in(\ref{bs7}) is expressed through the 
on-shell 
momenta $k_1,k_2$ as $p=k_1+k_2-\omega\tau$, in contrast to off-mass 
shell relation $p=l_1+l_2$. For the Bethe-Salpeter function 
$\Phi(l_1,l_2)$ parametrized in terms of the momenta $l_1,l_2$ the 
formula (\ref{bs7}) reads:   
\begin{equation} \label{bs8}                                                     
\psi(k_1,k_2,p,\omega \tau ) =\frac{(\omega\cd k_1 )(\omega\cd k_2 
)}{\pi (\omega\cd  p)}\int_{-\infty }^{+\infty }\Phi 
(l_1=k_1-\omega\tau/2+\omega\beta,l_2 
=k_2-\omega\tau/2-\omega\beta)d\beta       
\end{equation}                                                                  
In ordinary LFD, eqs.(\ref{bs7}) and (\ref{bs8})  correspond to the integration
over $dk_{-}$. This equation makes the  link between the Bethe-Salpeter
function $\Phi$ and the wave function  $\psi$ defined on the light front
specified by $\omega$.  It should be  noticed however that eq.(\ref{bs8}) is
not necessarily an exact solution of  eq.(\ref{eqwf3}), since, as a rule,
different approximations are made for the Bethe-Salpeter kernel and for the
light-front one. In the ladder approximation, for example, the Bethe-Salpeter
amplitude contains the box  diagram, including the time-ordered diagram with
two exchanged particles in the intermediate state, as indicated in fig.
\ref{box4} in appendix \ref{f-lf}. This contribution is absent in the 
light-front ladder kernel.

The quasipotential type equations for the light-front wave function derived by
restricting arguments of the Bethe-Salpeter amplitude to the light-front plane
$z+t=0$ and corresponding electromagnetic form factors  were studied  in refs.
\cite{gkmtf75,gm75}.

\section{Application to the Wick-Cutkosky model}                                
                                                                                
\subsection{Solution in the covariant formulation of LFD}                         
                                                                                
As a simple, but nevertheless instructive, example, we shall derive  in this 
section the light-front  wave function of a system consisting of two scalar
particles with  mass  $m$ interacting through the exchange of a massless scalar
particle   calculated  in the ladder approximation. This is the so-called 
Wick-Cutkosky model. The  diagrams that determine the kernel are shown in
fig.~\ref{fkern}.  The kernel is given by eq.(\ref{k1}) with $\mu=0$.   Going
over from the kernel ${\cal K}$ to the potential $V=-{\cal  K}/(4m^2)$,
introducing the constant $\alpha=g^2/(16\pi m^2)$, and  expressing (\ref{k1})
by means of the relations  (\ref{4.10}-\ref{4.7b})   in terms of
$\vec{k},\vec{k}',\vec{n}$, we obtain~\cite{karm80}:                           
\begin{equation}\label{k3}                                                      
V=-4\pi\alpha/\vec{K}^2,
 \end{equation}
where                                   
\begin{equation}\label{k4}                                                      
\vec{K}\,^2 = (\vec{k}\,' -
\vec{k}\,)^2                                         -
(\vec{n}\cd\vec{k}\,')(\vec{n}\cd\vec{k})                                           
\frac{(\varepsilon_{k'}-\varepsilon_k)^2}{\varepsilon_{k'}\varepsilon
_k}         +(\varepsilon_{k'}^2
+\varepsilon_k^2-\frac{1}{2}M^2)                           
\left|\frac{\vec{n}\cd\vec{k}\,'}{\varepsilon_{k'}}                                
-\frac{\vec{n}\cd\vec{k}}{\varepsilon_k}\right|\ .                                 
\end{equation}                                                                  

For $k,k' \ll m$, eq.(\ref{k3}) turns into the Coulomb potential 
in momentum space                                                    
\begin{equation}\label{k5}                                                      
V(\vec{k}\,',\vec{k})\simeq 
-\frac{4\pi\alpha}{(\vec{k}\,'-\vec{k})^2}\ . 
\end{equation}                                                                
For $\alpha\ll 1$, $|\epsilon_b|= |M-2m|=m\alpha^2/4\ll m$, the wave 
function is concentrated in the non-relativistic region of momenta.  
The non-relativistic wave function of the ground state in the Coulomb 
potential has the form:   
\begin{equation}\label{k6}                                                      
\psi(\vec{k})=\frac{8\sqrt{\pi m}\kappa^{5/2}} 
{(\vec{k}\,^2+\kappa^2)^2}\ , 
\end{equation} 
where $\kappa=\sqrt{m|\epsilon_b|}=m\alpha/2$ (it is normalized,  however,
according to (\ref{nor5}) with $\varepsilon_k\approx m$).   The  integral over
$d^3k'$ in (\ref{eqwf3}) is concentrated in the region  $k'\approx \kappa$. 
Therefore, at $k\gg \kappa$ the momentum  $\vec{k}\,'$ in
$V(\vec{k}\,',\vec{k},\vec{n},M^2)$ can be ignored,  and  from (\ref{eqwf3}) we
find:   
\begin{equation}\label{k7} 
\psi(\vec{k},\vec{n})=-\frac{mV(0,\vec{k},\vec{n},M^2)}                
{(2\pi)^3(\vec{k}\,^2
+\kappa^2)}\int\psi(\vec{k}\,')d^3k'\ .                     
\end{equation}                                                                  
Note, that in (\ref{k7}) 
$\int\psi(\vec{k}\,')d^3k'=(2\pi)^3\psi(r=0)$. The wave function
(\ref{k6}) is normalized to 1.  Substituting in the r.h.s. of eq.(\ref{k7}) the
expressions (\ref{k3},\ref{k4}) for $V$ and (\ref{k6})
for $\psi$, we obtain                
\begin{equation}\label{k8}                                                     
\psi(\vec{k},\vec{n})                                                           
=\frac{8\sqrt{\pi m}\kappa^{5/2}}{(\vec{k}\,^2+\kappa^2)^2 
\left(1+\frac{\displaystyle |\vec{n}\cd\vec{k}|}                                   
{\displaystyle\varepsilon_k}\right)}\ .                                         
\end{equation}                                                                  
This relativistic wave function of the ground state with zero  total angular
momentum is a good approximation of a more exact one in the range  $k>\kappa$.
Corrections of order $\alpha \log(\alpha)$ should be considered in  the range
$k<\kappa$ (see \cite{fft73}).
 Although the kernel (\ref{k3}), (\ref{k4})  contains  the modulus
$|\vec{n}\cd\vec{k}\,'/\varepsilon_{k'}  -\vec{n}\cd\vec{k}/\varepsilon_k|$,
one can show that the solution  has  no ``cusp'' at $\vec{n}\cd\vec{k}=0$. 
This cusp in (\ref{k8})  appears  due to our approximations.  

One can check on this simple example that it is the retardation of  the 
interaction that is the dynamical reason for the dependence of the  wave 
function on the variable $\vec{n}$. The non-relativistic Coulomb  expression
for the kernel (\ref{k5}) does not contain retardation and  does not depend on
$\vec{n}$ while the relativistic kernel (\ref{k3})  contains retardation and
depends on $\vec{n}$. This leads to the  dependence of the wave function on the
argument $\vec{n}$. It may seem  at first sight that the dependence of the wave
function on $\vec{n}$ is not due  to the contribution of the many-body sectors.
Indeed, the coupling of the  two-body sector with the other sectors must
contain a coupling  constant, whereas the parameter that determines the
dependence of the  wave function on $\vec{n}$ is the nucleon mass and does not
contain  the  coupling constant. This argument should be taken with some care
however. It can be seen from  eq.(\ref{eqwf3}) that the operator
$k^2+m^2-M^2/4$ acting on the  wave function is equivalent to the potential
acting on this same wave function.  It is thus proportional to the coupling
constant.  Such a result underlies the  disappearance of the $\vec{n}$
dependence of the integrand of eq.(\ref{AAA})  previously mentioned.

We emphasize that retardation leads to both the $\vec{n}$-dependence  and the
presence  of the carriers of the  interaction in the intermediate state, which
contribute to the many body sectors.  However,  these  two effects, being
important in full measure in a truly relativistic  system, can manifest
themselves in a different way in weakly bound  systems. Neglecting the
many-body sectors does not necessarily entails  to neglect the
$\vec{n}$-dependence of the  wave function at $k \approx m$. The above
mentioned $\alpha \log(\alpha)$  correction to eq.(\ref{k8}) for instance,
which originates from the last  $\vec{n}$-dependent term in (\ref{k4}), is
$\vec{n}$ independent and has no  counterpart from the many-body sector. As we
shall see also in chapter  \ref{nns}, it is  necessary to take into account  
the $\vec{n}$-dependence of the wave function even when one  restricts to the
two-nucleon sector. 

The wave  function of the 2p state can be found analogously. In the 
representation (\ref{sp1}), it has the form \cite{karm80}:                                                                  
\begin{eqnarray}\label{bs12}                                                   
&&\psi ^\lambda (\vec k,\vec n)= \frac{8\pi \kappa 
^{7/2}m^{1/2}}{\sqrt{6}}\frac {1}{\left(\vec k\,^2+\frac{1} 
{\displaystyle 4}\kappa ^2\right)^3\left(1+ \frac{\displaystyle |\vec 
n\cd\vec k|}{\displaystyle\varepsilon_k}\right)^2} 
\\ 
&&\times
\left\{kY_{1\lambda }(\vec k/k) +Y_{1\lambda}(\vec{n}) 
\left[\frac{(2\varepsilon_k-M)^2} {4\varepsilon_k M} 
(\vec n\cd\vec k)-\frac{(\vec k^2+\frac{\displaystyle 1}
{\displaystyle 4}\kappa ^2)} {2m}\left(\theta (-\vec n\cd\vec 
k)-\theta (\vec n\cd\vec 
k)\right)\right]\right\}.
\nonumber
\end{eqnarray} 
The wave function corresponding to the angular  momentum  $l=1$ contains the
spherical function $Y_{1\lambda }(n)$. This is  an illustration of the fact 
that the vector $\vec n$ participates in  the construction of the total angular
momentum on the same ground  as the relative momentum $\vec{k}$.   The
dynamical difference between the  solution with  $\vec{k} \| \vec{n}$ and
$\vec{k} \perp \vec{n}$ is obviously related  to the  property that some of the
components of the angular momentum  $\vec{J}$, before using the angular
condition, depend on the interaction.

\subsection{Solution in the Bethe-Salpeter approach}                            
\label{wcbs}                                                                    
The exact expression for the Bethe-Salpeter function in the Wick-Cutkosky model
is found in the form of the integral  representation \cite{wcm,nak69} and,
for zero angular momentum, reads:                       
\begin{equation}                                                                
\label{bs8p}                                                                    
\Phi (l,p)=-\frac i{\sqrt{4\pi }}\int_{-1}^{+1}\frac{               
g(z,M)dz}{(m^2-M^2/4-l^2-zp\cd l-i\epsilon)^3}\ .                                   
\end{equation}                                                                  
Substituting (\ref{bs8p}) in (\ref{bs7}), and making the change of 
variable 
$\beta \to \beta/\omega \cd p$, we find:                              
\begin{equation}                                                                
\label{bs9}                                                                     
\psi =-\frac{ix(1-x)}{2\pi ^{3/2}}\int_{-\infty }^{+\infty               
}d\beta \int_{-1}^{+1}\frac{g(z,M)dz}{[(\vec k\,^2+\kappa                       
^2)(1-zz_0)+\beta (z_0-z)-i\epsilon]^3}\ ,                                      
\end{equation}                                                                  
where $z_0=1-2x$, $\kappa^2=m^2-M^2/4.$                                         
                                                                                
For fixed $z\neq z_0$, the integrand,  as a function of $\beta $,   has one 
pole of third order, and therefore the integral is zero. For $z=z_0$,  the
integrand does not depend on $\beta $ and the integrand over  $\beta  $
diverges. To find the value of the integrand (\ref{bs9}), we first  calculate
the integral over $dz$ from $z_0-\epsilon $ to  $z_0+\epsilon  $ at $\epsilon
\rightarrow 0$, retaining only the term proportional  to  $\epsilon $, and then
over $d\beta $. As a result, one obtains  \cite{bjs}:         
\begin{equation}\label{bs10}
\psi =\frac{g(1-2x,M)}{2^5\sqrt{\pi }x(1-x) 
(\vec k\,^2+\kappa ^2)^2}\ .                                                                       
\end{equation}                                                                  
The spectral function $g(z,M)$ is determined by a differential                  
equation \cite {wcm,nak69} and has no singularity at $z=0$.  
The exact wave function (\ref{bs10}) is therefore analytic at $x=1/2$.  The 
approximate explicit solution found in \cite{wcm} for $g(x,M)$ has 
the 
form:  
\begin{equation}\label{bs11} 
g(z,M)=2^6\pi \sqrt{m}\kappa ^{5/2}(1-|z|)\ .  
\end{equation} 
Inserting (\ref{bs11}) in (\ref{bs8p}) and integrating over $z$,   one can
recover the approximate solution of the Bethe-Salpeter  equation given by
eq.(\ref{wcm02}). The discontinuity of the spectral function $g(z,M)$ at $z=0$
is again  the result of our approximation, since the solution (\ref{bs11}) 
corresponds to an asymptotically small binding energy. Substituting 
(\ref{bs11}) in (\ref{bs10}), we reproduce the expression (\ref{k8})  for the
wave function. The discontinuity of (\ref {bs11}) at $z=0$  results in the
``cusp'' of (\ref{k8}) at $\vec n \cd \vec k=0$. 

          
\section{Angular momentum and angular condition}\label{am-ac}                   
                                                                                
The wave function (\ref{bs12}) is consistent with the general 
structure of the         
light-front wave function of a system with total angular momentum 
equal         
to 1 (for spinless constituents):                                               
\begin{equation}\label{ac1}                                                     
\psi^{\lambda}(\vec{k},\vec{n})=f_1Y_{1\lambda}(\vec{k}/k)                      
+f_2Y_{1\lambda}(\vec{n})\ ,                                                    
\end{equation}                                                                  
with  $f_{1,2}=f_{1,2}(\vec{k}\,^2,\vec{n}\cd\vec{k})$.                            
The states with higher angular momentum are constructed similarly,              
using Clebsch-Gordan coefficients. In this case, it is clear that the              
angular momentum operator in the representation (\ref{sp1}) has the             
form:                                                                           
\begin{equation}\label{ac2}                                                     
\vec{J} =                                                                       
-i[\vec{k}\times \partial/\partial\vec{k}\,] -i[\vec{n}\times                   
\partial/\partial\vec{n}]\ .                                                    
\end{equation}                                                                  
The same expression for $\vec{J}$ was found in 
refs. \cite{fudaall}. It can 
be shown to be quite general for spinless particles. 
Acting on the wave function (\ref{k8}) for zero angular momentum and 
on 
any function depending on the scalars $\vec{k}\,^2$ and 
$\vec{n}\cd\vec{k}$, the operator (\ref{ac2}) gives zero.  The 
generalization for the case of constituents with non-zero spins is 
evident:                                                                        
\begin{equation}\label{ac3}                                                     
\vec{J} =                                                                       
-i[\vec{k}\times \partial/\partial\vec{k}\,] -i[\vec{n}\times                   
\partial/\partial\vec{n}]+\vec{s}_1+\vec{s}_2\ ,                                
\end{equation}                                                                  
where $\vec{s}_{1,2}$ are the spin operators of the constituents. 

As we explained in sect. \ref{sp}, the relative momentum $\vec{k}$ and the spin
operators $\vec{s}_1,\vec{s}_2$ have identical transformation properties.  In
the standard approach, the angular momentum operator obtains the same form 
after performing the Melosh transformation \cite{melosh,coester92}, once the
$\vec{n}$-dependent term is neglected in (\ref{ac3}).  
In addition, the factor  $\vec{n}\cd\vec{k}$ in
$f_{1,2}(\vec{k}\,^2,\vec{n}\cd\vec{k})$ turns  into $k_z$. At first glance,
the dependence of the wave function on  $k_z$ can be interpreted as a violation
of rotational invariance (see  ref.~\cite{ji}). Indeed, this wave function is
not eigenfunction of  the  operator $\vec{L}\,^2$, where $\vec{L}$ is the usual
angular momentum  operator: 
\begin{equation}\label{ac4}                                           
\vec{L}=-i[\vec{k}\times\partial/\partial\vec{k}]\ .                            
\end{equation}                                                                  
The rotations changing the position of the light-front plane  $t+z=0$ are 
dynamical, the corresponding angular momentum operator contains the 
interaction,  and does not therefore coincide with  eq. (\ref{ac4}). 

In the covariant formulation, the action of the dynamical angular 
momentum operator has to coincide with the action of the operator 
(\ref{ac2}).  This is provided by the so called angular condition 
(\ref{kt12}) derived in section \ref{trpr}.                                                             

As explained in section~\ref{trpr}, the solutions of eq.(\ref{eqwf3})  with
non-zero angular momentum are always degenerate if one does not  apply the
angular condition. This condition eliminates this  unphysical  degeneracy. This
can be seen easily in the example of the wave  function 
(\ref{ac1}).                         

For this aim, it is convenient to represent (\ref{ac1}) in the form:            
\begin{equation}\label{ac5}                                                     
\vec{\psi}(\vec{k},\vec{n}) = f_1\vec{k} + f_2\vec{n}\ .
\end{equation}                                                                  
By analogy with eq.(\ref{kt30}), one can construct the operator:                
\begin{equation}\label{ac6}                                                     
A=(\vec{n}\cd\vec{J})^2.                                                          
\end{equation}                                                                  
Due to eq.(\ref{ac2}), this operator can be represented in the form:  
$A=(\vec{n}\cd\vec{L})^2$.  Any $\vec{n}$-dependence of the kernel 
does not violate conservation of the projection of the operator 
$\vec{L}$ on the direction $\vec{n}$. Hence, the operator $A$ is 
conserved.  As a scalar, it commutes with the angular momentum 
operator 
$\vec{J}$:  $[\vec{J},A] =0$.  Therefore, any solution of 
eq.(\ref{eqwf3}) with definite angular momentum is characterized also 
by the eigenvalues of the operator $A$:                                                              
\begin{equation}\label{ac7}                                                     
A\psi_{\alpha}=\alpha\ \psi_{\alpha}\ .                                        
\end{equation}                                                                  
For $J=1$, there are two eigenstates with $\alpha=0$ and $\alpha=1$ 
respectively:                                     
\begin{equation}\label{ac8}                                                     
\vec{\psi}_0 =                                                                  
\vec{n}(\vec{n}\cd\vec{k})h_0(\vec{k}\,^2,\vec{n}\cd\vec{k})\ ,                       
\end{equation}                                                                  
\begin{equation}\label{ac9}                                                     
\vec{\psi}_1 = \left[\vec{k} -\vec{n}(\vec{n}\cd\vec{k})\right]                    
h_1(\vec{k}\,^2,\vec{n}\cd\vec{k})\ ,                                              
\end{equation}                                                                  
where $h_{0,1}$ are arbitrary scalar functions. In the case where  $h_{0,1}$ do
not depend on $\vec{n}\cd\vec{k}$, the states  $\vec{\psi}_{0,1}$ are
eigenstates of the usual angular momentum operator  squared:
$\vec{L}\,^2\vec{\psi}_{0,1} = 1(1+1)\vec{\psi}_{0,1}$.  Rewritten in terms of
the spherical functions, they still have no  definite projection of the usual
angular momentum on the $z$-axis. Thus,  $\vec{\psi}_0$ is proportional to
$\sum_m a_m Y_{1m}(\vec{k}/k)$  (with  $a_m= Y^*_{1m}(\vec{n})$) and, hence, is
a superposition of states  with different projections $m$. In the
non-relativistic limit $h_0=h_1$, and the degeneracy of the states (\ref{ac8}),
(\ref{ac9}) is nothing but the usual degeneracy relative to projections of the
angular momentum.

In the practical case of the approximate two-body equation  (\ref{eqwf3}), the
states (\ref{ac8}) and (\ref{ac9}) are not  degenerate but correspond to two
different energies.  Both of them  are  nonphysical.  It is difficult to
project explicitly the general  angular  condition (\ref{kt13}) on the two-body
sector. However, the model  solution (\ref{bs12}), found from the known
Bethe-Salpeter function,  does not contain any ambiguity, and hence, satisfies
the angular  condition. It has the form of eq.(\ref{ac5}), i.e., is a
combination  of  two states with $\alpha=0$ and $\alpha=1$ (neglecting at this
level of approximation the energy splitting). This example illustrates the 
result, proved in section \ref{trpr}, that the angular condition allows  to
construct the physical  solution from unphysical degenerate solutions.  Another
example is  given in  \cite{fudaall}. The  angular condition does not
eliminate the $\vec{n}$-dependence of the  wave function.  The angular 
condition in the ladder approximation  was  found in
\cite{karm82}.                                                    

                                                                                
\chapter{The nucleon-nucleon potential}\label{nnp}                              
The understanding of the $NN$ interaction should ultimately rely on  QCD. 
However, the non-perturbative character of this theory in the  low  energy
range makes this goal quite far away. For many purposes, using  effective
degrees of freedom such as mesons in the $t$-channel or  nucleons and baryon
resonances in the $s$-channel in order to describe  this interaction will still
be relevant, provided it is used in a  given  domain of energy and momentum. In
the present context, this approach  is natural  because {\it i)} the meson
exchange picture provides a reasonably good  description of the $NN$
interaction and {\it ii)} it allows one to  consider specific relativistic
corrections. The effects due to  a deeper understanding of hadronic
physics will have to be considered,  involving in the simplest case off-shell
effects, the effective  character of some meson exchanges or of hadronic form
factors.                                          

\section{Mesonic degrees of freedom in nuclei}\label{mdf} 
The relevance of mesonic degrees of freedom in describing the $NN$  interaction
stems from the study of peripheral partial waves phase  shifts at low energy.
These ones are dominated by the exchange of the  pion, due to both its low mass
and the low energy domain. In related  area, dealing for instance with
electromagnetic properties of light  nuclei, the relevance of the pion degree
of freedom relies on meson  exchange current contributions to the deuteron
electrodisintegration  near threshold \cite{hockert}, where it allows to
achieve  agreement  with experiment at momentum transfers around $Q^2 \sim
0.5$  (Gev/c)$^2$  \cite{auf}. As this contribution is not required by a pure 
phenomenological description of the $NN$ interaction, it is a clear  evidence
for  the role of the pion in the description of nuclear  forces. In both cases 
($NN$ interaction and deuteron  electrodisintegration cross-section), the role
of the pion is related  to the peculiar nature of this particle within QCD. The
pion is the  Goldstone boson associated to the spontaneous breaking of chiral 
symmetry which provides its low mass. The $\pi$-exchange is now a common  
component
of all the  models of the $NN$ interaction.  

Beyond single pion exchange, one expects contributions due to  $2,3,\ldots$
pion exchanges. They have shorter and shorter ranges, 1  $fm$ or less, and it
becomes more and more difficult to  identify  their  contribution to the $NN$
interaction as the number of exchanged pions  increases. At this point, models
differ from each other.                                                                        

Some models \cite{wir}, following the philosophy of the Reid Soft Core  (RSC)
potential \cite{reid}, simply parametrize this part in terms of  Yukawa
potentials, but  without precise relation to meson exchanges.  As  a support
for such an approach, one may argue that nucleons have a  radius of the order
of 0.8 $fm$ (as far as this radius  can be identified to the proton charge
radius) and consequently, below a  distance of 1.5 $fm$, nucleons began to
overlap, making the  description  of $NN$ interaction in this range in terms of
exchange of mesons quite  doubtful.  

A second approach relies on the use of dispersion relations \cite{cdr}  to
calculate the irreducible $2\pi$-exchange contribution to the $NN$  scattering 
amplitudes. The  interaction potential is obtained by removing the 
$2\pi$-exchange contribution resulting from the iteration of one 
$\pi$-exchange which is automatically generated by the dynamical  equation used
to calculate the two-body wave function. This allows one  to  extend the
potential to distances as low as 0.8 $fm$. Notice that the  full
$2\pi$-exchange scattering amplitude has covariance properties  that are lost
by the separate $\pi$-exchange and $2\pi$-exchange  contribution to the $NN$
potential. The short range part is largely  phenomenological, with the
consequence that the Lorentz structure of  the $NN$ amplitude may not be
correctly determined, quite similarly to  the previous approach. An important
point to be noticed is that this  approach relies on physical $\pi N$
scattering amplitudes with the  consequence that no form factor is introduced
at the $\pi NN$ ($\pi  N\Delta,$ $\ldots$) vertices. Off-shell effects, which
have a shorter  range, are incorporated in the phenomenological part. On the
other  hand, the $\pi N$ scattering amplitude satisfies properties expected 
from chiral symmetry, although a pseudoscalar $\pi NN$ coupling is  used. The
large contribution from the excitations of $N\bar{N}$ pairs
($Z$-diagrams) to the $\pi N$  scattering  amplitude expected in this case is
cancelled here by a contribution  involving mainly the $\Delta$ resonance
intermediate state (calculated  via dispersion relations). Finally, although it
does not incorporate  explicit $\rho$-exchange contribution, this approach can
account for  it, through the $2\pi$ interaction in a P-state in the
$t$-channel.                                                                    

A third approach is based on a field-theoretical description of the 
meson-nucleon interaction, including vertex form factors 
\cite{bonn,machl,nuj}.  In the simplest approximation, the $NN$  interaction
results from a sum of contributions involving single meson  exchanges: $\pi$
for the long range part, ''$\sigma$" giving  attraction  at intermediate
distances (1 $fm$) and accounting for the exchange of  $2\pi$ in a S-state,
$\rho$ accounting for the exchange of $2\pi$ in a  P-state, and $\omega$ for
the short-range repulsive part. Other well  known mesons such as
$\eta,a_1,\delta$ may also be considered, but are  quantitatively less
important. While some phenomenology is involved in  the vertex form factors or
through adjustments of coupling constants,  this approach has the great
interest to provide a parametrization of  the $NN$ interaction in terms of
explicit exchanges of mesons, which  are expected to be important, if not
dominant, in some channels.  The  fit to $NN$ scattering is quite satisfactory
and there is no real  indication at present that such an approach breaks down,
even at short  distances where it could fail. Because it keeps the full
Lorentz  structure attached to the different exchanged mesons, this approach
is  quite appropriate to the present
purpose.                                                                        

The above approach can be improved by considering two-meson exchanges. This is
especially important for a better account of the ''$\sigma$"  exchange
contribution, whose effective character is well known. As in  the dispersion
relations approach, the iterated one meson (pion in  practice) exchange has to
be removed from the full $2\pi$-exchange  contribution to the $NN$ amplitude. 
The above contribution also  involves $\Delta$ resonances (and other higher
resonances) in  intermediate states in a quite similar way as Van der  Waals
forces in atomic physics are generated. The non-local character attached to 
such contributions is likely to be accounted for only in an  approximate  way
by single meson exchanges (as ``$\sigma$'' and $\rho$
mesons).                     

\section{The non-relativistic NN potential}\label{nrp}                          
The non-relativistic NN potential is to be used with the Schr\"odinger 
equation. In momentum space, which is more appropriate for an 
extension 
to relativistic calculations, it reads:                                            
\begin{equation}\label{nrp1}                                                    
\frac{\vec{k}^2}{m}\psi(\vec{k}) +\int \frac{d^3k'} 
{(2\pi)^3} \langle\vec{k}\vert V\vert \vec{k}'\rangle
\psi(\vec{k}')  = E\psi(\vec{k})\ ,                                                               
\end{equation} 
where $\vec{k}$ represents one half of the relative momentum 
of the two nucleons, $\vec{k}={1\over 2}(\vec{k}_1-\vec{k}_2)$, and 
similarly for $\vec{k}'$.                                                                
In $r$-space, where many models were developed, the corresponding 
equation would be:                                                                       
\begin{equation}\label{nrp2}                                                    
-{\vec{\nabla}\,^2\over m}\psi(\vec{r}) +\int d^3r' 
\langle\vec{r}\vert 
V\vert\vec{r}\,'\rangle \psi(\vec{r}\,')=E\psi(\vec{r})\ .                      
\end{equation}                                                                  
Typically, in a meson exchange theory, $\langle\vec{k}\vert V\vert              
\vec{k}\,'\rangle$ takes the following form in the simplest case:               
\begin{equation}\label{nrp3}                                                    
V(\vec{k}, \vec{k}\,') \equiv \langle\vec{k}\vert V\vert 
\vec{k}\,'\rangle \propto \frac{g^2} {\mu^2          
+(\vec{k}-\vec{k}\,')^2}\ ,                                                     
\end{equation}                                                                  
while in $r$-space, one would get                                               
\begin{equation}\label{nrp4}                                                    
V(\vec{r}, \vec{r}\,') \equiv \langle\vec{r}\vert 
V\vert\vec{r}\,'\rangle \propto                             
g^2\delta^{(3)}(\vec{r}-\vec{r}\,') \frac{\exp(-\mu r)}{4\pi r}\ .              
\end{equation}                                                                  
where $\vec{r}$ represents the relative distance between the two 
nucleons, 
$\vec{r}=\vec{r}_1-\vec{r}_2$. The same interaction that is non-local 
 in momentum space, eq.(\ref{nrp3}), is local in $r$-space. This 
explains why the last one is often preferred in practical 
calculations.                      

%
\subsection{Spin structure of the non-relativistic $NN$ potential} 
\label{sst}  
Quite generally, the $NN$ non-relativistic Galilean invariant 
potential         
contains 5 terms with different spin structure (assuming that strong            
interactions conserve isospin). It may be written in momentum space 
as:  
\begin{eqnarray}\label{sst1}                                                    
V(\vec{k}, \vec{k}\,')  &=& V_C(\vec{k},\vec{k}\,') +      
V_{SS}(\vec{k},\vec{k}\,') \vec{\sigma}_1\cd\vec{\sigma}_2 +              
V_{SO}(\vec{k},\vec{k}\,'){i\over 2}(\vec{\sigma}_1+\vec{\sigma}_2)             
\cd [\vec{k}\times \vec{k}']                                               
\nonumber\\                                                                     
&&+V_T(\vec{k},\vec{k}\,')\left\{(\vec{k}-\vec{k}\,')^2                         
\vec{\sigma}_1\cd\vec{\sigma}_2 -                                         
3(\vec{k}-\vec{k}\,')\cd\vec{\sigma}_1\:                       
(\vec{k}-\vec{k}\,')\cd\vec{\sigma}_2\right\} \nonumber\\    
&&+ V_{SO2}(\vec{k},\vec{k}\,')                                                 
\vec{\sigma}_1\cd [\vec{k}\times\vec{k}\,']\:                  
\vec{\sigma}_2\cd [\vec{k}\times\vec{k}\,']\ ,                
\end{eqnarray}                                                                  
where $V_C(\vec{k},\vec{k}\,'),$ $ V_{SS}(\vec{k},\vec{k}\,'), $ $ 
V_{SO}(\vec{k},\vec{k}\,'),$ $ V_T(\vec{k},\vec{k}\,')$ and 
$V_{SO2}(\vec{k},\vec{k}\,')$ stand respectively for central, spin-spin, 
spin-orbit, tensor and quadratic spin-orbit forces respectively. They 
are predominantly scalar functions of $(\vec{k}-\vec{k}\,')^2$ (with 
further slight dependence on $\vec{k}\,^2$ and $\vec{k}\,'^2$) and 
contain isospin independent and dependent terms:                                                            
\begin{equation}\label{sst2}                                                    
V(\vec{k},\vec{k}\,') =V^0(\vec{k},\vec{k}\,')                                  
+V^1(\vec{k},\vec{k}\,')\vec{\tau}_1\cd\vec{\tau}_2\ .                     
\end{equation}                                                                  
Other structures, which have an off-energy shell character, are discarded on
the basis that their effect can be re-absorbed in the shorter range part of the
interaction model, mostly phenomenological. Such terms have been
considered in ref. \cite{amdes}.

In $r$-space, and assuming a dependence of $V(\vec{k},\vec{k}\,')$  on 
$(\vec{k}-\vec{k}\,')^2$ only, the above interaction takes the 
following form:  
\begin{eqnarray}\label{sst3}                                                    
V(\vec{r},\vec{r}\,')= \delta^{(3)}(\vec{r} 
-\vec{r}\,') \left\{V_C(r) +V_{SS}(r)\vec{\sigma}_1\cd\vec{\sigma}_2 
+V_{SO}(r)\frac{\vec{\sigma}_1 +\vec{\sigma}_2}{2}\cd\vec{l}\right.           
\nonumber\\                                                                     
\left. V_T(r)\left(\frac{3\vec{\sigma}_1\cd\vec{r}\;\; 
\vec{\sigma}_2\cd\vec{r}}{r^2} -\vec{\sigma}_1\cd\vec{\sigma}_2\right) 
+ V_{SO2}(r)\left( \frac{\vec{\sigma}_1\cd 
\vec{l}\;\vec{\sigma}_2\cd\vec{l} +\vec{\sigma}_2\cd\vec{l}\; 
\vec{\sigma}_1\cd\vec{l}}{2} + \cdots \right)\right\}, 
\end{eqnarray}
where the points indicate the existence of other terms involving 
$V_{SO2}$, but usually neglected. The potentials $V_C(r)$ and 
$V_{SS}(r)$ can have a linear $k^2$ dependent term \cite{nuj,paris}, 
which, by using an appropriate transformation \cite{paris}, can be 
dealt 
with in  coordinate space without too much difficulty. In  
momentum space, such terms 
can be kept to any order without further difficulty. These terms have 
quite different origins. Some simply come from the mathematical 
structure of Dirac spinors. Other ones come from the dependence on the 
$s$-variable arising from two-meson exchange ($\sqrt{s} $ is the 
energy 
of the system in its c.m. system). This $s$-dependence can be turned 
into a $\vec{k}\,^2$ dependence if one assumes $s\simeq 4(k^2+m^2)$.  
This procedure amounts  to neglect off-shell effects that can be 
incorporated phenomenologically through a fit to experimental data of 
the short range part of the $NN$ interaction.                                                        

Typically, the effects under consideration are of order  $\vec{k}^2/m^2$ and 
look like kinematical relativistic corrections. Knowing their origin  is 
however not sufficient to fix them in the $NN$ potential. The  equation,  which
the NN potential has to be associated with, has to be  precised in order to
have an unambiguous prescription. Thus, for the  Schr\"odinger equation (or
equivalent ones such as the  Lippmann-Schwinger equation for the $T$-matrix for
instance), the  appropriate  definition of $V(\vec{k},\vec{k}\,')$  is:        
\begin{equation}\label{sst4}                                                    
V_{\sigma_2\sigma_1}^{\sigma'_2\sigma'_1}(\vec{k},\vec{k}\,') =                 
\sqrt{{m\over \varepsilon_k}}\frac{\bar{u}^{\sigma_1}(\vec{k})                  
O_1u^{\sigma'_1}(\vec{k}\,') \;\;\bar{u}^{\sigma_2}(-\vec{k})                   
O_2u^{\sigma'_2}(-\vec{k}\,')} {4m^2(\mu^2 +(\vec{k}-\vec{k}\,')^2)}            
\sqrt{{m\over \varepsilon_{k'}}},                                               
\end{equation}                                                                  
where the spinors are normalized as $\bar{u}u=2m$, hence the factor  $4m^2$ in
the denominator in (\ref{sst4}), and $O$ represents the  vertex describing the
interaction of nucleons with the mesons of  interest. The factors
$\sqrt{m/\varepsilon_k}$ and  $\sqrt{m/\varepsilon_{k'}}$, that are neither 1,
as in the  non-relativistic limit nor $m/\varepsilon_k$ and
$m/\varepsilon_{k'}$  as one may expect from the normalization factors relative
to the two nucleons in the initial and final states, are required  to satisfy a
unitarity condition \cite{cdr}. Furthermore, their  omission in earlier
calculations of the $NN$ potential was responsible  for a large energy
dependence of the $2\pi$-exchange contribution  (obtained by removing the
iterated one pion exchange from the full  $2\pi$-exchange contribution to the
$NN$ scattering amplitude).  The  presence of the factors
$\sqrt{m/\varepsilon_k}$ and  $\sqrt{m/\varepsilon_{k'}}$ in (\ref{sst4}) is
important in comparing relativistic  calculations to non-relativistic ones.
They already include some sizeable relativistic kinematical corrections.                                                                    

Anticipating a comparison with the light-front equation for the $NN$ 
system, as given by eq.(\ref{eqwf3}) for instance, we can rewrite the 
Schr\"odinger equation (\ref{nrp1}) with eq.(\ref{sst4}) for the 
potential and $E\equiv k_0^2/m$, as follows: 
\begin{equation}\label{sst5}                                           
(k^2-k_0^2)\left(\sqrt{\frac{\varepsilon_k}{m}}\psi(\vec{k})\right) = 
-m^2\int\frac{d^3k'}{(2\pi)^3 \varepsilon_k\,'}\frac{\bar{u}(\vec{k}) 
O_1u(\vec{k}') 
\;\;\bar{u}(-\vec{k}) O_2 u(-\vec{k}\,')} {4m^2(\mu^2 
+(\vec{k}-\vec{k}\,')^2)} 
\left(\sqrt{\frac{\varepsilon_{k'}}{m}}\psi(\vec{k}\,')\right). 
\end{equation}  
Quite generally, the on-shell $NN$ amplitude, from which the $NN$ 
potential is derived, can be expressed as a sum of independent 
invariants. Their number, five, is in relation with the number of 
independent terms in the non-relativistic limit given by 
eq.(\ref{sst1}). Their choice is not unique however. They may be built 
from the different tensors $1(S)$, $\gamma_5(PS)$, $\gamma_{\mu}(V)$, 
$\gamma_{\mu}\gamma_5(A)$ and $\sigma_{\mu\nu}(T)$. Other choices, in 
closer relation to the $2\pi$-exchange contribution can also be made 
\cite{cdr}. They differ by off-shell effects, which may be accounted 
for in a phenomenological way by the fit of shorter range contributions 
to experimental data, as already mentioned.                                                     

\subsection{Physical inputs}\label{input}                                       
The $NN$ potential due to the exchange of single mesons which we  consider here
can be calculated from the following Lagrangian  densities  describing the
meson-nucleon couplings.\footnote{In the present paper  we  use the definition
of coupling constants corresponding to   $g_{\pi NN}^2/4\pi \approx 14$, in
contrast to  refs. \cite{ck-deut,ck-fsi,dkm95}, where the definition with  
$g_{\pi NN}^2  \approx 14$ has been
used.}                                                            

({\it i}) Pseudoscalar mesons ($\pi$, $\eta$) (PS coupling):                    
\begin{equation}\label{4.1a}                                                    
{\cal L}^{int} = i\ g\ \bar{\psi} \gamma_5 \psi\ \phi^{(ps)}\ ;               
\end{equation}                                                                  
                                                                                
({\it ii}) Scalar mesons ($\sigma$, $\delta$):                                  
\begin{equation}\label{4.1b}                                                    
{\cal L}^{int} = g\ \bar{\psi} \psi \phi^{(s)}\ ;                             
\end{equation}                                                                  
                                                                                
({\it iii}) Vector mesons ($\rho$, $\omega$):                                   
\begin{equation}\label{4.1c}                                                    
{\cal L}^{int} = \bar{\psi}\                                                    
 [g\gamma^{\mu}\phi^{(v)}_{\mu} + (f/4m)\sigma^{\mu\nu}(\partial_{\mu}          
\phi^{(v)}_{\nu} - \partial_{\nu}\phi^{(v)}_{\mu})]\ \psi \ .                      
\end{equation}  
For simplicity, the isospin degrees of 
freedom have been omitted.                                                                        

Other types of coupling may be considered, such as the pseudo-vector 
one for the $\pi$ (or $\eta$) meson for instance:                                       

({\it iv}) Pseudoscalar mesons ($\pi$, $\eta$) (PV coupling):                   
\begin{equation}\label{input1}                                                  
{\cal L}^{int} =                                                                
-\frac{g}{2m}\ \bar{\psi} \gamma^{\mu} \gamma_5\                              
\psi\ \partial_{\mu}\ \phi^{(ps)}\ .                                            
\end{equation}                                                                  
For a transition involving on-mass-shell nucleons described by  positive 
energy spinors, couplings ({\it i}) and ({\it iv}) are equivalent.   They
strongly differ  off-mass shell as it is well known. The first  one  can give
rise to larger effects due to the so-called $Z$-diagram it  leads to, while the
second one does not. The choice of the coupling is  deeply connected with the
way in which chiral symmetry is chosen to be 
realized.                                       

As far as nucleons described by positive energy spinors are retained  in  the
description of the $NN$ system, the potential is unique. As will  be  seen
later in section \ref{lfobep}, different potentials can be obtained on the light 
front depending on the choice of the particular coupling used at the 
meson-nucleon vertex (like for instance the PS or PV $\pi NN$ couplings). Using
the  standard approximation made in the meson propagator consisting in  taking
$k_0-k'_0=0$, neglecting therefore retardation effects, one  gets  in the NN
c.m.s.:            
\begin{eqnarray}\label{input2}                                                  
V(\vec{k},\vec{k}\,')&=& \sqrt{\frac{m}{\varepsilon_k}}\left\{ g_{\pi 
NN}^2 \frac{\left[\bar{u}(\vec{k})\gamma_5u(\vec{k}\,')\right]_1 \left[ 
\bar{u}(-\vec{k})\gamma_5u(-\vec{k}\,')\right]_2} {4m^2(\mu_{\pi}^2 
+(\vec{k}-\vec{k}\,')^2)}\;\vec{\tau}_1\!\cdot\!\vec{\tau}_2 \right.              
\nonumber\\                                                                     
&&-g_{\sigma NN}^2\frac{\left[\bar{u}(\vec{k})u(\vec{k}\,')\right]_1 
\left[ \bar{u}(-\vec{k})u(-\vec{k}\,')\right]_2} {4m^2(\mu_{\sigma}^2 
+(\vec{k}-\vec{k}\,')^2)}                                                       
\nonumber\\                                                                     
&&+\frac{g_{\omega NN}^2} {4m^2(\mu_{\omega}^2 
+(\vec{k}-\vec{k}\,')^2)} \bar{u}(\vec{k}) \left[(g+f)\gamma_{\mu}- 
{f\over 2m}(k_1+k'_1)_{\mu}\right]_1u(\vec{k}\,')      
\nonumber\\                                                                     
&&\left.\times\bar{u}(-\vec{k}) \left[(g+f)\gamma^{\mu}- {f\over 
2m}(k_2+k'_2)^{\mu}\right]_2u(-\vec{k}\,') \right\} 
\sqrt{\frac{m}{\varepsilon_{k'}}}\ ,                                            
\end{eqnarray}                                                                  
where $k_1,k_2$ $(k'_1,k'_2)$ are the initial (final)
 on-shell nucleon momenta.
Only a few significant exchanges have been displayed in 
(\ref{input2}). 
Other ones, $\eta$, $\delta$ or $\rho$ can easily be obtained by 
removing or inserting the appropriate isospin dependence. Hadronic 
form 
factors may also be introduced by giving to the meson-nucleon coupling 
constants a dependence on the meson four-momentum.                              

\subsection{Choice of the parametrization}                                      
\label{comp}                                                                    
There are various single meson exchange models in the literature. They are not
all equivalent, with the consequence that they should be used within a defined
scheme. One of the most ambitious model is the field theory motivated Bonn
model (Bonn-E) \cite{bonn}. Its energy dependence, which has a relationship
with retardation effects included here in the kernel of the interaction, has
prompted the authors to derive energy independent models (now denoted QA and RA)
that can be used more easily. The emphasis on reproducing the deuteron D state
properties of the original model has biased this derivation. The model is
unable to account for the observable mixing angle $\epsilon_1$ beyond 100 MeV.
These drawbacks have been corrected in later versions (QB, QC, RB) \cite{machl}
by relaxing the constraints on the deuteron D state properties that are known
to be model dependent
\cite{bd88}.                                                              

Other models differ by the values of the coupling constants, including 
the associated form factors. This is expected for short-range 
contributions which have an effective character anyway, but this also 
occurs for the $\pi NN$ coupling constant, which may be sensitive to 
isospin breaking effects \cite{nuj}. The last one is important for a 
fine description of the $NN$ interaction but unimportant in the 
present 
review. The difference with Bonn $QB$ and $QC$ models \cite{machl}, 
which motivation was shortly explained above, involves changes in both 
the $\pi NN$ coupling and those determining short range contributions.                                                                  

The difference between momentum and coordinate space models has  conceptually
an other origin. Working in  one or the other should  indeed be equivalent.  In
practice however, this is true as far as the  linearly $k^2$-dependent terms,
which are tractable in $r$-space  models, represent an accurate description of
the full momentum  dependence of the $NN$ interaction model. Recent
developments, dealing  with the well known $\pi$-exchange, show that off-shell
effects  involving fourth order terms in the nucleon momentum are  relevant for
describing the $NN$ interaction in  coordinate space \cite{amdes}.  Their 
neglect explains a large part of the differences in the bare  predictions
obtained from the Bonn $QB$ \cite{machl} and Paris  \cite{paris} models, which
reproduce the same $NN$ scattering data  otherwise. In particular, the deuteron
D-state probability of  coordinate space  models is expected to be
systematically between 0.5 \% and 1 \% higher  than in  momentum space
models.                                                       

Other differences may be mentioned. The genuine term, 
$\vec{\sigma}_1\cd [\vec{k}\times\vec{k}\,'] \; \vec{\sigma}_2\cd 
[\vec{k}\times\vec{k}\,'] $, in eq.(\ref{sst1}) is approximately 
replaced by the term $\vec{\sigma}_1\cd\vec{l} \; 
\vec{\sigma}_2\cd\vec{l} + \vec{\sigma}_2\cd\vec{l} \; 
\vec{\sigma}_1\cd\vec{l}$ in  coordinate space models. This can 
introduce some 
bias in fitting the potential parameters to $NN$ scattering data. The 
full $2\pi$-exchange contribution considered in the Paris model does 
not seem to bring significant improvement upon a single meson exchange 
model, probably because these ones contain enough parameters (coupling 
constants together with vertex form factors) to accurately account for 
these $NN$ data. This is indirectly confirmed by the Argonne model 
\cite{wir} which also does well with these data although it does not 
incorporate $k^2$ dependent terms as in the Bonn $R$, Paris or 
Nijmegen 
models and, on the other hand, fit the deuteron quadrupole moment to 
the observed one without any regards to the contribution of meson 
exchange currents.                                    

Altogether, it seems that $NN$ scattering data put strong constraints  on the
$NN$ interaction models and  within a given scheme do  not allow
for much freedom. Differences are closely related to the  scheme under
consideration and, most often, involve terms of order $(\vec{k}/m)^2$, that are
of relativistic origin. For some part, the present  models may be unitary
equivalent, implying the existence of corrections pertinent to the model under
consideration for any calculation of the physical observables. The choice of 
some particular model depends on a prejudice \cite{amdes}, models  that keep
track of these theoretically motivated $(\vec{k}/m)^2$ terms should deserve a
special  attention
however.                                                      

\section{Meson exchange interaction on the light front}                         
As already mentioned, the approach to the non-relativistic $NN$ interaction is
semi-phe\-no\-me\-no\-lo\-gi\-cal. This means that the meson-nucleon field 
theory is used to obtain a form of  the $NN$ interaction kernel (OBEP, as a
rule), and the parameters of this  kernel are found from fitting the
experimental data in the framework of the  Schr\"odinger equation. 

A similar approach can be based on the  relativistic quantum mechanics in the
front form. Besides the Hamiltonian, one  should introduce the interaction in
all the generators of the Poincar\'e group, which change the orientation of
the light-front plane \cite{dirac}. 

This approach has been developed in a number of studies (see, in particular, 
\cite{ls78,lev83,terent76,bt76,bt77,kt77,bkt79,lev85,lev95}). We do not
describe them here in details (see \cite{kp91,coester92} for a review). One
should mention that in this approach the full program  of fitting the phase
shifts and the relativistic kernel has not yet been  realized. 

The kernel in these references is considered as purely phenomenological, i.e.
it is not derived from light-front field theory, even in the  framework of
OBEP. It is assumed that it does not depend on the orientation of the
light-front plane. However, if one uses the field theory to construct  the
off-energy shell light-front kernel, this one always gets a dependence  on the
light front through the four-vector $\omega$ or its spatial part $\vec{n}$ (see
examples in sect.  \ref{lfobep}). It seems reasonable therefore to use this
kernel as an input  in the light-front version of relativistic quantum
mechanics. This cannot  be done by a simple replacement of the
$\vec{n}$-independent kernel in the above versions of the relativistic quantum
mechanics by the $\vec{n}$-dependent one. This would destroy the commutation
relations, found for the $\vec{n}$-independent mass operator. In order to
restore the covariance, one should find a solution of the Poincar\'e group
algebra with this new $\vec{n}$-dependent mass operator. Due to covariance, the
exact on-shell amplitudes in this approach will not depend on $\omega$ (with
the accuracy of numerical calculations). 

It is quite probable that in both  phenomenological light-front approaches
(with the $\omega$ - independent and dependent kernels), in spite of the
completely different behavior of the off-shell amplitudes, the phase shifts
and the binding  energies can be equally fitted, at least in the two-body
problem. This  implies that there exists an unitary transformation between
corresponding  sets of the Poincar\'e generators. However, this does not imply
their {\em practical} equivalence, since this unitary operator has to be
dynamical and highly complicated, since it has to change the  off-shell
behaviour of all the partial amplitudes. It hardly can be useful in practice. 

It seems therefore reasonable to develop a phenomenological scheme inspired
from  field theory and to start with the  OBEP kernel in the light-front
relativistic approach. This way seems rather promising, since it reconciles the
main features of the field-theoretical kernel on the light front with
relativistic covariance. It has not yet been developed in its full form.

%
\subsection{OBEP on the light front}\label{lfobep}                                           
The contribution due to single meson exchange, which is shown in 
fig. \ref{fkern}, can be obtained from the meson-nucleon interaction 
described by the Lagrangians given by eqs.(\ref{4.1a})-(\ref{input1}). 
The analytical expression for the pion exchange in the pseudo-scalar 
coupling, for instance, is a straightforward generalization of 
eq.(\ref{k1}). It has the form:             
\begin{eqnarray}\label{4.2}                                                     
V^{\sigma'_2\sigma'_1}_{\sigma_2\sigma_1}                                       
&=&\frac{g_{\pi NN}^2}{4m^2} \vec{\tau}_1\cd\vec{\tau}_2                           
\left\{\frac{\theta\left(\omega\cd (k'_1 - k_1)\right)F^2}{\mu^2- 
 (k'_1-k_1)^2 + 2\tau'\omega\cd (k'_1-k_1)-i\epsilon} \right. 
 \nonumber \\
&+ &\left. \frac{\theta\left(\omega\cd (k_1-k'_1)\right)F^2} {\mu^2 - 
 (k_1-k'_1)^2 + 2\tau\omega\cd (k_1-k'_1)-i\epsilon}\right\}  
\nonumber 
\\ &\times&\left[\bar{u}^{\sigma_2}(k_2)\gamma_5 
u^{\sigma'_2}(k'_2)\right] \;\left[\bar{u}^{\sigma_1}(k_1)\gamma_5 
u^{\sigma'_1}(k'_1)\right] \ ,                                                                         
\end{eqnarray}                                                                  
where $F$ is a phenomenological form factor, assumed to depend                  
on the same argument as the denominator of the kernel. Here and below 
it is introduced ``by hand".

The $\eta$ exchange contribution can be obtained from (\ref{4.2}) by 
removing the isospin factor and replacing $g_{\pi NN}^2$ by $g_{\eta 
NN}^2$.  The contributions due to $\sigma$- and $\delta$-exchanges are 
obtained from the previous ones by removing the $\gamma_5$ matrices 
and 
changing the overall sign. As for the vector meson exchange, it 
contains a peculiarity because of the   contact terms due to the 
vector 
meson propagator and the coupling with derivative. We therefore 
explain 
this point in more details. The starting expression for the vector 
meson exchange contribution, obtained by applying the rules of the 
graph technique to the diagram of fig.~\ref{fkern}, has the form:                                                 
\begin{eqnarray}\label{v1}                                                      
{\cal K}&=&\int \bar{u}(k_1)\left[g\gamma^{\alpha}+ 
\frac{f}{2m}\sigma^{\alpha'\alpha}(i)(k-\omega\tau_1)_{\alpha'}\right]          
u(k_1')                                                                         
\nonumber\\                                                                     
&&\times\left[-g_{\alpha\beta}+                                                 
\frac{(k-\omega\tau_1)_{\alpha}(k-\omega\tau_1)_{\beta}}{\mu^2}\right]          
\nonumber \\                                                                    
&&\times \bar{u}(k_2)\left[g\gamma^{\beta}+                                     
\frac{f}{2m}\sigma^{\beta'\beta}(-i)(k -\omega\tau_1)_{\beta'}\right] 
u(k_2')    
\nonumber\\                                                                     
&&\times\delta\left[(k_1'-k_1+\omega\tau_1-\omega\tau')^2-\mu^2\right]          
\theta(\omega\cd (k_1'-k_1))\frac{d\tau_1}{\tau_1-i\epsilon} 
\nonumber \\           
&+&\frac{1}{(2\pi)^2}\int \bar{u}(k_1)\left[g\gamma^{\alpha}+                   
\frac{f}{2m}\sigma^{\alpha'\alpha}(-i)(k-\omega\tau_1)_{\alpha'}\right]         
u(k_1')                                                                         
\nonumber\\                                                                     
&& \times\left[ -g_{\alpha\beta}+                                               
\frac{(k-\omega\tau_1)_{\alpha}(k-\omega\tau_1)_{\beta}}{\mu^2}\right]          
\nonumber \\                                                                    
&&\times \bar{u}(k_2)\left[g\gamma^{\beta}+                                     
\frac{f}{2m}\sigma^{\beta'\beta}(i) (k-\omega\tau_1)_{\beta'}\right] 
u(k_2')     
\nonumber\\                                                                     
&&\times\delta\left[(k_1-k_1'+\omega\tau_1-\omega\tau)^2-\mu^2\right]           
\theta(\omega\cd (k_1-k_1'))\frac{d\tau_1}{\tau_1-i\epsilon}\ ,                       
\end{eqnarray}                                                                  
where $k$ is the meson momentum. The factors $(i)$ and $(-i)$ at 
$(k-\omega\tau_1)$ are associated with outgoing and incoming lines,
according to sect.\ref{sp1-der}.  
The factor $\omega\tau_1$, everywhere in the difference 
$(k-\omega\tau_1)$ in the expression (\ref{v1}), incorporates the 
contact terms  for both the vector meson exchange and for the coupling 
with derivative.                                                                

After integrating over $\tau$ and introducing the form factor $F$, we get for 
the potential $V=-{\cal K}/4m^2$:  
\begin{eqnarray}\label{v2} 
V^{\sigma'_2\sigma'_1}_{\sigma_2\sigma_1}                                       
&=& \frac{F^2}{4m^2(\vec{K}^2+\mu^2)}\bar{u}^{\sigma_1}(k_1) \left\{ 
(g+f)\gamma_{\alpha}  - 
\frac{f}{2m}(k_1+k'_1)_{\alpha}\right.\nonumber\\                               
&+&\left.\frac{if}{2m}\sigma_{\alpha'\alpha}\omega^{\alpha'}                    
\left[\tau\theta(\omega \cd (k_1-k_1'))- \tau'\theta(\omega                        
\cd (k_1'-k_1))\right]\right\} u^{\sigma'_1}(k'_1)                                 
 \nonumber \\                                                                   
&&\times\bar{u}^{\sigma_2}(k_2)                                                 
\left\{(g+f)\gamma^{\alpha}- \frac{f}{2m}(k_2+k'_2)^{\alpha}\right.
\nonumber\\ 
&+&\left.\frac{if}{2m}\sigma^{\beta'\alpha}\omega_{\beta'}                      
\left[\tau\theta(\omega\cd (k_1'-k_1))-                                            
\tau'\theta(\omega\cd (k_1-k_1'))\right]\right\} u^{\sigma'_2}(k'_2)
\nonumber\\  
&-&\frac{ 
g^2}{4m^2\mu^2(\vec{K}^2+\mu^2)}\left[\bar{u}^{\sigma_1}(k_1)         
\hat{\omega}u^{\sigma'_1}(k'_1)\right] \left[                                   
\bar{u}^{\sigma_2}(k_2)                                                         
\hat{\omega}u^{\sigma'_2}(k'_2)\right]\tau\tau',                                
\end{eqnarray}                                                                  
where $\vec{K}\,^2$ is given by eq.(\ref{k4}).                                                      

The contribution of the isovector exchange ($\rho$-meson) is obtained  
by multiplying (\ref{v2}) by the factor $\vec{\tau}_1\cd\vec{\tau}_2$. 
The terms in (\ref{v2}) proportional to $\tau$, $\tau'$ and 
$\tau\tau'$ 
were omitted in ref.~\cite{ck-deut}.            

For reasons already explained, the pseudovector $\pi NN$ coupling is  sometimes
preferred to the pseudoscalar one. To illustrate the  treatment  of terms
corresponding to derivative couplings of the meson field, we  just give the
result here:                                                
\begin{eqnarray}\label{pv}                                                      
V^{\sigma'_2\sigma'_1}_{\sigma_2\sigma_1}                                       
&=&\frac{g_{\pi
NN}^2}{4m^2}\left\{\frac{                                       
\theta(\omega\cd (k'_1 - k_1))F^2}{\mu^2- (k'_1-k_1)^2 +  2\tau'\omega\cd
(k'_1-k_1)-i\epsilon} \right.\nonumber\\ &&\times 
\left[\bar{u}^{\sigma_2}(k_2)                                                   
\left(\gamma_5-\frac{\hat{\omega}\gamma_5\tau}{2m}\right)                                  
u^{\sigma'_2}(k'_2)\right]
\left[                                               
\;\bar{u}^{\sigma_1}(k_1)                                                       
\left(\gamma_5+\frac{\hat{\omega}\gamma_5\tau'}{2m}\right)                                 
u^{\sigma'_1}(k'_1)\right]                                                     
\nonumber
\\ &+&
\frac{                                                                     
\theta(\omega\cd (k_1-k'_1))F^2}{\mu^2 - (k_1-k'_1)^2 + 2\tau\omega\cd 
(k_1-k'_1)-i\epsilon} \nonumber\\ &&\left.\times 
\left[\bar{u}^{\sigma_2}(k_2)                                                   
\left(\gamma_5+\frac{\hat{\omega}\gamma_5\tau'}{2m}\right)                                 
u^{\sigma'_2}(k'_2)\right]
\left[                                               
\;\bar{u}^{\sigma_1}(k_1)                                                       
\left(\gamma_5-\frac{\hat{\omega}\gamma_5\tau}{2m}\right)                                  
u^{\sigma'_1}(k'_1)\right]\right\}                                             
\;\;                                                                            
\vec{\tau}_1\cd\vec{\tau}_2\ ,
\nonumber\\                                         
&&                                                                              
\end{eqnarray}                                                                  
The comparison with the expression obtained with the pseudo-scalar $\pi  NN$ 
coupling evidences differences which depend on the four-vector  $\omega_{\mu}$
and vanish in the limit of on-energy shell nucleons  ($\tau=\tau'=0$). 

It is also interesting to compare the interaction on  the light front,
eq.(\ref{4.2}), with the Bonn-$Q$ model  (\ref{input2}), which contains some
relativity. Differences involve  again terms depending on the four-vector
$\omega_{\mu}$, which vanish  on-energy shell. Further differences, due to the
time component of the  four-momentum of the meson in its propagator
(retardation effects), vanish on energy shell. The difference in the
normalization factors $\sqrt{m/\varepsilon_k}$ and $\sqrt{m/\varepsilon_k'}$,
discussed in sect. \ref{sst}, is related to the particular equation for the
wave function.

%
\subsection{Beyond OBEP on the light front}\label{beyond}     

The interaction due to single meson exchanges can 
produce contributions depending on the arbitrary four vector $\omega$. While 
this dependence vanishes for the Born amplitude (which is  calculated for
on-shell particles), there is no guarantee that  this
property holds to all orders. This implies an asymptotic scattering behavior,
and  therefore a scattering amplitude which depends on $\omega$. A quite
similar problem  arises in other approaches based on a description referring to
the plane  determined by $\sigma=\lambda x$ with the particular choice
$\lambda=(1,0,0,0)$,  but here the difficulty is more striking as results may
depend on the arbitrary orientation of the three-vector $\vec \omega (\vec n)$
while physical ones should not.
For example, the relativistic wave function originating from the
$\pi$-exchange,
calculated in a perturbative way in the first $1/m$ order, reads \cite{dkm95}:
\begin{equation}\label{bey3}
\delta \psi^{(1)\,\pi}(\vec{k},\vec{n})= -\frac{g^2_{\pi NN}}{8m^2}
\vec{\tau}_1\cd\vec{\tau}_2
\frac{m}{\vec{k}\,^2+\kappa^2} \int
\frac{d^3k'}{(2\pi)^3} \frac{(\vec{k}\,^2-\vec{k}\,'^2)}{m} 
\frac{[\vec{\sigma}_1\times\vec{\sigma}_2]\cd 
[(\vec{k}-\vec{k}\,')\times \vec{n}]}{\mu^2+(\vec{k}-\vec{k}\,')^2}
\psi^{(0)}(\vec{k}\,')\ .
\end{equation}
The quantity  $\kappa^2$ is positive in the
case of a bound state such as the deuteron, and it is negative 
for a scattering state.  

The defect of the wave function (\ref{bey3}) is that it has a
$\vec{n}$-dependent residue at the pole at $k=i\kappa$, that can contribute to
an observable amplitude. A similar defect in the scattering state wave function
appears in the dependence of asymptotical states on $\vec{n}$, i.e., in the 
on-energy-shell amplitude, in contradiction with the results of sect.
\ref{lfgt}.

Knowing that the sum of the contributions of all  irreducible diagrams to all
orders should reproduce results obtained from Feynman  diagrams, which are
independent of $\omega$, one can guess that higher order terms in the
interaction are essential for removing the above unpleasant  feature. In this
section, we explore their contributions and the role they may  play. As will be
shown below, the defect of the wave function (\ref{bey3}) is corrected by
incorporating the contact term in the $NN$ kernel \cite{dkm95p}.

\begin{figure}[hbtp]
\centerline{\epsfbox{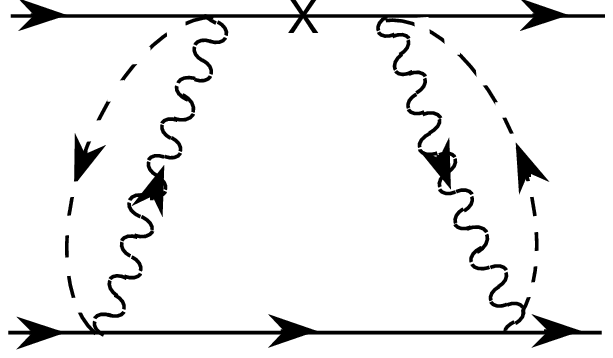}}
\figcap{Irreducible diagram with contact term.}
\label{soc}
\end{figure}

To give some insight about the relevance of these extra terms in the
interaction, we will concentrate on a selected one, where the contact term
discussed for the spin 1/2 particle (eq.(\ref{rul14})) is operating (see fig.
\ref{soc}). This will be done by retaining the relevant lowest $1/m$ order
terms, and for contributions in relation with the single $\pi$-exchange. At
this order, only $\sigma$- and $\omega$-exchanges contribute, so in fig.
\ref{soc} one wavy line corresponds to $\pi$-exchange and the other wavy line
corresponds to  $\sigma$ and $\omega$. The corresponding contribution to the
kernel has the form:
\begin{eqnarray}\label{bey1}
&&\delta V^{\pi,\sigma +\omega}(\vec{k},\vec{k}\,'',\vec{n})\equiv 
\delta V_1^{\pi,\sigma +\omega}+\delta V_2^{\pi,\sigma +\omega}=
-\frac{g_{\pi NN}^2}{8m^2} 
\vec{\tau}_1\cd\vec{\tau}_2\;[\vec{\sigma}_1\times\vec{\sigma}_2] 
\\
&&\times\int\frac{d^3k'}{(2\pi)^3}
\left[\frac{[(\vec{k}-\vec{k}\,')\times\vec{n}]} 
{\mu^2+(\vec{k}-\vec{k}\,')^2}V^{\sigma +\omega}(\vec{k}\,'-\vec{k}\,'')-
V^{\sigma +\omega}(\vec{k}-\vec{k}\,')
\frac{[(\vec{k}\,'-\vec{k}\,'')\times\vec{n}]} 
{\mu^2+(\vec{k}\,'-\vec{k}\,'')^2}\right]\ ,
\nonumber
\end{eqnarray}
where $V^{\sigma +\omega}$ represents the dominant contribution to the
interaction at zeroth order in $1/m$ \cite{dkm95p}.

The insertion of $\delta V^{\pi,\sigma +\omega}$ in an equation such as 
(\ref{2.16bb}) provides an extra, $\vec{n}$-dependent, contribution to the wave
function. In a perturbative calculation it is given by:
\begin{eqnarray}\label{bey2}
 &&\delta \psi^{\pi,\sigma +\omega}(\vec{k},\vec{n})\equiv
 \delta \psi_1^{\pi,\sigma +\omega}(\vec{k},\vec{n})+
 \delta \psi_2^{\pi,\sigma +\omega}(\vec{k},\vec{n})
\nonumber\\
 &&=-\frac{m}{\vec{k}\,^2+\kappa^2}\int \frac{d^3k''}{(2\pi)^3} 
 \left[\delta V_1^{\pi,\sigma +\omega}(\vec{k},\vec{k}\,'',\vec{n})+
 \delta V_2^{\pi,\sigma +\omega}(\vec{k},\vec{k}\,'',\vec{n})\right]
 \psi^{(0)}(\vec{k}\,'')\ ,
\end{eqnarray}
where $\psi^{(0)}(\vec{k}\,'')$ is a zeroth-order non-relativistic wave
function.

One should also take into account all the iterations of the above contributions
with the non-relativistic potential $V^{\sigma+\omega}$, since these iterations
remain in the first $1/m$ order. They are represented as
\begin{eqnarray}\label{bey4} 
\delta \psi_3^{\pi,\sigma +\omega}(\vec{k},\vec{n})=
&-&\frac{m}{\vec{k}\,^2+\kappa^2} \int\frac{d^3k'}{(2\pi)^3}
V^{\sigma+\omega}(\vec{k}-\vec{k}\,') 
\left[\delta \psi^{(1)\,\pi}(\vec{k}\,',\vec{n})
+\delta \psi_1^{\pi,\sigma +\omega}(\vec{k}\,',\vec{n})\right.
\nonumber\\
&+&\left.\delta \psi_2^{\pi,\sigma +\omega}(\vec{k}\,',\vec{n})
+\delta \psi_3^{\pi,\sigma +\omega}(\vec{k}\,',\vec{n})\right]\ , 
\end{eqnarray} 
where the integrand  incorporates the function $\delta
\psi_3^{\pi,\sigma +\omega}$ itself.  Therefore eq.(\ref{bey4}) is an
inhomogeneous equation for $\delta \psi_3^{\pi,\sigma +\omega}$.

It turns out (see below) that $\delta \psi_2^{\pi,\sigma +\omega}+ \delta
\psi_3^{\pi,\sigma +\omega}=0$. Hence, the full correction to the 
wave function reads:
\begin{equation}\label{beysum}
\delta\psi^{\pi}(\vec{k},\vec{n})=\delta \psi^{(1)\,\pi}+
\delta \psi_1^{\pi,\sigma +\omega}\ .
\end{equation}
It can be written as:
\begin{eqnarray}\label{bey5}
&&\delta \psi^{\pi}(\vec{k},\vec{n})=
-\frac{g^2_{\pi NN}}{8m^2}
\vec{\tau}_1\cd\vec{\tau}_2\frac{m}{\vec{k}\,^2+\kappa^2}
[\vec{\sigma}_1\times\vec{\sigma}_2]
\int\frac{d^3k'}{(2\pi)^3}
\frac{[(\vec{k}-\vec{k}\,')\times \vec{n}]}
{\mu^2+(\vec{k}-\vec{k}\,')^2}
\\
&&\times \left[\frac{\vec{k}\,^2}{m}\psi_0(\vec{k}\,')- 
\int
d^3k''\left(\frac{\vec{k}\,'^2}{m}\delta^{(3)}(\vec{k}\,'-\vec{k}\,'')
+\frac{1}{(2\pi)^3}V^{\sigma+\omega}(\vec{k}\,'-\vec{k}\,'')
\right)\psi_0(\vec{k}\,'')\right]\ .
\nonumber
\end{eqnarray}
Applying eq.(\ref{2.16bb}) at the lowest order considered here allows to
replace the integral in the last bracket by
$-\frac{\displaystyle{\kappa^2}}{\displaystyle{m}}\psi_0(\vec{k}\,')$. 
It is then easily seen that this term
adds to the first one in the bracket to cancel the front factor
$\frac{\displaystyle{m}}{\displaystyle{\vec{k}\,^2+\kappa^2}}
$ at the r.h.s.  of
(\ref{bey5}). One thus gets:
\begin{equation}\label{bey5p}
\delta \psi^{\pi}(\vec{k},\vec{n})
= -\frac{g^2_{\pi NN}}{8m^2}
\vec{\tau}_1\cd\vec{\tau}_2
\int\frac{d^3k'}{(2\pi)^3}
\frac{[\vec{\sigma}_1\times\vec{\sigma}_2]\cd 
[(\vec{k}-\vec{k}\,')\times \vec{n}]}{\mu^2+(\vec{k}-\vec{k}\,')^2}
\psi_0(\vec{k}\,')\ .
\end{equation}
The pole in eq.(\ref{bey3}) at $k=i\kappa$ ($k=k_0$ in continuous spectrum
case) has now disappeared. This result is important since it guarantees that
the observables, such as asymptotic normalization or phase shifts, do
not depend on the arbitrary direction $\vec{n}$, though the wave function 
\ref{bey5p} depends on it. 

In $r$-space eq.(\ref{bey5p}) takes a simpler form:
\begin{equation}\label{bey6}
\delta \psi^{\pi}(\vec{r},\vec{n})= 
i\frac{g^2_{\pi NN}}{8m^2}
\vec{\tau}_1\cd\vec{\tau}_2\;[\vec{\sigma}_1\times\vec{\sigma}_2]\cd
\left[\vec{\nabla} _r \left(\frac{\exp(-\mu r)}{r}\right)\times \vec{n}\right]
\psi_0(\vec{r})\ ,
\end{equation}
which evidences a radial part in close correspondence with that appearing in
the pair term contribution to meson exchange currents or also in relativistic
components in other approaches.
 
To calculate the contribution $\delta \psi_3^{\pi,\sigma+\omega}$
one has to substitute for $\delta
\psi^{(1)\,\pi}+\delta \psi_1^{\pi,\sigma +\omega}$ in the integrand in 
eq.(\ref{bey4})  the wave function
(\ref{bey5p}). The result of this substitution coincides with 
$-\delta \psi_2^{\pi,\sigma +\omega}$. Hence, the equation (\ref{bey4}) turns
into 
\begin{eqnarray}\label{bey4p} 
&&\delta \psi_2^{\pi,\sigma +\omega}(\vec{k},\vec{n})+
\delta \psi_3^{\pi,\sigma +\omega}(\vec{k},\vec{n})=
\nonumber\\
&&-\frac{m}{\vec{k}\,^2+\kappa^2} \int\frac{d^3k'}{(2\pi)^3}
V^{\sigma+\omega}(\vec{k}-\vec{k}\,')
\left[\delta \psi_2^{\pi,\sigma +\omega}(\vec{k}\,',\vec{n})
+\delta \psi_3^{\pi,\sigma +\omega}(\vec{k}\,',\vec{n})\right]\ . 
\end{eqnarray} 
This is the homogeneous equation relative to 
$\delta \psi_2^{\pi,\sigma +\omega}+ \delta \psi_3^{\pi,\sigma +\omega}$.
However it does not determine any eigenvalue, since the energy $E$ in
$\kappa^2=-mE$ is already fixed by the equation for the complete wave function.
Therefore its solution is zero:
\begin{equation}\label{bey7}
\delta\psi_2^{\pi,\sigma+\omega}(\vec{k},\vec{n})+
\delta \psi_3^{\pi,\sigma+\omega}(\vec{k},\vec{n})=0. 
 \end{equation}  
The significance of this result may be obtained from a comparison with other
relativistic approaches \cite{gross6982}. In these approaches, a term like
(\ref{bey6}) is associated with a pair of nucleons with positive and negative
energies. Their interaction due to $\sigma$ and $\omega$ exchanges produces a
correction proportional to $V^{\sigma}-V^{\omega}$ (instead of 
$V^{\sigma}+V^{\omega}$ as in eq.(\ref{bey4})). The second term in 
eq.(\ref{bey1}) has therefore the effect to remove the contribution that
should be absent in any case.

In this section, we have considered some of the second order $\vec  n$
dependent contributions to the irreducible interaction $V$. This has  been
done to lowest $1/m$ order. These contributions which are the first  ones of
an infinite series seem to play an important role to make the theory 
consistent, accounting in particular for the $\vec{n}$ independence of the 
asymptotic wave function, which are related to physical properties. They also
show  that solving the equation of motion without using the appropriate
interaction may lead  to wrong results. This consideration makes the
relationship with other relativistic approaches closer. The present study is,
to a large extent, exploratory. It suggests that the interaction should
fulfill some constraints whose determination may be helpful for further work.

                                                                                
\chapter{Applications to the nucleon-nucleon
system}\label{nns}                 
As a direct application of the formalism developed above, we consider in this
chapter the two nucleon system. Since its non-relativistic phenomenology has
been developed over more than twenty years, it allows a sensible discussion of
the new features brought about by relativity.

\section{The deuteron wave function}\label{dlfwf}                               
                                                                                
\subsection{Spin structure} \label{dwf}                                         
                                                                                
The construction of the deuteron wave function is a nice illustration of
how to construct a state with non-zero angular momentum in the covariant 
formulation of LFD.  For this nucleus, $J^p=1^+$ (pseudovector particle),  the
decomposition of the wave function ${\mit\Phi}^{\lambda}_{\sigma_2\sigma_1}$ in
independent spin  structures has the general form~\cite{ck-deut}:  
\begin{equation}\label{nz1}                                                     
{\mit\Phi}^{\lambda}_{\sigma_2\sigma_1}(k_1,k_2,p,\omega \tau)=
\sqrt{m}e^{\lambda}_{\mu}(p) 
\bar{u}^{\sigma_2}(k_2)\phi_{\mu}U_c\bar{u}^{\sigma_1}(k_1)\ ,        
\end{equation}                                                                  
with                                                                            
\begin{eqnarray}\label{nz2} \phi_{\mu}=
\varphi_1\frac{(k_1                      - k_2)_{\mu}}{2m^2}
+\varphi_2\frac{1}{m}\gamma_{\mu}                           
+\varphi_3\frac{\omega_{\mu}}{\omega\cd
p}                                       
+\varphi_4\frac{(k_1-k_2)_{\mu}\hat{\omega}}{2m\omega\cd
p}\nonumber\\            +\varphi_5\frac{i}{2m^2\omega\cd p}\gamma_5
\epsilon_{\mu\nu\rho\gamma}          
(k_1+k_2)_{\nu}(k_1-k_2)_{\rho}\omega_{\gamma}
+                                
\varphi_6\frac{m\omega_{\mu}\hat{\omega}}{(\omega\cd p)^2}\
,                      
\end{eqnarray}  
and $m$ is the nucleon mass.                                                                                                                        
The light-front deuteron wave function is determined by six invariant functions
$\varphi_{1-6}$, depending on two scalar variables, e.g.  $s=(k_1+k_2)^2$ and
$t=(p -k_1)^2$.  The extra spin structures in front of $\varphi_{3-6}$ are
constructed by means of the four-vector $\omega$. Other possible structures
(e.g., $e^{\lambda}_{\mu} 
\overline{u}_2\hat{\omega}\gamma_{\mu}U_c\overline{u}_1$, etc.)  are expressed
through the six structures given by (\ref{nz2}).  The wave function (\ref{nz1})
is kinematically transformed  under rotations and Lorentz transformations $g$
of its arguments  according to eq.(\ref{wfp6}).             

The transformation to the new representation (\ref{sp1}) has the            
form:  
\begin{eqnarray}\label{nz3}                                              
{\mit\Psi}^{\lambda}_{\sigma_2\sigma_1}(k_1,k_2,p,\omega \tau) & = &                                   
\sum_{\lambda',\sigma_1',\sigma_2'}                                             
D^{(1)*}_{\lambda\lambda'}\{R(L^{-1}({\cal P}), p)\} 
D^{(\frac{1}{2})}_{\sigma_1\sigma_1'}\{R(L^{-1}({\cal                  
P}),k_1)\} 
\nonumber \\ 
&\times & 
 D^{(\frac{1}{2})}_{\sigma_2\sigma_2'}\{R(L^{-1}({\cal                
P}),k_2)\} {\mit\Phi}^{\lambda'}_{\sigma_2'\sigma_1'}(k_1,k_2,p,\omega \tau), 
\end{eqnarray}           
where, e.g., $R(L^{-1}({\cal P}),p)$ is given by (\ref{rot}), in which
$g=L^{-1}({\cal P})$, and ${\cal P}$ is given by (\ref{calp}). The matrix
$D^{(\frac{1}{2})}_{\sigma\sigma'}$ in (\ref{nz3})  is defined by eq.
(\ref{nz4}).                                                                               

The transformation law of the wave  function (\ref{nz3}) is given by
eq.(\ref{sp4}). In this new  representation given by (\ref{nz3}), the rotation
operator is the same for all  the spin indices and coincides with the operator
rotating the variables  $\vec{k}, \vec{n}$ in eqs.(\ref{sc6}).  

In the variables  $\vec{k}$ and $\vec{n}$, the problem of constructing the wave
function is the  same as in the non-relativistic case:  we have to construct
the general  form of a pseudovector from the variables $\vec{k}, \vec{n}$, the
Pauli  matrices and two-component spinors.  Since $\sigma_1, \sigma_2 = 
\pm1/2$, the wave function is a $2\times 2$-matrix.  In this respect,  the only
difference from the non-relativistic case is the presence of  an additional
vector $\vec{n}$ that increases the number of  independent structures.  The
decomposition of the wave function  (\ref{nz3}) takes therefore the
form~\cite{karm81}:                                                             
\begin{equation}\label{nz7}                                                     
{\mit\Psi}^{\lambda}_{\sigma_2\sigma_1}(\vec{k},\vec{n}) = 
\sqrt{m}w^\dagger_{\sigma_2} \psi^{\lambda}(\vec{k},\vec{n})\sigma_y 
w^\dagger_{\sigma_1}\ , 
\end{equation} 
with 
\begin{eqnarray}\label{nz8} 
\vec{\psi}(\vec{k},\vec{n}) & = & f_1\frac{1}{\sqrt{2}}\vec{\sigma} + 
f_2\frac{1}{2}(\frac{3\vec{k}(\vec{k}\cd\vec{\sigma})}{\vec{k}^2} 
-\vec{\sigma}) + f_3\frac{1}{2}(3\vec{n}(\vec{n}\cd\vec{\sigma}) 
-\vec{\sigma}) \nonumber \\ & + & 
f_4\frac{1}{2k}(3\vec{k}(\vec{n}\cd\vec{\sigma}) + 
3\vec{n}(\vec{k}\cd\vec{\sigma}) - 2(\vec{k}\cd\vec{n})\vec{\sigma}) 
\nonumber        
\\ & + & f_5\sqrt{\frac{3}{2}}\frac{i}{k}[\vec{k}\times \vec{n}] +              
f_6\frac{\sqrt{3}}{2k}[[\vec{k}\times \vec{n}]\times\vec{\sigma}]\ ,            
\end{eqnarray} 
where $w$ is the two-component nucleon spinor                  
normalized to $w^\dagger w=1$. 
The relation between $\psi^{\lambda}$ and $\vec{\psi}$ is the same 
as the relation between the spherical function $Y^{\lambda}_1(\vec{n})$ and
$\vec{n}$.
 The scalar functions $f_{1-6}$ depend on the           
scalars $\vec{k}\,^2$ and $\vec{n}\cd\vec{k}$ or $k \equiv \vert \vec{k}           
\vert $ and $z=\cos(\widehat{\vec{n}\vec{k}})$.  For an isospin zero state,       
the Pauli principle results into:                                               
\begin{equation}\label{nz9} f_{1,2,3,5}(k,z)                                    
=f_{1,2,3,5}(k,-z),\;\; f_{4,6}(k,z) = -f_{4,6}(k,-z)\ .                        
\end{equation}                                                                  
The reduction of the number of variables can be summarized as follows.  
The initial wave function (\ref{wfp6}) depends on 4 four-vectors, i.e.  
on 16 variables.  Four mass-shell constraints reduce this number to 12.  
The four-dimensional conservation law (\ref{sc1}) reduces it to 8.  
Lorentz-invariance eliminates the dependence on the velocity $\vec{v}$ 
of the system of reference, i.e., reduces the number of variables to 5.  
The wave function (\ref{nz8}) just depends on the 5 independent 
components of two vectors $\vec{k}, \vec{n}$, ($\vec{n}\,^2 =1)$.  
Because the rotational invariance eliminates their dependence on the 
three Euler angles, the scalar functions $f_i$ depend only on the two 
variables ($k$ and $z$ or $\vec{R}_{\perp}^2$ and $x$).  

In the region $k \ll m$, the functions $f_{3-6}$ which are of relativistic 
origin become negligible, $f_{1,2}$ do not depend anymore on $z$ and  turn into
the S- and D-waves:  $f_1\approx u_S, f_2 \approx -u_D$. From the  general
decomposition (\ref{nz8}), one recovers the usual  non-relativistic wave
function:                                                                       
\begin{equation}\label{nz10} 
\vec{\psi}_{NR}(\vec{k}\,) = u_S(k)\frac{1}{\sqrt{2}} \vec{\sigma} 
-u_D(k)\frac{1}{2} \left[\frac{3\vec{k} 
(\vec{k}\cd\vec{\sigma})}{\vec{k}\,^2} - \vec{\sigma}\right] \ .                                                         
\end{equation}                                                                  
The representation (\ref{nz2}) is however convenient in the calculation of
physical  observables, like electromagnetic form factor for instance. The 
decomposition (\ref{nz8}) is more adapted for comparison with the 
non-relativistic limit and will be chosen in this chapter to present  numerical
results.  The relations between the functions $\varphi  _{1-6}$ and $f_{1-6}$
can be found by comparing  eqs.(\ref{nz1}),(\ref{nz2}) with
eqs.(\ref{nz7}),(\ref{nz8}) in the  reference system where
$\vec{k}_1+\vec{k}_2=0$.  Indeed,  both  representations coincide in this
system, since all the $D$-matrices in  (\ref{nz3}) turn into unit matrices. 
These relations are given in  appendix~\ref{dcomp}.  

The number of components of the wave function is a priori not the same  in 
different relativistic approaches.  For example, the Bethe-Salpeter  amplitude
of the deuteron is determined by eight  components~\cite{tjonconf} (see
sect.~\ref{bsd}), whereas the Gross wave  function~\cite{gross} contains four
components.  It does not follow,  however, that these components do not lead to
physical observable  consequences.  In a given framework, they are
unambiguously related to  measured cross sections. The convenience of one 
approach or another depends on the problem under consideration. The relation 
between
the Bethe-Salpeter and  light-front components of the deuteron functions is
given in section~\ref{bsd}. We emphasize that within the LFD the number of
components is the same both in the non-covariant approach and in the covariant
formulation. In the non-covariant approach, these components manifest
themselves as six independent matrix elements  of the wave function
$\psi^{m}_{\sigma_1\sigma_2}$ with $m=0,\pm 1$, $\sigma_{1,2}=\pm 1/2$ instead
of two matrix elements in the  non-relativistic case.  The representation of
the wave function (\ref{nz8}) in  terms of the well defined spin structures and
corresponding components $f_{1-6}$  is however much more transparent  than in
terms of those matrix elements.  

Note that  when a deuteron state at rest, and described from two positive
energy spinors for its constituents, is seen from a different inertial frame,
it automatically acquires four extra components depending on the velocity of
this frame.  This also holds for the $^1S_0$ scattering state (one extra
component in this case). The equality of the total number of six components 
thus obtained with the number of components in the light-front approach is not
a  coincidence, since it is related in both cases to the appearence of an extra
vector. The extra components in the light-front approach just represent the
dynamical counterpart arising from the boost of the system at rest to some
velocity $\vec v$, and the vector $\vec{n}$ is the counterpart of the direction
$\vec{v}$ when $v\rightarrow c$.

\subsection{Two-body contribution to the deuteron normalization}\label{norm}                                
For the contribution of the two-body Fock component, the 
normalization integral (\ref{nor3}) is rewritten as:
\begin{equation}\label{na3} 
N_2^{\lambda'\lambda}={1\over (2\pi)^3}             
\int\sum_{\sigma_1 \sigma_2} {\mit \Phi}^{*J\lambda'}_{j_2 \sigma_2j_1          
\sigma_1} {\mit \Phi}^{J\lambda}_{j_2 \sigma_2j_1 \sigma_1}                     
\delta^{(4)}(k_1+k_2-p-\omega\tau) {d^3k_1\over                                 
2\varepsilon_{k_1}}{d^3k_2\over 2\varepsilon_{k_2}}2(\omega\cd p) 
d\tau.
\end{equation} 
Substituting the wave function (\ref{wfp1}) into the integral 
(\ref{na3}), we get:  
$$N_2^{\lambda'\lambda}=e^{*\lambda'}_{\mu}(p)I^{\mu\nu} 
e^{\lambda}_{\nu}(p)$$ 
where 
\begin{equation}\label{na4} 
I^{\mu\nu}={m\over (2\pi)^3}\int Tr\{\phi^{\mu}(\hat{k}_2+m) 
\phi^{\nu}(\hat{k}_1-m)\}\delta^{(4)}(k_1+k_2-p-\omega \tau) 
{d^3k_1\over 2\varepsilon_{k_1}}{d^3k_2\over 
2\varepsilon_{k_2}}2(\omega\cd p) d\tau\ , 
\end{equation} 
and $\phi$ is given by eq.(\ref{nz2}). 

Substituting in (\ref{nor9}) the integral given by  (\ref{na4}) with the wave
function (\ref{nz2}), one can obtain the  normalization condition for the
deuteron wave function in terms  of the components $\varphi_{1-6}$.  Expressing
the functions $\varphi_{1-6}$ through $f_{1-6}$ by the formulae  (\ref{ba5}) -
(\ref{ba10}) given in appendix  \ref{dcomp}  we can  represent it in terms of
the $f_{1-6}$ functions.  It is however simpler to do this directly in terms of
$f_{1-6}$. For this goal, consider the integral (\ref{na3}) in the  reference
system where $\vec{k_1}+\vec{k_2}=0$. The deuteron wave  function is given by
(\ref{nz8}). We then get for $N_2^{\lambda'\lambda}$, in  the basis where
$\lambda',\lambda=x,y,z$:
\begin{equation}\label{norm5}
N_2^{ij}={m\over (2\pi)^3}\int Tr\{\psi^{\dag}_i(\vec{k},\vec{n}) 
\psi_j(\vec{k},\vec{n})\}{d^3k\over\varepsilon_k}\ .
\end{equation}
The general structure of the integral (\ref{norm5}) is given by 
(\ref{nor8b}) and $A_2$ is found from (\ref{nor9}):
\begin{equation}\label{norm7}
A_2=\frac{1}{3}\delta_{ij}N_2^{ij}\ .
\end{equation}
Substituting in (\ref{norm5}) the wave function (\ref{nz8}), we get:
\begin{equation}\label{norm4} 
A_2=\frac{m}{(2\pi)^2}\int_0^\infty \frac{k^2dk}{\varepsilon_k}\int_{-1}^1 
I\ dz\ ,
\end{equation}
where
\begin{eqnarray}\label{norm3} I&=&f_1^2 
+f_2^2 +f_3^2 +f_4^2(3+z^2) + 
f_5^2(1-z^2) +f_6^2(1-z^2) \nonumber\\ 
&&- f_2 f_3(1-3z^2) +4f_2f_4z+4f_3f_4z
\end{eqnarray}
and $z=\cos(\widehat{\vec{q}\vec{n}})$.  
 In the case where the kernel of the interaction, $V$, is independent of M,
one has $A_2=1$. Keeping the dominating contribution  in the region $k\ll m$,
where the functions $f_{3-6}$ are negligible  and $f_1=u_S,\,f_2=-u_D$, we
recover the non-relativistic normalization  condition for the deuteron wave
function:   $${1\over 2\pi^2}\int (u_S^2+u_D^2)k^2dk=1\ .$$

\subsection{Equation for the wave function}\label{eq} 
Keeping all the 
spin indices, the equation for the wave function 
${\mit\Phi}^{\lambda}_{\sigma_1\sigma_2}$ has the form already indicated in 
eq.(\ref{2.16}). In terms of the variables $\vec{k}$ and
$\vec{n}$,  this equation can be rewritten as:  
\begin{eqnarray}\label{2.16b} 
\left[4(\vec{k}\,^2 + m^2)-M^2\right] \vec{\psi}_{\sigma_1\sigma_2} 
(\vec{k},\vec{n}) &=& -\frac{m^2}{2\pi^3} \\ &\times& \sum \int 
\left[\vec{\psi}'(\vec{k}\,',\vec{n}) 
\sigma_y\right]_{\sigma'_1\sigma'_2} 
V^{\sigma'_1\sigma'_2}_{\sigma_1\sigma''_2} 
(\vec{k}\,',\vec{k},\vec{n},M) (\sigma_y)_{\sigma''_1\sigma_1} 
\frac{d^3k'}{\varepsilon_{k'}} 
\nonumber 
\end{eqnarray} 
According to our discussion of section \ref{toto}, the 
function $\vec{\psi}'_{\sigma_1\sigma_2}$ is obtained from \newline
$\mit\Phi^{\lambda}_{\sigma'_1\sigma'_2}(k'_1,k'_2,p,\omega\tau')$ simply 
by expressing the arguments $k'_1,k'_2,p,\omega\tau'$ through 
$\vec{k},\vec{k}',\vec{n}$, as given by eqs.(\ref{4.10}-\ref{4.7b}).  
Expressing the  function $\vec{\psi}$ through 
${\mit\Phi}^{\lambda}_{\sigma_1\sigma_2}$ with the help of the $D$-matrices, in 
eq.(\ref{nz3}), the wave function
$\vec{\psi}'(\vec{k}\,',\vec{n})$ can then be  obtained from 
$\vec{\psi}(\vec{k}\,',\vec{n})$ by a rotation of spins according to:                                                                 
\begin{equation}\label{3.11a} 
\vec{\psi}'= D^{(\frac{1}{2})\dag}\{R(L^{-1}({\cal 
P}'),k_{2}')\}\vec{\psi} D^{(\frac{1}{2})}\{R(L^{-1}({\cal 
P}'),k_{1}')\},                               
\end{equation}                                                                  
where the matrix $D^{(\frac{1}{2})}\{R(L^{-1}({\cal P}),k)\}$ is 
defined by (\ref{nz4}).  

\section{Connection with the Bethe-Salpeter amplitude}\label{bsd}
The Bethe-Salpeter function of the deuteron $\Phi (p,k_1,k_2)$ is a $ 
4\times 4$ matrix depending on the two off-shell nucleon momenta $k_1$ 
and $k_2$ and the on-shell deuteron momentum $p=k_1+k_2$ with $p^2=M^2$.  
Both nucleons in the vertex $d\rightarrow np$ are final, whereas, 
as a convention, the 
index $\alpha $  in the matrix $\Phi _{\alpha \beta }$ corresponds to the final
nucleon and the index $\beta $  corresponds to the initial nucleon.  It
is therefore convenient to separate  from $\Phi ^{BS}$ at the right the charge
conjugation matrix $U_c$:   
\begin{equation} \label{bsd1}
\Phi=\psi ^{BS}U_c 
\end{equation} 
In order to construct the Bethe-Salpeter function $\psi ^{BS}$, we will 
start with the ``initial'' expression $a\hat \xi +b(k\cd\xi )/m$ and 
then multiply it by $ \hat k_1$ and $\hat k_2$ from right and left
respectively and 
from both sides.  Here $\xi \equiv e^{\lambda}(p)$ is the deuteron 
polarization vector with  $p\cd\xi=0$, and $ k=(k_1-k_2)/2$. We thus 
obtain:  
\begin{equation}\label{bsd2} 
\psi ^{BS}=\left[a_1\hat \xi +b_1\frac{(k\cd\xi )}m\right] 
+\left[a_2\hat \xi +b_2 \frac{(k\cd\xi )}m\right] \frac{\hat 
k_1}m+\frac{\hat k_2}m\left[a_3\hat \xi +b_3\frac{ (k\cd\xi 
)}m\right]+\frac{\hat k_2}m\left[a_4\hat \xi +b_4\frac{(k\cd\xi 
)}m\right]\frac{\hat k_1}m 
\end{equation} 
The functions $a_i,b_i$ depend on the scalars $k\cd p$ and $k^2$.  One 
can see from this equation that the deuteron Bethe-Salpeter function, 
$\psi ^{BS}$, contains eight spin structures.  This fact agrees with 
another classification \cite{ztjonall,gammel,kubis} using the partial 
waves $^{2S+1}L^{\rho}_J$, where the index $\rho$ corresponds to the so 
called $\rho$-spin, which operates on the positive- and negative-energy 
states exactly in the same manner as usual spin operates on the usual 
spin states.  In these terms, the Bethe-Salpeter function 
of the deuteron is determined by the following eight coupled waves 
\cite{ztjonall}:  
\begin{equation}\label{clas} ^3S^+_1,\quad ^3D^+_1,\quad                
^3S^-_1,\quad ^3D^-_1,\quad ^1P^e_1,\quad ^3P^o_1,\quad ^1P^o_1,\quad           
^3P^e_1.  
\end{equation} 
The indices $``+"$ and $``-"$ mean ``up" ($++$) and ``down" ($--$) 
$\rho$-states of both nucleons, the indices $e$ (even) and $o$ (odd) 
mean the triplet and singlet $\rho$-spin states.  From the Pauli 
principle it follows that the function $\Phi $ satisfies the symmetry 
relation:  $\Phi _{\alpha \beta }(k_1,k_2,p)=\Phi _{\beta \alpha 
}(k_2,k_1,p)$.  Hence, $\psi ^{BS}(1,2)=-U_c[\psi ^{BS}(2,1)]^t U_c$.  
To provide the symmetry relation in a simplest way, we transcribe 
(\ref{bsd2}) in the form:                                                                    
\begin{eqnarray} \label{bsd3}
\psi ^{BS}(k_1,k_2,p)= g_1 \frac{(k\cd\xi )}m+ g_2\hat \xi + 
g_3\frac{(\hat \xi \hat k_1-\hat k_2\hat \xi +2m\hat \xi )}{m} + 
g_4\frac{(k\cd\xi )}{m^2}(\hat k_1-\hat k_2+2m) 
\nonumber\\ 
+g_5\frac{(\hat \xi \hat k_1+\hat k_2\hat \xi )}{m}+ \frac{(\hat 
k_2-m)}m\left[g_6\hat \xi +g_7\frac{(k\cd\xi )}m\right] \frac{(\hat 
k_1+m)}m +g_8\frac{(k\cd\xi ) }{m^2}(\hat k_1+\hat k_2)\ . 
\end{eqnarray}  
The symmetry implies:  $g_i(k\cd p,p^2)= g_i(-k\cd p,p^2)$ for $i=1 
-4,6,7$ and \newline $g_{5,8}(k\cd p,p^2)=-g_{5,8}(-k\cd p,p^2)$.  

The relation between the Bethe-Salpeter and the light-front wave 
function is given by eq.(\ref{bs8}).  We show below that 
(\ref{bs8}) together with the deuteron Bethe-Salpeter function (\ref{bsd3}), 
coincides with the deuteron light-front wave function (\ref{nz2}).  

We substitute (\ref{bsd3}) in (\ref{bs8}). The 
momenta $k_1,k_2$ in eq.(\ref{bsd3}) are off the mass shell. 
However, after the following substitution in (\ref{bs8}):
$$k_1\rightarrow l_1=k_1-\omega\tau/2+\omega\beta,\quad
k_2\rightarrow l_2=k_2-\omega\tau/2-\omega\beta
$$
we have to put $k_1,k_2$ on mass shell. Due to that, we may use the 
Dirac equation $\bar u(k_2)(\hat{k}_2-m)= (\hat{k}_1+m)U_c\bar 
u(k_1)=0$.  After applying the Dirac equation, the function $\psi^{BS}$ 
in the integrand of (\ref{bs8}) writes:  
\begin{eqnarray}\label{bsd5}                                                    
\psi^{BS}(k_1-\omega\tau/2+\beta\omega, k_2-\omega\tau/2-\beta\omega,p) 
\rightarrow 
g_1\frac{1}{m}[(k\cd\xi)+(\omega\cd\xi)\beta] 
+g_2\hat{\xi} 
\nonumber\\ 
+g_3\frac{1}{m}[2(\omega \cd\xi)\beta
-\frac{1}{2}(\hat{\xi}\hat{\omega}-\hat{\omega}\hat{\xi})\tau]
+g_4\frac{1}{m^2}[(k\cd\xi)+(\omega\cd\xi)\beta]2\beta\hat{\omega} 
\nonumber\\ 
+ g_5\frac{1}{m}[(\hat{\xi}\hat{\omega}-\hat{\omega}\hat{\xi})\beta
-(\omega\cd\xi)\tau]
-g_6\frac{2(\omega\cd\xi)}{m^2}\hat{\omega}(\beta^2-\tau^2/4) 
-g_8\frac{1}{m}[(k\cd\xi)+(\omega\cd\xi)\beta]\hat{\omega}\tau\ .  
\end{eqnarray} 
The function $g_7$ does not contribute. One can see that the structures 
at $g_{1,2,4,6,8}$ in (\ref{bsd5}) coincide with the corresponding 
structures at $\varphi_{1-4,6}$ in (\ref{nz2}).  The expression 
$(\hat{\xi}\hat{\omega}-\hat{\omega}\hat{\xi})$ at $g_3$ and $g_5$ in 
(\ref{bsd5}) can be transformed to the form 
$\gamma_5e_{\mu\nu\rho\gamma}\xi_{\mu} p_{\nu}k_{\rho}\omega_{\gamma}$ 
at $\varphi_5$ in (\ref{nz2}).  For this aim we decompose 
$\hat{k}_2(\hat{\xi}\hat{\omega} -\hat{\omega}\hat{\xi})\hat{k}_1$ in 
the complete set of $4\times 4$ matrices:  
\begin{equation}\label{bsd6}         
M\equiv \hat{k}_2(\hat{\xi}\hat{\omega} 
-\hat{\omega}\hat{\xi})\hat{k}_1 = A +B\gamma_5 + C^{\mu}\gamma_{\mu} + 
D^{\mu}i\gamma_{\mu}\gamma_5 + E^{\mu\nu}i\sigma_{\mu\nu}\ . 
\end{equation} 
 The coefficients are given by:  
$A=\frac{1}{4}Tr(M)$, $B=\frac{1}{4}Tr(\gamma_5M)$, $2E_{\mu\nu} 
=\frac{1}{4}Tr(i\sigma_{\mu\nu}M)$ and $C_{\mu}=D_{\mu}=0$.  
Sandwiching eq.(\ref{bsd6}) between the spinors $\bar u(k_2)$ and 
$U_c\bar u(k_1),$ and using again the Dirac equation, we
find:                                          
\begin{eqnarray}\label{bsd7}                                                    
\bar u_2(\hat{\xi}\hat{\omega} -\hat{\omega}\hat{\xi})U_c\bar u_1 =             
\frac{4}{s}\bar u_2[i\gamma_5e_{\mu\nu\rho\gamma}                               
\xi_{\mu}k_{1\nu}k_{2\rho}\omega_{\gamma} -(k\cd\xi)(\omega                       
\cd p)-\hat{\xi}m(\omega \cd p)  \nonumber\\ 
+\frac{1}{2}(s-M^2)(x-\frac{1}{2})(\omega\cd\xi) 
+\frac{1}{2}(s-M^2)m\frac{\hat{\omega}(\omega\cd\xi)}{(\omega\cd p)}] 
U_c\bar u_1 \ ,
\end{eqnarray} 
where, as usual, $s=(k_1+k_2)^2=4(\vec{k}^2+m^2)$. This structure 
contributes to $\varphi_5$ and to other structures. By this way, we  
reproduce exactly the six structures in the light-front deuteron wave function 
(\ref{bsd5}). Comparing their coefficients, one can easily find the 
expressions for $\varphi_i$ in terms of the integral of $g_i$ over $\beta$.  The
arguments of the functions $g_i=g_i(k^2,p\cd k)$ are replaced  in the integrand 
by: 
\begin{eqnarray*}
 k^2&\rightarrow&  -\vec{k}\,^2+2m^2(x-\frac{1}{2})\beta',
 \\
p\cd k&\rightarrow &-\frac{1}{2}(x-\frac{1}{2})(s-M^2)+m^2\beta',  
\end{eqnarray*} 
where we introduce  the dimensionless variable $\beta'=\beta(\omega \cd
p)/m^2$, and, hence, depend on $\vec{k}\,^2$,
$\vec{n}\cd\vec{k}=-2\varepsilon_k (x-1/2)$. This  provides the explicit link
between Bethe-Salpeter and light-front wave  functions.  

\section{Two nucleon wave function in the $J^{\pi}=0^+$ scattering              
state}                                                                          
\label{1s0} 

This relativistic scattering state wave function has been            
calculated in ref. \cite{ck-fsi}.  Before coming to the relativistic case        
we remind some definitions for the non-relativistic continuous spectrum         
wave function in coordinate and momentum spaces.                                
\subsection{Non-relativistic wave function}\label{fsi}                          
                                                                                
We start with the known S-wave continuous spectrum wave function
in  coordinate space, $\psi_p(r)$, having the usual
asymptotical behavior:                    
\begin{equation}\label{fsi1}                                                    
\psi_p(r)\vert_{r\rightarrow\infty}\approx\psi_p^A(r) ={\sin(pr)\over           
pr} + f\frac{e^{ipr}}{r} = e^{i\delta}\frac{\sin (pr+\delta)}{pr}\ ,             
\end{equation}                                                                  
where $f=e^{i\delta}\sin\delta/p$ is the scattering                             
amplitude.  The corresponding energy is $E_p=p^2/m$.{                       
                                                                                
The Fourier transform\footnote{We use the same notation both for the            
wave function $\psi_p(r)$ in coordinate space and for its Fourier            
transform $\psi_p(k)$.  The difference is indicated by the arguments.}        
$\psi_p(k)$ of $\psi_p(r)$ is expressed through the off-shell scattering    
amplitude $f(k,p)$:                                                                      
\begin{equation}\label{fsi4} \psi_p(k)=\int                                  
\psi_p(r)\exp(-i\vec{k}\cd\vec{r})d^3r = \frac{2\pi^2}{p^2} \delta(k-p)+        
\frac{4\pi f(k,p)}{k^2-p^2-i\epsilon}\ ,                                        
\end{equation}                                                                  
where the S-wave off-shell amplitude $f(k,p)$ satisfies the Lippmann-Schwinger
equation. The variable $k$ is the argument of the Fourier transform, whereas
$p$ is related to the eigenvalue $E_p$.        

Because of the asymptotical behavior (\ref{fsi1}), valid for any short-range      
interaction, the momentum space wave function (\ref{fsi4}) has a                
singular term at $k=p$ and a quickly decreasing off-energy shell                
dynamical part $f(k,p)$. 

It is convenient to extract the global phase factor from the wave 
function:                            
\begin{equation}\label{fsi2} 
\psi_p(r)=e^{i\delta(p)}\varphi_p(r)              
\end{equation} 
and from the amplitude:
\begin{equation}\label{fsi5}                                                    
f(k,p)=e^{i\delta(p)}g(k,p)\ .                                                  
\end{equation} 
The function $\varphi_p(r)$ is then real and has the asymptotical form:  
\begin{equation}\label{fsi3} \varphi_p(r)\vert_{r                 
\rightarrow \infty} \approx \varphi_p^A(r)= \frac{\sin(pr+\delta)}{pr}\ .       
\end{equation} 
The function $g(k,p)$ is real with $g(p,p)=\sin\delta/p$.                                                
                                                                                
The Fourier transform of $\varphi_p^A$ thus gives:                                  
\begin{equation}\label{fsi6} \varphi_p^A(k)=\int                                
\varphi_p^A(r) \exp(-i\vec{k}\cd\vec{r})d^3r =\frac{2\pi^2}{p^2} 
\delta(k-p)+ \frac{4\pi f}{k^2-p^2-i\epsilon}\ ,                                             
\end{equation}                                                                  
where $f= f(p,p)$ is the on-shell scattering amplitude given above.  
Subtracting (\ref{fsi6}) from (\ref{fsi4}), one finds:  
\begin{equation}\label{fsi7}         
g(k,p)=\frac{\sin\delta}{p}+ (k^2-p^2)\int_0^{\infty}(\varphi_p(r)-             
\varphi_p^A(r))\frac{\sin(kr)}{k}r dr\ .                                        
\end{equation}                                                                  
By the relations (\ref{fsi4}) and (\ref{fsi5}), $g(k,p)$ determines the 
continuous spectrum wave function in momentum space.  The difference, 
$\varphi_p(r)-\varphi_p^A(r)$, decreases faster  for $r$ greater than 
the interaction range (exponentially for the Yukawa-type potentials) 
and the integral (\ref{fsi7}) can be easily computed.  

\subsection{Spin structure of the relativistic wave function in LFD}                                                     
This wave function is the relativistic light-front generalization of 
the non-relativistic $np$ wave function in the $^1S_0$ state.  The 
partial scattering amplitude (scattering through $J^{\pi}=0^+$ state) can be 
factorized as follows:                       
\begin{equation}\label{3.2} {\cal                                               
F}^{\alpha_1\alpha_2}_{\sigma_1\sigma_2}= e^{i\delta} {\cal                     
A}_{\sigma_1\sigma_2} \chi^{\dag}_{\alpha_1\alpha_2}(^1S_0)\ , \end{equation}      
where $\chi(^1S_0)$ is the final $^1S_0$-state spin function on energy          
shell normalized to 1 :  \begin{equation}\label{3.3}                            
\chi_{\alpha_1\alpha_2}(^1S_0)=\frac{1}{2\sqrt{2}m}
 \overline{u}^{\alpha_1}(p_1)      
\gamma_5 U_c\overline{u}^{\alpha_2}(p_2)\ , 
\end{equation}                      
                                                                                
In the non-relativistic limit in the center-of-mass 
system of the final nucleons, $\chi(^1S_0)$ turns into its non-relativistic 
counterpart:                                                                               
\begin{equation}                                                                
 \chi(^1S_0) \approx \frac{i}{\sqrt{2}}w^{\dagger}_1\sigma_yw^{\dagger}_2 =                     
 C^{00}_{\frac{1}{2}\sigma_ 1\frac{1}{2}\sigma _2}                              
 w^{\dagger}_{\sigma _1}(1)w^{\dagger}_{\sigma _2}(2)\ ,                                        
 \label{eld5}                                                                   
 \end{equation}                                                                 
where $w_\sigma$ represents the two-component nucleon spinor (see appendix
\ref{nota}), 
and $C^{00}_{\frac{1}{2}\sigma_1\frac{1}{2}\sigma_2}$ is the Clebsh-Gordan 
coefficient.                                                                 

The amplitude ${\cal A}$ does not depend on the spin projections of the 
final nucleons $\alpha_1$ and $\alpha_2$.  We have separated the phase 
factor $e^{i\delta}$, where $\delta$ is the $^1S_0$ phase shift.  We 
expand the off-energy shell amplitude ${\cal A}$ for forming the 
$0^+$-state in the most general invariant amplitudes~\cite{ck-fsi}:  
\begin{equation}\label{3.4} 
{\cal A}=\overline{u}_1\gamma_5 U_c 
\overline{u}_2 a_1 + \overline{u}_1\left[\frac{2m\hat{\omega}}{\omega 
\cd p}- \frac{4m^2}{s}\right] \gamma_5 U_c \overline{u}_2 a_2\ , 
\end{equation} 
where $s=(k_1+k_2)^2$, $\overline{u}_{1,2}= 
\overline{u}^{\sigma_{1,2}}(k_{1,2})$ and $a_{1,2}$ are scalar 
functions.  In the representation (\ref{nz3}) (without the matrix 
$D^{1*}$), the amplitude (\ref{3.4}) can be decomposed as 
follows~\cite{ck-fsi}:  
\begin{equation}\label{fsi9} 
G_{\sigma_1\sigma_2}= \frac{1}{\sqrt{2}} w_{\sigma_1}^{\dagger} \left(g_1+ 
\frac{i\vec{\sigma}\cd [\vec{k}\times \vec{n}]}{k}g_2\right) 
\sigma_yw_{\sigma_2}^{\dagger}\ , 
\end{equation} 
where the scalar functions $g_{1,2}$ depend on $k$ and 
$z=\cos\widehat{\vec{n}\vec{k}}$.  This procedure is completely 
analogous to the one used above for the deuteron wave function. 
The relativistic amplitude (\ref{fsi9}) contains not only the 
$^1S_0$ state with spin zero (the first item), but also a state with 
spin 1 (the second item).  Therefore, it can be misleading to use the 
notation $^1S_0$. We refer here to the $J^{\pi}=0^+$ state.  From the 
Pauli principle it follows:  
\begin{equation}\label{3.9b} 
g_{1,2}(k,z)=g_{1,2}(k,-z) \ 
.\end{equation} 
The relations between the invariant amplitudes in the representations 
(\ref{3.4}) and (\ref{fsi9}) are the following:  
\begin{equation}\label{3.9a} 
a_1=\frac{\sqrt{2}}{\pi}g_1,\,\,\, a_2= 
\frac{\sqrt{2}\varepsilon_k^2}{\pi mk}g_2\ .  
\end{equation} 

\subsection{Equation for the wave function} 
The inhomogeneous equation          
for the relativistic light-front amplitude has a form  similar to
that determining the bound state wave function            
(\ref{2.16}):                                                                   
\begin{eqnarray}\label{fsi10}
  &&{\cal F}(p_1,p_2;k_1,k_2,\omega\tau) = 
{\cal K}(p_1,p_2;k_1,k_2,\omega\tau) 
+ \frac{1}{(2\pi)^3}\int {\cal F}(p_1,p_2;k_1',k_2',\omega\tau')
\nonumber\\ 
&&\times 
{\cal K}(k_1',k_2',\omega\tau';k_1,k_2,\omega\tau) 
\delta^{(4)}(k_1'+k_2'-\omega\tau'-k_1-k_2+\omega\tau)                          
\frac{d\tau'}{(\tau'-i\epsilon)}
\frac{d^3k_1'}{2\varepsilon_{k_1'}}         
\frac{d^3k_2'}{2\varepsilon_{k_2'}}\ .  
\nonumber\\
&&
\end{eqnarray} 
The delta-function, together with the integration on one of the momenta, is
here kept for symmetry.                         
Substituting here the amplitude (\ref{3.2}) and using the variables 
$\vec{k}, \vec{n}$ in the system of reference where $\vec{{\cal P}}=0$ 
we find:  
\begin{equation}\label{3.10} 
\frac{1}{\sqrt{2}}(g_1+ \frac{i\vec{\sigma}\cd 
[\vec{k}\times\vec{n}]}{k}g_2)= 
-\frac{m^2}{4\pi\varepsilon_k}e^{-i\delta}\chi V 
-\frac{m^2}{32\pi^2\varepsilon_k} \int\frac{{\cal 
A}V}{(k'^2-p^2-i\epsilon)}\frac{d^3k'}{\varepsilon_{k'}}\ .                     
\end{equation} 
with $V=-{\cal K}/4m^2$.  The products $\chi V$ and 
${\cal A}V$ contain summation over spin indices.  

\section{Numerical results}\label{numer-d}                                      
                                                                                
                                                                                
In order to have a first estimate of the relativistic
corrections to the deuteron wave function, one can calculate the
components $f_1-f_6$ perturbatively, starting from the non-relativistic 
solution $u_S$ and $u_D$:
\begin{equation}\label{3.11}                                                    
\vec{\psi}_{\sigma_1\sigma_2}(\vec{k},\vec{n}) =                                
\frac{-m^2}{2\pi^3(4(\vec{k}\,^2 +m^2)-M^2)}                                    
\sum_{\sigma'_1\sigma'_2}\int                                                   
(\vec{\psi}'_{NR}\sigma_y)_{\sigma_1'\sigma'_2}                                 
V^{\sigma'_1\sigma'_2}_{\sigma_1\sigma''_2}                                     
(\vec{k}\,',\vec{k},\vec{n},M) (\sigma_y)_{\sigma''_2\sigma_2}                  
\frac{d^3k'}{\varepsilon_{k'}}, 
\end{equation} 
where $\vec{\psi}'_{NR}$ is given by eq.(\ref{3.11a}) with the replacement 
of $\vec{\psi}$ by $\vec{\psi}_{NR}$ defined by (\ref{nz10}).  
The approximation (\ref{3.11}) is based on the fact that due to the small 
deuteron binding energy, the deuteron wave function is concentrated at 
non-relativistic values of $k'$ and the integral in the r.h.s.
of (\ref{2.16b}) is dominated by the same domain.  From eq.(\ref{3.11}) 
one can find $f_{1-6}$ (see ref.~\cite{ck-deut} for more details).  

The comparison of eq.(\ref{3.11}) with the equation to be used with the 
Bonn-Q models (as written in eq.(\ref{sst5}) for instance) indicates that they
are  identical in the limit where $\vec{n}$ dependent terms and boost 
effects under the integral are omitted.  In this limit, the light-front 
wave function should be equal to 
$\sqrt{\varepsilon_k/m}\ \psi^{Bonn-Q}(k)$.  This one could therefore 
provide an approximate zeroth order in solving eq.(\ref{3.11}).  

The same approximate solution can be found for the $^1S_0$ scattering state
\cite{ck-fsi}. In this case, one replaces $\cal A$ in the integrand of
eq.(\ref{3.10}) by:
\begin{equation}\label{fsi11}  {\cal A}=\frac{\sqrt{2}}{\pi}
\bar{u}'_1\gamma_5 U_c  \bar{u}'_2 g_1(k',p)
\end{equation}  
with $g_1(k',p)=(\varepsilon_{k'}/m) g(k',p)$ and $g(k',p)$ given in 
(\ref{fsi7}). 

We use the  one-boson-exchange interaction with the parameters of the Bonn-QA
model  \cite{bonn}. This one was used in earlier calculations
\cite{ck-deut,ck-fsi} and in view of the exploratory character of the present
work, we have not  used a more realistic model (section \ref{beyond}). The
non-relativistic deuteron wave function was also taken from  ref. \cite{bonn}.
For the $^1S_0$ scattering  state,  we use the Paris potential \cite{paris} and
the wave function was taken from \cite{lois}.  

\begin{figure}[hbtp] 
\epsfxsize = 10.cm 
\centerline{\epsfbox{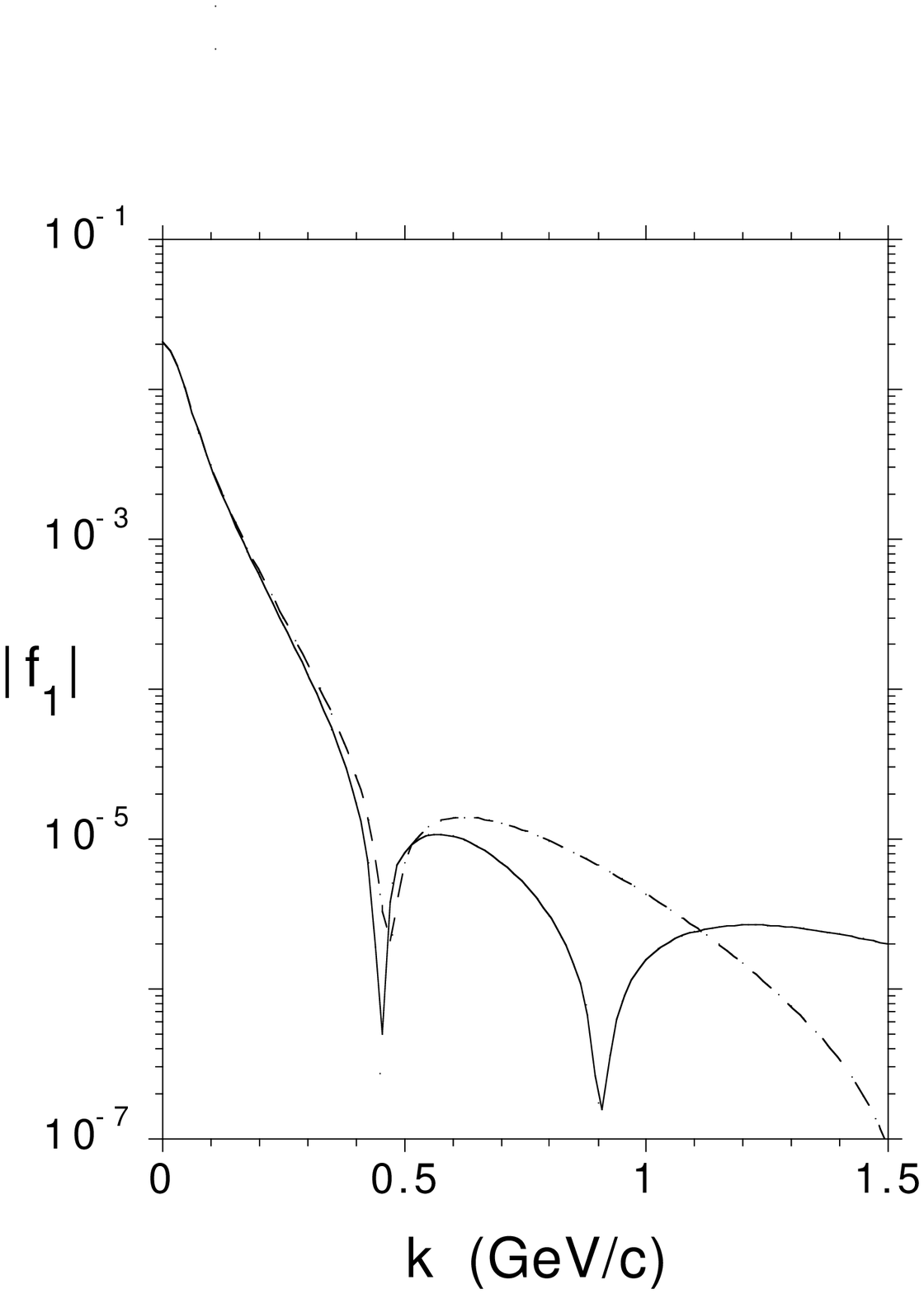}}
\figcap{ The function $f_1$, in (MeV/c)$^{-3/2}$, in absolute value,  as a
function of  $k$ at a fixed value of
$z=\cos\widehat{\vec{n}\vec{k}}=0.6$.  $f_1$ is positive
at the origin and changes sign twice. The dashed line is the non-relativistic
S wave function calculated with the Bonn potential, and the solid line 
is the relativitic result calculated in perturbation theory (see text).}
\label{ck-f1}
\end{figure}

Results for the deuteron components \cite{ck-deut} are displayed in
figs.\ref{ck-f1}-\ref{fckd4}. All the wave functions  are normalized by
$\sqrt{A_2}$, as given by (\ref{norm4}). In figs.\ref{ck-f1}  and \ref{ck-f2}
the functions $f_1(k,z)$ and $f_2(k,z)$ in  (MeV/c)$^{-3/2}$ are shown for a
fixed value of $z=0.6$, with $z=\cos(\widehat{\vec{k}\vec{n}})$, in the region 
$0\leq k \leq 1.5$ GeV/c.  They  are compared to their non-relativistic
counterparts (dashed lines). All the functions are shown together in  fig.
\ref{fckd4} as a function of $k$.

\begin{figure}[hbtp] 
\epsfxsize = 10.cm 
\centerline{\epsfbox{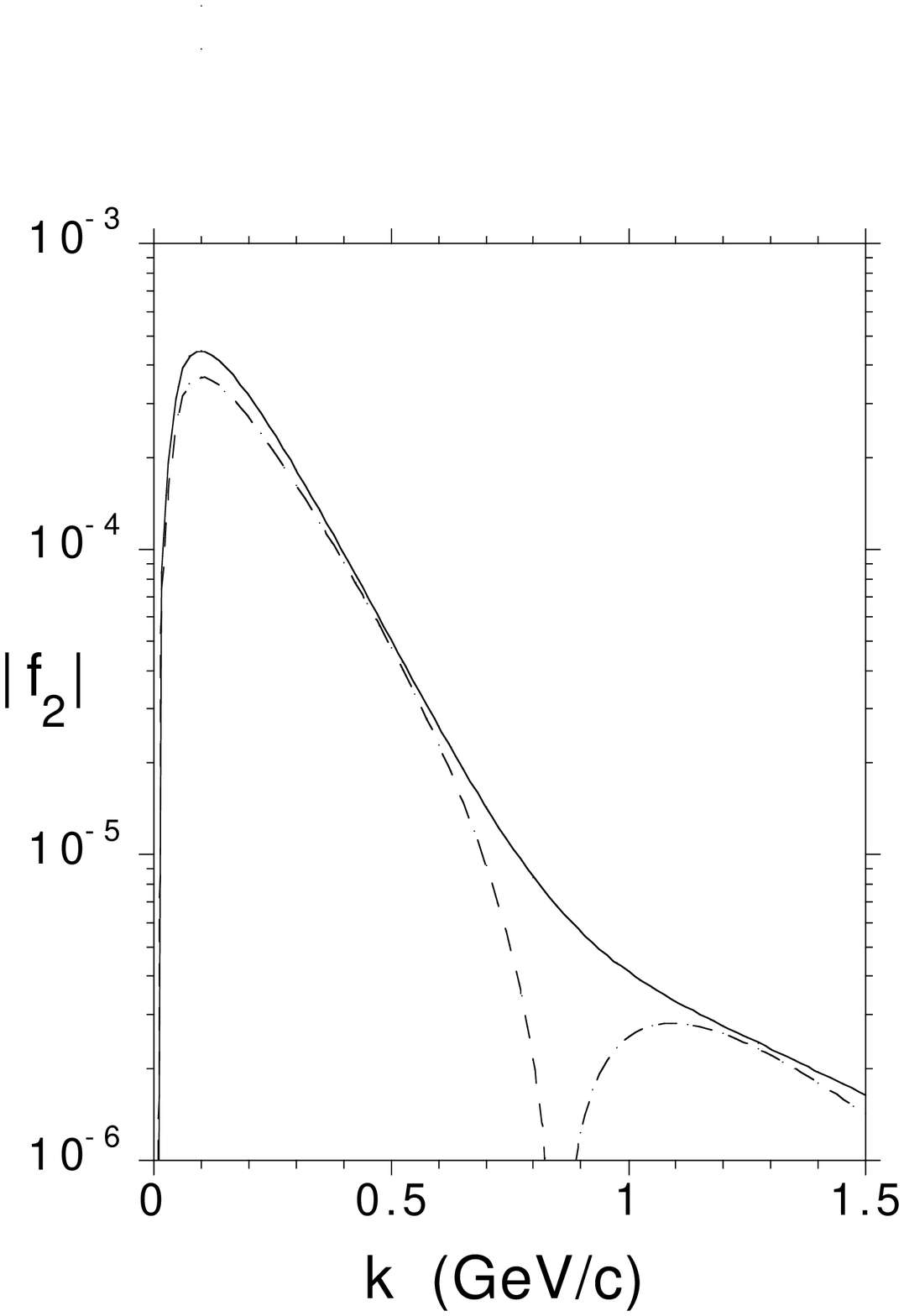}}
\figcap{The same as in fig. \protect{\ref{ck-f1}}, 
but for the function $f_2$.}
\label{ck-f2}
\end{figure}

\begin{figure}[hbtp] 
\epsfxsize =12.cm 
\centerline{\epsfbox{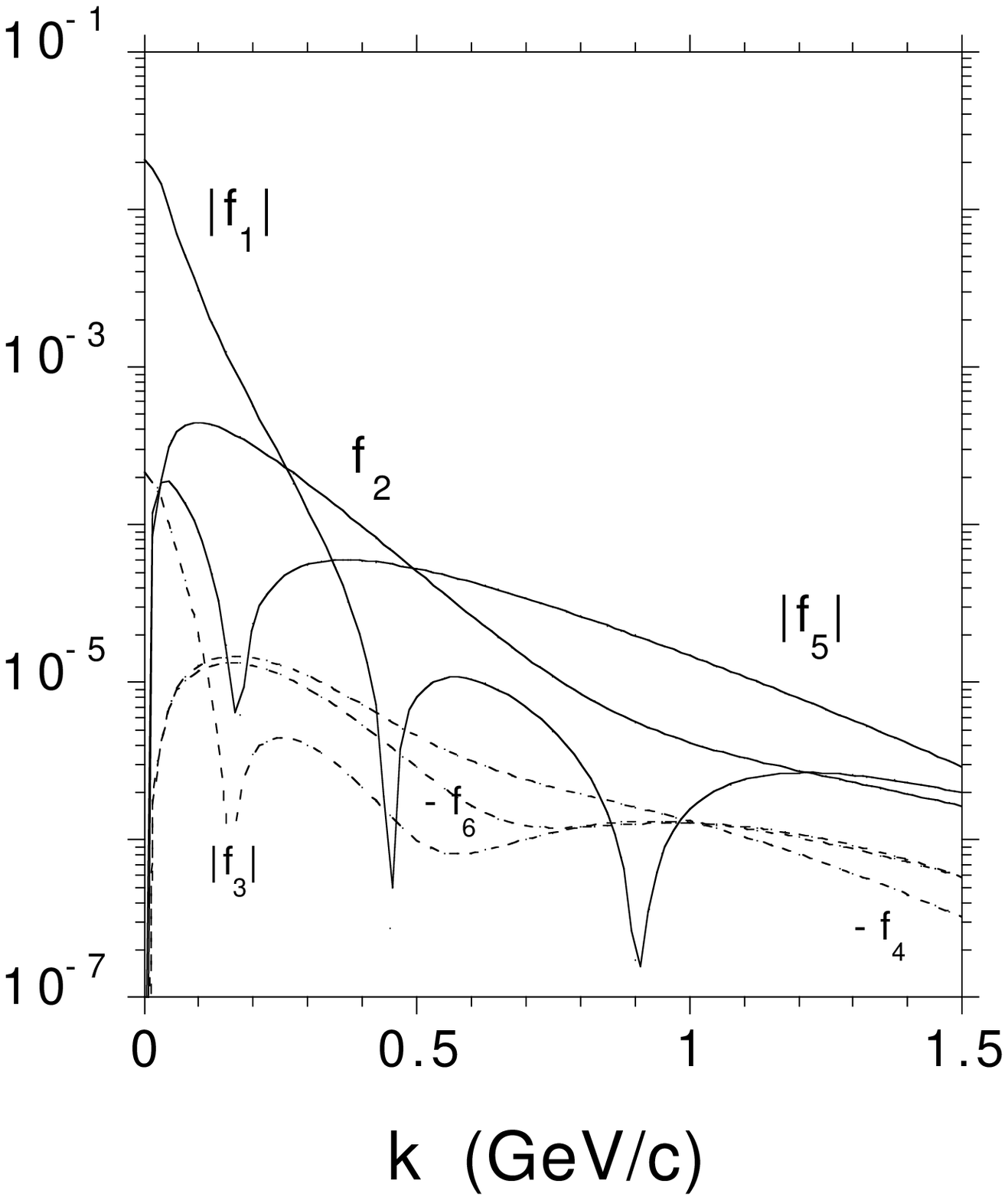}}
\figcap{The relativistic components $f_1-f_6$, in (MeV/c)$^{-3/2}$, as a 
function of $k$, at $z=0.6.$}
\label{fckd4}
\end{figure}

The wave functions $g_{1,2}$ \cite{ck-fsi} for the $^1S_0$ scattering state, 
for $E_p=1.5$ MeV ($p=37.52$ MeV/c), in (MeV/c)$^{-1}$, are shown in 
fig. \ref{fckd5}.

\begin{figure}[hbtp] 
\epsfxsize = 10.cm 
\centerline{\epsfbox{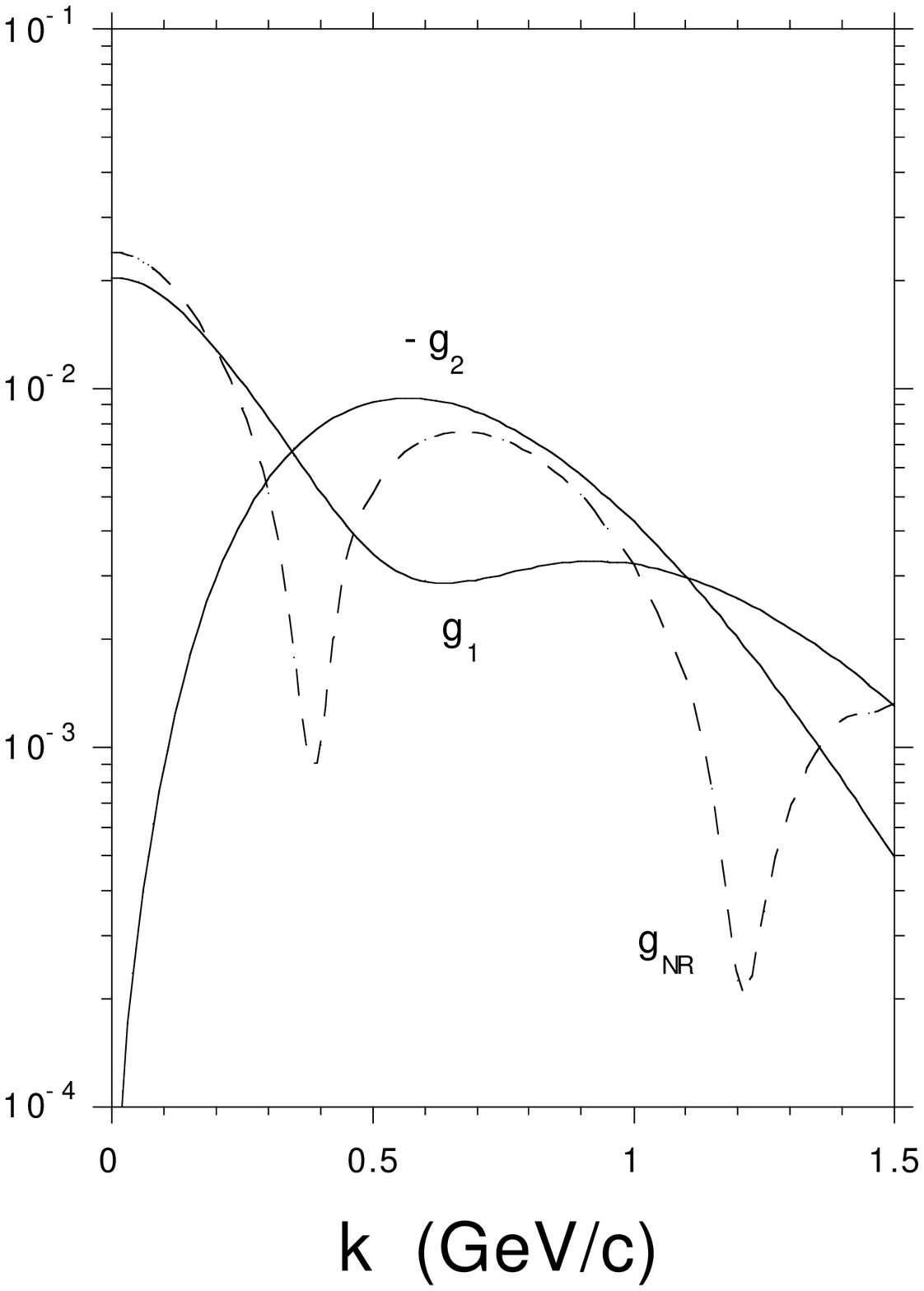}}
\figcap{The relativistic 
wave functions $g_{1,2}$ of the $^1S_0$ scattering state, in (MeV/c)$^{-1}$,
for $E_p=1.5$ MeV, at $z=0.6$ (solid line). The dashed line is the
non-relativistic $^1S_0$ scattering state wave function.}
\label{fckd5}
\end{figure}

In view of the results given in refs. \cite{ck-deut,ck-fsi}, one can make 
the following remarks.  
\begin{enumerate} 
\item 
In the region $k\geq$ 0.4-0.5 GeV/c, the  components $f_5$ of the deuteron wave
function, and $g_2$ of the $^1S_0$ scattering state are comparable to the
non-relativistic components. They are dominant above $k=0.5$ GeV/c. These
components are calculated here with the pseudo-scalar representation at the
$\pi NN$ vertex. We shall come back to this point in the next
chapter.                      

\item
The functions $f_1, -f_2$ and $g_1$ are close to  the input non-relativistic
wave functions in the region $0<k<0.5$ GeV/c. They are however strongly
modified  for larger momenta.  This indicates that relativistic corrections are
important in this region. Some of the differences at low $k$, for $f_2$ for
instance, may disappear once the low energy $NN$ scattering data or the
asymptotic normalization, $A_D$ and $A_S$, are required to be produced. 

\item Calculating the phase shift by means of the relativistic 
function, $g_1(p,p)=\sin\delta/p$, we get $\delta=48^\circ$ in 
comparison with the value $61^\circ$ found from the non-relativistic 
input Paris wave function.  The deviation of $13^\circ$ indicates that 
the relativistic effects on the low-energy phase shift are sizeable and 
advocate for a new fit of the parameters in the relativistic kernel.  This 
deviation shows the actual uncertainties of the relativistic kernels.  

\item
The relative contribution  of the
relativistic components $f_{3-6}$ to the normalization integral is of the order 
of 1\%.        
                                                                                
\item 
At $k$ of the order of 1 GeV/c  all the components of the
wave function strongly depend on $z$.  This is in contrast with the
components $f_1, f_2$ and $g_1$ in the non-relativistic, small $k$, region.
                                                                                
\item The $\vec{n}$ dependence of the wave function is a property of the wave
function  related to the off-energy shell amplitude.  At $k=p$ the scattering
amplitude  should not depend on $\vec{n}$, otherwise it would violate
rotational  invariance.  In practice this dependence could appear due to 
approximations.  It was found in \cite{ck-fsi} that, at $k=p\approx 
37.5$ MeV/c, $g_2$ is negligible compared to $g_1$ and the 
$z$-dependence of $g_1$ is of the order of $10^{-3}$.  These facts 
indicate that in this respect these calculations are self-consistent.            
\end{enumerate}                                                                 
                                                                                
The existence of the relativistic extra components $f_5$ and $g_2$ which
dominate already at moderate momenta may seem surprising given the success of
the non-relativistic phenomenology in the two nucleon systems.  This fact is
indeed related to the dominance of meson exchange currents in the deuteron 
electrodisintegration amplitude. It shows that the calculation of the 
light-front wave function omitting its $n$-dependence cannot be considered as 
realistic. We shall come back to this point in  section
\ref{lfdmec}, where the physical significance of these components will be
investigated in leading order in a $1/m$ expansion.


\chapter{The electromagnetic amplitude}
\label{ema}                               
The physical amplitude and electromagnetic form factors should not depend on a
particular choice of orientation of the light-front plane. This is however only
the case in a given order of perturbation theory or in an exact calculation. In
any approximate calculation, the light-front electromagnetic amplitude depends
on the orientation of the  light front, i.e., on the four-vector $\omega$. This
dependence is non-physical for any on-shell amplitude. We will show in this
chapter, for systems with spins 0, 1/2, 1 and transitions between them, how to
extract the physical form factors from the $\omega$-dependent electromagnetic
vertex. The covariant formulation of LFD will prove to be very powerfull in 
resolving the  ambiguities arising from the light-front orientation. 

\section{Factorization of the electromagnetic amplitude}\label{facema}                        
                                                                                
The form factors are experimentally extracted from the  scattering amplitude of
a  charged particle (usually the electron) from a bound system. One of the 
contribution to the amplitude is shown graphically in  fig. \ref{f1-ks92-a}.

\begin{figure}[hbtp]
\centerline{\epsfbox{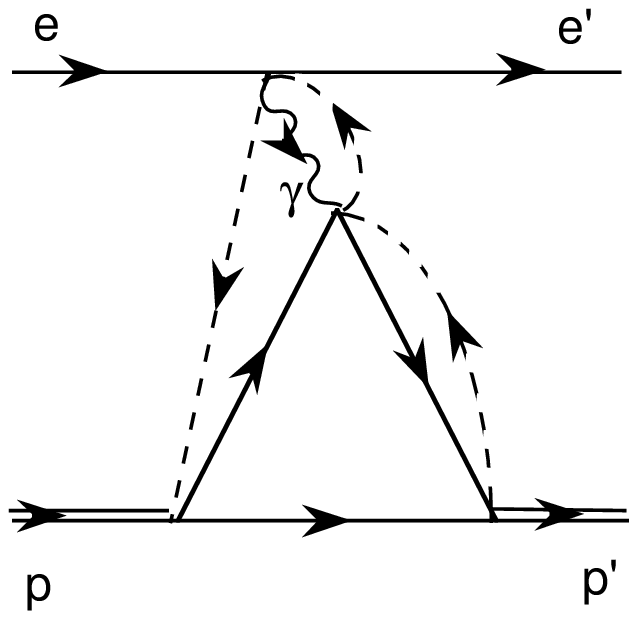}}
\figcap{Particular amplitude corresponding to the scattering of an electron from 
a bound system.}
\label{f1-ks92-a}
\end{figure}

For simplicity, we consider a two-body system to  start with, but all our
results are obviously applicable to many-body  systems. In the Feynman
approach, the amplitude of this process is  factorized in the product of the
electromagnetic vertex of the  projectile, the propagator $1/q^2$ of the
virtual photon and the  electromagnetic vertex of the bound system. This latter
is then expressed  through the physical form  factors. The complete set of
light-front diagrams should, of  course, be factorized and reproduces this
structure. However, the separate  amplitude represented in fig. \ref{f1-ks92-a}
does not in general factorize in  this formalism. Indeed, the expression for
this amplitude, corresponding to  the photon propagator and the adjoining
spurion lines has the form:                                                    

\begin{equation}\label{fac1}                                                    
M=\int (\cdots) \theta\left(\omega\cd (p'-p)\right)\delta\left((p'-p
+\omega\tau_3   -\omega\tau_2)^2\right)
\frac{d\tau_2}{\tau_2-i\epsilon}                        
\frac{d\tau_3}{\tau_3-i\epsilon}\ .                                             
\end{equation}                                                                 
The dots represent the parts of the amplitude which contains the matrix
element of the current $J_{\rho}$. They will be discussed in the next section. 

For definiteness, we suppose that $\omega\cd (p'-p)>0$.  Integrating
eq.(\ref{fac1}) over $d\tau_3$ by means of the  $\delta$-function, we get, with 
$q\equiv p'-p$:       
\begin{equation}\label{fac2}                                                    
M=\int(\cdots) \frac{d\tau_2} {\left[2(\omega                                   
\cd q)\tau_2-q^2-i\epsilon\right](\tau_2-i\epsilon)}\ .                            
\end{equation}                                                                  
This amplitude is not a product of the $1/q^2$ term with                        
the form factors.                                                               

\begin{figure}[htbp]
\centerline{\epsfbox{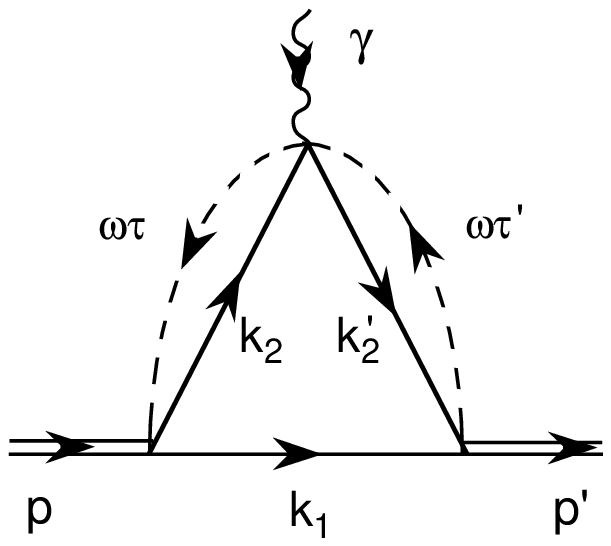}}
\figcap{Electromagnetic vertex of a bound system.}
\label{f1-ks92-b}
\end{figure}
                                                                                                    
However, as seen from (\ref{fac2}), the factorization can be obtained  under
the special condition 
\begin{equation}\label{fac3} 
\omega\cd (p'-p)=\omega\cd q=0\ , 
\end{equation} 
implied as the limit $\omega \cd  q\rightarrow +0$. Under this condition, the
amplitude (\ref{fac2}) obtains the form $M \sim f(q)/q^2$. The integration over
$d\tau_2$ in (\ref{fac2}) is absorbed by the form factors. The corresponding 
electromagnetic vertex can then be represented by fig. \ref{f1-ks92-b} with the
off-mass-shell photon momentum $q^2\neq 0$. In the standard formulation of LFD
with $\omega =(1,0,0,-1)$, the condition (\ref{fac3}) turns into the usual
condition $q_+=0$. This  condition can only be achieved for $q^2\le 0$.  

\section{Extracting the physical form factors}\label{pff}  

The problem of extracting the physical form factors from the elementary 
electromagnetic amplitude in the LFD is rather general  and appears for a
system with any spin.  As it was discussed in chapter  \ref{cov-graph}, the sum
of all the on-energy-shell amplitudes in a  given order of perturbation theory 
does not depend on the four-vector  $\omega$, whereas any given amplitude can
be $\omega$-dependent. This is so also for the on-energy shell electromagnetic
amplitudes.  The amplitude corresponding to a  separate diagram can indeed 
depend on
$\omega$. The diagram of  fig. \ref{f1-ks92-a} for instance is not the only one
contributing to the  form factor. Other contributions are shown in fig.
\ref{f2-ks92}.

\begin{figure}[hbtp]
\epsfxsize = 14.cm
\epsfysize = 5.cm
\centerline{\epsfbox{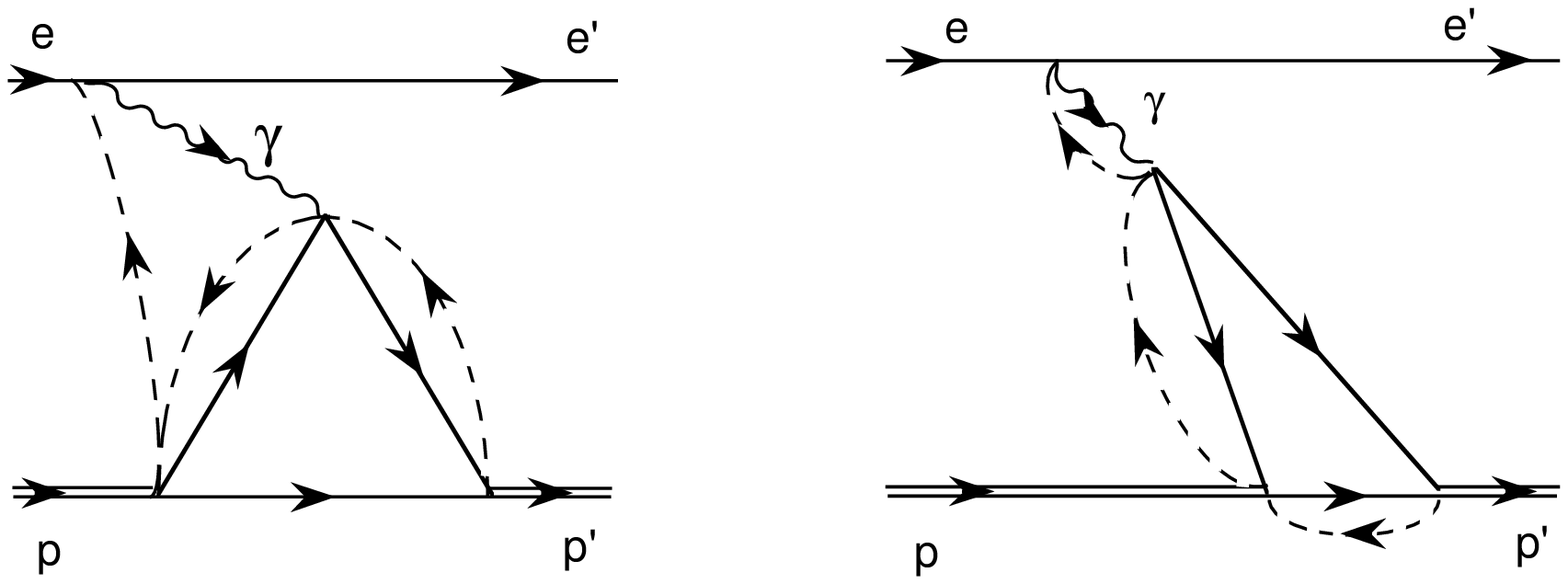}}
\figcap{Amplitude which cannot be expressed through the bound state wave 
function.}
\label{f2-ks92} 
\end{figure}

  The  matrix element of the exact current, $J_{\rho}=\langle
p'|J_{\rho}(0)|p\rangle$, can be represented as the sum of  all possible
contributions:
\begin{equation}\label{eq2}
 J_{\rho} = J^{(1)}_{\rho} + J^{(2)}_{\rho} +\cdots +
J^{(n)}_{\rho} +\cdots,
\end{equation}
where the superscripts is a short notation for indicating the various levels of 
approximation. 
In the case of the deuteron form factors for instance, the terms in 
(\ref{eq2}) correspond to the impulse approximation, meson exchange 
current contribution, etc. Any contribution in (\ref{eq2}) 
consists of the sum of two parts: a $\omega$ independent one denoted by
$F^{(n)}_{\rho}$, and a $\omega$ dependent one denoted by 
$B^{(n)}_{\rho}(\omega)$: 
\begin{equation}\label{eq3a}
J^{(n)}_{\rho} = F^{(n)}_{\rho} +B^{(n)}_{\rho}(\omega)\ .
\end{equation}
The explicit form of eq.(\ref{eq3a}) for  systems with spin 0, 1/2 and 1
is given below in sect. \ref{lfem}. 

Since the matrix element of the exact current does not depend on 
$\omega$, the $\omega$-dependent parts cancel each other:
\begin{equation}\label{eq4p}
B^{(1)}_{\rho}(\omega) +B^{(2)}_{\rho}(\omega) +\cdots 
+B^{(n)}_{\rho}(\omega) +\cdots =0\ ,
\end{equation}
and, hence, the matrix element is determined by the sum of the 
$\omega$-independent parts only:
\begin{equation}\label{eq5a}
 J_{\rho} =  F^{(1)}_{\rho} +  F^{(2)}_{\rho} +\cdots +
 F^{(n)}_{\rho} +\cdots\ .
\end{equation}

It follows that {\it the problem of restoring the independence of any physical 
amplitude on $\omega$ is reduced to the separation, and omission, of 
the ``pure" $\omega$-dependent terms}.

We emphasize that this definition of the form factors, after separating  out
$\omega$-dependent contributions, still leads  to approximate form  factors. 
However, our procedure guarantees that the form factors do  not receive any
spurious, non-physical contributions.  Since these  spurious,
$\omega$-dependent, contributions can be as large as the  physical ones, in
particular at high momentum transfer, our procedure  is the most adequate when
one wants to draw any conclusion on a  particular model, or to compare various
relativistic approaches. Moreover, it leads to an unambiguous determination
of the form factors,  independently of the orientation of the light front or of
the choice of  the component of the current which is used to calculate the
amplitude. This is  not the case in the usual formulation on the light-front
plane $t+z=0$, as it is already known for the deuteron form factors
\cite{ks92,ks94}. 

In the case of spin 0 and 1 systems, the criterion of irreducibility  for the
$\omega$-dependent contribution is rather evident (though with a subtle point
discussed in sect. \ref{ff}). It can be  read directly from the decomposition
of the current: the terms  proportional to $\omega$ do not contribute to the
physical form  factors. In the case of spin 1/2 particle, the unambiguous
separation  of the purely $\omega$-dependent part requires some care and is 
discussed in section \ref{ver}.

In a more general language, the extraction of the physical form factors 
can be explained as follows.  The physical electromagnetic amplitude 
$J_\rho \equiv \langle p'|J_{\rho}(0)|p\rangle $ does not depend on the 
hypersurface where the state vector $|p\rangle$ is defined and, hence, 
does not contain any term proportional to $\omega_{\rho}$, provided the 
current operator $J_{\rho}(0)$ is the full operator. This current 
operator has to contain the interaction.  Indeed, the commutation 
relation between $J_{\rho}(0)$ and the four-dimensional angular 
momentum $\hat{J}_{\alpha\beta}$ is similar to (\ref{kt25}), since it is 
universal for any four-vector:  
\begin{equation}\label{emv10}
\frac{1}{i}[J_{\rho}(0),\hat{J}_{\alpha\beta}] =g_{\rho\beta}J_{\alpha}(0)
-g_{\rho\alpha}J_{\beta}(0)\ . 
\end{equation}
Since $\hat{J}_{\alpha\beta}$ in (\ref{emv10}) contains the interaction, 
the current $J_{\rho}(0)$ has to contain the interaction too.  The 
consistency of the transformation properties of the current and of the 
state vector ensures the independence of the matrix element 
of the current $J_{\rho}(0)$ on the quantization plane orientation.

In practical calculations however, this consistency is violated twice:  
{\it i}) the diagram of fig. \ref{f1-ks92-b} corresponds to the free 
current and {\it ii}) the state vector is approximated by its two-body 
component.  Due to these reasons, the vertex $J_{\rho}$ obtains spurious
contributions proportional to $\omega_{\rho}$.  They are however
completely explicited in our covariant formulation, while they are hidden in
the usual formulation of the LFD with $t+z=0$.

\section{Light-front electromagnetic vertex}\label{lfem}                                    
                                                                                                                                      
The calculation of any amplitude with composite systems in the initial and
final states is done according to the rules of the graph technique discussed in
section \ref{lfgt}, with the coupling constants at the vertices replaced by the
vertex functions $\Gamma$ and $\Gamma^{\dagger}$ introduced in section
\ref{toto} for the final and initial systems respectively. We detail in this 
section how to
perform the calculations for spin 0, 1/2 and 1 bound states. 

\subsection{Spin 0 system}\label{sp0s}
Let us  consider the structure of the vertex of fig. \ref{f1-ks92-b} 
for a system of spin 0. It has the general form:                                    
\begin{equation}\label{emv1}                                                    
\tilde{J}_{\rho}=(p+p')_{\rho}F(Q^2) + \omega_{\rho} \frac{(p+p')^2} 
{2\omega           
\cd p} B_1(Q^2)\ ,  
\end{equation} 
in which the factor $(p+p')^2/(2\omega\cd p)$ is separated for convenience. The
tilde on top of the electromagnetic amplitude $\tilde{J}_{\rho}$  here and
below indicates that this amplitude is an approximate one and contains
$\omega$-dependent non-physical contributions, as compared to the exact
physical amplitude with no  tilde. As usual, we have $Q^2=-q^2 \equiv
-(p'-p)^2$. The invariant  functions $F$ and $B_1$ could depend in  principle
on $\omega \cd p$ and $\omega\cd p'$.  However, the  four-vector $\omega$ is
defined up to an arbitrary number, and, hence,  the theory is invariant
relatively to the replacement $\omega\rightarrow  \alpha\omega$, where $\alpha$
is a number. The form factors $F$ and  $B_1$ can therefore depend only on the
ratio $\omega \cd p'/\omega\cd  p$. But since $\omega\cd q=0$, we have $\omega
\cd p'/\omega\cd p =1$, and the functions $F$ and $B_1$ (and similar
functions for the  case of spin 1/2  and 1) depend on $Q^2$
only.                 

The general physical electromagnetic amplitude of a spinless system is  given
by: 
\begin{equation}\label{emv2}                                                
J_{\rho}\equiv\langle p'|J_{\rho}(0)|p\rangle =(p+p')_{\rho}F(Q^2)\ .                         
\end{equation}                                                                  
The main difference of the amplitude (\ref{emv1}) with respect to  (\ref{emv2})
is the presence of an additional contribution, proportional to 
$\omega_{\rho}$. Note that the expression (\ref{emv1}) satisfies the 
conservation of the current $\tilde{J}_{\rho}q^{\rho}=0$ 
since $\omega \cd  q=0$.  This
conservation is automatically fulfilled in this particular  case due to the
simplicity of the structure (\ref{emv1}), but it cannot be  imposed {\it a
priori}.  We will see below that for a system with spin  1, some
$\omega$-dependent terms violate indeed current
conservation.                   

In the simple case where the constituents are also spinless 
(Wick-Cutkosky model), the amplitude indicated in fig. \ref{f1-ks92-b} 
reads:                    
\begin{eqnarray}\label{emv3}                                                    
\tilde{J}_{\rho} &=&\int \left[(p+\omega\tau_2-k_1)_{\rho}+ 
(p'+\omega\tau_1-k_1)_{\rho}\right] \theta\left(\omega\cd 
(p-k_1)\right) \delta\left((p+\omega\tau_2-k_1)^2-m^2\right)\nonumber\\ 
&\times &\theta\left(\omega\cd (p'-k_1)\right) 
\delta\left((p+\omega\tau_1-k_1)^2-m^2\right) \theta(\omega\cd k_1) 
\delta(k_1^2-m^2)
\nonumber\\
&\times &\Gamma_1\Gamma_2\frac{d\tau_1}{(\tau_1-i\epsilon)}         
\frac{d\tau_2}{(\tau_2-i\epsilon)}\ \frac{d^4k_1}{(2\pi)^3} \ ,
\end{eqnarray} 
where the vertex functions $\Gamma_1,\Gamma_2$ depend on the momenta in 
the vertices 1 and 2. Integrating by means of the $\delta$-functions 
over $d\tau_1\ d\tau_2\ dk_0$, and expressing $\Gamma$ through $\psi$ 
by means of eq.(\ref{eqwf2}), we get:  
\begin{equation}\label{emv4}                                                    
\tilde{J}_{\rho}=\frac{1}{(2\pi)^3}\int\frac{(p+p'+\omega\tau_1 +\omega\tau_2 
-2k_1)_{\rho}}{(1-\omega\cd k_1/\omega\cd p)^2}\ \psi' \psi\ 
\theta\left(\omega\cd (p-k_1)\right) \frac{d^3k_1}{2\varepsilon_{k_1}}\ 
,          
\end{equation}                                                                  
where $\tau_1=\left[m^2-(p-k_1)^2\right]/(2\omega\cd (p-k_1))$ and                   
$\tau_2=\left[m^2-(p'-k_1)^2\right]/(2\omega\cd (p'-k_1))$                           

From (\ref{emv4}) one clearly sees that $J_{\rho}$ contains terms  proportional
to $\omega_{\rho}$ and, hence, has the structure of  eq.(\ref{emv1}). However,
to avoid any misunderstanding, we emphasize  that the contribution
$\omega\tau_1 +\omega\tau_2$ is not the only source of  $\omega$-dependence.
Even in the absence of this term and {\em in the case  where the wave function
$\psi$ does not depend on $\vec{n}$}, the term  proportional to $\omega_{\rho}$
survives because of momentum conservation at the  electromagnetic vertex.  

In the spinless case, the prescription to extract the physical contribution  to
the form factor, $F(Q^2)$, from  the $\omega$-dependent amplitude 
$\tilde{J}_\rho$ can be obtained immediately by multiplying both sides of
eq.(\ref{emv1}) by $\omega_{\rho}$. We thus get:                       

\begin{equation}\label{emv9} 
F(Q^2)=\frac{\tilde{J} \cd\omega}{2\omega\cd p}\ .          
\end{equation}                                                                  
With (\ref{emv4}), (\ref{emv9}) we obtain:
\begin{equation}\label{ff0}
F(Q^2)= \frac{1}{(2\pi)^3} \int \psi(\vec{R}_{\perp}^2,x)
\psi((\vec{R}_{\perp}-x\vec{\Delta})^2,x)\frac{d^2R_{\perp}dx}{2x(1-x)}.
\end{equation}
We have represented here, and in the following, the four-momentum transfer $q$ 
by $q=(q_0,\vec{\Delta},\vec{q}_{\|})$ with 
$\vec{\Delta}\cd\, \vec{\omega}=0$ and $\vec{q}_{\|}$ is parallel to 
$\vec{\omega}$. Since $\omega \cd q=0$, we have $Q^2=-q^2=\vec{\Delta}^2$.

In the usual light-front formulation, with $\omega
=(1,0,0,-1)$, eq.(\ref{emv9}) corresponds to expressing the form factor
through  the $\tilde{J}_+$ component. This is well known, and 
eq.(\ref{ff0}) has been found in ref. \cite{gbb72}. However, 
this procedure cannot be extended to the calculation of physical 
form factors of systems with total spin 1/2 and 1, as we shall see in
the next sections.

One can similarly calculate for comparison the non-physical form factor
$B_1$:
\begin{eqnarray}\label{ffb1}
B_1(Q^2)=\frac{2}{(2\pi)^3}\int \frac{(\vec{R}_{\perp}^2+m^2-M^2(1-x)^2
-x\vec{R}_{\perp}\cd\vec{\Delta}+Q^2/2)}{(4M^2+Q^2)}
\nonumber\\
\times\psi(\vec{R}_{\perp}^2,x)
\psi((\vec{R}_{\perp}-x\vec{\Delta})^2,x)\frac{d^2R_{\perp}dx}{2x(1-x)^3}.
\end{eqnarray}
Here and below we use $M$ for 
the total mass of the system, $m$ being used for the mass of its 
constituents.

For a system with small binding energy $|\epsilon_b|\ll m$, the wave function 
is mainly concentrated at $x \sim 1/2$, $\vec{R}_{\perp}^2\sim m|\epsilon_b|$.
For such a system, the function $B_1$ at $|q^2|\ll m^2$ has smallness 
$\sim 
|\epsilon_b|/m$ in comparison with $F$ or $\sim Q^2/M^2$, if $Q^2/M^2
> |\epsilon_b|/m$. However, at $Q^2\ge m^2$  the
function $B_1$ becomes comparable with $F$. We will illustrate 
this in  
section \ref{wcm} in the Wick-Cutkosky model.

\subsection{Spin 1/2 system}\label{ver} 

The electromagnetic physical amplitude for a spin 1/2 fermion is determined by 
the  two standard form factors:  \begin{equation}\label{ver1} J_{\rho} =
F_1(Q^2)\bar{u}'\gamma_{\rho}u
+\frac{iF_2(Q^2)}{2M}\bar{u}'\sigma_{\rho\nu}q^{\nu}u \equiv
\bar{u}'\Gamma_{\rho}u\ .  \end{equation} Here 
$\bar{u}'=\bar{u}^{\sigma'}(p')$,  $u=u^{\sigma}(p)$ are the  spin 1/2 fermion
bispinors and $M$ is the fermion mass. The charge ($G_E$) and magnetic ($G_M$) 
form factors are expressed through $F_1$ and $F_2$ as follows:   
\begin{equation}\label{ver1p}
G_E=F_1-\frac{ Q^2}{4M^2}F_2\ ,\quad G_M=F_1+F_2\ .
\end{equation}

The covariant light-front spin 1/2 electromagnetic vertex was firstly 
considered in ref. \cite{ks92} and investigated in detail in ref. \cite{km96}
(see also ref. \cite{weberxu}).

Like in the case of spin 0, the dependence of the 
corresponding amplitude on the extra four-vector $\omega$ increases 
the number of independent terms in the decomposition of the amplitude.  
Besides the two  structures entering in eq.(\ref{ver1}), one can 
construct the  following three structures:
\begin{equation}\label{eq1p}
\frac{1}{\omega\cd p}\bar{u}'\hat{\omega}u\ P_{\rho}\ , \quad
\frac{1}{\omega\cd p}\bar{u}'u\ \omega_{\rho}\ ,\quad 
\frac{1}{(\omega\cd p)^2}\bar{u}'\hat{\omega}u\ \omega_{\rho}\ ,
\end{equation}
where $P=p+p'$, $\hat{\omega}=\omega_{\mu}\gamma^{\mu}$. 

Other structures, like  non-gauge-invariant ones: $\bar{u}uq_{\rho}$,
$\bar{u}\hat{\omega}uq_{\rho}$, $\bar{u}'\sigma_{\rho\nu}u\ \omega^{\nu}$ and
also $i\bar{u}'\gamma_5u\ e_{\rho\mu\nu\gamma}  P^{\mu}q^{\nu}\omega^{\gamma}$ 
(not independent from the previous three), contradict $T$-invariance.

At first glance, it seems that  all the structures in (\ref{eq1p})
exhibit  no $\omega$-independent parts, since, like in the previous
section,  they are proportional to $\omega/\omega\cd p$ in the first and
second  degree. This is indeed true for the last two structures which are 
explicitly proportional to $\omega_\rho/\omega\cd p$.

This is however not the case for the first structure 
$\bar{u}'\hat{\omega}uP_{\rho}/\omega\cd p$ which is directly 
proportional to $P_{\rho}$, as one of the physical amplitudes (after 
Gordon decomposition).  It could in principle contain a  part which has 
to contribute to the physical form factors and therefore cannot be 
simply omitted together with the full structure.  This becomes evident 
in the particular situation where $p'=p$. In this case, due to the 
identity $\bar{u}(p) \gamma_{\mu}u(p)= \bar{u}up_{\mu}/M$, we get 
$\bar{u}\hat{\omega}uP_{\rho}/\omega\cd p =\bar{u}uP_{\rho}/M$, i.e., 
a non-zero $\omega$-independent contribution only. It indicates that
this part exists also  when $p' \neq p$. 

In order to construct the appropriate structure, we demand that it is 
orthogonal to the physical contribution given in (\ref{ver1}). Since 
there are only two physical contributions, any other contribution {\it 
proportional to} $P_{\rho}$ {\it and orthogonal to them} is completely 
independent and therefore non-physical.  Thus, eliminating this 
structure from the vertex, we shall not loose any physical contribution 
to the form factors.

For the two structures $\bar u' U^{\mu} u$ and $\bar u' V^{\nu} u$, we 
define here the orthogonality by the following condition:
\begin{equation}\label{ortho}
Tr[(\hat{p}'+M)U^{\rho}(\hat{p}+M)\overline{V}_{\rho}]=0
\end{equation}
with $\overline{V}_{\rho}=\gamma_0 V_{\rho}^{\dagger}\gamma_0$.  This  
definition
is motivated by the procedure, detailed in the next subsection,  to calculate
the form factors.

The only alternative to the structure $\bar{u}'\hat{\omega}u/\omega\cd  p$ is
the term $\bar{u}'(\hat{\omega}/\omega\cd p-a)u$ with arbitrary  $a$, since all
the possible scalar structures sandwiched with the spinors are  proportional
either to $\bar{u}'\hat{\omega}u$ or to $\bar{u}'u$.  The value of $a$ is then
determined by the orthogonality conditions. This gives:  
\begin{equation}\label{la} 
a=\frac{4M}{P^2}=\frac{1}{(1+\eta)M},\ \  \mbox{where}\ \  
\eta=\frac{Q^2}{4M^2}\ .  
\end{equation}

Indeed, at this value of $a$,{O the following two orthogonality conditions 
are simultaneously satisfied: 
\begin{eqnarray}\label{eq13}
Tr\left[(\hat{p}'+M)\gamma^{\rho}(\hat{p}+M)
\left(\frac{\hat{\omega}}{\omega\cd p} 
-\frac{4M}{P^2}\right)\right] P_{\rho}=0\ ,
\nonumber\\
Tr\left[(\hat{p}'+M)i\sigma^{\rho\nu}q_{\nu}(\hat{p}+M)
\left(\frac{\hat{\omega}}{\omega\cd p} 
-\frac{4M}{P^2}\right)\right]P_{\rho}=0\ .
\end{eqnarray}
The reason why a unique value of $a$ satisfies two conditions is a 
consequence of the fact that the expression $(\hat \omega/\omega\cd p 
-1/((1+\eta)M))$, being substituted in (\ref{eq13}), is proportional to 
$\hat e$, where e is the following four-vector:  
\begin{equation} 
e=\frac{\sqrt{P^2}}{2\omega\cd p}\left(\omega-2P\frac{\omega\cd p}
{P^2}\right) \ ,
\end{equation}
orthogonal to all available four momenta:
\begin{equation}\label{eqn1pp}
e\cd p=e\cd p'=e\cd q=0\ .
\end{equation}
For convenience, we shall normalize $e$ by the condition $e^2=-1$.
Note also that the term $\bar{u}'(\hat{\omega}/\omega\cd 
p-1/((1+\eta)M))u$ is proportional to $\bar{u}'\hat{e}u$.

The two last structures in (\ref{eq1p}) are not orthogonal to 
$\gamma_{\rho}$ and $i\sigma_{\rho\nu}q^{\nu}$. However, they are 
proportional to $\omega_{\rho}$, and, hence, do not contribute to the 
$\omega$-independent structures associated with $F_1,F_2$ and vice 
versa.  In this respect, the situation does not differ from the spin 0 
and 1 cases. We emphasize that in this way the structures proportional 
to $\omega 
_{\rho}/\omega \cd p$ and $P_{\rho}$ are separated by different formal 
criterions. 

Another way to separate the physical and non-physical structures is 
the following. One can simply subtract the two physical structures with 
undefinite coefficients from 
$\bar{u}'\hat{\omega}u\ P_{\rho}/(\omega\cd p)$ and then multiplying
at the left by $u'$ and at the right by $\bar{u}$ and taking sum over
polarizations get a $4 \times 4$ - matrix 
$M_{\rho}=(\hat{p}'+M)U_{\rho}(\hat{p}+M)$. This matrix can be decomposed in 
the full set of $4 \times 4$ -matrices, similarly to (\ref{bsd6}):
$$M_{\rho}=A_{\rho}+B_{\rho}\gamma_5+C_{\rho\mu}\gamma^{\mu}+ 
D_{\rho\mu}i\gamma_5\gamma^{\mu} + E_{\rho\mu\nu}i\sigma^{\mu\nu}.$$
Now the matrix $M_{\rho}$ can be analyzed in terms of the tensors $A_{\rho},
B_{\rho\mu}$, etc. We require them to be irreducible or proportional to
$\omega$. The tensor $A_{\rho}$ turns out to be proportional to 
$P_{\rho}$, and therefore is set to zero. The tensor $B_{\rho}$ equals 
to zero. The trace of
$C_{\rho\mu}$ is set to zero in order to construct an irreducible 
tensor. The antisymmetric part of it is proportional to $\omega$.
These two conditions fix the coefficients, and we obtain the final structure 
$U_{\rho}$ which exactly coincides with 
$[\hat{\omega}/\omega\cd p-1/((1+\eta)M)]P_{\rho}$. 
With these coefficients, the 
tensor $D_{\rho\mu}$ turned out to be proportional to $\omega$, and the
tensor $E_{\rho\mu\nu}$ is irreducible and proportional to  $\omega$ 
for the antisymmetric part. In this analysis, all the three nonphysical 
structures in 
(\ref{eq1p}) are treated on an equal footing, since the tensors obtained
after decomposition of the last two are proportional to $\omega_{\rho}$
and therefore are non-physical. 

It should  also be mentioned that in the Breit system, where 
$\vec{P}=0$, the term $\bar{u}'\hat{e}u$ obtains a particularly 
transparent form. In this system, the spatial part of $e$ is the unit 
vector directed along $\vec{\omega}$:  
$\vec{e}=\vec{\omega}/\omega_0=\vec{n}$, whereas $e_0=0$.  Therefore, 
the term $\bar{u}'\hat{e}u$ becomes proportional to $\vec{n}$:  
\begin{equation}\label{breit} 
\bar{u}'\hat{e}u=- \bar{u}'(p')\vec{n}\cd\vec{\gamma}u(p) \ ,
\end{equation}
with 
$\vec{n}$ orthogonal to both $\vec{p}$ and $\vec{p}\,'=-\vec{p}$. 

In an arbitrary system of reference  one can find in $\bar{u}'\hat{e}u$ an 
$\omega$-independent part.   Omitting this full structure, we inevitably omit
also this  $\omega$-independent part. Its form depends on the criterions we
used  above (orthogonality and irreducibility). These criterions, strictly
speaking, are however a matter of convention. They are imposed as  natural and
reasonable, but are not mathematically derived from some more general
principles.

The above consideration shows that we achieve (with the  structure
$\bar{u}'\hat{e}uP_{\rho}$) optimal separation of nonphysical and physical
contributions.

Collecting the various possible contributions, we obtain the following 
expression for the spin 1/2 light-front electromagnetic  amplitude\footnote{A
similar expression for $\tilde{\Gamma}_{\rho}$ was given in  \cite{ks92}. In
this reference however, as well as in \cite{weberxu}, the  structure
$\hat{\omega} P_{\rho}/\omega\cd p$ was used instead of the  structure
proportioanl to $B_1$
in  (\ref{eq12}). In this case, the form factor $G_E(0)$ does not coincide 
with the normalization integral.} \cite{km96}:
\begin{eqnarray}\label{eq12}
\tilde{J}_{\rho}&= &\bar{u}'\tilde{\Gamma}_{\rho}u\ ,
\nonumber\\
\tilde{\Gamma}_{\rho}&=&F_1\gamma_{\rho}
+\frac{iF_2}{2M}\sigma_{\rho\nu}q^{\nu}
+B_1\left(\frac{\hat{\omega}}{\omega\cd p} -\frac{1}{(1+\eta)M}
\right)P_{\rho}
+B_2\frac{M}{\omega\cd p}\omega_{\rho}
+B_3\frac{M^2}{(\omega\cd p)^2}\hat{\omega}\omega_{\rho}\ .
\nonumber\\
& &
\end{eqnarray}
The electromagnetic vertex (\ref{eq12}) is  
gauge invariant since $\tilde{J}_{\rho}q^{\rho}=0$ (with $\omega \cd q =0$). As 
mentioned above, the 
possible non-gauge-invariant terms are  forbidden by  $T$-invariance.

\subsubsection{The physical form factors}
We can now express the physical form factors $F_1$ and $F_2$ 
through 
the full vertex function $\tilde{\Gamma}_{\rho}$. We multiply 
$\tilde{J}_{\rho}$ by
$ [\bar{u}^{\sigma'}(p')\gamma^{\rho}u^{\sigma}(p)]^*$,
$[\bar{u}^{\sigma'}(p')i\sigma^{\rho\nu}q_{\nu}/(2M) 
u^{\sigma}(p)]^*$, etc. and sum over polarizations.  We thus 
obtain the following quantities: 

\begin{eqnarray}\label{eq14}
&&c_1=Tr[O_{\rho}\gamma^{\rho}]\ ,
\quad
c_2=Tr[O_{\rho}i\sigma^{\rho\nu}q_{\nu}]/(2M)\ ,
\quad
c_3=Tr[O_{\rho}(\hat{\omega}/\omega\cd p -1/(1+\eta)M)]P^{\rho}\ ,
\nonumber\\
&&c_4=Tr[O_{\rho}]\omega^{\rho}M/\omega\cd p\ ,
\quad
c_5=Tr[O_{\rho}\hat{\omega}]\omega^{\rho}M^2/(\omega\cd p)^2\ ,
\end{eqnarray}
where
\begin{equation}\label{eq14p}
O_{\rho}=(\hat{p}'+M)\tilde{\Gamma}_{\rho}(\hat{p}+M)/(4M^2)\ .
\end{equation}
By this way, we get a linear system of five equations for 
$F_1,F_2,B_{1-3}$ with the inhomogeneous part determined  by
$c_{1-5}$.   The way of calculating $c_{1-5}$ through traces with the matrix 
$O_{\rho}$ just corresponds to the operation used in the definition of  the
orthogonality condition (\ref{ortho}).
The values of $c_{1-5}$ can  be calculated in specific models for the structure
of the composite system, as shown in the next chapter.

Solving the above mentioned system of equations relative to $F_1$ and 
$F_2$, we find: 
\begin{eqnarray}\label{eq19}
F_1&=\frac{\displaystyle{1}}{\displaystyle{4(1+\eta)^2}}
&[(c_3+4c_4-2c_1) (1+\eta)+2(c_1+c_2)-2c_5(1+\eta)^2] \ , \\ 
F_2&=\frac{\displaystyle{1}}{\displaystyle{4\eta 
(1+\eta)^2}}&[(c_3+4c_4-2c_1)(1+\eta) +2(c_1+c_2)-2(c_5+c_4) 
(1+\eta)^2]\ .  
\label{eq20} \end{eqnarray} 
In spite of $\eta$ in the denominator in eq.(\ref{eq20}) and below, 
there is no singularity at $Q^2=0$. 

From eqs.(\ref{ver1p}), we can now easily obtain $G_E$ and 
$G_M$:  
\begin{eqnarray}\label{eq21}
G_E&=&c_4/2\ , \\
\label{eq21p}
G_M&=&\frac{1}{4\eta (1+\eta)}[(c_3+2c_4-2c_1) 
(1+\eta)+2(c_1+c_2)-2c_5(1+\eta)^2 ]\ ,
\end{eqnarray}
with $c_{1-5}$ given in (\ref{eq14}).
These formulae are quite general in the light-front approach. They are
applicable to any model of the spin 1/2 fermion structure. 

\subsubsection{Comparison with other approaches}
In the usual formulation of LFD on the plane $t+z=0$,  the  form 
factors of spin 1/2 systems are found from  the plus-component of the current 
(see, e.g., 
\cite{weberxu,keist94,capstic,salme95}), i.e., with our notation, from the 
contraction of $\tilde{J}_{\rho}$ in eq.(\ref{eq12}), with $\omega_{\rho}$. 
{\it 
This contraction eliminates the contributions of $B_{2,3}$, but does 
not eliminate the term with $B_1$}.  The form factors $F_1'$ and 
$F_2'$ 
deduced in this way are thus given by:
\begin{eqnarray}
\label{eq22} \tilde{J} \cd \omega&=&\bar{u}'[F_1\hat{\omega}
+\frac{iF_2}{2M}\sigma_{\rho\nu}\omega^{\rho}q^{\nu}
+2B_1(\hat{\omega}-\frac{\omega\cd p}{(1+\eta)M})]u \nonumber\\
&\equiv&\bar{u}'[ F_1'\gamma_{\rho}
     +\frac{iF_2'}{2M}\sigma_{\rho\nu}q^{\nu}]u\ \omega^{\rho}\ .
\end{eqnarray}
where
\begin{equation}\label{eq23}
F_1'=F_1+\frac{2\eta B_1}{1+\eta}\ ,\quad 
F_2'=F_2+\frac{2}{1+\eta}B_1\ .
\end{equation}
Equation (\ref{eq22}) shows that the structure of 
$\tilde{J} \cd \omega$ (or $\tilde{J}_+$) indeed coincides with the standard 
representation of the nucleon electromagnetic vertex, eq.(\ref{ver1}). 
However, $F_1'$ and $F_2'$ in (\ref{eq22}) are not the physical 
form 
factors, but their superposition (\ref{eq23}) with the  non-physical 
contribution $B_1$.

The expression for $B_1$ can be found from the above mentioned system of
equations, leading to: 
\begin{equation}\label{eq24}
B_1=-\frac{1}{8\eta (1+\eta)}[(c_3+4c_4-2c_1)(1+\eta)+2(c_1+c_2)-4c_5
(1+\eta)^2]\ .
\end{equation}
Together with eqs.(\ref{eq23}), one has for $F_1'$ and $F_2'$:
\begin{equation}\label{eq25}
F_1'=c_5/2,\quad F_2'=(c_5-c_4)/(2\eta)\ .
\end{equation}
From eqs.(\ref{eq25}) and (\ref{ver1p}) one can find the corresponding 
$G_E'$ and $G_M'$:
\begin{eqnarray}
G_E'&=&G_E\ ,\\
\label{eq26}
G_M'&=&G_M+2B_1=[c_5(1+\eta)-c_4]/(2\eta)\ .
\end{eqnarray}
The difference between the two approaches -- the first one separating 
out all the $\omega$-dependent terms to define the physical form 
factors, and the second based on the component $\tilde{J}_+$ of the 
electromagnetic current and incorporating non-physical item with $B_1$ 
-- is in the magnetic form factor. This difference is of relativistic 
origin and has to disappear for form factors of extremely 
non-relativistic systems (i.e. in the limit where the average internal 
momenta go to zero) at finite $Q^2$.  It also disappears in an exact 
calculation leading to all $B_i=0$. We shall estimate this difference
in a simple quark model for the nucleon structure in the next chapter.

\subsubsection{Off-energy shell effects in electromagnetic form 
factors}\label{offenff}

Though all the four-momenta are on the mass shell, this does not mean  that
there is no off-shell effects in the form factors at all.  If, for example, the
nucleon form factors are used in the calculation of the deuteron form factors, 
they
are part of a larger diagram and contain external spurion lines.  They are
off-energy shell. Such a form factor is shown in  fig. \ref{offshellff}.

\begin{figure}[hbtp]
\centerline{\epsfbox{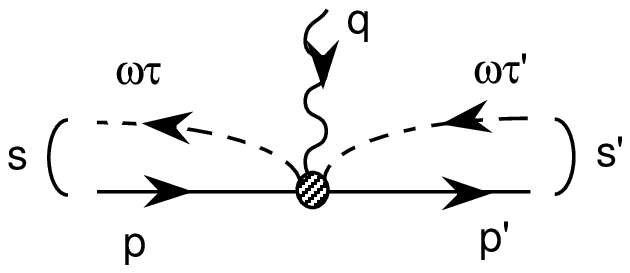}}
\figcap{Off-energy shell form factor.}
\label{offshellff}
\end{figure}

 In the
standard formulation of LFD this corresponds to the fact that the
minus-components of the nucleon momenta and the photon momentum are not related
by the conservation law. Hence, the off-energy shell form  factor can depend 
on the minus-components of the momenta, in addition to its $Q^2$-dependence. In
the  covariant formulation of LFD the graph for the off-energy shell form
factor contains external spurion lines. The off-shell effects are parametrized
in terms of the corresponding spurion  four-momenta. Besides $Q^2$, the
off-shell form factor $F$ depends therefore on two  variables \cite{karm78}:
\begin{equation}\label{off1}
F=F(Q^2,s,s'),\quad \mbox{where}\quad
 s=(p-\omega\tau)^2,\; s'=(p'-\omega\tau')^2. 
 \end{equation}
Since $\omega \cd q=0$, no other independent 
scalar products can be constructed. 

The spin structure of the off-energy shell electromagnetic vertex 
differs from  (\ref{eq12}) by  three  extra items:
\begin{eqnarray}\label{off2}
\Gamma^{off}_{\rho}=&&F_1\gamma_{\rho}
+\frac{iF_2}{2M}\sigma_{\rho\nu}q^{\nu}
+B_1\left(\frac{\hat{\omega}}{\omega\cd p} -\frac{1}{(1+\eta)M}
\right)P_{\rho}
+B_2\frac{M}{\omega\cd p}\omega_{\rho}
+B_3\frac{M^2}{(\omega\cd p)^2}\hat{\omega}\omega_{\rho}
\nonumber\\
&&+B_4\frac{(s'-s)}{M^3}q_{\rho}+
B_5\frac{(s'-s)}{M^2(\omega\cd p)}\hat{\omega}q_{\rho}+
B_6\frac{i(s'-s)}{2M(\omega\cd p)}\sigma_{\rho\nu}\omega^{\nu},
\end{eqnarray}
where all the  form factors depend on
$Q^2,s,s'$ and are symmetric relative to the permutation $s \iff s'$.  The last
three spin structures have been already mentioned after eq.(\ref{eq1p}). Their
opposite $T$-parity is corrected in (\ref{off2}) by the factor $(s'-s)$,
changing the sign due to permutation of initial and final momenta. We emphasize
that, in contrast to (\ref{eq12}), the $\omega$-dependent items in
(\ref{off2}) appear not because of approximation, but due to the off-energy 
shell continuation. Of course, in an approximate calculation of the off-energy
shell form factors these item appear due to both reasons.  The form factors
$B_{1-3}$ in an exact electromagnetic amplitude turn into zero on the energy 
shell
$\tau=\tau'=0\rightarrow s=s'=M^2$. 

The off-shell electromagnetic vertex 
of a spin-zero particle differs from eq.(\ref{emv1}) by the extra term
proportional to $(s'-s)q_{\rho}/M^2$. 

The extra form factors and the dynamical dependence of all the form factors  on
the off-shell variables $s,s'$ are usually unknown and are neglected in almost
all calculations.

\subsection{Spin 1 system}\label{ff}                                            
                                                                                
The electromagnetic form factors of a system with spin 1 (the deuteron, 
for instance) are defined by the following decomposition \cite{gj57} of 
the electromagnetic vertex:           
\begin{eqnarray}\label{ff1}                                                     
\langle\lambda'|J_\rho|\lambda\rangle &=& 
e_{\mu}^{*\lambda'}(p')\left\{P_{\rho}\left[{\cal F}_1(q^2) 
g^{\mu\nu} + {\cal F}_2(q^2){q^{\mu}q^{\nu}\over 2M^2}\right]
\right.                                                                                                 
+ \left.{\cal G}_1(q^2) (g^{\mu}_{\rho}q^\nu - 
g^{\nu}_{\rho}q^\mu)\right\}e_{\nu}^{\lambda}(p)\nonumber\\ 
&\equiv& 
e_{\mu}^{*\lambda'}(p')T^{\mu\nu}_{\rho}e_{\nu}^{\lambda}(p)\
.                
\end{eqnarray}                                                                  
Here $e_{\mu}^{\lambda}(p)$ is the deuteron polarization vector, $p$ 
and $p'$ are the initial and final deuteron momenta, 
$\lambda$ and $\lambda'$ are the corresponding helicities, 
$P=p+p'$ and $q = p' - p$. Charge 
($F_{C}$), magnetic ($F_{M}$) and quadrupole ($F_{Q}$) form factors are 
expressed through ${\cal F}_1$, ${\cal F}_2$ and ${\cal G}_1$ as 
follows:                         
\begin{eqnarray}\label{ch}                                                      
F_{C} &=& - {\cal F}_1 -{ 2\eta \over{3}}[{\cal F}_1 + {\cal G}_1               
- {\cal F}_2(1 + \eta)]\ ,\\                                                    
\label{mag}                                                                     
F_{M} &=& {\cal G}_1\ ,\\                                     
\label{q}                                                                       
F_Q &=& -{\cal F}_1 - {\cal G}_1 + {\cal F}_2(1 + \eta)\ ,                      
\end{eqnarray}                                                                  
where $\eta = Q^2/4M^2$. 

In the absence of polarization, the cross section of electron-deuteron 
scattering has the form:                                         
\begin{equation}\label{ff2}                                                    
\frac{d\sigma}{d\Omega}=\left(\frac{d\sigma}{d\Omega}\right)_0                  
\left(A(q^2)+\tan^2\frac{1}{2}\theta\; B(q^2)\right)\ , 
\end{equation}                                                                  
where $\theta$ is the electron scattering angle in the laboratory 
system, and $(d\sigma/d\Omega)_0$ is the Mott cross section:                            
\begin{equation}\label{eld8}                                                    
\left( \frac{d\sigma}{d\Omega}\right)_0 =                                       
\frac{\alpha ^2\cos^2(\theta/2)}{4E^2\sin^4(\theta/2)}\ .       
\end{equation}                                                                  
The functions $A(q^2)$, $B(q^2)$ are expressed through the form factors 
as follows: 
\begin{eqnarray}                                         
A(q^2)&=&F^2_{C}(q^2)+\frac{8}{9}\eta^2 F^2_Q(q^2)+\frac{2}{3}                           
\eta F^2_{M}(q^2)\ , \\                                                 
B(q^2)&=&\frac{4}{3}\eta(1+\eta) F^2_{M}(q^2)\ .  \label{ff3b}                                 
\end{eqnarray}                                                                  
                                                                                
Unpolarized experiments are able to separate $A$ and $B$, but do not allow one
to extract all three form factors. To do this, one should measure a 
polarization observable. Due to its spin one, the deuteron can have, besides
vector polarization,   quadrupole polarization. The corresponding observable is
$T_{20}$ given by: 
\begin{eqnarray}\label{t20}
&&T_{20}\left(A(q^2)+\tan^2\frac{1}{2}\theta\; B(q^2)\right)=
\nonumber\\
&&-\frac{1}{\sqrt{2}}\left[\frac{8}{3}\eta F_C F_Q +\frac{8}{9}\eta^2 
F_Q^2 +\frac{1}{3}\eta\left(1+2(1+\eta)  
\tan^2\frac{1}{2}\theta\right)F_M^2\right].
\end{eqnarray} 
The deuteron electromagnetic properties are reviewed in ref. \cite{lour94}.

The decomposition (\ref{ff1}) is valid for the full  electromagnetic 
current operator. Considering the one-body current $J_\rho^0$, 
as indicated in
fig. \ref{f1-ks92-b}, we get the following analytical 
expression:                   
\begin{eqnarray}                                                   
&&\langle \lambda ' \vert J_\rho^0 \vert \lambda \rangle
=                       e_{\mu}^{*\lambda'}(p')
\tilde{J}^{\mu\nu}_{\rho}e_{\nu}^{\lambda}(p)\ ,
\nonumber\\              
&&\tilde{J}^{\mu\nu}_{\rho}=\frac{m}{(2\pi)^3}\int                                      
Tr[\phi'^{\mu}(\hat{k}'_2+m)                                                    
\Gamma_{\rho}(\hat{k}_2+m)\phi^{\nu}(\hat{k}_1-m)]                              
\frac{\theta(\omega\cd k_1)\theta(\omega \cd k_2)\theta(\omega\cd
k'_2)}                {(1-\omega\cd k_1/\omega\cd
p)^2}\frac{d^3k_1}{2\varepsilon_{k_1}}\ , 
\nonumber\\ 
&&
\label{ff3a} 
\end{eqnarray}                                                                 
where $\Gamma_{\rho}$ is the electromagnetic vertex of the nucleon  given by
eq.(\ref{ver1}) (with $M$ replaced by $m$). The matrix $\phi^{\mu}$  is given
by eq.(\ref{nz2}), while the matrix $\phi'^{\nu}$ is obtained from
$\phi_{\nu}$  by the replacement $k_2\rightarrow k'_2$. This expression is
derived using the graph technique presented in chapter \ref{cov-graph}. It is
similar to the calculation indicated in sect. \ref{sp0s} for spinless
particles. The difference with (\ref{emv3}), (\ref{emv4}) is due to the
non-zero spins.  We explain below, how eq.(\ref{ff3a}) is derived. 

According to the remark in the end of sect. \ref{rule-spin}, we start in the
graph \ref{f1-ks92-b} from the outgoing deuteron line (i.e., from the  vertex
$NN\rightarrow d'$), follow the line in the direction opposite to the 
orientation of any nucleon line  and
reach its end, i.e., the vertex $d\rightarrow NN$.  Then we start again from
the vertex  $NN\rightarrow d,$ and follow another  nucleon line, also in the
direction opposite to its orientation. We thus obtain:
\begin{eqnarray}\label{ffe1}
&&e_{\mu}^{*\lambda'}(p')\left\{[\gamma_0\phi'^{\mu}U_c\gamma_0]^{\dagger}
(\hat{k}'_2+m)
\Gamma_{\rho}(\hat{k}_2+m)\phi^{\nu}U_c\right\}_{\beta\alpha}
e_{\nu}^{\lambda}(p)
\nonumber\\
&&\times (\hat{k}_1+m)_{\beta\alpha}.
\end{eqnarray}
The first row in (\ref{ffe1}) corresponds to the upper nucleon line in the
diagram, the second one to the lower line.
We attach the deuteron wave function to the upper nucleon line.
The first row contains the factor $[\gamma_0\phi'^{\mu}U_c\gamma_0]^{\dagger}$
 which originates from the 
complex conjugated final wave function of the deuteron: 
$$[e^{\lambda'}_{\mu}(p') \bar{u}(k'_2)\phi'^{\mu}U_c\bar{u}(k_1)]^*=
e_{\mu}^{*\lambda'}(p')u(k_1)[\gamma_0\phi'^{\mu}U_c\gamma_0]^{\dagger}
 u(k'_2).$$
We keep in (\ref{ffe1}) the matrix indices $\alpha$ and $\beta$ explicitly. 
Since both nucleon 
lines are equivalent, the order of indices $\beta,\alpha$ is the same.
This means that one of the factors in (\ref{ffe1}) (we take the second one) 
is the transposed matrix. With the wave function (\ref{nz2}) we get:
$$[\gamma_0\phi'^{\mu}U_c\gamma_0]^{\dagger}=-U_c\phi'^{\mu}$$
and, hence, obtain the factor: 
$$-U_c(\hat{k}_1+m)^tU_c=(\hat{k}_1-m),$$
that gives the trace in eq.(\ref{ff3a}).

Like in the case of a spinless system, the light-front vertex 
(\ref{ff3a}) depends on the four-vector $\omega$.  Its 
expansion in terms of form factors contains therefore extra terms.
 This expansion 
has the general form \cite{ks92}:                                               
\begin{equation}\label{ff4} 
\langle\lambda'|\tilde{J}_\rho|\lambda\rangle =
\frac{\displaystyle{1}}{\displaystyle{2\omega\cd p}}
e_{\mu}^{*\lambda'}(p')\tilde{J}^{\mu\nu}_{\rho}e_{\nu}^{\lambda}(p),
\quad \mbox{where} \quad
\tilde{J}^{\mu\nu}_{\rho}= T^{\mu\nu}_{\rho} + 
B^{\mu\nu}_{\rho}(\omega) \ ,
\end{equation}
where $T^{\mu\nu}_{\rho}$ has the structure given by eq.(\ref{ff1})
and is  determined by the physical form factors,
${\cal  F}_1$, ${\cal F}_2$ and ${\cal G}_1$.
We recall that tilde is used to distinguish approximate amplitudes and matrix
elements, containing $\omega$-dependent non-physical contributions, 
 from the $\omega$-independent ones. The tensor
$B^{\mu\nu}_{\rho}$ contains the $\omega$  dependent 
terms:                                                                          
\begin{eqnarray}\label{ff5}                                                     
B^{\mu\nu}_{\rho} &=& {M^2\over2(\omega\cd
p)}\;\omega_{\rho}\left[B_1 g^{\mu\nu} +
B_2{q^{\mu}q^{\nu}\over M^2} +  B_3 M^2
{\omega^{\mu}\omega^{\nu}\over (\omega \cd p)^2}
+  B_4 {q^{\mu}\omega^{\nu} -
q^{\nu}\omega^{\mu}\over2\omega\cd p}  \right]       
\nonumber\\                                                                     
&+& B_5 P_{\rho} M^2 {\omega^{\mu}\omega^{\nu}\over (\omega \cd p)^2 } 
+ B_6 P_{\rho} {q^{\mu}\omega^{\nu} - q^{\nu}\omega^{\mu} \over 
2\omega\cd p} + B_7 M^2 {g^{\mu}_{\rho}\omega^{\nu} + 
g^{\nu}_{\rho}\omega^{\mu} \over \omega \cd p} 
\nonumber\\                                                                
&+& B_8 q_{\rho} {q^{\mu}\omega^{\nu} + q^{\nu}\omega^{\mu}\over 
2\omega \cd p} \
.\end{eqnarray}                                                          
Here $P = p + p'$ and $B_1,...,B_8$  are invariant functions.   The  tensor
$B^{\mu\nu}_{\rho}$ as well as $T^{\mu\nu}_{\rho}$ is  symmetrical with
respect to the simultaneous permutations $p  \rightleftharpoons p'$ $(q
\rightarrow -q)$ and $\mu \rightleftharpoons  \nu$. The structures at
$B_7$ and $B_8$ in  eq.(\ref{ff5}) do  not satisfy the conservation of current,
that is  $q^{\rho}\tilde{J}^{\lambda'\lambda}_{\rho} \neq 0$.  

The total number of independent spin structures can be calculated  without
their explicit construction. Owing to the $\omega$-dependence,  the helicity
indices in the vertex are not constrained by the equality  $\lambda = \lambda_1
+ \lambda_2$ ($\lambda$ is the photon helicity,  $\lambda_1, \lambda_2$ are the
hadron helicities).  Therefore we have  3x3x3 = 27 initial matrix elements.
$P$-invariance reduces this number  down to 14.  $T$-invariance (or
$C$-invariance in annihilation channel)  eliminates in addition 4 structures
and finally 10 independent  structures remain.  This calculation automatically
takes into account  the conservation of the electromagnetic current which would
lead to a  relation between $B_7$ and $B_8$. However, $\omega$-dependent  terms
in  eq.(\ref{ff5}) do not obligatory satisfy this condition. The  functions
$B_7$ and $B_8$  are therefore independent and we obtain 11  structures  (or
invariant form factors): three of them are in $T^{\nu\mu}_{\rho}$  and the
remaining eight ones are in $B^{\nu\mu}_{\rho}$.  

From the point of view of the usual light-front approach with $t+z=0$,  the
dependence of the vertex (\ref{ff4}) on the orientation of the light  front
means lack of relativistic covariance. Hence, the formula  (\ref{ff1}), based
on this covariance, is not any longer valid. 
  Therefore, the  number of independent matrix elements in
eq.(\ref{ff4}) is not reduced  to 3, but is larger. The increase of the number
of matrix elements just  corresponds to the increase of the number of
independent spin  structures in eq.(\ref{ff5}). In the formulation of the
light-front with $t + z = 0$, the  contribution of these structures is not
separated out from the physical  form factors.  In the covariant formulation,
the dependence on the light front is parametrized explicitly by
eqs.(\ref{ff4}),({\ref{ff5}). 

However, like in the spin 1/2 case, and in the absence of explicit evaluation
of the $\omega$-dependent terms, 
one can see some ambiguity in the procedure: what has to be considered as 
a non-physical contribution? Consider for example the term proportional to
$B_5$: $P_{\rho} e_{\mu}^{*\lambda'}(p')\omega_{\mu} \omega_{\nu}
e_{\nu}^{\lambda}(p)/(\omega\cd p)^2$. 
Multiplying it by $e',e^*$ and summarizing
over $\lambda',\lambda$, we get:
\begin{equation}\label{ex}
P_{\rho} \sum_{\lambda',\lambda}e_{\mu}^{\lambda'}(p')
e_{\mu'}^{*\lambda'}(p')\frac{\omega_{\mu'} \omega_{\nu'}}
{(\omega\cd p)^2}e_{\nu'}^{\lambda}(p)
e_{\nu}^{*\lambda}(p)=
P_{\rho}\left(\frac{\omega_{\mu} \omega_{\nu}}
{(\omega\cd p)^2}
-\frac{p'_{\mu} \omega_{\nu}}
{(\omega\cd p)M^2}-
\frac{\omega_{\mu} p_{\nu}}
{(\omega\cd p)M^2}+
\frac{p'_{\mu} p_{\nu}}{M^4}\right).
\end{equation}
It contains, apart from $\omega$-dependent contributions, the term $p'_{\mu} 
p_{\nu}/{M^4}$.
This one cannot be removed 
from (\ref{ex}) and be included in the physical part, since it is
required by the transversality of the tensor (\ref{ex}).  The orthogonality
condition, similar to the spin-1/2 case (obtained by the generalization of
eq.(\ref{ortho})), results in the charge form factor $F_C$ which does not
coincide at $q=0$ with the
normalization integral, in contrast to the spin-1/2 case. We therefore do not 
impose 
this condition in the spin-1 case, and define the non-physical
structures as proportional to the first and higher degrees of $\omega$,
including the terms where $\omega$ is contracted with the  polarization vector.
After that, the physical  form factors can be unambiguously separated from the
spurious  $\omega$-dependent contributions. This allows one to obtain the
contribution to the physical form  factors in terms of the electromagnetic
amplitude in an explicit form by a generalization of the formula (\ref{emv9}). 
This approach is confirmed by comparison with the Bethe-Salpeter solution in
the Wick-Cutkosky model (see sect. \ref{wcm1}).

\subsubsection{Contributions to physical form factors}

The explicit expressions for the form factors are obtained by solving 
eq.(\ref{ff4}) relative to  ${\cal F}_1$, ${\cal F}_2$ and 
${\cal  G}_1$.  The procedure is described in ref. \cite{ks92}.  The
result is given 
below:                                                          

\begin{eqnarray}\label{ff6}                                                     
{\cal F}_1 &=& \tilde{J}^{\mu\nu}_{\rho}{\omega^{\rho} \over{2\omega\cd
p}} \left[g_{\mu\nu} - {q_{\mu}q_{\nu}
\over{q^2}} - {{P_{\mu}\omega_{\nu} +
P_{\nu}\omega_{\mu}} \over{2\omega\cd p}}+
P^2{{\omega_{\mu}\omega_{\nu}} \over{4(\omega \cd
p)^2}}\right] \ ,\\                
\label{ff7}                                                                     
{{\cal F}_2 \over{2M^2}}& =& -\tilde{J}^{\mu\nu}_{\rho}{\omega^{\rho} 
\over{2(\omega \cd p)
q^2}}                                                                        
\left[g_{\mu\nu} \right. 
\nonumber\\ 
 &&\left.- 2{{q_{\mu}q_{\nu}}
\over{q^2}} - {{P_{\mu}\omega_{\nu} +
P_{\nu}\omega_{\mu}} \over{2\omega\cd p}} +
M^2{{\omega_{\mu}\omega_{\nu}} \over{(\omega \cd
p)^2}}  - {{q_{\mu}\omega_{\nu} -
q_{\nu}\omega_{\mu}} \over{2\omega  \cd p}}\right]
 \ ,\\                                                                 
\label{ff8}                                                                     
{\cal G}_1 &=& {1 
\over{4}}\tilde{J}^{\mu\nu}_{\rho}\left\{2{{g_{\mu}^{\rho}q_{\nu} -      
g_{\nu}^{\rho}q_{\mu}} \over{q^2}} + {{g_{\mu}^{\rho}\omega_{\nu}
+ g_{\nu}^{\rho}\omega_{\mu}} \over{\omega\cd
p}}\right.                             
\nonumber\\                                                                     
&&+ {\omega^{\rho} \over{\omega\cd
p}} \left[-
P^2{{q_{\mu}\omega_{\nu}                                                
- q_{\nu}\omega_{\mu}} \over{2(\omega\cd p
)q^2}} + {{q_{\mu}P_{\nu} -
q_{\nu}P_{\mu}} \over{q^2}} +
P^2{{\omega_{\mu}\omega_{\nu}} \over{2(\omega\cd
p)^2}} -{{P_{\mu}\omega_{\nu} +
P_{\nu}\omega_{\mu}} \over{2(\omega\cd p)}}\right]                                                                    
\nonumber\\                                                                     
&&\left.+ P^{\rho}\left[{{q_{\mu}\omega_{\nu}
- q_{\nu}\omega_{\mu}} \over{(\omega\cd
p) q^2}} -                                   
{{\omega_{\mu}\omega_{\nu}} \over{(\omega\cd p)^2}}\right]
-  q^{\rho}{{q_{\mu}\omega_{\nu} +
q_{\nu}\omega_{\mu}} \over{(\omega \cd p)
q^2}}\right\} \ .                                                               
\end{eqnarray}                                                                  
These expressions determine the electromagnetic form factors. 
In spite of the fact that $\omega$ enters the r.h.s.
of eqs.(\ref{ff6}-\ref{ff8}), these expressions do not depend on 
$\omega$.  

The form factor ${\cal F}_1$, eq.(\ref{ff6}) which coincides at $Q^2=0$ with 
$F_C(0)$, does not coincide with the normalization integral 
(\ref{nor9}). This is due to the fact that the normalization is obtained on the
basis of  the decomposition (\ref{nor8}), where the non-physical structure
proportional to
$E$,   contains $g^{\mu\nu}$ besides $\omega^{\mu}\omega^{\nu}$. This is 
required by the
irreducibility of this structure and is compatible with the angular condition. 
In the 
decomposition of the electromagnetic vertex, we have not separated out this
term from the  terms proportional to $\omega^\mu \omega^\nu$, as we explained
above. We believe that the
concordance of $F_C(0)$ with the normalization integral is beyond the
accuracy of the current calculations which omits all the Fock components except
for the two-body one. In practical calculations of the deuteron elastic form
factors, the difference is very small. 

\subsubsection{Comparison with other approaches}\label{fff}
 
The deuteron electromagnetic form factors were already calculated in 
LFD in refs. 
\cite{chung88,fs79,glaz83,gk84,fgks89,ffs93,gian94}.
More generally,
spin-1 form factors were considered in refs.
\cite{ks92,ks94,bh92,card95,karm96}. Using the exact
decomposition (\ref{ff1}), one can express
the  matrix elements  $<\lambda'|J_+|\lambda>\equiv J_{\lambda'\lambda}$ in
terms of the form factors:
\begin{eqnarray}                                                     
J_{11}&=&-{\cal F}_1 +\eta {\cal F}_2
\label{iig1}\\  
J_{1-1}&=&-\eta {\cal F}_2
\label{iig2}\\
J_{10}&=&-\sqrt{2\eta}({\cal 
F}_1 -\eta {\cal F}_2 +{\cal G}_1/2)  
\label{iig3}\\
J_{00}&=&-(1-2\eta){\cal F}_1 -2\eta^2 {\cal F}_2 +2\eta 
{\cal G}_1\ .\label{iig4}
\end{eqnarray}
The four matrix elements in (\ref{iig1}-\ref{iig4}) can be  expressed 
through three form factors and, hence, are not independent from each 
other. They satisfy the relation \cite{gk84}:  
\begin{equation}\label{ac0}
(1+2\eta)J_{11}+J_{1-1}-2\sqrt{2\eta}J_{10}-J_{00}=0\ ,
\end{equation}
which can be checked by direct substitution. This condition is called
``angular  condition" (do not confuse with the angular condition
for the state vector discussed in  sect.\ref{ac}). 

In principle, with the exact matrix elements $J_{\lambda'\lambda}$
one can calculate any triplet of  matrix elements and 
find the form factors. It is not so however for the approximate ones 
$\tilde{J}_{\lambda'\lambda}$. The following triplets of the matrix elements 
were used in the literature:  $\tilde{J}_{11},\tilde{J}_{1-1},\tilde{J}_{00}$ 
and 
$\tilde{J}_{11},\tilde{J}_{1-1},\tilde{J}_{10}$ in ref. \cite{gk84} (the so 
called solutions A
and  B respectively); $\tilde{J}_{1-1},\tilde{J}_{10},\tilde{J}_{00}$ in 
ref. \cite{bh92}. For example, taking the
triplet  $\tilde{J}_{11},\tilde{J}_{1-1},\tilde{J}_{00}$ and solving the system 
of equations  (\ref{iig1}),(\ref{iig2}) and (\ref{iig4}) (with
$J_{\lambda'\lambda}\rightarrow \tilde{J}_{\lambda'\lambda}$)  relative to $
{\cal F}_1, {\cal F}_2,{\cal G}_1$, one finds:  
\begin{eqnarray}\label{gia} 
{\cal F}_1^{GK-A}&=&-\tilde{J}_{11}-\tilde{J}_{1-1},
\nonumber\\
{\cal F}_2^{GK-A}&=&-\tilde{J}_{1-1}/\eta,
\nonumber\\
{\cal G}^{GK-A}&=&
-\frac{1}{2\eta}\left[(1-2\eta)\tilde{J}_{11}+\tilde{J}_{1-1}-\tilde{J}_{00}
\right]\ .
\end{eqnarray} 
In refs. \cite{chung88,ffs93} the form factors were expressed through
different  combinations of four matrix elements.

The matrix elements in the r.h.s. of 
eqs.(\ref{gia}) calculated in LFD 
are however not given by the decomposition (\ref{ff1}), but
should be calculated according to eqs.(\ref{ff4}), (\ref{ff5}). The matrix
elements of (\ref{ff4}) have the  form:
\begin{eqnarray}\label{iigc}
\tilde{J}_{1,1}&=&-{\cal F}_1 +\eta {\cal F}_2\ ,
\nonumber\\  
\tilde{J}_{1,-1}&=&-\eta {\cal F}_2\ , 
\nonumber\\
\tilde{J}_{1,0}&=&-\sqrt{2\eta}({\cal F}_1 -\eta {\cal F}_2               
+{\cal G}_1/2)+\sqrt{\eta/2}B_6\ ,
\nonumber\\
\tilde{J}_{0,0}&=&-(1-2\eta){\cal F}_1 -2\eta^2 {\cal F}_2 
+2\eta {\cal G}_1-2\eta B_6 +B_5+B_7\ .                                                                     
\end{eqnarray}
The matrix elements $\tilde{J}_{11},\tilde{J}_{1-1}$ have the same form as 
$J_{11},J_{1-1}$, whereas $\tilde{J}_{10},\tilde{J}_{00}$ differ from 
$J_{10},J_{00}$  by the items containing the nonphysical form factors 
$B_5,B_6,B_7$.  Other nonphysical form factors $B_{1-4}$ and $B_8$ do not 
contribute to these matrix elements.

The matrix elements $\tilde{J}_{\lambda'\lambda}$ do not satisfy the 
condition (\ref{ac0}). Substituting $\tilde{J}_{\lambda'\lambda}$ in 
eq.(\ref{ac0}) instead of $J_{\lambda'\lambda}$, we get:
\begin{equation}\label{ac1p}
\Delta \equiv 
(1+2\eta)\tilde{J}_{11}+\tilde{J}_{1-1}-2\sqrt{2\eta}\tilde{J}_{10}-
\tilde{J}_{00}=
-(B_5+B_7)\ .
\end{equation}
This function was calculated for the deuteron in ref. \cite{gk84} and for 
$\rho$-meson in refs. \cite{keist94,card95}. Note that $\Delta$ in(\ref{ac1p}) 
should not be confused with the perpendicular component of the momentum transfer 
$q$.

The substitution in (\ref{gia}) of the matrix elements 
$\tilde{J}_{\lambda'\lambda}$ from (\ref{iigc}) gives:
\begin{eqnarray}\label{ffa}
{\cal F}^{GK-A}_1&=&{\cal F}_1,\quad                              
{\cal F}^{GK-A}_2={\cal F}_2\ ,                                 
\nonumber\\
{\cal G}_1^{GK-A}&=&{\cal G}_1-B_6-\frac{1}{2\eta}\Delta,
\end{eqnarray}
This solution gives ${\cal F}_1,{\cal F}_2$
without any nonphysical contribution (though still approximate), but ${\cal 
G}_1^{GK-A}$  contains the nonphysical form factor $B_6$ and, in 
addition, the function $\Delta=-(B_5+B_7)$, eq.(\ref{ac1p}), 
responsible for the violation of the condition (\ref{ac0}).

The different solutions were compared numerically in 
ref. \cite{card95} (for the case of $\rho$-meson form factor) and 
analytically in ref. \cite{karm96}. 
The results are summarized in  table \ref{tab1}.
Calculations by analytical formulae from this table
 exactly coincide with the numerical calculations \cite{card95}.

\begin{table}
\begin{center}
\begin{tabular}{|c|l||l||l||l||}
\hline
No. & Version
& Form factor ${\cal F}_1$  & Form factor ${\cal F}_2$  
&Form factor ${\cal G}_1$ \\
\hline \hline
1 & GK-B \protect{\cite{gk84}} & ${\cal F}_1$  & ${\cal F}_2$  
& ${\cal G}_1-B_6$ \\ 
\hline
2 & GK-A \protect{\cite{gk84}} & ${\cal F}_1$  & ${\cal F}_2$  
& ${\cal G}_1-B_6-\frac{\displaystyle{1}}
{\displaystyle{2\eta}}\Delta$ \\ 
\hline
3 & BH \protect{\cite{bh92}} & ${\cal F}_1+ 
\frac{\displaystyle{1}}{\displaystyle{1+2\eta}}\Delta$ 
& ${\cal F}_2$
& ${\cal G}_1-B_6-\frac{\displaystyle{2}}
{\displaystyle{1+2\eta}}\Delta$ \\
\hline
4 & D & ${\cal F}_1+\Delta$
& ${\cal F}_2+\frac{\displaystyle{1}}{\displaystyle{\eta}}\Delta$
& ${\cal G}_1-B_6$ \\
\hline
5 & CCKP \protect{\cite{chung88}} 
&${\cal F}_1+\frac{\displaystyle{1}} 
{\displaystyle{2(1+\eta)}}\Delta$
& ${\cal F}_2$
& ${\cal G}_1-B_6-\frac{\displaystyle{1}}
{\displaystyle{1+\eta}}\Delta$\\
\hline
6 & FFS \protect{\cite{ffs93}}
&${\cal F}_1$
& ${\cal F}_2-\frac{\displaystyle{1}} 
{\displaystyle{2(1+\eta)^2}}\Delta$
& The same as for CCKP\\
\hline
\end{tabular}
\end{center} 
\caption{The set of form factors ${\cal F}_1, {\cal F}_2, 
{\cal G}_1$ calculated in six versions discussed in the section 
\protect{\ref{fff}}.}  
\label{tab1}
\end{table}
From the table \ref{tab1}, one can see that all the methods give 
superpositions of the physical form factors with the nonphysical ones. 
Most of them contain the nonphysical contribution $\Delta$ and differ from 
each other by its coefficient. This is so not only for
the methods considered above, but also for any other method based on the 
$\tilde{J}_+$ component. If $\Delta=0$, the form factors do not depend on the 
prescription, as it was pointed out in \cite{keist94,card95}. But
${\cal G}_1$ still contains $B_6$.

Note that the terms  $B^{\mu\nu}_{\rho}$ contribute to                          
the matrix elements of the $J^{(0)}_+$ current component if at least            
one of the projections $\lambda,\lambda'$ equals to zero. In ref.               
\cite{glaz83} it was proposed to avoid using the zero projections of           
$\lambda,\lambda'$, but consider the matrix elements with $\lambda,             
\lambda'=\pm 1$ of the current components $J^{(0)}_+$ and                       
$\vec{J}^{(0)}_{\perp} =\{J^{(0)}_x,J^{(0)}_y\}$. Taking 
into account           
that $\omega_+=0$ for $\omega= (1,0,0,-1)$, one gets:                           
$$                                                                              
e_{\nu}^{*\lambda'}(p') B^{\nu\mu}_+e_{\mu}^{\lambda}(p)=0,\quad                
e_{\nu}^{*\lambda'}(p') B^{\nu\mu}_{\perp}e_{\mu}^{\lambda}(p)=0\qquad          
\mbox{for }\lambda, \lambda'=\pm 1\ .                                           
$$                                                                              
We see that this method indeed gives the solution for the form          
factors, in which the nonphysical contributions vanish entirely.                                                                 

In both methods the complete separation of the physical form factors 
from nonphysical ones is realized at the expense of incorporating, 
besides $\tilde{J}_+$, other components of the current.  A similar situation 
takes place with the form factors of a spin 1/2 system \cite{km96}.  
These other components can receive contributions not contained in $\tilde{J}_+$ 
(for example from the instantaneous part of the fermion propagators, which do 
not 
contribute to $\tilde{J}_+$ due to $\gamma_+^2=0$). The instantaneous
contributions (contact terms) have been  already taken into account in
the deuteron electrodisintegration  \cite{dkm95}.
As we shall see in the following section, these  contributions have a
very important physical interpretation in terms of
meson exchange currents in the non-relativistic approach. They can
therefore be  interpreted also as a two-body current. Considering
the  $\tilde{J}_+$-components only, these contributions would be lost. This
also shows that the consideration of other components of the current,
beside $\tilde{J}_+$, is inevitable if one wants to keep all physical
contributions, and only these, to the form factors.

\subsection{The transitions $1^+ - 0^+$ and $1^- - 0^-$} 
\label{eld} 

The pseudo-vector -- scalar transition $1^+ - 0^+$ corresponds to  the deuteron
electrodisintegration amplitude  near threshold (a few MeV in the
center-of-mass system of the $np$  pair). In this case, the final state is
dominated by the singlet S-wave   and the amplitude does not depend on the
direction of the relative  np-momentum. The amplitude of the vector --
pseudoscalar transition  $1^- - 0^-$, corresponds, for example, to the $\rho -
\pi$ transition,  and has the same structure. For definiteness we will discuss
here the deuteron  electrodisintegration.

In the impulse approximation, LFD
was applied to the deuteron electrodisintegration amplitude in 
ref.~\cite{keister88} and, in its explicitly covariant form, in 
ref.~\cite{adk94}. The corresponding amplitude  can be written as:  
\begin{equation} 
M_{2 \rightarrow 3} =\frac{4\pi\alpha}{q^2} 
\bar{u}(k'_e)\gamma^{\rho}u(k_e)e^{\lambda}_{\mu}(p)F^{\mu}_{\rho} 
\chi(^1S_0)\ ,  
\label{eld3} 
\end{equation}  
where
$F_{\mu\rho}$ describes the vertex $d + \gamma^* \rightarrow\; ^1S_0$.  
The function $\chi(^1S_0)$ is the spin function of the final nucleons 
in the $^1$S$_0$-state we discussed in section \ref{1s0}.  The vertex 
$F_{\mu\rho}$ can be represented as follows:   
\begin{equation} \label{eld6} 
F_{\mu\rho} = 
\frac{1}{4m^2}e_{\rho\mu\nu\gamma}q^{\nu}{P}^{\gamma}A\ , 
\end{equation} 
where  ${P} = p + p'$, $p' = p_1 + p_2$ and $A$ is a dimensionless 
scalar function: $A = A(Q^2,\nu)$.  The tensor~(\ref{eld6}) is the only 
one which has the appropriate properties expected from covariance, 
transversality and parity conservation to describe the vertex 
$d\gamma^* \rightarrow np(^1 S_0)$. Let us emphasize that the vertex 
(\ref{eld6}) automatically satisfies the transversality condition 
$F_{\mu\rho}q^{\rho} = 0$. In this particular case, the Lorentz 
invariance leads to gauge invariance for the amplitude.  This is a 
consequence of the particular $^1$S$_0$ final state.  The cross-section of 
inclusive electron-nucleus scattering is usually represented in terms 
of the structure functions $W_1$ and $W_2$ (see, e.g.,~\cite{martino}):   
\begin{equation} 
\frac{d\sigma}{d\Omega_edE'} = \left( \frac{d\sigma}{d\Omega} 
\right)_0(W_2 + 2\tan^2(\frac{\theta}{2})W_1)\ .  
\label{eld7} 
\end{equation}  
The structure 
functions $W_2$ and $W_1$ can be expressed through the unique scalar
amplitude $A$ \cite{adk94}:    
\begin{equation} 
W_2 = \frac{p^*Q^2}{3\cd 2^6\pi^2m^4}| A|^2\ , 
\label{eld9} 
\end{equation} 
\begin{equation} 
W_1 = (1 + \nu^2/Q^2)W_2\ .  
\label{eld10} 
\end{equation} 
Here $p^*$ is the c.m.-momentum of the final nucleons.
All the dynamical information about the process is therefore contained 
in the transition form factor $A$.  We stress that this result 
follows only from the assumption that the final $^1$S$_0$ state 
dominates and is valid independently of the presence or absence of 
final state interaction, meson exchange currents and isobar 
configurations.  Their contribution only influences the amplitude $A$. 
 
The particular amplitude for the vertex $d\gamma^* \rightarrow np(^1 S_0)$ in 
any approximate
calculation, denoted by $\tilde{F}_{\mu\rho}$,  does not coincide, however, with 
$F_{\mu\rho}$  in (\ref{eld6}). The tensor 
$\tilde{F}_{\mu\rho}$ depends in that case on $\omega$. Following the arguments 
presented in section \ref{pff}, we can decompose $\tilde{F}_{\mu\rho}$ 
on the general invariant amplitudes. It has the form:  
\begin{eqnarray}\label{am4} 
\tilde{F}_{\mu\rho}=\frac{1}{2m^2}e_{\rho\mu\nu\gamma}q^{\nu}p^{\gamma}A
+ 
e_{\rho\mu\nu\gamma}q^{\nu}\omega^{\gamma}B_1+ 
e_{\rho\mu\nu\gamma}p^{\nu}\omega^{\gamma}B_2\\ 
\nonumber 
+(V_{\mu}q_{\rho}+ V_{\rho}q_{\mu})B_3+(V_{\mu}\omega_{\rho}+ 
V_{\rho}\omega_{\mu})B_4+
\frac{1}{2 m^2\omega\cd p}(V_{\mu}p_{\rho}+ V_{\rho}p_{\mu})B_5\ , 
\end{eqnarray} 
where $V_{\mu}=e_{\mu\alpha\beta\gamma}\ \omega^{\alpha}\ q^{\beta}\ 
p^{\gamma}$. The decomposition (\ref{am4}) contains the symmetrical 
structures in front of the functions $B_{3,4,5}$. Corresponding 
antisymmetric terms (like $V_{\mu}q_{\rho}- V_{\rho}q_{\mu}$) are not 
independent and can be expressed through the first three contributions. We 
emphasize that $\tilde{F}_{\mu\nu}$ depends on $\omega$ even for the components 
of the wave function which do not depend on $\omega$. In the latter 
case, $\omega$  enters through the rules of the graph technique.  Similarly to 
the case of the deuteron 
form factors, the decomposition (\ref{am4}) enables us to separate 
the $\omega$-independent parts from nonphysical 
$\omega$-dependent ones so that one can unambiguously extract the 
physical form factors from the initial tensor $\tilde{F}_{\mu\rho}$. From 
eq.(\ref{am4}), we immediately find:  
\begin{equation}\label{am5} 
A=-\frac{m^2}{Q^2\ (\omega\cd 
p)}e^{\mu\rho\nu\gamma}q_{\nu}\omega_{\gamma}\tilde{F}_{\mu\rho}\ .  
\end{equation} 
This simple formula is the analogue of formulae 
(\ref{ff6}) - (\ref{ff8}) for the deuteron form factors. 

\subsubsection{Comparison with other approaches} 
The transition form factor $A$ given by (\ref{am5}) cannot be reduced 
to the 
$\rho=+$ component of the amplitude 
$\tilde{F}_{\mu\rho}$. For example, in the reference system where $q_0=0$,
and $\vec{q}$ is parallel to the $x$-axis, eq.(\ref{am5}) is reduced to:
\begin{equation}\label{tr1}
A=\frac{-m^2q_x}{Q^2p_+}(\tilde{F}_{y+}-\tilde{F}_{+y})\ .
\end{equation}
This formula includes $\tilde{F}_{+y}$ with $\rho=y$. 

The contribution of the plus-component is given, as usual, by 
the contraction $\tilde{F}_{\mu\rho}\omega^{\rho}$. If $\tilde{F}_{\mu\rho}$ 
would coincide
with  $F_{\mu\rho}$ in eq.(\ref{eld6}), i.e., would not contain any 
non-physical contributions, we indeed could find the form factor from the 
plus-component by the equation:
\begin{equation}\label{tr2}
A=-\frac{2m^2}{Q^2(\omega\cd p)^2}e^{\alpha\mu\nu\gamma}
\omega_{\alpha}q_{\nu}p_{\gamma}(F_{\mu\rho}\omega^{\rho})\ .
\end{equation}
However, applying this equation to $\tilde{F}_{\mu\rho}$, we get:
\begin{equation}\label{tr3}
A'=-\frac{2m^2}{Q^2(\omega\cd p)^2}e^{\alpha\mu\nu\gamma}
\omega_{\alpha}q_{\nu}p_{\gamma}(\tilde{F}_{\mu\rho}\omega^{\rho})=
A-B_5\ .
\end{equation}
Like in the cases considered above, we get a superposition of the 
physical form factor $A$ and nonphysical one $B_5$. Rewritten in 
terms of the 
components, the contraction (\ref{tr3}) obtains the form:
\begin{equation}\label{tr4}
A'=\frac{-2m^2q_x}{Q^2p_+}\tilde{F}_{y+}\ .
\end{equation}
The difference with eq.(\ref{am5})  is reduced to the replacement  of
$\tilde{F}_{y+}$ in  (\ref{tr4}) by the antisymmetric combination
$(\tilde{F}_{y+}-\tilde{F}_{+y})/2$  in  (\ref{tr1}). This seems very natural
since the physical amplitude (\ref{eld6}) is antisymmetric relative to
permutation of the indices  $\rho,\mu$, and, hence, any symmetrical
contribution is a spurious one and has to be excluded (independently of the
fact that it depends on $\omega$ or not). This is another reason to separate
it. We emphasize that  the difference  between the amplitudes $A$ and
$A'$ is of relativistic origin  and disappears  in the non-relativistic
limit.

\subsection{The transitions $0^- - 1^+$ and $0^+ - 1^-$} 
\label{pspv} 

The pseudo-scalar -- pseudo-vector transition $0^- - 1^+$ corresponds,
for example, to $\pi - A_1$ and to $K - K_1$ transitions. The 
scalar -- vector transition amplitude $0^+ - 1^-$ has the same form.

The general transition amplitude reads:
\begin{equation}\label{pspv1}
\langle PS(p')\vert J_{\rho}\vert PV(p,\lambda)\rangle
=[F_1((p\cd q)g_{\rho}^{\mu}-p_{\rho}q^{\mu})
+F_2(q_{\rho}q^{\mu}-q^2g_{\rho}^{\mu})]e_{\mu}^{\lambda}(p)\ .
\end{equation}
It is determined by two form factors $F_1$ and $F_2$.

In an approximate calculation, this amplitude should incorporate the 
$\omega$-dependent contributions. These are given by:
\begin{equation}\label{pspv2}
\langle PS(p')\vert \tilde{J}_{\rho}\vert PV(p,\lambda)\rangle
=\tilde{G}_{\rho}^{\mu}e_{\mu}^{\lambda}(p)\ ,
\end{equation}
where
\begin{eqnarray}\nonumber
\tilde{G}_{\rho\mu}&=&F_1((p\cd q)g_{\rho\mu}-p_{\rho}q_{\mu})
+F_2(q_{\rho}q_{\mu}-q^2g_{\rho\mu})
\nonumber\\
&+&A'p_{\mu}(p'_{\rho}-q_{\rho}(q\cd p')/q^2)
+C p'_{\mu}q_{\rho}+C'p_{\mu}q_{\rho}
\nonumber\\
&+&B_1q^2 \omega_{\mu} p'_{\rho}/\omega\cd p
+B_2p'_{\mu}\omega_{\rho}+B'_2 p_{\mu}\omega_{\rho}
+B_3\omega_{\mu}q_{\rho}+B_4\omega_{\mu}\omega_{\rho}\ .
\label{pspv3}
\end{eqnarray}
The terms  $A',C',B'_2$ do not contribute to eq.(\ref{eq2}), since 
$p^{\mu}e_{\mu}^\lambda(p)=0$. They appear however before contraction  with
$e_{\mu}^\lambda(p)$.  The term $C p'_{\mu}q_{\rho}$ appears, since an
approximate amplitude is not gauge invariant. It should be also  separated. Its
appearance has nothing to do with the  $\omega$-dependence or LFD. In the
Feynman  approach its separation is equivalent to the construction, by hand, 
of a gauge invariant amplitude by the following subtraction:  
$G_{\rho\mu}\rightarrow  G_{\rho\mu}-(G_{\rho'\mu}q^{\rho'})q_{\rho}/q^2$.

One can construct other gauge non-invariant terms, proportional to 
$p'_{\mu}p_{\rho}$, $q_{\mu}p_{\rho}$ and $q_{\mu}p'_{\rho}$, but they
are expressed through the structures included in (\ref{pspv3}). The term 
$V_{\mu}V_{\rho}$ with $V_{\mu}=e_{\mu\alpha\beta\gamma}\omega^{\alpha}
q^{\beta}p^{\gamma}$ is also expressed through them.

Solving eq.(\ref{pspv3}) relative to $F_1,F_2$ one gets:
\begin{equation}\label{pspv4}
F_1=\frac{\tilde{G}_{\rho\mu}}{q^2}
\left[-\frac{q^{\mu}\omega^{\rho}}{(\omega\cd p)}
+(m_{ps}^2-m_{pv}^2-q^2)\frac{\omega^{\mu}\omega^{\rho}}{2(\omega\cd
p)^2} \right],
\end{equation}
\begin{eqnarray}\nonumber
F_2&=&\frac{\tilde{G}_{\rho\mu}}{(q^2)^2}\left\{
[(m_{ps}^2-m_{pv}^2)^2-q^2(2m_{ps}^2-q^2)]
\frac{\omega^{\mu}\omega^{\rho}}{2(\omega\cd p)^2}
+(m_{pv}^2-m_{ps}^2+q^2)\frac{q^{\rho}\omega^{\mu}}{2(\omega\cd p)
}\right.
\nonumber\\
&+&\left.
(m_{pv}^2-m_{ps}^2)\frac{\omega^{\rho}q^{\mu}}{(\omega\cd p)
} +q^2\frac{\omega^{\mu}p^{\rho}+\omega^{\rho}p'^{\mu}}{(\omega\cd p)
} +q^{\mu}q^{\rho}-q^2g^{\mu\rho}
\right\}.
\label{pspv5}
\end{eqnarray}
Here $m_{ps}$ and $m_{pv}$ are the masses of pseudo-scalar
and pseudo-vector mesons.

\subsubsection{Comparison with other approaches} 

In ref. \cite{az90}
the form factors are found from the component $\rho=+$. 
Contracting (\ref{pspv1}) with $\omega_{\rho}$, we get:  
\begin{equation}\label{pspv6} 
\langle PS(p')\vert \tilde{J}_{\rho}\vert 
PV(p,\lambda)\rangle\omega^{\rho} =[F_1((p\cd q)\omega^{\mu}-(\omega\cd 
p)q^{\mu}) -F_2q^2\omega^{\mu}]e_{\mu}^{\lambda}(p). 
\end{equation} 
Contracting (\ref{pspv2}) with $\omega_{\rho}$, we reproduce the equation
(\ref{pspv6}) with new form factors $F'_1,F'_2$
given by: 
$$F'_1=F_1,\quad F'_2=F_2-B_1\ .$$
Hence, the $\tilde{J}^+$ component provides $F_1$ without any spurious 
contributions, while the form factor  $F'_2$ is determined by:
\begin{equation}\label{pspv8}
F'_2=F_2- B_1=\frac{\tilde{G}_{\rho\mu}\omega^{\rho}}{q^2
(\omega\cd p)}
\left[m_{pv}^2\frac{\omega^{\mu}}{\omega\cd p}-p^{\mu}\right].
\end{equation}
It includes the spurious contribution $B_1$.

                                                                                
\chapter{Electromagnetic observables}\label{emob}                                          

The formalism and methods presented above are applied to different physical
systems and processes: pion, nucleon and deuteron electromagnetic form factors,
$\rho-\pi$ and $ed-enp$ transitions. With standard assumptions and
approximations, we easily reproduce results given in the literature.  Some of
these results, however, like the asymptotic behavior of the pion form factor
and meson-exchange current contributions in $ed-enp$ amplitude, obtain a new and 
clear physical
interpretation in our approach, since they directly reflect the presence of
extra relativistic components in the corresponding wave functions.

\section{Electromagnetic form factors in the Wick-  Cutkosky model}
\label{wcm} 
The Wick-Cutkosky model \cite{wcm} allows to get
 the explicit form of both the 
Bethe-Salpeter functions and the light-front wave functions
 for  states with angular momentum $J=0$ and $J=1$. 
This gives us the opportunity to compare the electromagnetic form factors
calculated by two methods: by the standard one, with the use
 of the Feynman rules and
through the Bethe-Salpeter amplitude, and by means of the light-front
graph technique through the light-front wave function. This comparison is 
especially non-trivial in the $J=1$ case, where 
the physical form factors should be calculated after separating out the 
non-physical ($\omega$-dependent) amplitude from the physical one.  

\subsection{Spin 0}\label{wcm0}
For a spin-0 system consisting of  spinless constituents, the form factor is 
expressed through the light-front wave 
function by eq.(\ref{ff0}). Its expression through the Bethe-Salpeter
function $\Phi(k,p)$ corresponding to the Feynman graph of fig. \ref{fev} reads:

\begin{figure}[hbtp]
\centerline{\epsfbox{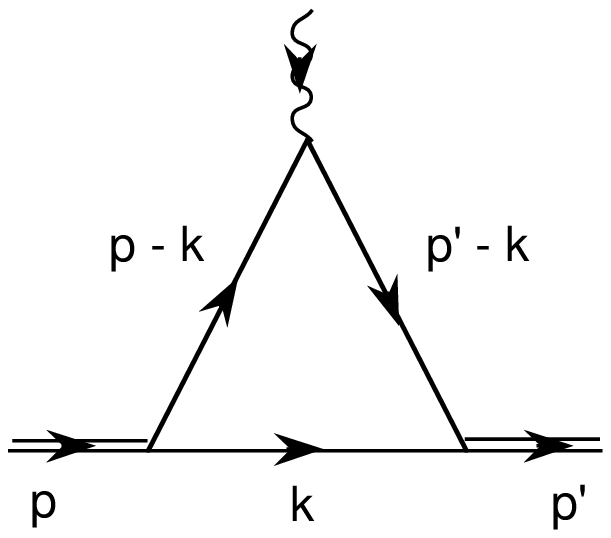}}
\figcap{Feynman electromagnetic vertex of a two-body system.}
\label{fev}
\end{figure}

\begin{equation}\label{wcm01}
(p+p')_{\rho}F(t)=i\int (p+p'-2k)_{\rho}\Phi(\frac{1}{2}p-k,p)
\Phi(\frac{1}{2}p'-k,p')(m^2-k^2)\frac{d^4k}{(2\pi)^4}.
\end{equation}

The Bethe-Salpeter function is given by eq.(\ref{bs8p}). 
Substituting here the spectral 
function (\ref{bs11}), one can find the explicit form of the 
Bethe-Salpeter amplitude:
\begin{equation}\label{wcm02}
\Phi(k,p)=-ic\left[\left(m^2-\frac{1}{2}M^2-k^2\right)
\left(m^2-(\frac{1}{2}p+k)^2-i0\right)
\left(m^2-(\frac{1}{2}p-k)^2-i0\right)\right]^{-1},
\end{equation}
where $c=2^5\sqrt{\pi m\kappa^5}$ with $\kappa=\sqrt{m|\epsilon_b|}=
m\alpha /2$. 

The analytical (in the asymptotical region) and numerical comparison of the
form factors  calculated through the light-front wave function, eq.(\ref{ff0}),
and through the Bethe-Salpeter amplitude, eq.(\ref{wcm01}), was carried out in
ref. \cite{ks92}. The form factors calculated by these two approaches are
compared in table  \ref{tabff}.

\begin{table}
\begin{center}
\begin{tabular}{||r||c|c|c||}
\hline
  & \multicolumn{3}{|c||}{$F(t)$}\\
\cline{2-4}
$|t|/m^2 $ & BS & LF & NR  \\ 
\hline 
  0   & $1.00E+00$ & $9.67E-01$ & $1.00E+00$ \\
\hline
0.01 & $5.14E-01$ & $4.98E-01$ & $5.17E-01$ \\
\hline
1    & $6.02E-04$ & $5.75E-04$ & $6.23E-05$ \\
\hline
25   & $1.01E-06$ & $9.76E-07$ & $1.05E-06$ \\
\hline
100  & $6.40E-08$ & $6.20E-08$ & $6.55E-08$ \\
\hline
225  & $1.28E-08$ & $1.24E-08$ & $1.29E-08$ \\
\hline
\end{tabular}
\end{center} 
\caption{Electromagnetic form factor of a system with spin $J=0$ in the 
Wick-Cutkosky model for $\alpha=0.08$. Columns BS, LF and NR represent the 
results of the calculation
using the Bethe-Salpeter amplitude (\protect{\ref{wcm02}}), the
light-front wave function (\protect{\ref{k8}}) and non-relativistic wave 
function (\protect{\ref{k6}}) respectively.}
\label{tabff} 
\end{table} 

Both approaches give the same asymptotical behavior of the form factors
at $|t|\gg m^2$:
\begin{equation}\label{wcm03}
F(t)\approx \frac{16\alpha^4m^4}{t^2}\left[1+
\frac{\alpha}{2\pi}\log\left(\frac{|t|}{m^2}\right)\right],
\end{equation}
where $\alpha=g^2/(16\pi m^2)$, $g$ is the coupling constant in the 
Wick-Cutkosky model. 

In contrast to ref. \cite{ks92} the LF form factor in  table 
\ref{tabff} has not been renormalized at $t=0$. In the light-front approach,
the form factor at low $t$, as well as the normalization, is smaller than the
Bethe-Salpeter one by $4\alpha/3\pi$. For the normalization, the discrepancy
has its origin in the contribution of the derivative of potential to the norm
operator, as indicated in eq.(\ref{AAA}). For the form factor, this corresponds
to omitted contributions indicated in fig. \ref{f4-ks92} (the so-called recoil
contributions).
If corrections to the form factor and to the norm (which coincide with each
other at $Q^2=0$) are included, then the renormalization of the light-front
result, as done in ref. \cite{ks92}, is not required.

At high $t$, the results obtained in LFD 
and in the Bethe-Salpeter approach are very similar. Non-relativistic 
calculation is
 also close
to them, when $(\alpha/2\pi)\log (|t|/m^2) \ll 1$. 
This is a peculiarity of the Wick-Cutkosky model. In the asymptotical region 
where 
$(\alpha/2\pi)\log (|t|/m^2)> 1$ LFD and Bethe-Salpeter form factors are still
close to each other, while they differ from non-relativistic one. 

The asymptotics of the non-physical form 
factor $B_1$, which appears in the decomposition (\ref{emv1}) and is 
given 
by eq.(\ref{ffb1}), has the form:
\begin{equation}\label{wcm04}
B_1(t)\approx \frac{32\alpha^4 m^4}{t^2}.
\end{equation}
In the intermediate region, $B_1$ exceeds $F$ : 
$B_1(t)/F(t)\approx 2$ while it is negligible in the 
non-relativistic region. 
According to \cite{ks92}, $B_1$ becomes comparable with the physical
form factor $F(t)$ at $|t|\sim m^2$. Therefore its 
separation is important. In the spin-0 case it is separated automatically, 
if one proceeds from the $\rho=+$ component of the current. However, this is 
not so for the spin-1 case.

\begin{figure}[htbp]
\epsfxsize = 15.cm
\epsfysize = 15.cm
\centerline{\epsfbox{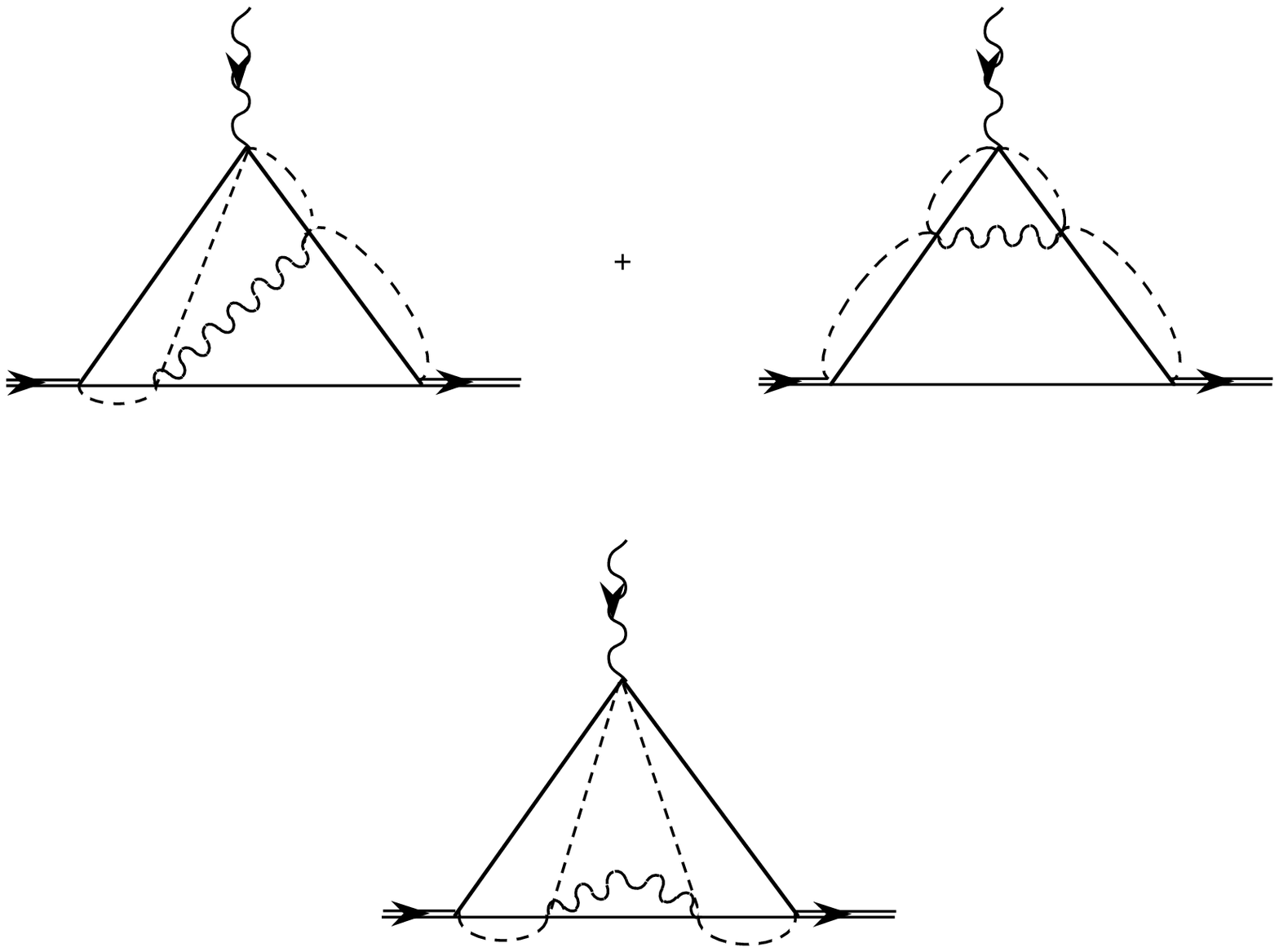}}
\figcap{Amplitudes corresponding to admixtures of the three-body 
sector in the state vector. These amplitudes correspond to retardation 
effects.}\label{f4-ks92}
\end{figure}

\subsection{Spin 1}\label{wcm1}
For a system with angular momentum $J=1$ made out from two spinless 
constituents, 
the light-front matrix 
elements $\tilde{J}^{\mu\nu}_{\rho}$ of the current are obtained from the 
spin-0 case equation
 (\ref{emv4}) with the replacement of the spin-0 wave functions 
$\psi',\psi$ by the spin-1 wave functions $\psi'_{\mu},\psi_{\nu}$. 
In four-dimensional notations, $\psi_{\mu}$ has the form:
\begin{equation}\label{wcm11}
{\mit\Phi}^{\lambda}(k_1,k_2,p,\omega \tau)=
e^{(\lambda)}_{\mu}(p)\psi^{\mu},\quad 
\psi^{\mu}=\varphi_1\frac{(k_1-k_2)^{\mu}}{2}+ 
\varphi_2\frac{\omega^{\mu}}{\omega\cd p}\ .
\end{equation}
The explicit form of the components $\varphi_{1,2}$ in (\ref{wcm11}) 
can be found from the comparison of (\ref{wcm11}) with (\ref{bs12}) in the
reference frame where $\vec{k}_1+\vec{k}_2=0$.
Substituting $\tilde{J}^{\nu\mu}_{\rho}$ in eqs.(\ref{ff6})-(\ref{ff8}),
in order to separate the nonphysical contributions,
we find three form factors ${\cal F}_1,{\cal F}_2,{\cal G}_1$.

On the other hand, in the Bethe-Salpeter approach, the electromagnetic
vertex is directly related to the physical form factors by
the decomposition (\ref{ff1}) and can be obtained from (\ref{wcm01}) by the 
replacement of the spin-0 Bethe-Salpeter amplitude by the spin-1
function. The latter has the following form:
\begin{eqnarray}
\Phi^{\lambda}(k,p)&=&e^{\lambda}_{\mu}(p)k^{\mu}\Phi_1(k,p)\ ,
\nonumber \\
\Phi_1(k,p)&=&-ic_1\left[\left(m^2-\frac{1}{2}M^2-k^2\right)^2
\left(m^2-(\frac{1}{2}p+k)^2-i0\right)\right. \nonumber \\
&&\left.\left(m^2-(\frac{1}{2}p-k)^2-i0\right)\right]^{-1}\ ,
\label{wcm12}
\end{eqnarray}
where $c_1=2^7\sqrt{\pi m}\kappa_1^{7/2}$, $\kappa_1=\sqrt{m|\epsilon_b|}=
m\alpha/4$. 

\begin{figure}[htbp]
\epsfxsize=8.0 cm
\centerline{\epsfbox{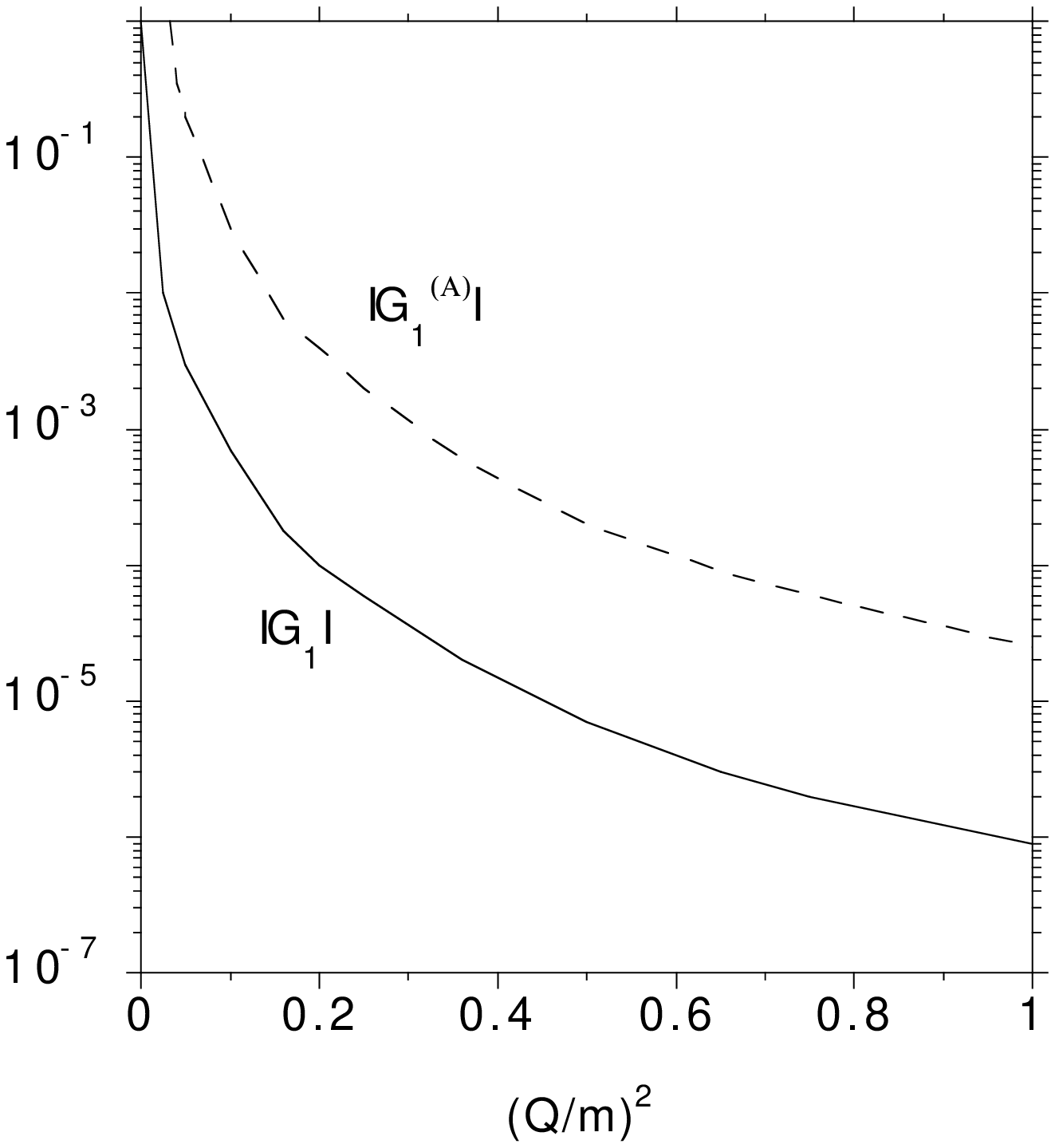}}
\figcap{The form factor ${\cal G}_1$
in the Wick-Cutkosky model. The solid line represents the calculation with the 
Bethe-Salpeter function (\protect{\ref{wcm12}}) and by eq.(\protect{\ref{ff8}}) 
with the light-front wave function (\protect{\ref{bs12}}). Dashed line 
illustrates the form factor incorporating the non-physical contribution by 
eq.(\protect{\ref{ffa}}) (solution A from ref. \protect{\cite{gk84}}).}
\label{wcg1}
\end{figure}

The form factor ${\cal G}_1$ calculated in the two approaches in ref.
\cite{ks94}  is shown by the solid line in fig. \ref{wcg1}. The two
calculations give the curves indistinguishable from each other. In order to
show the importance of a proper separation  of the non-physical contribution we
show also in fig. \ref{wcg1}  the form factor corresponding to the solution A
from ref. \cite{gk84} (dashed line). It is given by eq.(\ref{ffa}) and contains
the non-physical contributions $\Delta$ and $B_6$. It considerably differs
from  the correct form factor. According to ref. \cite{ks94}, this difference
in the model is mainly due to $\Delta$, whereas the $B_6$ contribution is
negligible.

In the asymptotical 
region, these Bethe-Salpeter and light-front form factors are given by the 
following analytical expressions \cite{ks92}:
\begin{eqnarray}\label{wcm13}
{\cal F}_1\approx&& -\frac{\alpha^6 m^6}{|t|^3}\left[1+\frac{\alpha}{4\pi}
\log\left(\frac{|t|}{m^2}\right)\right],
\\
{\cal F}_2\approx&& -\frac{48\alpha^6 m^8}{t^4}\left[1+\frac{\alpha}{4\pi}
\log\left(\frac{|t|}{m^2}\right)\right],
\\
{\cal G}_1\approx&& \frac{\alpha^6m^6}{|t|^3}.
\end{eqnarray}

We emphasize that, generally speaking, using two-body Bethe-Salpeter function 
does not correspond to two-body Fock component, but incorporates 
implicitly many-body sectors of the state vector.

\section{Applications to the quark model}
We shall investigate in this section a few simple examples on the application to 
the quark model of 
the formalism developed in this review. We shall discuss 
in some details the pion wave function and form factors, and give some results 
for other observables.

\subsection{The pion form factor}

\subsubsection{Pion wave function}                                              

The spin structure of the pion wave function as a system consisting of 
quark and antiquark in the $J^{\pi}=0^-$ state is identical to the $np$ 
wave function in $J^{\pi}=0^+$ state, eq.(\ref{3.4}). The negative parity 
is generated automatically by opposite internal parities of the quark and 
antiquark and therefore does not change the spin structure. The pion 
wave function has thus the form:                       
\begin{equation}\label{pion}                                                    
\psi=\frac{1}{\sqrt{2}}\bar{u}(k_2)\left[A_1\frac{1}{m}+ 
A_2\frac{\hat{\omega}}{\omega\cd p}\right]
\gamma_5 v(k_1)=\bar{u}(k_2)Ov(k_1)\ ,
\end{equation}
where $O=(A_1/m+ A_2 \hat{\omega}/\omega\cd p)\gamma_5$, and 
$m$ is the quark mass. For simplicity, we do not take into account
 isospin.

The representation of this wave function in terms of the variables 
$\vec{k}$ and $\vec{n}$ is almost identical to (\ref{fsi9}):
\begin{equation}\label{eq3}
\psi=\frac{1}{\sqrt{2}}w_2^t\left(g_1+\frac{i\vec{\sigma}\cd [\vec{n}\times 
\vec{k}]}{k}g_2\right)w_1\ ,
\end{equation} 
with the following relations between the invariant functions:
\begin{eqnarray}\label{eq3p}
A_1&=&-\frac{m}{2 \varepsilon_k} (g_1+\frac{m}{k} g_2), \quad 
A_2=\frac{\varepsilon_k}{k} g_2\ .
\end{eqnarray}
The normalization condition is a particular case of (\ref{nor3}):
\begin{equation}\label{eq4}
N_2=\int \sum_{\lambda_1\lambda_2}\psi_{\lambda_1\lambda_2}^*
\psi_{\lambda_1\lambda_2}D\ ,
\end{equation}
where
\begin{equation}\label{eq5}
D=\frac{1}{(2\pi)^3}\frac{d^3k_1}{(1-x)2\varepsilon_{k_1}}
=\frac{1}{(2\pi)^3}\frac{d^3k}{\varepsilon_k}
=\frac{1}{(2\pi)^3}\frac{d^2R_{\perp}dx}{2x(1-x)}\ .
\end{equation}
Substituting in eq.(\ref{eq4}) the wave function represented by (\ref{pion}) 
and ({\ref{eq3}), we get:
\begin{equation}\label{eq6}
N_2=
\int 
\left[\frac{ (m^2+\vec{R}_{\perp}^2)}{m^2x(1-x)}A_1^2+4 A_1 A_2 
+4x(1-x)A_2^2\right]D=
\int [g_1^2+(1-z^2)g_2^2]D\ ,
\end{equation}
where $z=\cos(\widehat{\vec{k}\vec{n}})$. Obviously,  in the case where the 
interaction potential
between quarks is independent of  $M$, as often assumed in phenomenological
approaches, $N_2$ should be equal to one.

\subsubsection{Pion form factor}
The diagrammatical representation of the  pion electromagnetic current in the
impulse approximation almost
coincides with fig. \ref{f1-ks92-b}. For  the interaction of a virtual photon
with  a quark, the horizontal line has to be replaced by the double line of
antiquark. This current has the form: 
\begin{equation}\label{eq9}
\tilde{J}^{\gamma q}_{\rho}=\int Tr[-\overline{O'}(\hat{k}_2'+m)j_{\rho}
(\hat{k}_2+m)O(m-\hat{k}_1)]\frac{1}{(1-x)}D\ ,
\end{equation}
where
$O$ depends on the initial momenta, and $O'$ depends on the final momenta.
The quark current $j_{\rho}$ is taken as:
\begin{equation}\label{eq10a}
j_{\rho} =  f_1\gamma_{\rho} 
+\frac{i f_2}{2m}\sigma_{\rho\nu}\ q^{\nu}\ .
\end{equation}

In this equation $ f_1$ and $ f_2$ are the quark form factors. The
minus sign in  $Tr[-\overline{O'}\cdots]$ is from the fermion loop. We remind
that both lines in the loop, single and double, corresponding to the quark and
antiquark, are in the same direction. But since, according to the rules of the
graph technique,  they are followed in the same and opposite orientations
respectively, one makes a
loop  when going through them.

The pion form factor  for the $\gamma q$ interaction is given by:
\begin{equation}\label{eq9p}
F^{\gamma q}(Q^2)=\int
Tr[-\overline{O'}(\hat{k}_2'+m)(\omega\cd j)(\hat{k}_2+m) 
O(m-\hat{k}_1)]\frac{1}{2\omega\cd p(1-x)}D\ .
\end{equation}

The pion form factor  for the $\gamma \bar{q}$ interaction is found similarly:
\begin{equation}\label{eq10}
F^{\gamma \bar{q}}(Q^2)=\int Tr[-O(m-\hat{k}_2)
(\omega\cd j)(m-\hat{k}'_2)
\overline{O'}(m+\hat{k}_1)]\frac{1}{2\omega\cd p(1-x)}D\ .
\end{equation}  
The momenta $k_2,k'_2$ in (\ref{eq9}) are the quark momenta, whereas they are 
the antiquark momenta  in 
(\ref{eq10}).

After a trivial generalization for the case of different masses of 
the quark and antiquark, these formulas can be applied to the $K$-meson form 
factors. The final form factor is expressed through the momentum 
transfer $\vec{\Delta}$ (with $Q^2=\vec{\Delta}^2$) and the integration 
variables $\vec{R}_{\perp},x$ with the use of the invariants indicated in  
appendix \ref{kr}.

Let us for a moment put $A_2=0$. We thus have:  
\begin{equation}\label{eq11} 
F^{\gamma 
q}(Q^2)=2\int \left[(m^2+\vec{R}_{\perp} 
\vec{R}_{\perp}')f_1 -\frac{\vec{\Delta}^2}{2}xf_2\right] 
\frac{A_1 A'_1}{x(1-x)m^2}D\ .
\end{equation}

With $A_1$ expressed through $g_1$ by (\ref{eq3p}) the form factor (\ref{eq11}) 
exactly 
coincides with the expressions
given in ref. \cite{card95-pirho}.  We emphasize that the calculation of
(\ref{eq11}),  and more generally of any form factor or transition amplitude is, 
due to the covariance of our approach, a simple routine for  analytical
computer calculations. 

When $A_2$ is different from zero,
 but for  pointlike quarks, i.e. with $  f_1=1$, $ f_2=0$ 
in the current (\ref{eq10a}), the calculation of (\ref{eq9p}) gives:
\begin{eqnarray}\label{eq11pp}
&& F^{\gamma q}(Q^2)=\int \left[\frac{m^2+\vec{R}_{\perp}^2-x\vec{R}_{\perp}
\cd\vec{\Delta}}
{x(1-x)m^2}A_1 A'_1+2(A_1A'_2+A'_1 A_2)+4x(1-x)A_2A'_2\right]D\ ,
\nonumber \\
&&
\end{eqnarray}
where $A'_{1,2}=A_{1,2}(\vec{R}_{\perp}-x\vec{\Delta},x)$.
We shall use this expression in the following to evaluate the
asymptotical behaviour of the pion form factor.

Note that for the interaction of the virtual photon with the antiquark, in 
eq.(\ref{eq10}), we get: $F^{\gamma \bar{q}}=-F^{\gamma q}$. 
Hence, the full form factor of a system consisting  of identical $q\bar{q}$ is
zero:  $F^{\gamma q}(t)+ F^{\gamma \bar{q}}=0$.  We get
opposite sign of  $F^{\gamma \bar{q}}$  automatically, {\em without
introducing negative charge of  antiquark}. This negative charge is
automatically taken into account in this formalism. It can be obtained
from the antiquark contribution, multiplying it by a sign factor depending on
the process under consideration.  

For the $\rho-\pi$ transition, the quark and antiquark contributions
have the same sign. The transition amplitude $\tilde{F}_{\mu\rho}$ can be 
obtained in this case from the $\pi$ elastic 
amplitude
 $J_{\rho}$ by  replacing in eqs.(\ref{eq9})
the initial pion wave function $O$  by the wave function of the vector 
meson $\phi_{\mu}$. The form of this wave function coincides with the deuteron 
wave function (\ref{nz1}). 

\subsubsection{Asymptotical behaviour of the pion form factor}\label{pionas}

It is well known that QCD provides a $1/Q^2$ asymptotical behaviour  for the
pion  form factor \cite{brodsky}. As it was shown above, the pion wave function
contains in general two components. The first one  leads  to the usual
non-relativistic wave function, while the second one is of purely relativistic
origin and $\omega$-dependent. We shall show in this section that the
asymptotical $1/Q^2$ behaviour of the pion  form  factor is entirely determined 
by this last $\omega$-dependent component of the  pion wave function
\cite{bd-piff}.  This piece is essential in order to explain the difference
with the form factor of the $J=0$ state in the Wick-Cutkosky model, which
asymptotical behaviour is $1/Q^4$.

The asymptotical behaviour of the form factor can be related to the asymptotical 
behaviour of the wave 
function. We shall calculate the latter in a $q\bar{q}$ model with
a one-gluon exchange kernel. 
The equation for the wave function corresponding to the diagram indicated 
in fig. \ref{feq}
has the form:
\begin{eqnarray}\label{as1}
&&\psi(k_1,k_2,p,\omega\tau) = -\frac{1}{s-M^2}\int
\frac{1}{(2\pi)^3}\delta^{(4)}(p+\omega\tau'-k'_1-k'_2)  
2(\omega\cd p)d\tau'
\nonumber\\
&&\times\bar{u}(k_2)\gamma_{\mu}(\hat{k}'_2+m)
\theta(\omega\cd k'_2)\delta(k_2'^2-m^2)d^4 k'_2
O'(m-\hat{k}'_1)\theta(\omega\cd k'_1)\delta(k_1'^2-m^2)d^4k'_1
\gamma^{\mu}v(k_1)
\nonumber\\
&&\times {\cal K}(k'_1,k'_2,\omega\tau';k_1,k_2,\omega\tau).
\end{eqnarray}
Here $M$ is the pion mass. We use the gluon propagator in the Feynman gauge and 
${\cal K}$ is the scalar part of the gluon
propagator (after separation of $-g_{\mu\nu}$). It coincides with the scalar 
$t$-channel exchange amplitude, given by eq.(\ref{k1}) (where one should 
put $\mu=0$). The matrix $O'$ is
the ``inner" part of the pion wave function as defined in
eq.(\ref{pion}). Prime means that it depends on the momenta in the loop.
For simplicity, we omit here any color degrees of freedom which are 
irrelevant for our consideration. 

To find the asymptotical behavior of the wave function, we shall use the 
iterative 
procedure already explained in chapter \ref{nns} to find the deuteron 
wave function. We suppose that $O'$ is concentrated in a finite region of 
the quark momenta and the integral (\ref{as1}) is dominated by this domain. 
These momenta are negligible relative to the asymptotical ones we are 
interested in. This means that we can 
evaluate all the factors except for $O'$ and the delta-functions at zero
relative quark momentum. This corresponds
to $\vec{R}\,'_{\perp}=0$ and $x'_1=x'_2=1/2$. In that case, the quark 
four-momenta are equal to each other: 
$$k'_1=k'_2=(p+\omega\tau_0)/2,$$
where $\tau_0=(4m^2-M^2)/(2\omega\cd p)$. We thus get the following 
replacement in (\ref{as1}):
\begin{eqnarray}\label{as2}
&&\int \frac{1}{(2\pi)^3}\delta^{(4)}(p+\omega\tau'-k'_1-k'_2) 
2(\omega\cd p)d\tau'
\nonumber\\
&&\times\theta(\omega\cd k'_2)\delta(k_2'^2-m^2)d^4k'_2\;\, O'\;\,
\theta(\omega\cd k'_1)\delta(k_1'^2-m^2)d^4k'_1
\;\Rightarrow \; O_0,
\end{eqnarray}
where $O_0$ is the integrated value of the wave function.

We substitute in (\ref{as1}) the one-component wave function obtained from 
(\ref{pion}) with $A_2=0$. Together with the replacement (\ref{as2}),  this
gives in (\ref{as1})  $O'=A_0\gamma_5/m$, where $A_0$ is a constant. The 
second  component $A_2$ of the wave function is then automatically generated by
eq.(\ref{as1}).

We thus find for the pion wave function:
\begin{equation}\label{as4}
\psi=-\frac{2A_0}{s-M^2}\bar{u}(k_2)(\hat{p}+\hat{\omega}\tau_0-4m)
\gamma_5v(k_1){\cal K}\ .
\end{equation}
 
Substituting here $p$ from the relation $p+\omega\tau=k_1+k_2$ and using the 
Dirac equation, we finally obtain:
\begin{equation}\label{as5}
\psi=\frac{2A_0}{s-M^2}\bar{u}(k_2)(\hat{\omega}(\tau-\tau_0)+2m)
\gamma_5v(k_1){\cal K}\ .
\end{equation}
The  pion wave function (\ref{as5}) has the form of eq.(\ref{pion}) with the 
following 
two components:
\begin{equation}\label{as6}
A_1=\frac{4m^2}{s-M^2}A_0{\cal K},\quad A_2\approx A_0{\cal K}\ 
\end{equation}
with $s=\frac{\displaystyle{\vec{R}_{\perp}^2+m^2}}
{\displaystyle{x(1-x)}}$. The kernel ${\cal K}$ given in (\ref{k1}) 
can be rewritten in terms of the variables $\vec{R}_{\perp},x$.
For $x \leq x'$, it reads:
\begin{eqnarray}\label{kernrx}
{\cal K}(\vec{R}'_{\perp},x';\vec{R}_{\perp},x,M^2)=&&
g^2\left[\mu^2+\frac{x'}{x}\left(1-\frac{x}{x'}\right)^2m^2\right.
+\frac{x'}{x}\left(\vec{R}_{\perp}-\frac{x}{x'}\vec{R}'_{\perp}\right)^2
\nonumber\\
&+&\left.(x'-x)\left(\frac{m^2+\vec{R}_{\perp}^{\prime 2}}{x'(1-x')}-M^2\right)
\right]^{-1} \ ,
\end{eqnarray}
with $\mu=0$.
For $x \geq x'$ $K(\vec{R}'_{\perp},x';\vec{R}_{\perp},x,M^2)$ is obtained from
(\ref{kernrx}) by the replacement: $x \leftrightarrow x'$, $\vec{R}_{\perp}
\leftrightarrow \vec{R}'_{\perp}$.
As indicated above, one should then put $\vec{R}'_{\perp}=0,x'=1/2$. 

We will see below that this wave function enters in the form factor at 
$x=1/2$. Then $s \approx 4 \vec{R}^2_{\perp}$. As follows from 
(\ref{kernrx}), the kernel ${\cal K}$ at $\vec{R}'_{\perp}=0,x'=1/2,x=1/2$ 
obtains the simple form: ${\cal K}=g^2/\vec{R}^2_{\perp}$. The components of the 
wave function which dominate in the asymptotical region
are then given by:
\begin{equation}\label{as8}
A_1=\frac{g^2m^2A_0}{\vec{R}^4_{\perp}},\quad 
A_2=\frac{g^2A_0}{\vec{R}^2_{\perp}}.
\end{equation}
In this region, the component $A_2$ decreases more 
slowly than $A_1$.

The form factor is now easily calculated through the components $A_1,A_2$ 
according to eq.(\ref{eq11pp}). The integral for the form factor is dominated
by  the region where one of the two wave functions (initial or final ones)
takes its non-relativistic value,
whereas the other wave function corresponds to the asymptotical one. This
is equivalent to the standard  calculation discussed in the literature
in the usual formulation of LFD ~\cite{brodsky}.  
Hence, similarly to eq.(\ref{as1}), we can replace $A'_1,A'_2$ in 
(\ref{eq11pp}) by
$$
A'_1= A_0\gamma_5\delta^{(2)}(\vec{R}_{\perp})\delta(x-1/2),\quad A'_2=0.
$$
and take the asymptotical values (\ref{as8}) for $A_1$ and $A_2$.
We thus find:
\begin{equation}\label{as7}
F(Q^2)=\frac{A_0}{\pi^3}(2A_1 + A_2)\approx 
\frac{A_0}{\pi^3} A_2.
\end{equation}
The substitution of $A_2$ from (\ref{as8}) in (\ref{as7}) 
with $\vec{R}_{\perp}=\vec{\Delta}/2$ and $\vec{\Delta}^2=Q^2$ 
gives:
\begin{equation}\label{as9}
F(Q^2) \approx \frac{4 g^2 A_0^2}{\pi^3}\frac{1}{Q^2}.
\end{equation}

From (\ref{as7}) and (\ref{as9})
we see that the asymptotical behaviour $\propto 1/Q^2$ of the pion form 
factor is indeed determined
by the extra component $A_2$ of the pion wave function, related to the
$\omega$-dependent spin structure. This originates from the slower  decrease of
$A_2\propto 1/\vec{R}^2_{\perp}$ in  comparison to $A_1\propto 
1/\vec{R}^4_{\perp}$ (see eq.(\ref{as8})).
The enhancement of $A_2$ is determined by the factor 
$\tau(\omega\cd p)=(s-M^2)/2\approx 2\vec{R}^2_{\perp}$ in 
(\ref{as5}). This extra factor originates from the part $\omega\tau$ in 
the conservation law $p+\omega\tau=k_1+k_2$ used to derive eq.(\ref{as5})
and therefore accompanies namely the $\omega$-dependent 
spin structure. 


\subsection{The nucleon form factors}\label{nucleonff}
As we have already seen in the case of the two-body systems, it is convenient to
parametri\-ze the covariant light-front wave function  in terms of the following
three sets of variables:

({\it i}) The four-momenta $k_i$ of each constituent. In these 
variables, the wave function of any three-quark system has the following 
form:  
\begin{equation}\label{wf2} 
\psi=\psi(k_1,k_2,k_3,p,\omega\tau;\sigma_1,\sigma_2,\sigma_3,\sigma)\ 
, 
\end{equation} 
where $\sigma$ is the nucleon spin projection and $\sigma_{1,2,3}$ are 
the quark spin indices, with the following 
conservation law \cite{karm88}: 
\begin{equation}\label{wf1}
k_1+k_2+k_3=p+\omega\tau\ .
\end{equation}

({\it ii}) The three-vector variables $\vec{q}_i$ which are identical, in the
c.m.s., to the constituent momenta. They are constructed similarly to the
two-body case, (\ref{sc4}):
\begin{equation}\label{wf3}
\vec{q}_i=L^{-1}({\cal P})\vec{k}_i\ , 
\end{equation}
where ${\cal P}=p+\omega\tau=k_1+k_2+k_3$ and $L^{-1}({\cal P})$ is the Lorentz
boost operator defined in (\ref{sc4}). The sum of $\vec{q}_i$ gives zero, as
it should be for the relative momenta: $\vec{q}_1+\vec{q}_2+\vec{q}_3=0$.

({\it iii}) The four-vectors: $R_i=k_i-x_ip$, 
where $x_i=\omega\cd k_i/\omega\cd p$. They can be represented as 
$R_i=(R_{i0}, \vec{R}_{i\perp},\vec{R}_{i\parallel})$, where 
$\vec{R}_{i\perp} \cd \vec{\omega}=0$, $\vec{R}_{i\parallel}$ is 
parallel to $\vec{\omega}$. We thus have the following relations:
$$
\vec{R}_{i\perp}^2=-R_i^2\ ,$$
$$\vec{R}_{1\perp}+\vec{R}_{2\perp}+\vec{R}_{3\perp}=0\ ,$$
$$x_1+x_2+x_3=1\ .$$

We do not pretend here to a calculation of the form 
factors in a realistic model for the nucleon. In order to illustrate 
our approach and estimate the effect of eliminating the unphysical 
$B_1$ contribution from the physical  form factors, we consider a 
simple model in which two neutral ``quarks"  labeled 2 and 3 are 
coupled to zero total angular momentum, so that the spin of  the 
nucleon is determined by the charged ``quark" labeled 1.  We put the 
masses of all the quarks equal to each other.  The 
corresponding wave function reads: 
\begin{equation}\label{wf5} 
\psi(k_1,k_2,k_3,p,\omega\tau;\sigma,\sigma_1,\sigma_2,\sigma_3)=
[\bar{u}^{\sigma_2}(k_2)\gamma_5U_c\bar{u}^{\sigma_3}(k_3)]
[\bar{u}^{\sigma_1}(k_1)u^{\sigma}(p)]\psi_0(k_1,k_2,k_3,p,\omega\tau)\ ,
\end{equation}
where $U_c=\gamma_2\gamma_0$ is the charge conjugation matrix.  The 
spinor structure of (\ref{wf5}) coincides with one of the invariants 
given in \cite{weber}. In order to show the method more clearly, we 
do  not symmetrize the wave function (\ref{wf5}) and do not introduce 
isospin. So, we have only one (charged) nucleon -- ``proton".  In 
eq.(\ref{wf5}) $\psi_0$ is a scalar function.

\begin{figure}[htbp]
\centerline{\epsfbox{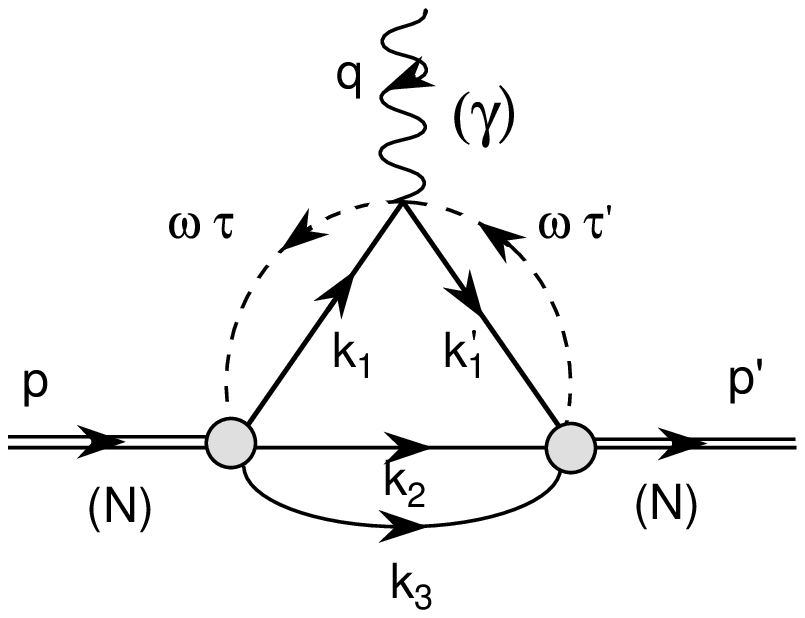}}
\figcap{Nucleon electromagnetic amplitude in the three-quark 
model.}
\label{nem}
\end{figure}

We consider the charged quark 1 as a structureless particle coupled to 
the electromagnetic  current:  $j_{\rho} 
=\bar{u}(k_1')\gamma_{\rho}u(k_1)$. The electromagnetic nucleon vertex 
(\ref{eq12}) is shown in fig. \ref{nem} and has the form: 
\begin{equation}\label{qff1}
\tilde{J}_{\rho}=\bar{u}'\tilde{\Gamma}_{\rho}u=\int[\bar{u}(p')(\hat{k}_1'+m)
\frac{\gamma_{\rho}}{x_1}
(\hat{k}_1+m) u(p)]Tr[(\hat{k}_2+m) (\hat{k}_3+m)] 
\psi_0\psi_0'D(R_i,x_i)\ .
\end{equation}
Here $\psi_0$ and $\psi_0'$ depend on the initial and final momenta
respectively. Details of the calculation can be found in ref. \cite{km96}.

At zero momentum transfer, the difference 
$G_M'(0)-G_M(0)$ obtains the simple form:
$$
G_M'(0)-G_M(0)=2B_1(0)=-\frac{\langle \vec{k}\,^2\rangle}{3m^2}.
$$
Here $\langle \vec{k}\,^2\rangle$ is the average quark 
momentum. The difference between $G_M'$ 
and $G_M$ is of course of relativistic origin.

\begin{figure}[htbp]

\vspace*{1cm}

\centerline{\epsfbox{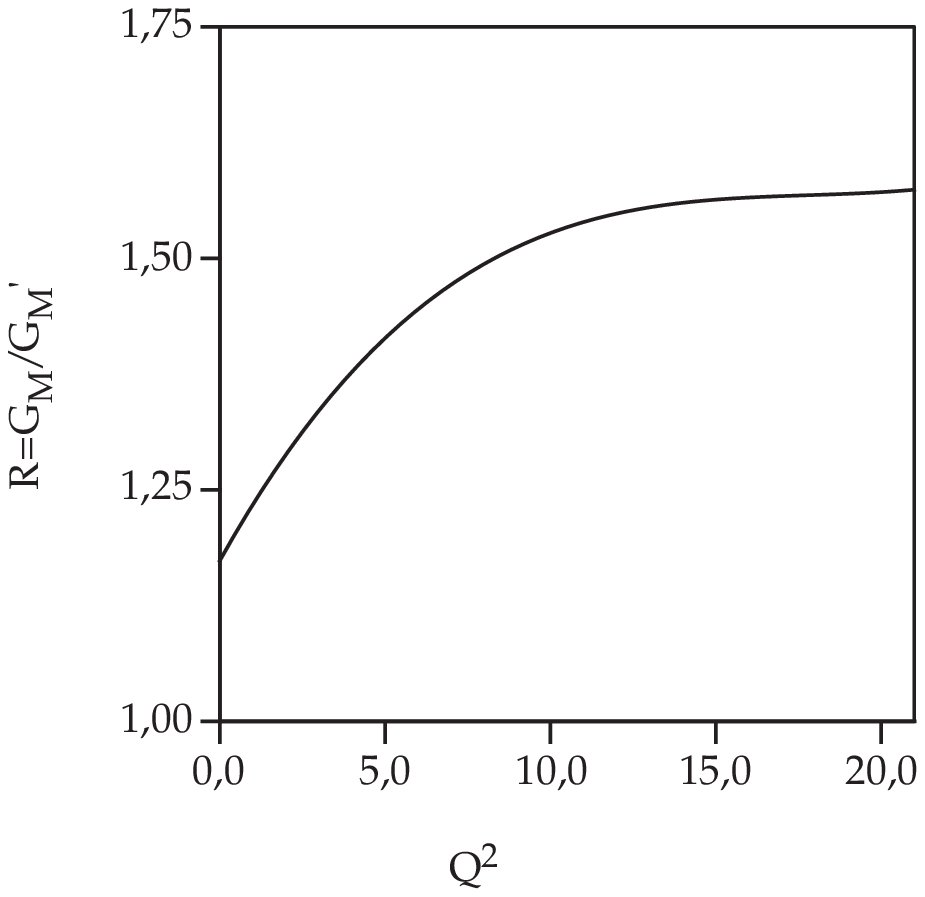}}
\figcap{The ratio $G_M/G'_M$ as a function of $Q^2$ in $(GeV/c)^2$\cite{km96}.}
\label{ratg}

\vspace*{1cm}

\end{figure}

We indicate in fig. \ref{ratg} the ratio $R =
G_M/G'_M$ of the physical form  factors $G_M$ obtained
with our formalism and the form factor $G'_M$  which would be obtained with the
usual procedure (using the $J_+$  component of the current), as discussed in
section 3.3. Both form factors are  calculated with a simple harmonic
oscillator radial wave function \cite{km96}.  We recall that  the electric form
factor $\tilde{G}_E$ is the same in both approaches. 

The difference between $G_M$ and $G'_M$ is of relativistic
origin. It  is already sizable at $Q^2=0$ for the magnetic moment 
(enhancement of about 15\%). The two form factors largely differ at  high
momentum transfer, where corrections of the order of 50\% are to  be expected. 
The physical form factor $G_M$, i.e. the form factor free  from any
ambiguity from the position of the light front,  is always larger 
than $G'_M$ in this model. According to fig. \ref{ratg}, the magnetic radius  is
larger for $G_M$ than for $G'_M$.

\section{Application to the nucleon-nucleon systems}\label{anns}                     
\subsection{Light-front dynamics and meson-exchange  
currents }\label{lfdmec} 
\subsubsection{Non-relativistic phenomenology}
The qualitative, and to a large extent also quantitative, understanding of 
electromagnetic observables in the few-body systems in terms of meson-exchange 
currents is now well established \cite{jfm}. This is particularly 
transparent for isovector magnetic transitions like the deuteron 
electrodisintegration cross-section near threshold. These two-body currents 
appear in that case already at order $1/m$, while they are of higher order for 
charge and isoscalar magnetic transitions. Last but not least, these currents 
are also required by the low-energy soft-pion theorems.

In order to extend these non-relativistic analyses to a wide range of momenta, 
let say for $Q^2>M^2$, a relativistic framework is necessary. In order to be 
fully consistent, the relativistic
analyses  have to give, as a limiting case, the well-known results in terms of
meson-exchange currents. While  genuine two-body currents like the mesonic 
contribution indicated
in fig. \ref{genuineMEC}} has  to be considered explicitly in both LFD and
non-relativistic formalisms, some  of them may appear differently depending on
the underlying framework. Physical  amplitudes should however be the same.

\begin{figure}[htbp]

\vspace{1cm}

\centerline{\epsfbox{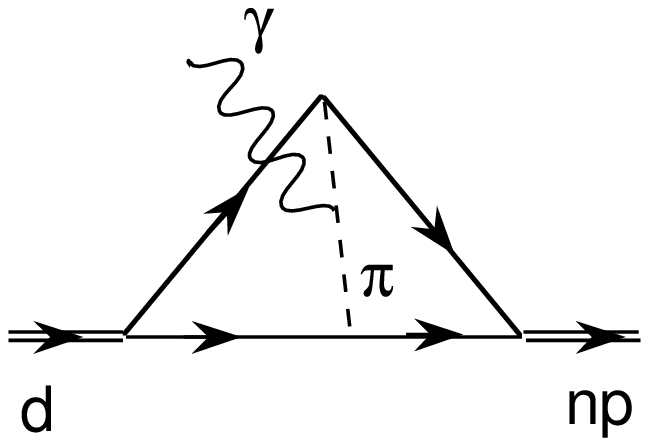}}
\figcap{Two-body meson exchange current: the mesonic current.}
\label{genuineMEC}

\vspace{1cm}

\end{figure}

This is the case for the so-called pair term which, in the non-relativistic 
approach, is indicated on fig. \ref{pairMEC} (we assume here that the $\pi NN$
vertex is of  the pseudo-scalar type, we shall come back to this point at the
end of this  section). This contribution cannot survive on the light front
because of the absence of vacuum fluctuation diagrams, together with the
particular  constraint $\omega \cd q=0$. A carefull analyzis of the deuteron 
electrodisintegration amplitude near threshold in LFD is therefore of
particular  interest.

\begin{figure}[hbtp]

\vspace{1cm}

\centerline{\epsfbox{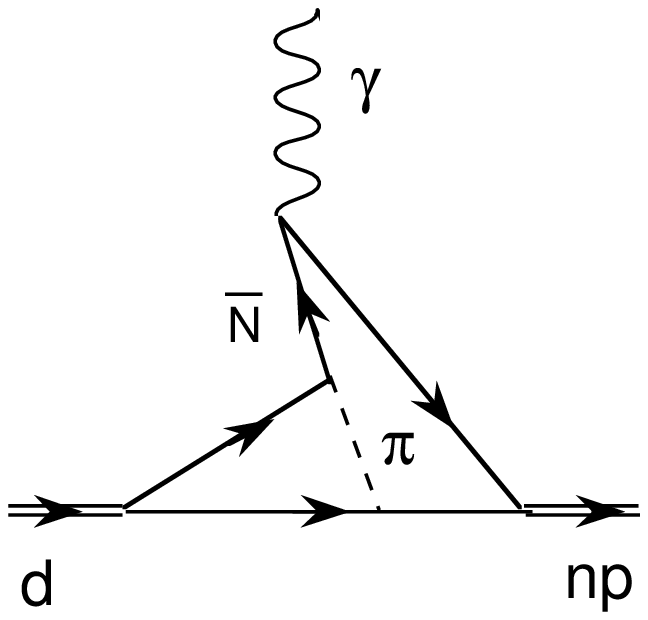}}
\figcap{Two-body meson exchange current: the pair term.}
\label{pairMEC}

\vspace{1cm}

\end{figure}

\subsubsection{Relativistic impulse approximation}

The vertex $d+\gamma^*\rightarrow np(^1S_0)$ is represented graphically 
in fig. \ref{f2-adk}. It corresponds to the following amplitude:  
\begin{figure}[htbp]

\vspace{1cm}

\centerline{\epsfbox{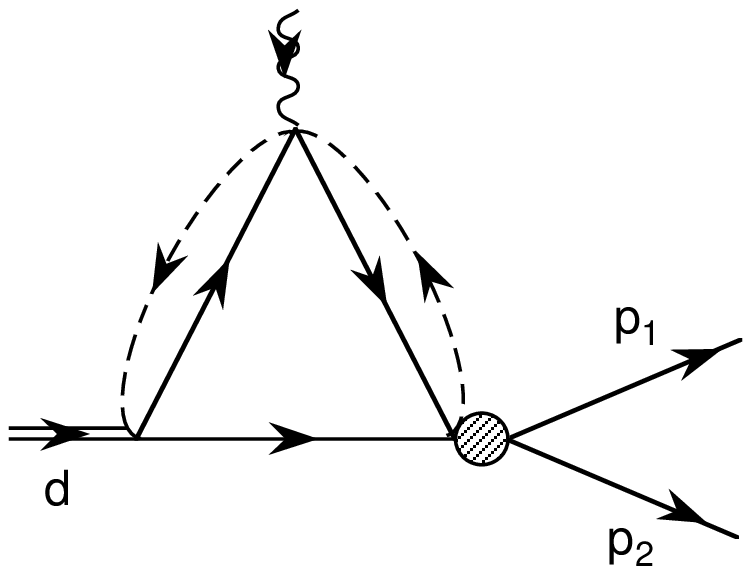}}
\figcap{The electromagnetic vertex 
$\gamma^*d\rightarrow np(^1S_0$).}
\label{f2-adk} 

\vspace{1cm}

\end{figure}

\begin{equation}\label{am1} 
{\tilde F}_{\mu\rho} = c\int \frac{Tr[\gamma_5(\hat{k}'_2 + m)\Gamma_{\rho}^V 
(\hat{k}_2 + m)\phi_{\mu}(\hat{k}_1 - m)]} {(1 - \omega\cd 
k_1/\omega\cd p)^2}\varphi_f \frac{d^3k_1}{\varepsilon_{k_1}}\ ,                                            
\end{equation}                                                                 
where $c =i2^{-11/2}\pi^{-3}\sqrt{m}$, $\phi_{\mu}$ is the deuteron wave 
function (\ref{nz2}), $\varphi_f$ is the final state wave function and 
$\Gamma_{\rho}$ is the nucleon electromagnetic vertex (\ref{ver1}).  
The superscript $V$ refers to the isovector part of the form factors 
which is the only one which contributes here.  In the plane wave approximation, 
we simply have:  
\begin{equation}\label{am3}                                                     
{\tilde F}_{\mu\rho}= i\sqrt{2m}Tr[\gamma_5(\hat{k}_f+m) 
\Gamma_{\rho}^V(\hat{k}_2+m)\phi_{\mu}(\hat{k}_1-m)]\ .                         
\end{equation}                                                                  
                                                                                
The expression of the electromagnetic current used here involves the Pauli and 
Dirac form factors $F_1$ and $F_2$. They differ off-energy-shell
from the  Sachs form factors used in ref. \cite{adk94}.
The present one for its $F_1$ part is consistent 
with the minimal substitution in the Dirac equation.  Sachs form 
factors may be used, but extra terms have then to be introduced in the 
current to get the exact result.  

The amplitude $A$, determining the cross section (\ref{eld7}) is found by the
formula (\ref{am5}). In the plane wave approximation and for zero c.m.-energy
in the final state, the amplitude $A$  calculated with the deuteron wave
function (\ref{nz2}) has the form:                       
\begin{equation}\label{am5p}                                                    
A=\frac{m^{5/2}2^{3/2}}{\varepsilon_k}(2G_M^V\varphi_2-G_E^V\varphi_5) 
\end{equation}                                                                  
with $k=\Delta/2$. We introduce in (\ref{am5p}) the charge and magnetic 
nucleon form factors according to (\ref{ver1p}). The deuteron 
components $\varphi_{1,3,4,6}$ do not contribute to (\ref{am5p}). 
Expressing $\varphi_2$, $\varphi_5$ through $f_1,f_2,f_5$ by eqs.(\ref{ba6}), 
(\ref{ba9}) (and neglecting the small contributions from
 $f_3$ and $f_4$) we get:                                            
\begin{equation}\label{am8}                                                     
A=2m^{3/2} 
\left\{G_M^V[u_S(\Delta/2)+\frac{1}{\sqrt{2}} u_D(\Delta/2)]-                   
G_E^V\frac{\sqrt{3}\ m}{\Delta}f_5(\Delta/2)\right\},                           
\end{equation}                                                                  
where $\Delta=\sqrt{Q^2}$. 

Neglecting $f_5$ in 
eq.(\ref{am8}), we recover the usual expression in  the plane wave 
approximation.                                                      

\subsubsection{Meson exchange currents and 1/m expansion}\label{extr-dis}                                                       
 
In order to make the connection with  the non-relativistic phenomenology, it
is  instructive to calculate the $f_5$ component in leading order in a $1/m$ 
expansion. Following the iterative procedure used in refs.
\cite{ck-deut,ck-fsi},  and recalled in section   \ref{numer-d}, the $f_5$
component has the following  form in momentum space and in $1/m$ order
\cite{dkm95}:

 \begin{eqnarray}\label{3.8} 
f_5[\vec{k}\times \vec{n}]_i/k= -
\frac{3g_{\pi NN}^2\sqrt{2}}{8\sqrt{3}\pi^2m^2} \int
\frac{(\vec{k}\,^2-\vec{k}\,'^2)}{(\vec{k}\,^2+\kappa^2)}
\frac{d^3k'}{\mu^2+(\vec{k}-\vec{k}\,')^2}\nonumber \\ \times
\left[\frac{1}{\sqrt{2}}u_S(k')\delta_{ij}-\frac{1}{2}
(\frac{3k_i'k_j'}{\vec{k}\,'^2}
-\delta_{ij})u_D(k')\right][(\vec{k}-\vec{k}\,')\times\vec{n}]_j\ .
\end{eqnarray} 
In the limit where the momentum $k'$ is restricted by the integration domain to 
low momenta, one can take $(\vec{k}^2-\vec{k}'^2)/(\vec{k}^2-\kappa^2) \simeq 
1$. We shall come back to this approximation later on.
                                                                          
The coordinate space integral for the $f_5$ component has the form:                                                  
\begin{equation}\label{f5}                                                      
f_5(k)=-\frac{3g_{\pi NN}^2}{8\sqrt{3\pi}m^2}\int_0^{\infty}                       
\frac{\exp(-\mu r)}{r}(\mu r+1)j_1(kr)\left[u(r)+\frac{1}{\sqrt{2}}             
w(r)\right] dr \ ,                                                                 
\end{equation} 
where $u$ and $w$ are the S and D state deuteron wave functions in coordinate 
space. The  extra component $g_2$ of the final  
state wave function can be calculated simply from isospin arguments, leading to 
the replacement $-3g_{\pi NN}^2 \rightarrow  
-(3+1)g_{\pi NN}^2= -4g_{\pi NN}^2$ in the amplitude.  

The amplitude A has thus the following form, in coordinate space:             
\begin{eqnarray}\label{3.10p}                                                   
A=4\pi^{1/2}m^{3/2}\left\{ G_M^V 
\int_0^{\infty}\left[u(r)j_0( \frac{r\Delta} {2})- 
\frac{1}{\sqrt{2}}w(r)j_2(\frac{r\Delta}{2})\right] rdr\right.                  
\nonumber \\ +\left.\frac{G_C^Vg_{\pi NN}^2}{m\Delta}\int_0^{\infty}               
\frac{\exp(-\mu r)}{r}(\mu r+1)\left[u(r)                                       
+\frac{1}{\sqrt{2}}w(r)\right] j_1(\frac{r\Delta}{2})dr\right\}.                
\end{eqnarray}                                                                  
                                                                                
The structure of the second term is that of a two-body contribution of order 
$g_\pi^2/4\pi$. It is similar to the one given by the pair term in the 
non-relativistic MEC analyses \cite{jfm,hockert,mat84,riska}. However, its 
strength is smaller by a factor of 2.                                                                 
                                                                                
\subsubsection{Contribution from the contact term}\label{cted} 

The one to one correspondence with the non-relativistic phenomenology in terms 
of MEC is indeed recovered if one collects all the contributions of order 
$g_{\pi NN}^2/4\pi$ to the light-front amplitude. The missing contribution
originates  in this particular example from the contact term which appears once
we have  singled out the one-pion exchange mechanism in generating the
$f_5(g_2)$  component of the deuteron ($^1S_0$) wave function. It is indicated
in  fig. \ref{ct4}. There are four diagrams corresponding to the contact terms.
Two of them, \ref{ct4}(a) and (b), correspond to the contact interaction at the
left of the photon line. Two other diagrams, \ref{ct4}(c) and (d), correspond to 
the 
right contact  term with the two possible time orderings for the
 lower vertex. 
Each diagram can be calculated in terms of the kinematical invariants
given in  section \ref{twoloop} of appendix  \ref{kr}.
 In leading order
in $1/m$, the contact term contribution is exactly equal to  the contribution
from the $f_5(g_2)$ components, and therefore the total  amplitude  of the
deuteron electrodisintegration near threshold is strictly equivalent in  LFD
and in the non-relativistic phenomenology including meson-exchange currents 
(more precisely the 
pair term ) \cite{dkm95}.

\begin{figure}[hbtp]

\vspace{1cm}

\centerline{\epsfbox{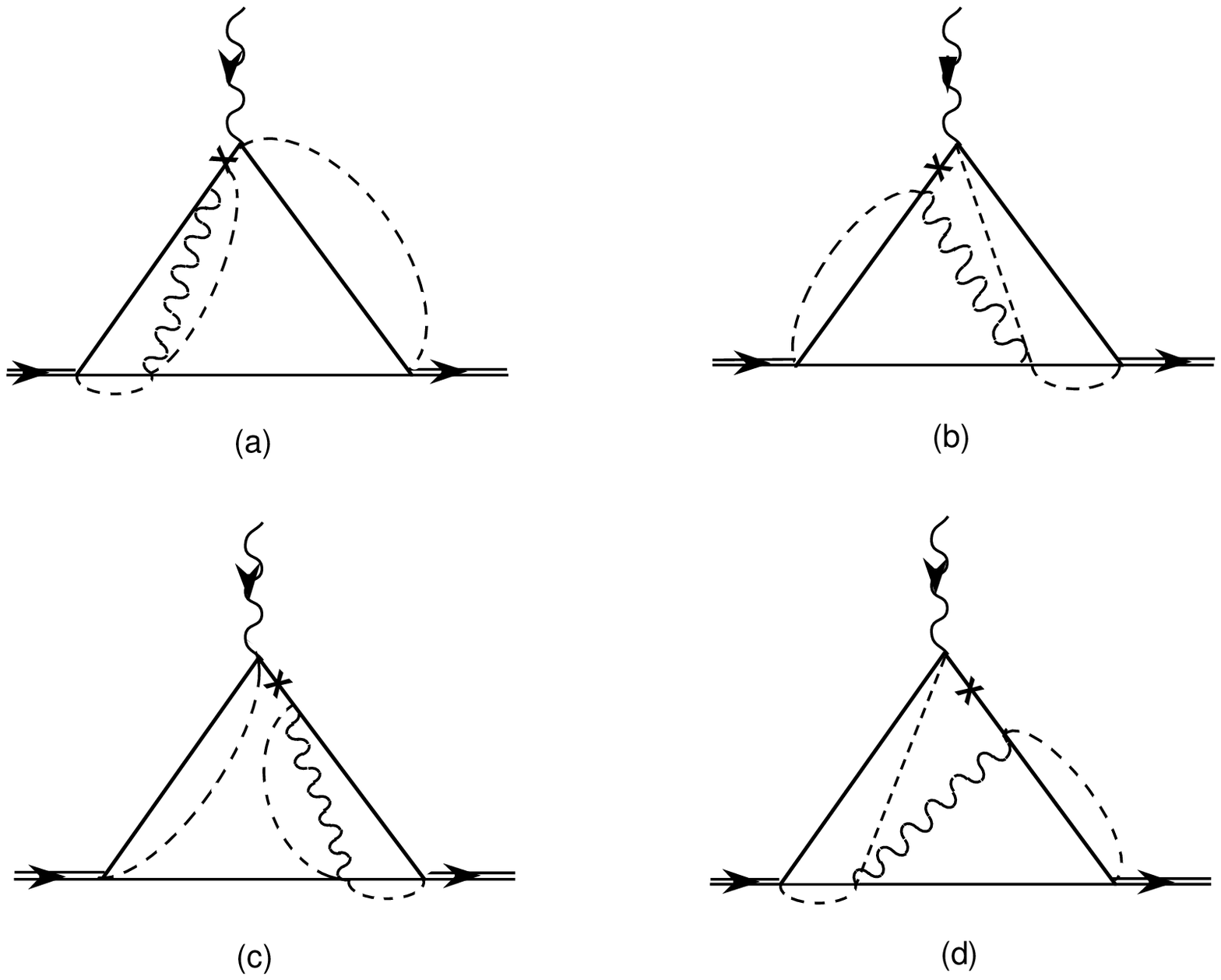}}
\figcap{Contact terms contribution to the electromagnetic interactions with the
deuteron.} 
\label{ct4}

\vspace{1cm}

\end{figure}

This analysis can even be extended to the whole domain of momenta in 
eq.(\ref{3.8}), i.e. without neglecting the factor 
$(\vec{k}^2-\vec{k}'^2)/(\vec{k}^2+\kappa^2)$ in the integrand. This term
gives  in fact a wrong asymptotical behaviour of the wave function.  As shown
above in sect. \ref{beyond} and in ref. \cite{dkm95p}, incorporation of the
complete contact interaction which arises with the other  components of the
kernel (NN potential) like $\sigma$ and $\omega$ exchange  and of a higher
order iterations, gives a correction which turns the factor 
$(\vec{k}^2-\vec{k}'^2)/(\vec{k}^2+\kappa^2)$ into 1.

For the deuteron electrodisintegration, the dominant component $\varphi_5$ of
the deuteron wave function does not contribute to the amplitude corresponding
to the contact term.  For the elastic $ed$ scattering with the representation
(\ref{ver1}) for the nucleon electromagnetic vertex $\Gamma_{\rho}$, the contact
terms contribute to  the deuteron electromagnetic form factor ${\cal G}_1$ only
and do not contribute to ${\cal F}_1$ and ${\cal F}_2$.

\subsubsection{The relevance of $\pi$ exchange. Numerical results}

The 1/m analysis of the relativistic contributions to the deuteron 
electrodisintegration amplitude is based on the $\pi$-exchange interaction with 
the pseudo-scalar $\pi NN$ vertex. It is also well known that, in leading $1/m$ 
order, the pair 
contribution associated with the pseudo-vector $\pi NN$ coupling is zero.  
After minimal substitution in the $\pi NN$ vertex, generating a genuine 
$NN\pi\gamma$ current, the deuteron electrodisintegration amplitude is 
however equivalent to the one given in the pseudo-scalar 
representation.  In LFD, the equivalence between 
the two representations is realized in the following way. In the 
pseudo-vector representation with $\pi$-exchange only, the $f_5$  
component in the deuteron wave function  is strictly zero in leading 
order. This is due to the off-shell  condition at the $\pi NN$ vertex. 
The contribution from the light-front  contact term is also of higher 
$1/m$ order for this representation. On  the other hand, the current 
originating from the direct coupling of the  photon to the pseudo-vector 
$\pi NN$ vertex has its analogue in LFD, 
providing the equivalence between relativistic  and non-relativistic 
formulations in leading $1/m$ order.

\begin{figure}[htbp]
\epsfxsize=6cm
\centerline{\epsfbox{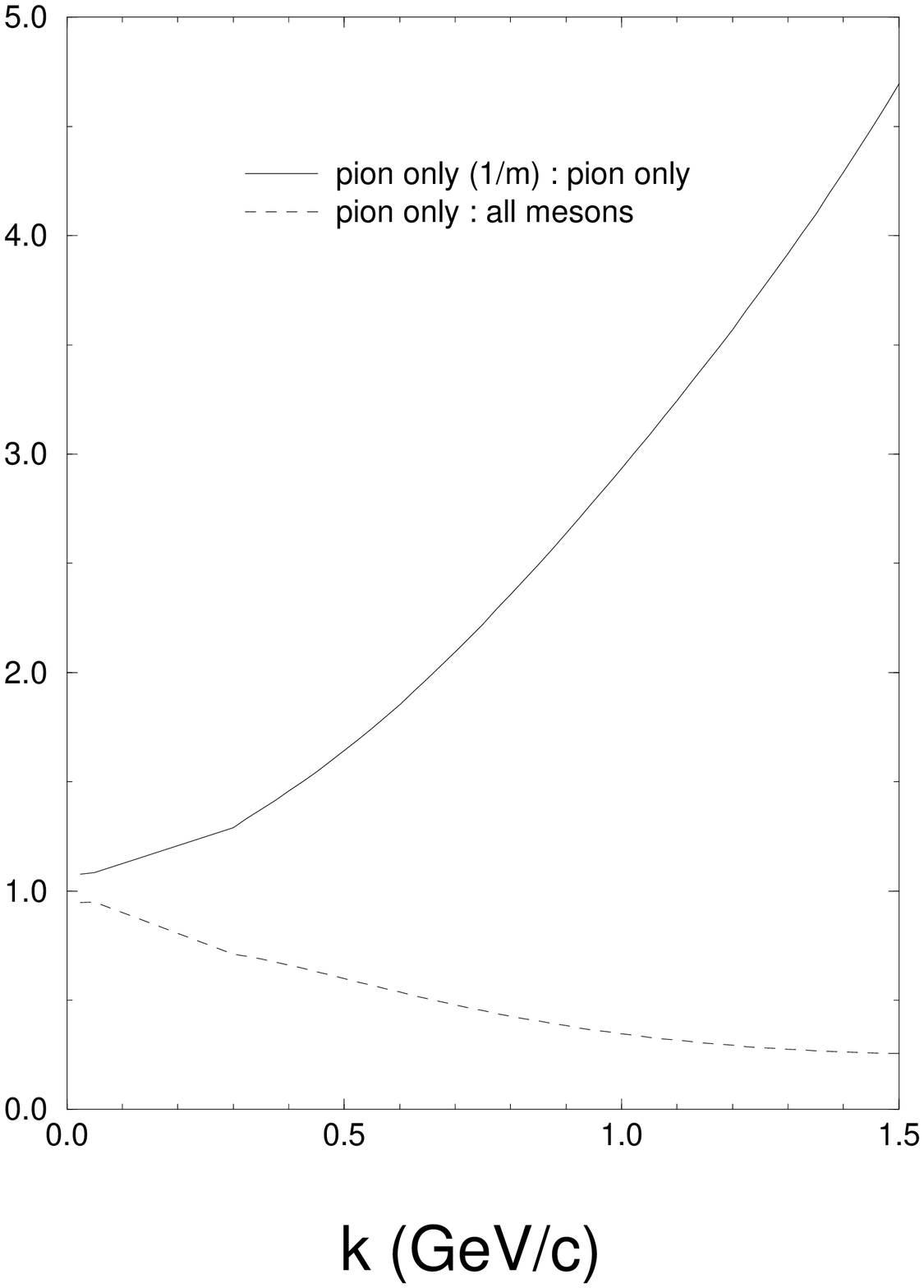}}
\vspace{6.cm}
\figcap{The solid curve is the ratio of $f_5$'s in $1/m$  approximation and
beyond, both for the pion exchange only. The dashed curve is the ratio of
$f_5$ for pion only to $f_5$ incorporating all the  meson exchanges, both
beyond $1/m$ approximation.}
\label{ratf5}
\end{figure}

The accuracy of the $1/m$ approximation and the contribution of the pion
exchange relative to all mesons are shown in fig. \ref{ratf5}. The solid curve
in fig. \ref{ratf5} is the ratio of $f_5$ for the pion  exchange only, 
calculated in the $1/m$ approximation by eq. (\ref{3.8}), to the $f_5$, also
for the pion exchange only, but calculated beyond the $1/m$ approximation. It
shows the accuracy of the $1/m$ approximation, which is very good in the
non-relativistic region of $k$, but changes the result by a large factor for
higher $k$. The dashed curve is the ratio of $f_5$ for pion only to $f_5$
incorporating all the  meson exchanges, both beyond the $1/m$ approximation.
The pion exchange contributes  about 60 per cent in the region $k\approx 0.4$
GeV/c, where $f_5$ dominates. 

The cross section of  deuteron electrodisintegration 
is shown in fig.\ref{edenp1}.  
\begin{figure}[htbp]
\epsfxsize=6cm
\centerline{\epsfbox{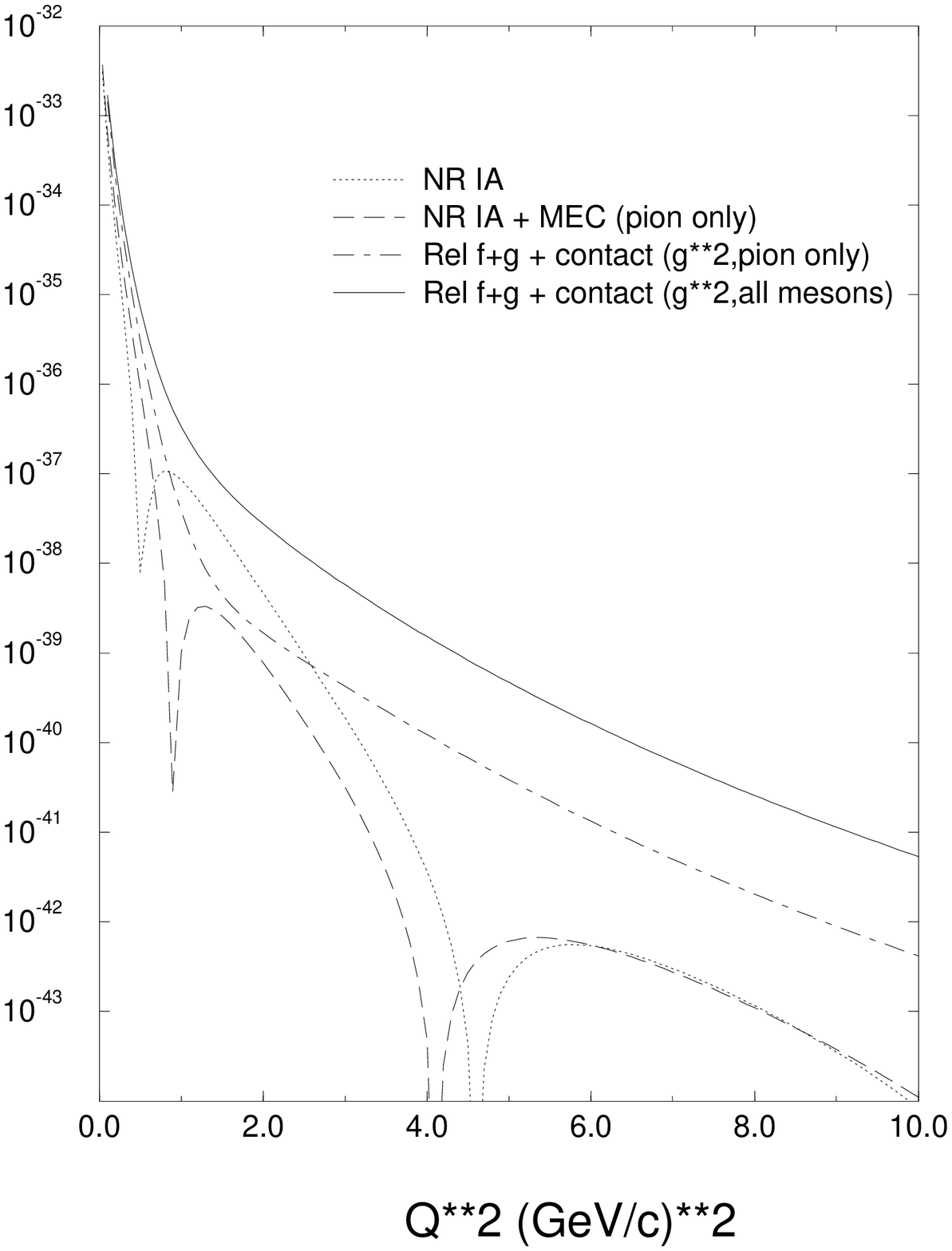}}
\vspace{2cm}
\figcap{The cross section $\frac{\displaystyle{d\sigma}}
{\displaystyle{d\Omega_e dE'}}$ of  deuteron  electrodisintegration
cross-section in $cm^2$/MeV/$sr$ for $\theta_e=155^{\circ}$. The dotted curve
is the non-relativistic impulse approximation with the Bonn deuteron wave
function \protect{\cite{bonn}} and with the Paris scattering state wave
function \protect{\cite{lois}}. The dashed curve includes the dominant
contribution to meson exchange currents (the standard pair term), calculated
with $\pi$-exchange only. The dot-dashed curve is the relativistic impulse
approximation with the light-front deuteron wave function ($f_1,f_2,f_5$) and
the scattering state one ($g_1,g_2$) and includes the contact term. The extra
components $f_5$, $g_2$ and the contact term have been calculated with pion
exchange only. It does not include the product  $f_5\times g_2$ and extra
components of  the wave functions in the calculation of the contact term. The
solid curve differs from the dot-dashed one by the contribution of all mesons 
both in $f_5,g_2$ and in the contact term.}
\label{edenp1} 
\end{figure}
As it was explained above, there is equivalence, in $1/m$ order, between the
relativistic impulse approximation (including all components of the wave
functions, together with the contact interaction), and non-relativistic
calculations including the dominant contribution to meson-exchange current (the
so-called pair term).
The comparison between meson exchange
contributions (pair term) and the  light-front approach is shown in fig.
\ref{edenp1}. The dotted line is the non-relativistic  impulse approximation
without meson exchange currents. The dashed line is the sum of the
non-relativistic impulse  approximation and of the contribution from the pair
term  for $\pi$-exchange only. The dot-dashed line is the relativistic impulse 
approximation (with $f_5$ and $g_2$) together with the  contribution from the 
contact term. The relativistic components $f_5, g_2$ and the contact term are 
calculated
with $\pi$-exchange only and with the pseudo-scalar $\pi NN$ vertex, according 
to
the iterative procedure indicated in sect.\ref{numer-d}. To be consistent with
this iterative procedure, we  omit in the amplitude terms of higher order than 
$g_{\pi NN}^2$. The
solid curve differs from the dot-dashed one by the contribution of all mesons 
both in $f_5,g_2$ and in the contact term. For the nucleon electromagnetic form
factors the dipole fit \cite{lapidus} was used. 

The deviation between the
dotted  and dashed lines shows the well known influence of meson-exchange
currents due to the pion
pair term. The deviation  between the dashed and  dot-dashed lines indicates
that  the influence of relativistic effects to both the deuteron and the 
scattering
state wave function, and to the electromagnetic operator (contact term) are 
important. The comparison
between dash-dotted and solid curves shows that the contribution of other
mesons, in addition to the pion, considerably influence the cross section at
large $Q^2$.

The relativistic calculations in fig.
\ref{edenp1} are higher than non-relativistics ones. This difference is related
to a deviation of relativistic scattering state wave function $g_1$ from
non-relativistic one (see fig. \ref{fckd5}). As seen in fig. \ref{fckd5}, the 
non-relativistic scattering state wave function in momentum space, in contrast
to the non-relativistic one, changes sign. On the other hand, one can expect
that the recoil-type diagram, indicated on fig. \ref{f4-ks92}, partially cancels 
the
difference between both calculations. These calculations should therefore be
considered as an estimate of relativistic  effects and are not compared with
experimental data. 

The difference  between the non-relativistic impulse approximation (dotted
curve) and the relativistic impulse approximation, calculated with the same
wave functions (neglecting $f_5$ and $g_2$), is negligible. This was already 
noticed in
ref. \cite{keister88,adk94}.

\subsection{The elastic deuteron form factors}                                             
The deuteron form factors can be calculated according to  
eqs.(\ref{ff6})-(\ref{ff8})
which express them through the tensor $\tilde{J}^{\mu\nu}_{\rho}$.
This tensor is expressed through the deuteron wave 
function  by means of the rules of the graph technique given in chapter
\ref{cov-graph}. In the impulse approximation this expression is given by
eq.(\ref{ff3a}). It includes the full deuteron wave function $\phi^{\mu}$,
with its six  components. We shall take into account below the dominant
ones, $f_1,f_2$ and $f_5$. The contribution of the contact term to the tensor
$\tilde{J}^{\nu\mu}_{\rho}$ is obtained by calculating the amplitude
corresponding to the graph of fig. \ref{ct4}. All relevant kinematical 
invariants
are given in appendix \ref{kr}. 

Like in the case of the electrodisintegration cross-section, the component
$f_5$ of the deuteron wave function was calculated by an iterative procedure,
keeping the first degree of the  relativistic kernel (of the order of $g^2$).
Relativistic contributions of order $g^4$, like terms in $f_5^2$, have been
removed from the calculation of the form factor.

\begin{figure}[htbp]

\vspace{1cm}

\epsfxsize=6cm
\centerline{\epsfbox{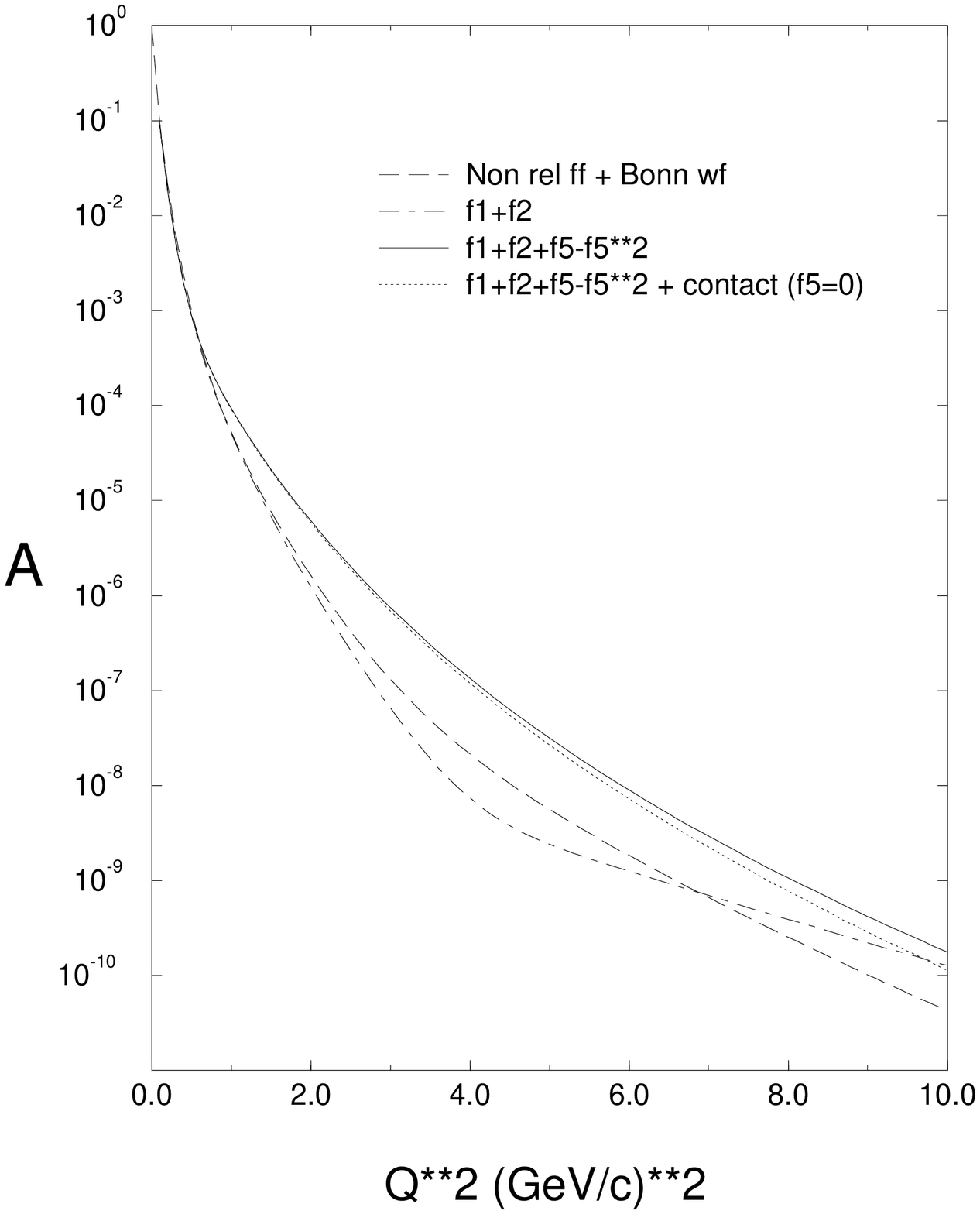}}

\vspace{1.5cm}

\figcap{The structure function $A(Q^2)$ for the deuteron. The dashed line is 
non-relativistic impulse approximation calculated with the Bonn wave function.  
The
dot-dashed curve is the relativistic impulse approximation  with relativistic
deuteron components $f_1$ and $f_2$. The solid curve is the same as the
dot-dashed one, but including the $f_5$ component (in  first degree only).
The dotted curve incorporates, in addition, the contact term. The nucleon form
factor were taken in the dipole form \cite{lapidus}.}
\label{A1}
\end{figure}

The influence of 
relativistic effects on the structure function $A(Q^2)$ is shown in fig. 
\ref{A1}.
 The dashed curve
corresponds to the non-relativistic impulse approximation \cite{gourdin,elias69} 
with the S-
and D-waves of the Bonn-QA wave function  \cite{bonn}.  The dot-dashed line is
calculated in the light-front formalism with relativistic deuteron components
$f_1$ and $f_2$  (solid lines in figs. \ref{ck-f1} and \ref{ck-f2}), but
without $f_5$. The solid curve is similar to the dot-dashed one, but
includes the $f_5$ component (in  first degree only). The dotted curve
incorporates, in addition, the contact term (where the contributions of 
order  $g^2$  are taken into account only).

\begin{figure}[htbp]

\vspace{1cm}

\epsfxsize=6cm
\centerline{\epsfbox{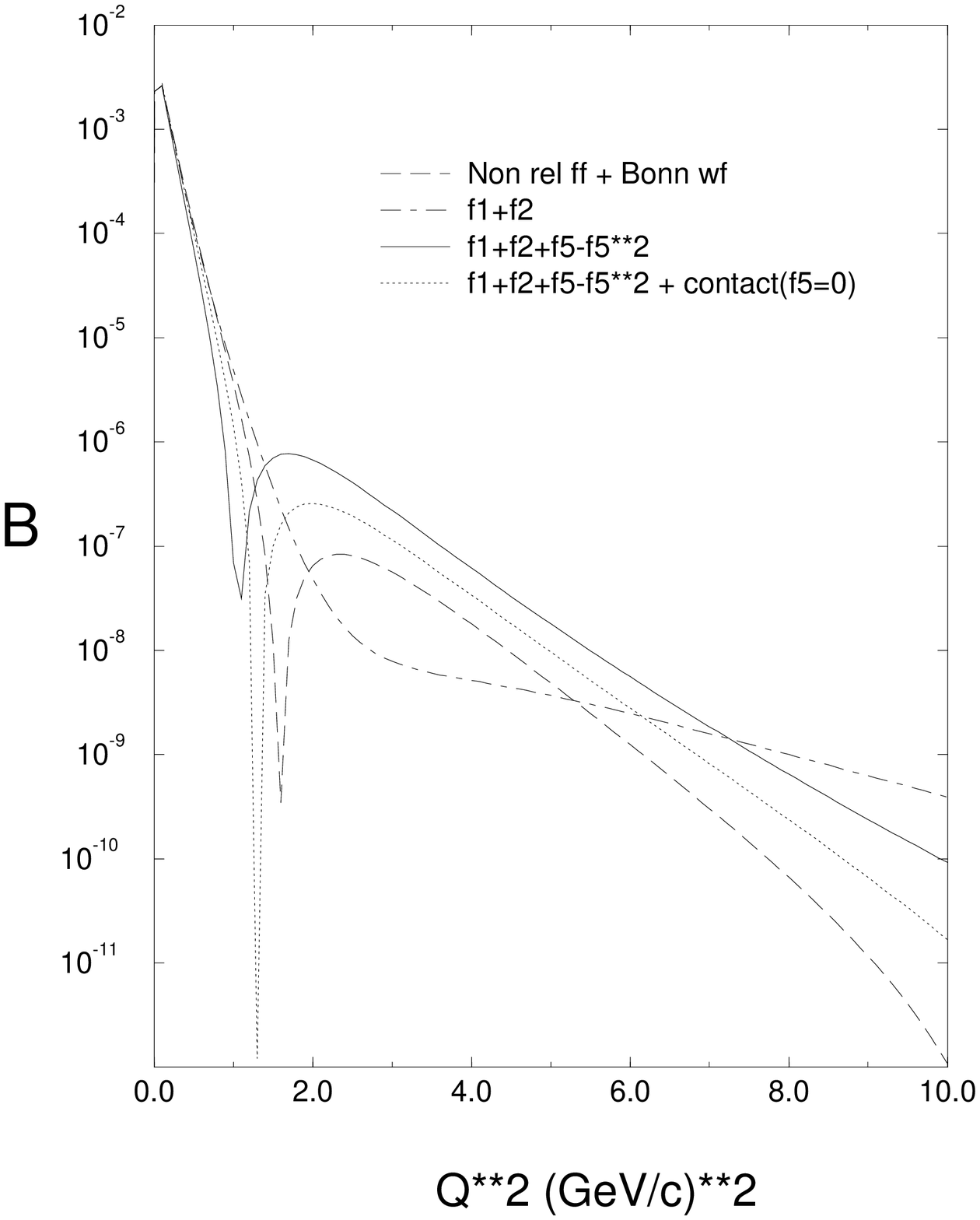}}
\vspace{1.5cm}
\figcap{Same as fig. \protect{\ref{A1}} but for the $B(Q^2)$ 
form factor of the deuteron.}
\label{B1}

\vspace{1cm}

\end{figure}

Figure \ref{B1} the
$B(Q^2)$ deuteron form factor. One can see that the function $B(Q^2)$ is more 
sensitive to the different
approximations and contributions than $A(Q^2)$. The comparison between the 
dot-dashed
and solid curves shows that the influence of the extra component $f_5$ of the
deuteron wave function is important. The minimum of $B(Q^2)$ in this figure
is the consequence of the $f_5$ component.

\begin{figure}[htbp]
\vspace{1cm}
\epsfxsize=6cm
\centerline{\epsfbox{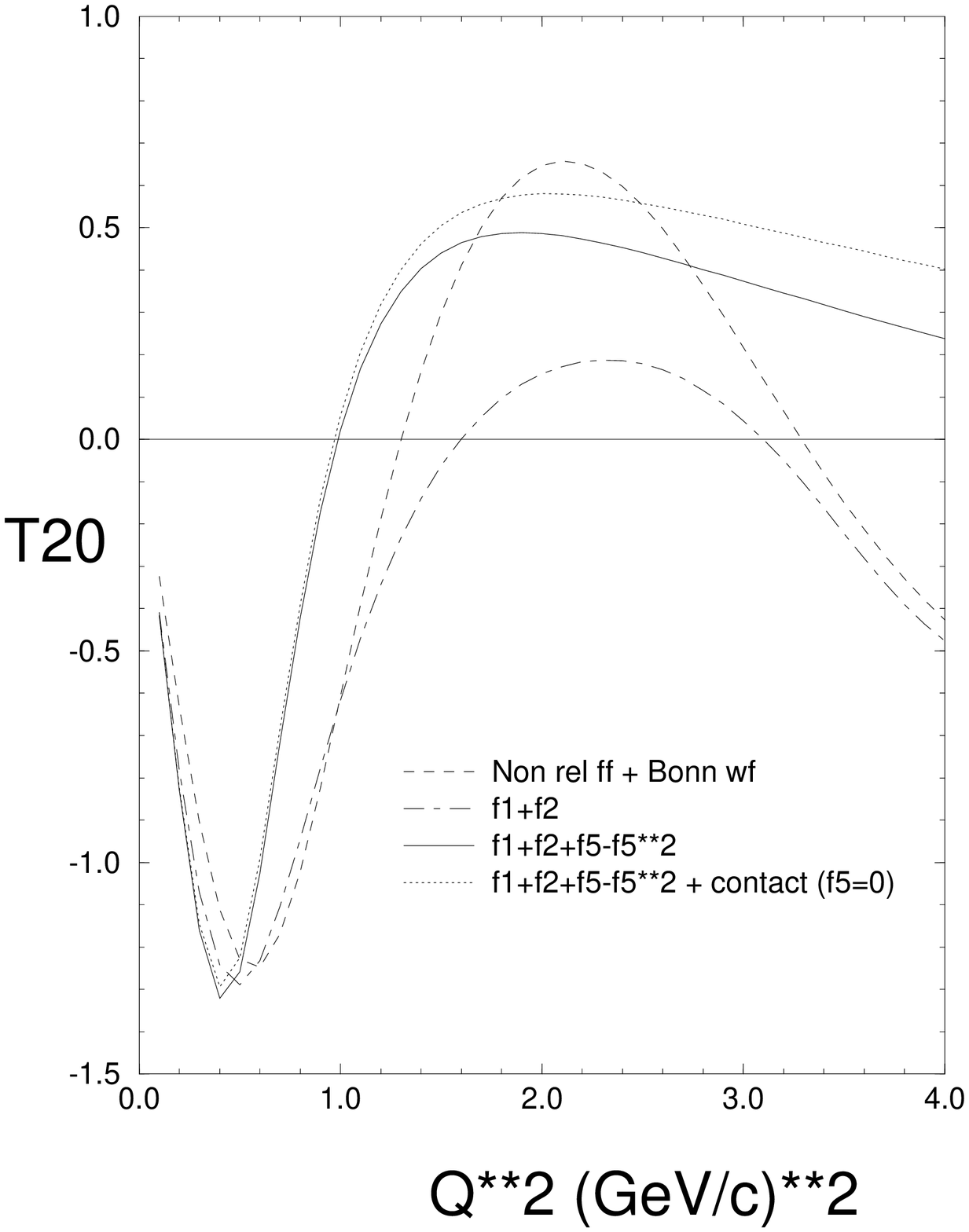}}
\vspace{1.5cm}
\figcap{The deuteron tensor polarization $T_{20}$ at $\theta=70^{\circ}$. 
calculated by relativistic formulas for form factors.  The designations of the
curves are the same as in figs. \protect{\ref{A1}} and \protect{\ref{B1}}}.
\label{T20}
\end{figure}

Figure \ref{T20} shows the influence of different approximations to
$T_{20}$. The effect of incorporating the component $f_5$ is qualitatively 
similar to that
obtained when adding the contribution of the pair current in non-relativistic
calculations \cite{sriska91}.

\begin{figure}[htbp]

\vspace{1cm}

\epsfxsize=6cm
\centerline{\epsfbox{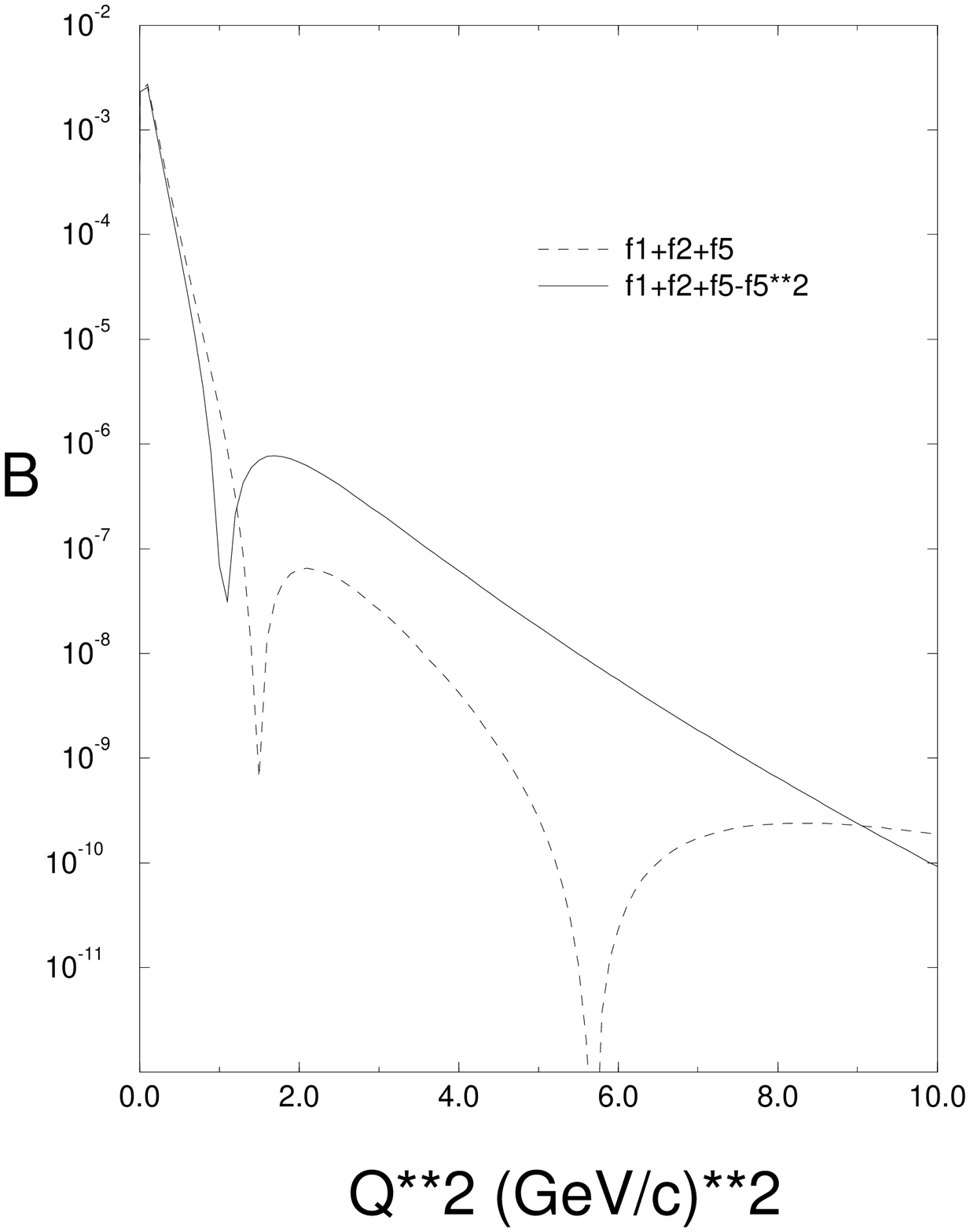}}

\vspace{1.5cm}

\figcap{The structure function $B(Q^2)$. The solid curve is the same as in
fig. \protect{\ref{B1}}. The dashed curve incorporates the quadratic terms in
$f_5$.}
\label{B2}

\vspace{1cm}

\end{figure}

Figure \ref{B2} illustrates the influence of $B(Q^2)$ of the higher order
terms than $g^2$. Their importance indicates the need to carry out the
calculations beyond the iterative procedure. The function $A(Q^2)$ is less
sensitive to this contribution.

The results presented in this section should be considered only as indicative.
They can be improved in many respects. An exact solution of the equation for
the deuteron and scattering state wave functions is indeed feasible. It implies
the redefinition of the parameters of the NN kernel. However, we believe a
realistic estimate of relativistic corrections to the deuteron can be performed
along the lines presented in this section. It necessitates a careful choice of
the NN potential to generate the kernel (energy-dependent OBE potential for
instance), and a good starting point for the non-relativistic wave functions
(with kinematical relativistic corrections indicated in sect. \ref{nrp}). This
may imply also the use of a PV $\pi NN$ coupling to be consistent with chiral
symmetry constraints in leading order. This necessitates also a careful
treatment of second order contributions, as discussed in sect.\ref{beyond}.
Finally, the calculation of electromagnetic observables should include
retardation effects as indicated in fig.\ref{f4-ks92}.

                                                                                
\chapter{Concluding remarks}\label{concl}    

We have reviewed the general properties of the covariant formulation of  LFD, 
and its application to relativistic few-body systems. We have first detailed
the  general  transformation properties of  the state vector, both kinematical
and dynamical,  and  emphasized the most important features brought about by
the covariant  formulation of  the  state vector on  the plane $\omega \cd
x=0$.  The explicit covariance provides similar advantages over the standard
formulation, as the Feynman graph technique in comparison to old fashioned
perturbation theory. We have presented a covariant light-front graph technique
to calculate  physical amplitudes. 

The covariance enables us also to  construct  bound 
and scattering states of definite angular momentum, separating kinematical and
dynamical aspects of this problem. The angular momentum of a system is a 
property of the wave
function relative to transformation of the coordinate system and therefore has
in our formalism a purely kinematical nature. Dynamics is involved in the
composition of this angular momentum from the spins of its constituents. 

We illustrate our  formalism  with  the most simple and adequate system, the
two nucleon system, and construct  explicitly  the deuteron wave function and
the $^1S_0 $ scattering state. In both cases, the dominant relativistic 
components of the wave function originating from its dependence on the light 
front position (the variable $\vec{n}$) has been found.  We also  extend our 
formalism to the $\pi$
and nucleon wave function in terms of valence quarks and  antiquarks.

This formulation is  essential in order to extract the physical amplitude
for all  electromagnetic transitions. This is mandatory in all practical
calculations in  order  to  avoid non-physical contributions coming from the
definite choice of  $\omega$. We show how this applies to the electromagnetic
form factors  of spin 0,  1/2 and 1 systems, as well as for the deuteron
electrodisintegration. We show  also in  this  last example how one can recover
easily, in a $1/m$ expansion, the well known  results  obtained in the
non-relativistic phenomenology in terms of meson-exchange  currents  (the 
so-called pair term). The dominant $\vec{n}$-dependent components of the wave 
functions are essential in order to reproduce these results. This shows that the 
calculations which omit this dependence cannot be considered as realistic. 

These developments have been made particularly simple, and intuitive, by the 
three- dimensional nature of the formalism, and the absence of vacuum
fluctuations.  This  enables   a transparent link with the non-relativistic
framework developed over  the  last  twenty years in the microscopic structure
of few-nucleon systems. 

We expect many new developments in the near future along the lines presented
in  this  review, both in nuclear and particle physics.
One of them concerns the particular way to implement the independence of
various results on the light-front orientation. Some restricted examples
(determination of the $NN$-interaction, normalization of the wave function,
$\ldots$)
have been considered in this review to show how this canbe done, but more 
extensive work is
required. Another related aspect, which is related, concerns the construction of 
the
generators satisfying, in the truncated Fock space, the commutation relation  
of the Poincar\'e group algebra, that also provides independence of observables
on the light-front orientation.

This feature of LFD -- the dependence of  approximate on-shell amplitudes on
$\omega$, which has to be excluded from  observables, -- has, however, a
positive aspect. It gives a quantitative measure of the incompleteness of a
given approximation. The model calculations in other approaches where this
feature  is absent are often still incomplete, but this does not manifest itself 
explicitly.  

In nuclear  physics, the exact  calculation of  the two-body wave function is
worth to be done. This however necessitates to  reparametrize  the NN potential
starting from the same relativistic formalism. This is   essential in 
order to have quantitative predictions for electromagnetic observables in the 
few  GeV  range. In particle physics, the structure of the pion and nucleon can
be  investigated  easily  in this framework, and electromagnetic form factors
calculated in the high  momentum  region. Application to virtual compton
scattering can also be considered, as  well  as the extension to axial
transitions. One other particularly interesting  application is the  structure
of heavy quarkonium, like the $J/\psi$, in order to investigate the  importance
of  relativistic dynamical corrections, i.e. the importance of higher Fock
states in  the  $J/\psi$  wave function.

Last but not least, the application of the covariant  formulation of LFD to 
field theory should 
be investigated.

\section*{Acknowledgement} 
One of the authors (V.A.K.) is sincerely grateful for the warm hospitality of
the theory group at the Institut des Sciences Nucl\'{e}aires, Universit\'e
Joseph Fourier, in Grenoble and of Laboratoire de Physique Corpusculaire,
Universit\'e Blaise Pascal, in Clermont-Ferrand, where  part of this work was
performed.

                                                                                
\appendix                                                                       
\chapter{Notations}\label{nota}                                                 
                                                                                
We use the standard covariant/contravariant notation, with implicit summation
over repeated indices, according to:
\begin{equation}
x^\mu=(t,{\vec x}),\quad x_\mu=g_{\mu\nu} x^\nu, \quad
g_{\mu\nu}= \left(\begin{array}{rrrr} 1&0&0&0 \\ 0&-1&0&0 \\ 0&0&-1&0 \\ 
0&0&0&-1
\end{array}\right).
\end{equation}
The Dirac matrices satisfy the anticommutation relation:
\begin{equation}
\gamma_\mu \gamma_\nu + \gamma_\nu \gamma_\mu = 2g_{\mu\nu}.
\end{equation}
The standard representation for the 
Dirac matrices and spinors is used:
\begin{equation}                                                                
\gamma_0 =                                                                      
\left(\begin{array}{rr} 1 & 0 \\ 0 & -1 \end{array}\right),\quad              
\vec{\gamma}= \left(\begin{array}{rr}0&\vec{\sigma} \\ -\vec{\sigma}            
 & 0
\end{array}\right),\quad
\gamma_5 =-i\gamma_0\gamma_1\gamma_2\gamma_3 =
\left(\begin{array}{rr}0&-1\\-1&0\end{array}\right),
\label{b2}                                                              
\end{equation}                                                                  
where $\vec{\sigma}$ are the usual Pauli matrices. Note the minus sign for the
definition of $\gamma_5$.

The tensor $\sigma_{\mu\nu}$ is:
\begin{equation}
\sigma_{\mu\nu}=\frac{i}{2} (\gamma_\mu\gamma_\nu- \gamma_\nu\gamma_\mu). 
\label{b2p}                                                                      
\end{equation} 
Throughout the review the following convention is taken: 
\begin{equation} \hat{k} = k_\mu \gamma^\mu.
\end{equation}

The Dirac spinors  $u^\sigma (k)$ and $v^\sigma (k)$ for spin 1/2 fermion and
antifermion respectively satisfy the following Dirac equations:
\begin{equation}
(\hat{k}-m)u^\sigma (k)=\bar u^\sigma (k)(\hat{k}-m)=0,\quad
(\hat{k}+m)v^\sigma (k)=\bar v^\sigma (k)(\hat{k}+m)=0,
\end{equation}
with the normalization:
\begin{equation}
\bar u^\sigma (k) u^{\sigma'} (k)=2m\delta_{\sigma\sigma'}
=-\bar v^\sigma (k)v^{\sigma'} (k).
\end{equation}
They are given by:
\begin{equation} 
u^{\sigma}(k)= \sqrt{\varepsilon_k+m} \left(\begin{array}{c} 
1\\
\frac{\displaystyle\vec{\sigma}\cd\vec{k}}
{\displaystyle(\varepsilon_k+m)}
\end{array}\right) w^{\sigma}, \quad
v^{\sigma}(k)= \sqrt{\varepsilon_k+m} \left(\begin{array}{c} 
\frac{\displaystyle\vec{\sigma}\cd\vec{k}}
{\displaystyle(\varepsilon_k+m)}\\
1
\end{array}\right) w^{\sigma}.
\end{equation}
where $w^{\sigma}$ are the two-component spinors and                            
$\varepsilon_k = \sqrt{\vec{k}\,^2+m^2}$.
The summation over spin indices writes in that case:
\begin{equation}
\sum_{\sigma} u_{\alpha}^\sigma (k) \bar u_{\beta}^\sigma (k)= 
(\hat{k}+m)_{\alpha\beta},\quad
\sum_{\sigma}  v_{\alpha}^\sigma (k) \bar v_{\beta}^\sigma (k)= 
(\hat{k}-m)_{\alpha\beta}.
\end{equation}

The charge conjugation matrix is defined as: $U_c =  \gamma_2 \gamma_0$.

We use the following properties:  $\sigma_y\vec{\sigma}^t
=  -\vec{\sigma}\sigma_y$, $U_c\gamma_{\mu}^t=-\gamma_{\mu}U_c$, where
the superscript $t$ means transposition.                                           

The polarization vector $e^{(\lambda)}_{\mu}(p)$ of a spin 1 particle of mass
$m$ satisfies the following general relations:
\begin{equation}\label{ba4p}
p^{\mu}e^{(\lambda)}_{\mu}(p)=0,\quad
e^{*\lambda'}_{\mu}(p)e^{\lambda}_{\mu}(p)=-\delta_{\lambda'\lambda},
\quad
\sum_{\lambda}e^{\lambda}_{\nu}(p)e^{*\lambda}_{\mu}(p)=
-g_{\nu\mu}+\frac{p_{\nu}p_{\mu}}{m^2}.
\end{equation}

\chapter{Relation to other techniques}\label{other}                             
                                                                                
\section{Relation to the Weinberg rules}\label{lf-w}                            
As we already mentioned, the approach considered in this report is a 
generalization of the usual LFD obtained by the 
replacement of the particular value of the light-front vector $\omega = 
(1,0,0,-1)$ by the most general one $\omega = (\omega_0, \vec{\omega})$ 
with the only constraint $\omega^2 =0$. Hence, the graph technique developed in 
section~\ref{lfgt} is the direct generalization of the Weinberg 
rules~\cite{weinberg}. These rules have been obtained from old 
fashioned perturbation theory in the infinite momentum frame. We shall 
show below for the case of a spinless particle how to transform the 
amplitudes calculated in the covariant light-front graph technique to 
the Weinberg amplitudes.  This explicit transformation was found 
in~\cite{karm78}. 
\par                                                          
Any diagram of the covariant light-front graph technique can be 
represented as a sequence of intermediate states, each of them containing, 
besides 
the particles,  one spurion line and only one. An example of 
such contribution is indicated in fig.~\ref{intermed}. 

\begin{figure}[hbtp]
\centerline{\epsfbox{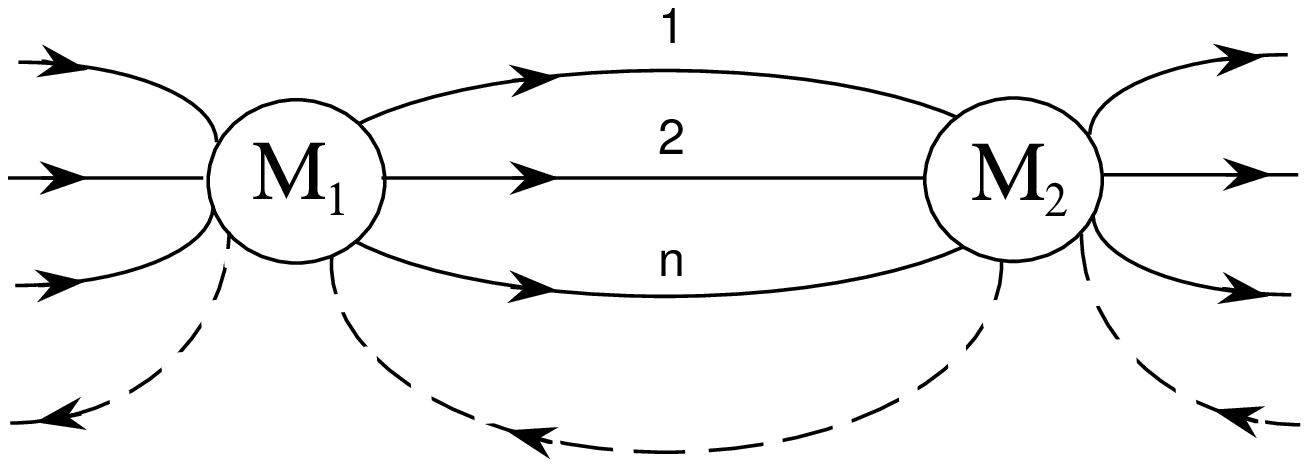}}
\figcap{Amplitude with one intermediate state shown explicitly. Other
intermediate states, if any, are absorbed by the amplitudes $M_1$ and
$M_2$.}
\label{intermed}
\end{figure}

\noindent
The expression 
for the corresponding amplitude has the form:                                                                           
\begin{eqnarray}\label{w1}                                                      
M&=&\int M_1 M_2\  \delta^{(4)}(\sum p_i-\sum k_i+\omega\tau)                   
\ \delta^{(4)}(\sum p'_i-\sum k_i+\omega\tau) \nonumber  \\                     
&&\times\frac{d\tau}{2\pi(\tau-i\epsilon)}\prod_{j=1}^n \theta(\omega\cd k_j)          
\ \delta(k_j^2-m_j^2)\ d^4k_j\ .                                                
\end{eqnarray}                                                                  
If a particle passes through a vertex without interaction, (like in
fig.~\ref{f1-ks92-b} the particle in the lower part of the diagram  propagates
without interaction in the upper vertex), the intermediate states can 
still be factorized. This is ensured by the following  representation of
the propagator:  
\begin{eqnarray}\label{w1a} 
\int (\cdots)\  \theta 
(\omega \cd k)\ \delta(k^2-m^2)\ d^4k = 
\int (\cdots)\ (2\omega\cd k_1)\
\theta (\omega\cd k_1)\ \delta(k_1^2-m^2)\ d^4k_1  
\nonumber \\ 
 \times\delta^{(4)}(k_1 +\omega\tau -k_2)\ d\tau\ \theta(\omega \cd k_2)\ 
\delta(k_2^2-m^2)\ d^4k_2\ , 
\end{eqnarray} 
which reduces the amplitude with the non-interacting particle to the form 
of eq.(\ref{w1}).

From the conservation equation $\sum p_i+\omega\tau=\sum k_i$ in \ref{w1}, it 
follows that:  $$\tau =\left[(\sum k_i)^2-(\sum p_i)^2\right]/(2\omega 
\cd\sum p_j)\ .  $$ Integrating over $dk_{0j}$ we get:                                                                            
\begin{equation}\label{w2}                                                      
M=\delta^{(4)}(\sum p_i-\sum p'_i)\int\frac{M_1M_2(-2\omega\cd\sum p_i)          
\delta^{(4)}(\sum p_i-\sum k_i+\omega\tau)d\tau} {2\pi[(\sum                    
p_i)^2-(\sum k_i)^2+i\epsilon]}\prod_j\frac{d^3k_j}{2\varepsilon_j}\ .          
\end{equation}                                                                  
In order to transform the expression (\ref{w2}) to a form analogous to 
old fashioned perturbation theory in the infinite momentum frame, we 
introduce the variables:                                     
\begin{eqnarray}\label{w3}                                                      
R_i^{ext}=p_i-y_i\sum p_j,\quad y_i=\omega\cd p_i/(\omega\cd\sum p_j)\ 
, \nonumber \\ R_i=k_i-x_i\sum p_j,\quad x_i=\omega \cd k_i/(\omega 
\cd\sum p_j)\ ,                                                                         
\end{eqnarray}                                                                  
satisfying $\sum y_i=\sum x_i=1$ and $R_i\cd\omega=R_i^{ext} 
\cd\omega=0$ with $R=(R_0,\vec{R}_{\perp}, 
\vec{R}_{\|})$, where $\vec{R}_{\perp}\cd\vec{\omega}=0$ and 
$\vec{R}_{\|}$ is parallel to $\vec{\omega}$. Taking into account that 
$R^2=-\vec{R}^2_{\perp}$, we can write the denominator of eq.(\ref{w2}) 
into the form:                                     
\begin{equation}\label{w4} \left[\sum                                           
\frac{(\vec{R}^{ext}_{i\perp})^2+m^2_i}{y_i}- \sum                              
\frac{\vec{R}_{i\perp}^2+m^2_i}{x_i}+i\epsilon\right]\ .                        
\end{equation}                                                                  
This form coincides with the one given by the Weinberg 
rules~\cite{weinberg}. The factor $d^3k_j/2\varepsilon_j$ in these 
variables turns into $d^2R_jdx_j/2x_j$, the domain of integration over 
$dx_j$ being from 0 to 1. The expression $$ (2\omega \cd\sum 
p_i)\delta^{(4)}(\sum p_i-\sum k_i+ \omega\tau)d\tau $$ turns into $$ 
2\delta^{(2)}(\sum \vec{R}_{i\perp})\ \delta(\sum x_i-1)\ . $$ By this 
way, one can transform the expression of any intermediate state.  The 
equivalence between the expression for any amplitude in the covariant 
formulation of light front dynamics and the one given by the Weinberg rules is
thus  realized. The vectors $\vec{R}_{\perp}$ play here the role of the 
momenta transverse to the infinite momentum, and the variables 
$x_i,y_i$ are analogous to the fractions of the particle momenta 
with respect to the infinite momentum.                                      

Note that the factorization formula (\ref{w1a}) is generalized for the 
propagator of a spin $1/2$ particle  as follows:  
\begin{eqnarray}\label{1}                                                       
\int (\ldots)(\hat{k}+m) \delta(k^2-m^2)d^4k = 
\int(\ldots)                     
\frac{\omega\cd k_1}{m}
\theta (\omega\cd k_1) \delta(k_1^2-m^2)(\hat{k}_1+m)d^4k_1
\nonumber \\                                                                    
\times\delta^{(4)}(k_1+\omega\tau-k_2)d\tau                 
\theta(\omega\cd k_2)                                                             
\delta(k_2^2-m^2) (\hat{k}_2-\hat{\omega}\tau +m) d^4k_2\ .                     
\end{eqnarray}                                                                  
The part $-\hat{\omega}\tau$ in (\ref{1}) corresponds to the instantaneous 
interaction which appears in light front dynamics, since the fermion 
line with the momentum $k_2$ now ``extends over a single time 
interval'' (see section \ref{lfgt}). 

\section{Relation between the Feynman amplitudes and the                        
Weinberg rules}\label{f-w} 

It is also instructive to derive the Weinberg 
rules directly from the Feynman diagrams. This allows one to avoid some 
subtle problems of light-front quantization. The direct relation 
between Feynman and the light-front amplitudes was shown for a few 
cases in ref.~\cite{schmidt} and investigated in detail in 
ref.~\cite{lbak}. The problem of renormalization was investigated
in ref. \cite{lbak2}. We illustrate this relation by the simplest case
of the triangle Feynman graph shown in fig.~\ref{fev}. This 
example has been considered also in ref.~\cite{sawicki}.  
Up to a multiplicative factor, 
the corresponding 
Feynman amplitude for scalar particles of mass m has the
form:                                                 
\begin{equation}\label{tri1}                                                    
F_{\triangle}= \int\frac{d^4k}{(2\pi)^4}                                        
\frac{1}{(k^2-m^2+i\epsilon)                                                    
((k-p)^2-m^2+i\epsilon)((k-p')^2-m^2+i\epsilon)}\ .                             
\end{equation}                                                                  
Introducing the new light front variables:                                      
\begin{equation}\label{fw2}                                                     
k_+=\frac{k_0+k_3}{\sqrt{2}},\quad k_-=\frac{k_0-k_3}{\sqrt{2}},\quad           
\vec{k}_{\perp}=(k_1,k_2) \ ,                                                   
\end{equation}                                                                  
we rewrite eq.(\ref{tri1}) in the form:                                         
\begin{equation}\label{tri2}                                                    
F_{\triangle}=\int\frac{dk_+d^2k_{\perp}}{(2\pi)^3}M_{\triangle},\quad          
M_{\triangle}=\int \frac{dk_-}{2\pi\phi_3}\frac{1}{(k_--H_1) (k_--H_2)          
(k_--H_3)},                                                                     
\end{equation}                                                                  
where $\phi_3=8k_+(k_+-p_+)(k_+-p'_+)$ and                                      
\begin{eqnarray}\label{tri3}                                                    
H_1&=&\frac{\vec{k}_{\perp}\,^2+m^2-i\epsilon}{2k_+} \nonumber \\
H_2&=&p_-+\frac{(\vec{k}_{\perp}-\vec{p}_{\perp})^2+m^2-i\epsilon}                
{2(k_+-p_+)} \nonumber \\                                                    
H_3&=&p'_-+\frac{(\vec{k}_{\perp}-\vec{p}\,'_{\perp})^2+m^2-i\epsilon}            
{2(k_+-p'_+)}\ .                                                                
\end{eqnarray}                                                                  
Let $p^2=p'^2=m^2$ and $q^2=(p-p')^2 < 0$. For simplicity we put                
$q_+=0$, i.e., $p_+=p'_+$. At $k_+<0$ and $k_+>p_+$ all the poles in            
$k_-$ of the integrand in eq.(\ref{tri2}) are above or below the             
real axis. Hence, the integral differs from zero $0<k_+<p_+$ only.           
Application of the residue theorem gives:                                       
\begin{equation}\label{tri4}                                                    
M_{\triangle}=-\frac{i}{\phi_3}\frac{1}{(H_1-H_2)(H_1-H_3)}\ .                   
\end{equation}                                                                  
Introducing $x=k_+/p_+$, we get:                                                
\begin{equation}\label{tri5}                                                    
F_{\triangle}(t)=-\frac{i}{(2\pi)^3}\int\frac{\displaystyle                     
1}{\displaystyle\left[\frac{\vec{k}^2_{\perp}+m^2}                              
{x(1-x)}-m^2\right]\left[\frac{(\vec{k}_{\perp}-x\vec{\Delta})^2 + m^2}         
{x(1-x)}-m^2\right]}\frac{d^2k_{\perp}dx}{2x(1-x)^2}\ ,                           
\end{equation}                                                                  
where $t=q^2=-\vec{\Delta}^2$.                                                  
The form (\ref{tri5}) reproduces the amplitude given by the Weinberg          
rules.                                                                          

\begin{figure}[htbp]
\centerline{\epsfbox{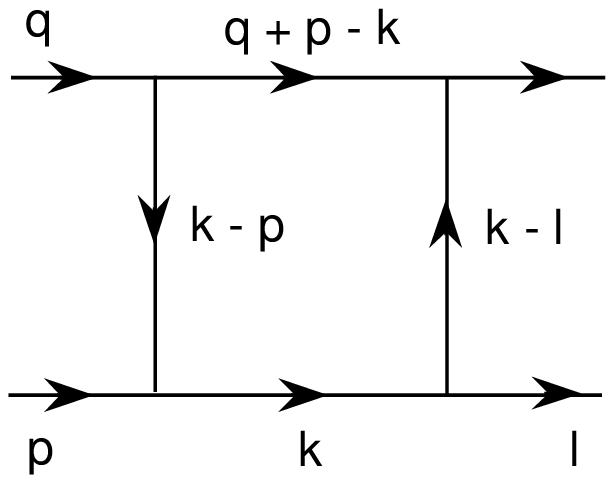}}
\figcap{Box Feynman graph.}
\label{box}
\end{figure} 

The next step is the box diagram of fig.~\ref{box}. This example
is considered in ref.~\cite{lbak}. The amplitude has the
form:                     
\begin{equation}\label{tri6}                                                    
F_{\Box }=\int D(k_1,m)D(k_2,\mu)D(k_3,m)D(k_4,\mu)\frac{d^4k}{(2\pi)^4}\ ,                 
\end{equation}                                                                  
with $D(k,m)=i/(k^2-m^2+i\epsilon)$ and $k_1=k$, $k_2=k-l$, $k_3=p+q-k$,          
$k_4=k-p$.  Particles are supposed to have mass $m$ or $\mu$ depending on their 
appearance in the $s$ or $t$ channel.                                                                   
This amplitude can be represented analogously to (\ref{tri2}):  
\begin{equation}\label{tri7}                                                    
F_{\Box }=\int\frac{dk_+k^2_{\perp}}{(2\pi)^3}M_{\Box}\ , \quad                 
M_{\Box }=\int\frac{dk_-}{2\pi\phi_4}\frac{1}{\prod_{i=1}^4(k_- - H_i)},        
\end{equation}                                                                  
where $\phi_4=16k_+(k_+-l_+)(k_+-p_+-q_+)(k_+-p_+)$ and $H_{1-4}$           
are analogous to (\ref{tri3}):                                               
\begin{eqnarray}\label{tri8} H_1  & =                                           
& \frac{\vec{k}\,^2_\perp+m^2-i \epsilon }{2 k_+}\ , \nonumber  \\              
H_2  & = &  l_- + \frac{(\vec{k}_\perp - \vec{l}_\perp)^2+\mu^2-i                 
\epsilon}{2(k_+ - l_+)}\ , \nonumber \\ 
H_3  & = & p_- + q_- +                  
\frac{(\vec{k}_\perp - \vec{p}_\perp - \vec{q}_\perp)^2 + m^2-i                 
\epsilon} {2 (k_+ -p_+-q_+)}\ , \nonumber  \\ 
H_4 & = & p_- +                   
\frac{(\vec{k}_\perp - \vec{p}_\perp)^2 + \mu^2-i \epsilon}{2 (k_+-p_+)}\ .       
\end{eqnarray}                                                                  
Let $p_+>l_+$.  Then at $k_+<0$ and at $k_+>p_+ +q_+$ the imaginary  parts
$ImH_{1-4}$ have the same signs and, hence, the integral for  $M_{\Box}$ is
zero.  At $0<k_+<l_+<p_+ <p_++q_+$ there is  one pole in the variable $k_-$ at
the down  half plane relative to the real axis and three  poles are at upper
half plane.  The residue reproduces the expression given by  the Weinberg rules
for the diagram analogous to fig.~\ref{box1}, as in  the case of the triangle
graph indicated  in fig.~\ref{fev}.

\begin{figure}[htbp]
\centerline{\epsfbox{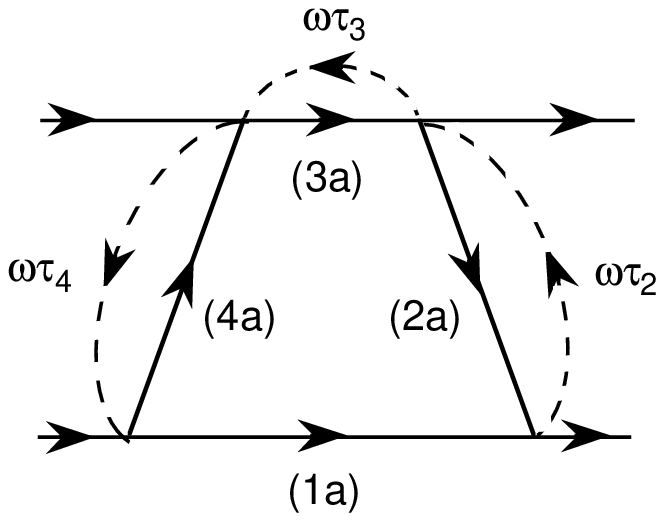}}
\figcap{Time-ordered box graph.}
\label{box1}
\end{figure}  

At $k_+>p_+>l_+$, but $k_+<p_+ +q_+$, the amplitude corresponds
to a diagram similar to fig. \ref{box1}, but up side down. Note that if the 
poles have imaginary parts of the same
signs, the corresponding lines of the time-ordered graph have the same
(clockwise  or counter clockwise) directions. For the poles with opposite
imaginary  parts, the directions of lines are
opposite.                           

New feature appears at $l_+<k_+<p_+$, when two pairs of poles are at  the
opposite sides of the real axis. In this case, any given residue  does not
reproduce any time-ordered diagram. However, the sum of them  can be
transformed~\cite{lbak} to a form consisting of two other parts,  each of them
coinciding with the amplitudes given by the Weinberg rules  for the diagrams
analogous to those shown below in figs.~\ref{box3},  \ref{box4}.  We explain 
this below in more detail  in the case of the covariant formulation of LFD. 
The general algorithm is developed in  ref.~\cite{lbak}.

\section{Relation between the Feynman amplitudes and the cova\-riant 
light-front graph tech\-nique}\label{f-lf} 
\subsection{Spin 0} 
Like in the case of the Weinberg rules, the Feynman approach and the 
covariant light-front graph technique give different ways of 
calculating one and the same $S$-matrix and, hence, give one and the 
same on-shell amplitude. However, these amplitudes are represented in 
different forms. It is therefore instructive to transform explicitly 
the Feynman amplitudes to the form of covariant light-front amplitudes.  

We will explain this transformation in the simplest case of spinless 
particles for the box diagram indicated in fig.~\ref{box}. The 
analytical expression for this diagram is given by (\ref{tri6}).  For 
 this transformation, we first represent the Feynman propagator in the 
following form~\cite{amn1}:                                                     
\begin{eqnarray}\label{flf3}                                                    
iD(k,m)&=&-\frac{1}{k^2-m^2+i\epsilon} = \int                                       
\frac{\delta^{(4)}(\omega\tau+k-p)}                                             
{\tau-i\epsilon}\ \theta(\omega\cd p)\ \delta(p^2-m^2)\ d^4p\
d\tau                
\nonumber \\ 
&&+\int \frac{\delta^{(4)}(\omega\tau-k-p)}{\tau-i\epsilon}\
\theta(\omega\cd p)\ \delta(p^2-m^2)\ d^4p\ d\tau\ .            
\end{eqnarray}                                                                  
Substituting (\ref{flf3}) in (\ref{tri6}), we get:                              
\begin{eqnarray}\label{flf4}                                                    
F_{\Box}=                                                                       
\int \frac{d^4k}{(2\pi)^4}                                                      
&\displaystyle \biggl\{&\frac{\delta^{(4)}(\omega\tau_1+k-p_1)} 
{\tau_1-i\epsilon}\ \theta(\omega\cd p_1)\  \delta(p_1^2-m^2)\biggl|_{=(1a)}
\nonumber \\                                                                    
&+\displaystyle 
&\frac{\delta^{(4)}(\omega\tau_1-k-p_1)}{\tau_1-i\epsilon}\ 
\theta(\omega \cd p_1) \ 
\delta(p_1^2-m^2)\biggr|_{=(1b)}\biggr\}d^4p_1d\tau_1                   
\nonumber \\                                                                    
&\times\displaystyle \biggl\{&\frac{\delta^{(4)}(\omega\tau_2+k-l-p_2)} 
{\tau_2-i\epsilon}\ \theta(\omega\cd p_2)\ 
\delta(p_2^2-\mu^2)\biggl|_{=(2a)}       
\nonumber \\                                                                    
&+\displaystyle &\frac{\delta^{(4)}(\omega\tau_2-k+l-p_2)} 
{\tau_2-i\epsilon}\ \theta(\omega \cd p_2) \ 
\delta(p_2^2-\mu^2)\biggr|_{=(2b)}\biggr\}d^4p_2d\tau_2 
\nonumber \\ 
&\times\displaystyle 
\biggl\{&\frac{\delta^{(4)}(\omega\tau_3+k-p-q-p_3)} 
{\tau_3-i\epsilon}\ \theta(\omega\cd p_3)\ 
\delta(p_3^2-m^2)\biggl|_{=(3a)}       
\nonumber \\                                                                    
&+\displaystyle &\frac{\delta^{(4)}(\omega\tau_3-k+p+q-p_3)} 
{\tau_3-i\epsilon}\ \theta(\omega \cd p_3) \ 
\delta(p_3^2-m^2)\biggr|_{=(3b)}\biggr\}d^4p_3d\tau_3                   
\nonumber \\                                                                    
&\times\displaystyle \biggl\{&\frac{\delta^{(4)}(\omega\tau_4+k-p-p_4)} 
{\tau_4-i\epsilon}\ \theta(\omega\cd p_4)\ 
\delta(p_4^2-\mu^2)\biggl|_{=(4a)}       
\nonumber \\                                                                    
&+\displaystyle 
&\frac{\delta^{(4)}(\omega\tau_4-k+p-p_4)}{\tau_4-i\epsilon}\ 
\theta(\omega \cd p_4)\ 
\delta(p_4^2-\mu^2)\biggr|_{=(4b)}\biggr\}d^4p_4d\tau_4\ .                 
\end{eqnarray}                                                                  
Let us first consider the product of the first items in the four braces 
in eq.(\ref{flf4}). Using schematic notations, we have:                         
\begin{eqnarray}\label{flf5}                                                    
\int (1a)(2a)(3a)(4a)\ldots = \int \frac{d^4k}{(2\pi)^4} 
\frac{\delta^{(4)}(\omega\tau_1+k-p_1)}{\tau_1-i\epsilon} 
\frac{\delta^{(4)}(\omega\tau_2+k-l-p_2)}{\tau_2-i\epsilon} 
\nonumber \\        
\times\frac{\delta^{(4)}(\omega\tau_3+k-p-q-p_3)}{\tau_3-i\epsilon}             
\frac{\delta^{(4)}(\omega\tau_4+k-p-p_4)}{\tau_4-i\epsilon}                     
(\cdots)d\tau_1 d\tau_2 d\tau_3 d\tau_4\ .                                      
\end{eqnarray}                                                                  
The functions $\theta(\omega \cd p_i)$, $\delta(p_i^2-\mu^2)$ and all 
other factors in (\ref{flf5}) are absorbed in $(\cdots)$. To exclude 
$\tau_1$ from the numerator, we make the following change of variables 
$$k+\omega\tau_1= k',\ \tau_2-\tau_1=\tau_2',\ 
\tau_3-\tau_1=\tau_3',\ \tau_4- \tau_1=\tau_4',$$ 
after which we find:                                                            
\begin{equation}\label{flf6}                                                    
\int (1a)(2a)(3a)(4a)\ldots=\int 
\frac{(\cdots)d\tau_1}{(\tau_1-i\epsilon) 
(\tau_2'+\tau_1-i\epsilon)(\tau_3'+\tau_1-i\epsilon) 
(\tau_4'+\tau_1-i\epsilon)}=0\ .                                                 
\end{equation}                                                                  
The numerator $(\cdots)$ in (\ref{flf6}) does not depend on $\tau_1$.  
All the poles of the integrand in (\ref{flf6}) relative to $\tau_1$ are 
above the real axis.  Therefore the integral (\ref{flf6}) equals to 
zero.  Similarly we get $\int (1b)(2b)(3b)(4b)=0$.  

We suppose that all the external momenta are on the mass shells and,  for
convenience, $\omega \cd p >\omega \cd l$.  The expression  (\ref{flf4})
contains 16 items.  However, only three of them differ  from zero, namely,
$\int (1a)(2b)(3b)(4b)$, $\int (1a)(2a)(3b)(4a)$  and\- $\int (1a)(2a)(3b)(4b)$. 
In all other items, except for the ones considered above $\int
(1a)(2a)(3a)(4a)$ and $\int (1b)(2b)(3b)(4b)$,  we get the products of  the
$\theta$-functions with mutually incompatible restrictions on the  scalar
products of $\omega$ with four-momenta. Let assign to the items  $(a)$ the
arrow counter clockwise in the loop and to the items $(b)$ --  the clockwise
arrow.  Put on the external lines with the momenta $p$  and $q$ the arrows
incoming to the box and to the lines with momenta  $l$ and $p+q-l$ -- the
outcoming arrows. Then the items which disappears due to the $\theta$-functions
correspond to the diagrams containing at  least one vertex with all incoming or
outcoming lines, which are  interpreted as vacuum fluctuations. One of these
graphs, corresponding  to the product $\int (1b)(2a)(3b)(4b)$ is shown in
fig.~\ref{vacbox}. 

\begin{figure}[hbtp]
\centerline{\epsfbox{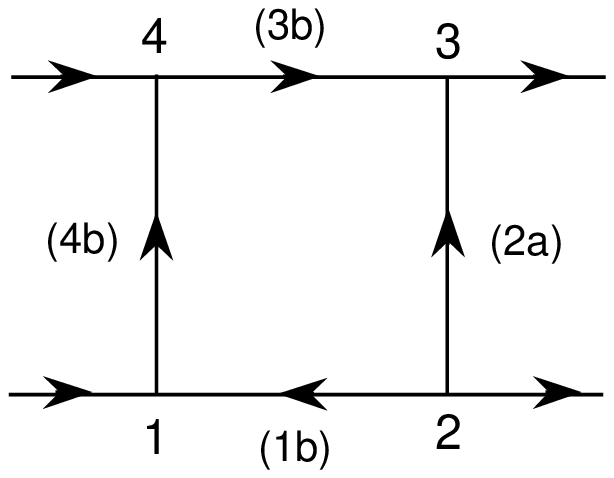}}
\figcap{Time-ordered box graph with the vacuum vertex No. 2.}
\label{vacbox}
\end{figure}

\noindent
The  vertex 2 in fig.~\ref{vacbox} corresponds to the
creation of particles  from the vacuum. The items $\int (1a)(2a)(3a)(4a)$ and
$\int  (1b)(2b)(3b)(4b)$, which are zero, correspond to the loops with all the 
lines directed clockwise or counter clockwise.  Now consider the item 
\begin{eqnarray}\label{flf7}  \int (1a)(2b)(3b)(4b)\ldots &=& \int 
\frac{\delta^{(4)}(\omega\tau_1+k-p_1)}{\tau_1-i\epsilon}                       
\frac{\delta^{(4)}(\omega\tau_2+l-k-p_2)}{\tau_2-i\epsilon} 
\nonumber\\                                                                              
&\times& \frac{\delta^{(4)}(\omega\tau_3+p+q-k-p_3)}{\tau_3-i\epsilon} 
\frac{\delta^{(4)}(\omega\tau_4+p-k-p_4)}{\tau_4-i\epsilon}                     
(\cdots)\ d\tau_1 \ . \nonumber\\
&&                                                           
\end{eqnarray}                                                                  
Again after replacing
$k+\omega\tau_1=k'$,                                      
$\tau_{2,3,4}+\tau_1=\tau'_{2,3,4}$ we get:                            
\begin{equation}\label{flf7a}                                                   
\int
(1a)(2b)(3b)(4b)\ldots=\int                                                
\frac{(\cdots)d\tau_1}{(\tau_1-i\epsilon)                                       
(\tau'_2-\tau_1-i\epsilon)(\tau'_3-\tau_1-i\epsilon)                              
(\tau'_4-\tau_1-i\epsilon)}\
,                                                   
\end{equation}                                                                  
where $(\cdots)$ does not depend on $\tau_1$. Calculating the
residue            at $\tau_1=i\epsilon$ and deleting the prime over $\tau$'s
we find:

\begin{eqnarray}\label{flf8}                                                    
\int  (1a)(2b)(3b)(4b)\ldots&=&i\int\theta(\omega\cd k)\delta(k^2-m^2)
\nonumber \\
&&\times\theta(\omega\cd (l-k))\ \delta((l-k +         
\omega\tau_2)^2-\mu^2)\nonumber \\                                                
&&\times\theta(\omega\cd (p+q-k))\ \delta((p+q-k+\omega\tau_3)^2-m^2)                   
\nonumber \\ 
&&\times\theta(\omega\cd (p-k))\ \delta((p-k+\omega\tau_4)^2-\mu^2) 
\nonumber \\ 
&&\times\frac{d\tau_2}{(\tau_2-i\epsilon)}\frac{d\tau_3}{(\tau_3-i\epsilon)} 
\frac{d\tau_4}{(\tau_4-i\epsilon)}\frac{d^4k}{(2\pi)^3}\ .                  
\nonumber\\                                                                     
&&                                                                              
\end{eqnarray}                                                                  

The expression (\ref{flf8}), after separating the factor $i$  (due to its
presence in the definition (\ref{rul8}) of the amplitude),  exactly corresponds
to the light-front amplitude  indicated by the diagram in fig.~\ref{box1}. The
integral $\int (1a)(2a)(3b)(4a)$ is calculated analogously (by integrating over
the variable  $\tau_3$ instead of $\tau_4$) and corresponds to the  diagram 
having the form of the overturned trapezium (relative to fig. \ref{box1}).

\begin{figure}[htbp]
\centerline{\epsfbox{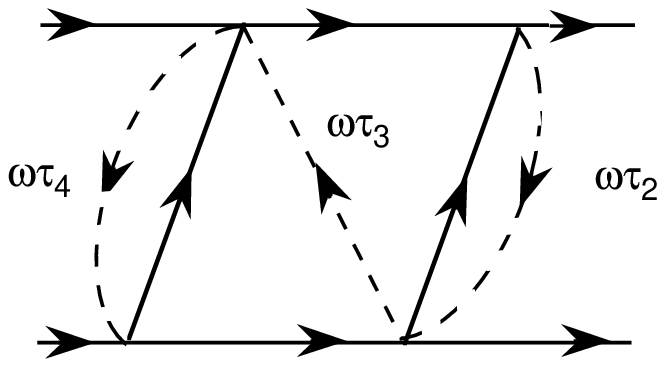}}
\figcap{Time-ordered box graph.}
\label{box3}
\end{figure}

\begin{figure}[hbtp]
\centerline{\epsfbox{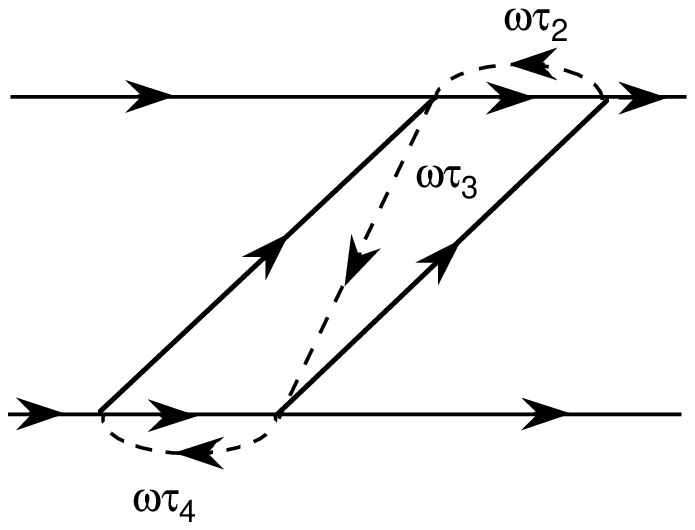}}
\figcap{Time-ordered box graph.}
\label{box4}
\end{figure}

There are  two other diagrams in the light-front dynamics shown in 
figs.~\ref{box3} and~\ref{box4}.  We can find both of them in the item $\int 
(1a)(2a)(3b)(4b)$. After similar transformations we get:  
\begin{equation}\label{flf9} 
\int (1a)(2a)(3b)(4b)\ldots =\int 
\frac{(\cdots)d\tau_1}{(\tau_1-i\epsilon) 
(\tau_2+\tau_1-i\epsilon)(\tau_3-\tau_1-i\epsilon)                              
(\tau_4-\tau_1-i\epsilon)}\ ,                                                   
\end{equation}                                                                  
Taking the sum of the two residues at $\tau_1=+i\epsilon$ and                   
$\tau_1=-\tau_2+i\epsilon$, we obtain two items:                                
\begin{equation}\label{flf10}                                                   
\int (1a)(2a)(3b)(4b)\ldots                                                     
=\int                                                                           
\left[\frac{1}{\tau_2\tau_3\tau_4}-\frac{1}{\tau_2(\tau_2+\tau_3)               
(\tau_2+\tau_4)}\right](\cdots)\ .                                              
\end{equation}                                                                  
Each of them does not correspond to any diagram. However,                       
(\ref{flf9}) indeed gives the sum of the diagrams in                            
figs.~\ref{box3} and~\ref{box4}.                                                
                                                                                
To transform it to the appropriate form, we shall use the simple                
formula: \begin{equation}\label{flf13}                                          
\frac{1}{\alpha\beta} =\frac{1}{\alpha+\beta} (\frac{1}{\alpha} +               
\frac{1}{\beta})\ .                                                             
\end{equation}                                                                  
This formula should be applied recursively to pairs of factors in the 
integrand having the poles in the integration variables with the
imaginary parts of  opposite signs. For example, applying (\ref{flf13}) to 
$1/(\beta_2\beta_3\beta_4)$, first to $1/\beta_2\beta_3$ and then again 
to $1/\beta_2\beta_4$, we get:
\begin{equation}\label{flf14} 
\frac{1}{\beta_2\beta_3\beta_4} = \frac{1}{(\beta_2+ \beta_3)                   
\beta_3\beta_4} + \frac{1}{(\beta_2 + \beta_3) (\beta_3+ \beta_4)               
\beta_4} +\frac{1}{(\beta_2 +\beta_3) (\beta_2 +\beta_4)\beta_2}\ .             
\end{equation}                                                                  
We apply (\ref{flf14}) to the integrand of (\ref{flf9}), i.e.,  substitute
$\beta_2=\tau_2 +\tau_1 -i\epsilon$, $\beta_3 =\tau_3- \tau_1 -i\epsilon$ and
$\beta_4 = \tau_4 -\tau_1 -i\epsilon$.  The factors $1/\beta_2$ and $\beta_3$
in $1/\beta_2\beta_3$ have the poles relative to the integration variable
$\tau_1$  with the  imaginary parts of opposite signs. The same is true for the
factors  $\beta_2$ and $\beta_4$ in $1/\beta_2\beta_4$. By this way, we
transform the integrand of (\ref{flf9}) to the form  which after integration
over $d\tau_1$ and standard replacement of variables leads to the
expression:                                              
\begin{eqnarray}\label{flf12}                                                   
\int (1a)(2a)(3b)(4b)\ldots &=&i\int\theta(\omega\cd k)\delta(k^2-m^2)
\nonumber \\
&&\times\theta(\omega\cd (k-l))                         
\delta((k-l+\omega \tau_2 - \omega\tau_3)^2-\mu^2)
\nonumber \\           
&&\times\theta(\omega\cd (p+q-k)) \delta((p+q-k+\omega\tau_3)^2-m^2)                   
\nonumber \\
&&\times\theta(\omega\cd (p-k))                                       
 \delta((p-k+\omega\tau_4)^2-\mu^2) \nonumber\\                                              
 &&\times\frac{d\tau_2d\tau_3d\tau_4}                                                   
 {(\tau_2-i\epsilon)(\tau_3-i\epsilon)(\tau_4-i\epsilon)}                       
 \frac{d^4k}{(2\pi)^3}                                                  
 \nonumber \\                                                                   
&+&i\int\theta(\omega\cd k)\delta(k^2-m^2)
\nonumber \\
&&\times \theta(\omega\cd (k-l)) \delta((k-l+\omega\tau_3-\omega\tau_4))^2 
-\mu^2)
\nonumber \\                                                       
&&\times\theta(\omega\cd (p+q-k)) \delta((p+q-k +\omega\tau_2 + \omega\tau_4 
-\omega \tau_3)^2 -m^2)
 \nonumber \\ 
&& \times\theta(\omega\cd (p-k)) 
\delta((p-k+\omega \tau_4)^2 -\mu^2) \nonumber \\
&&\times\frac{d\tau_2d\tau_3d\tau_4} 
 {(\tau_2-i\epsilon)(\tau_3-i\epsilon)(\tau_4-i\epsilon)}                       
 \frac{d^4k}{(2\pi)^3}\ .                                                 
\end{eqnarray}                                                                  
Two items in eq.(\ref{flf12}) exactly correspond to the amplitudes for          
the diagrams of figs.~\ref{box3} and~\ref{box4}.                                
The method can be generalized to more complicated diagrams. The general         
case is considered in ref.\cite{lbak} and the renormalization 
in ref. \cite{lbak2}.                                          
                                                                                
\subsection{Spins 1/2 and 1}\label{nonzs}                                                    
For spin 1/2, one should use the following representation of the                
propagator~\cite{amn2}:                                                         
\begin{eqnarray}\label{flf15a}                                                  
iS(k) &=& -\frac{\hat{k} +m}{k^2-m^2+i\epsilon} 
\nonumber \\ 
&=&                                 
\quad\int \frac{\delta^{(4)} (\omega\tau+k - p)} {\tau - i\epsilon}                  
(\hat{p} -\hat{\omega} \tau +m)                                                 
\theta(\omega\cd p)\                                                              
\delta(p^2-m^2)\ d^4p\ d\tau \nonumber \\                                       
&&-\int \frac{\delta^{(4)} (\omega\tau-k - p)} {\tau - i\epsilon}               
(\hat{p} -\hat{\omega} \tau -m)                                                 
\theta(\omega\cd p)\                                                              
\delta(p^2-m^2)\ d^4p\ d\tau  \nonumber \\                                                                                                                 
&=&\quad\int \frac{\delta^{(4)} (\omega\tau+k - p)} {\tau -                          
i\epsilon}(\hat{p} +m)                                                          
\theta(\omega\cd p)\                                                              
\delta(p^2-m^2)\ d^4p\ d\tau \nonumber\\                                        
&&+\int \frac{\delta^{(4)}                                                      
(\omega\tau-k - p)} {\tau - i\epsilon}(m-\hat{p})                               
\theta(\omega\cd p)\                                                              
 \delta(p^2-m^2)\ d^4p\ d\tau \nonumber \\                                      
&&-\frac{\hat{\omega}}{2(\omega\cd k)}\ . \label{flf15b}                          
\end{eqnarray}                                                                  
In the representation given by eq.(\ref{flf15b}), the propagator contains  the
spin parts $(\hat{p}+m)$ and $(m-\hat{p})$ with the on-shell four-momentum
$p^2=m^2$ and the extra contact term $-\hat{\omega}/2(\omega\cd k)$, like in
the rules of the graph technique, sect. \ref{rule-spin}.  At
$\omega=(1,0,0,-1)$ this contact term coincides with the one given in
refs.~\cite{brodsky73,drell70,chang73}.                                      

For the propagator with spin 1, we have the representation:                      
\begin{eqnarray}\label{flf16a}                                                  
iD_{\mu\nu}(k) &=& -\frac{-\displaystyle g_{\mu\nu}                               
+\frac{\displaystyle k_{\mu}k_{\nu}}                                            
{\displaystyle m^2}} {k^2-m^2+i\epsilon}                                       
\nonumber \\                                                                    
&=&\quad\int \frac{\delta^{(4)} (\omega\tau+k - p)} {\tau - i\epsilon}                
 \left(-g_{\mu\nu}+\frac{(p-\omega\tau)_{\mu}                                    
(p-\omega\tau)_{\nu}}{m^2}\right)                                               
\theta(\omega\cd p)\                                                              
\delta(p^2-m^2)\ d^4p\ d\tau \nonumber \\                                       
&&+\int \frac{\delta^{(4)} (\omega\tau-k - p)} {\tau - i\epsilon}               
\left(-g_{\mu\nu}+\frac{(p-\omega\tau)_{\mu}                                     
(p-\omega\tau)_{\nu}}{m^2}\right)                                               
\theta(\omega\cd p)\                                                              
\delta(p^2-m^2)\ d^4p\ d\tau \nonumber\\&& \\                                   
&=&\quad\int \frac{\delta^{(4)} (\omega\tau+k - p)}                                  
{\tau - i\epsilon}                                                              
\left(-g_{\mu\nu}+\frac{p_{\mu} p_{\nu}}{m^2}\right)                             
\theta(\omega \cd p)\                                                              
\delta(p^2-m^2)\ d^4p\ d\tau                                                    
\nonumber\\ 
&&+\int \frac{\delta^{(4)} (\omega\tau-k - p)} {\tau -              
i\epsilon}                                                                      
\left(-g_{\mu\nu}+\frac{p_{\mu} p_{\nu}}{m^2}\right)                             
\theta(\omega\cd p)\                                                              
\delta(p^2-m^2)\ d^4p\ d\tau \nonumber \\                                       
&&-\frac{k_{\mu}\omega_{\nu}+k_{\nu}\omega_{\mu}} {2(\omega\cd k) m^2}             
- \frac{\omega_{\mu}\omega_{\nu}}{4(\omega\cd k)^2m^2} (k^2-m^2)\ .               
 \label{flf16b}                                                                 
\end{eqnarray}                                                                  
The item in the last line of (\ref{flf16b}) is the vector contact term.         
                                                                                
Let now the particle 1 in fig.~\ref{box} has spin 1/2. Then the contact 
term for the particle 1 does not contribute to the amplitude indicated 
in fig.~\ref{box1}. Indeed, this amplitude was calculated in 
eq.(\ref{flf7a}), as the residue at $\tau_1 =i\epsilon$ in the pole of 
the propagator of particle 1.  Replacing this propagator by the contact 
term, we see that the only pole above the real axis disappears and the 
amplitude turns to zero.  However, the contact terms for the particles 
2,3 and 4 contribute to the diagram of fig.~\ref{box1}. According to 
(\ref{flf15a}) they can be incorporated by replacing the spin part of 
the propagators $(\hat{p}\pm m)$ by $(\hat{p}-\hat{\omega}\tau\pm m)$.                                              

These facts agree with the general rule given in section \ref{lfgt}:  the
contact terms have to be associated with these particles which lines  are
directly connected by a spurion line. This also coincides with the rule  given
in ref.~\cite{brodsky73}: contact terms modify only the  propagators
corresponding to the ``lines extending over a single time  interval''.  The
lines of the particles 2,3,4  in fig.~\ref{box1}  extend over single time
intervals and are associated with contact  terms, while the line for the
particle 1 extends over three time  intervals and is not associated with any
contact term.  Another example  is given by the diagram of fig.~\ref{box3}.  In
this case, the contact term contributes for the exchanged particles (if they
have spin) and does not contribute for the particles shown by the horizontal
lines. On the contrary, in fig.~\ref{box4} the contact term contributes for the
horizontal internal lines and does not contribute for the exchanged 
particles.                                              

\chapter{Relation between the deuteron components}                              
\label{dcomp}                                                                   
We give here for completeness the relations between the six invariant 
 functions coming in the decomposition of the wave function 
(\ref{nz1}), (\ref{nz2}) and (\ref{nz9}).  These two representations are 
connected by eq.(\ref{nz3}) and coincide with each other in the system 
of reference  where $\vec{{\cal P}} =0$.  In this system, we have 
$\vec{k_1} = -\vec{k}_2 =\vec{k}$.                       

The deuteron polarization vector $e^{(\lambda)}_{\mu}(p)$ in the 
representation with $\lambda = j = x,y,z$ in the rest system has the 
form: $e^{(j)}_i(0)=\delta^j_i$, $e^{(j)}_0(0)=0$. In an arbitrary 
system where $\vec{p}\neq 0$, it can be obtained by the Lorentz 
transformation (cf.(\ref{sc4})):                                                               
\begin{equation}                                                                
\vec{e}\,^{(j)}(p) = \vec{e}\,^{(j)}(0) + 
\frac{\left(\vec{e}\,^{(j)}(0)\cd 
\vec{p}\,\right)\vec{p}}{M(M+p_0)},\;\;\;\; e^{(j)}_0(p) 
=\frac{\vec{e}\,^{(j)}(0)\cd\vec{p}}{M}\ .                            
\label{ba3}                                                                      
\end{equation}                                                                  
In the system of reference where $\vec{{\cal P}}=0$, we have:                     
$$p_0=\frac{4\varepsilon_k^2 + M^2}{4\varepsilon_k},\;\;\;                              
\vec{p}= -\vec{n}\frac{\varepsilon_k^2 - m^2}{\varepsilon_k}\ .$$                     
 With these expressions                  
for $p$ we get from eq.(\ref{ba3}):                                              
\begin{equation}\label{ba4}                                                                
e^{(j)}_i(p)= \delta^j_i + \frac{(2\varepsilon_k - M)^2}                         
{4M\varepsilon_k}n_i n_j,                                                       
\;\;\;e^{(j)}_0(p)= -\frac{4\varepsilon_k^2-M^2}{4M\varepsilon_k}n_j\ .          
\end{equation}                                                                  
The polarization vector (\ref{ba4}) satisfies the general relations 
(\ref{ba4p}).

Substituting in eqs.(\ref{nz1},\ref{nz2}) the expressions for the 
nucleon spinors and Dirac matrices given in appendix \ref{nota} and the 
deuteron polarization vector given above, we obtain the structure of 
the wave function given in eq.(\ref{nz8}). Comparing the coefficients 
of identical spin structures, we can find the expressions of 
$\varphi_i$ in terms of $f_i$. In the approximation where $M\approx 2m$,
they have the form:                                                  
\begin{eqnarray}                                                                
\varphi_1 &=& \frac{m^2(2\varepsilon_k + m)}{4\varepsilon_k k^2}f_2             
+\frac{m^2}{4\varepsilon_k(\varepsilon_k + m)}(\sqrt{2}f_1 - f_3 + zf_4        
- \sqrt{3}zf_6)\ ,                                                              
\label{ba5}\\                                                                    
\varphi_2 &=& \frac{m}{4\varepsilon_k}(\sqrt{2}f_1 - f_2 -f_3 -2zf_4)\ ,        
\label{ba6}\\                                                                    
\varphi_3 &=& -\frac{\sqrt{2}k^5}{4\varepsilon_k^2(\varepsilon_k + 
m)^3}zf_1 -\frac{(2\varepsilon_k + 
m)k^3}{4\varepsilon_k^2(\varepsilon_k + m)^2}zf_2 
+\frac{(\varepsilon_k^2 + 4m\varepsilon_k + m^2)k} 
{4\varepsilon_k^2(\varepsilon_k + m)}zf_3                                       
\nonumber \\                                                                    
&&+\frac{3m}{2k}\left(1-\frac{z^2k^6}                                            
{6m\varepsilon_k^2(\varepsilon_k + m)^3}\right)f_4                              
+\frac{\sqrt{3}m}{2k}\left(1 +                                                  
\frac{z^2k^6}{2m\varepsilon_k^2(\varepsilon_k + m)^3}\right)f_6\ ,              
\label{ba7} \\                                                                   
\varphi_4 &=& -\frac{3m}{2k}f_4 + \frac{\sqrt{3}m}{2k}f_6\ ,                     
\label{ba8}\\                                                                    
\varphi_5 
&=&\frac{1}{2}\sqrt{\frac{3}{2}}\frac{m^2}{k\varepsilon_k}f_5\ ,      
\label{ba9} \\                                                                   
\varphi_6 &=&                                                                   
\frac{k^4}{2m\varepsilon_k(\varepsilon_k+m)^2}(\sqrt{2}f_1-f_2                  
+zf_4-\sqrt{3}zf_6)                                                            
-\frac{(\varepsilon_k^2+4m\varepsilon_k+m^2)}{2m\varepsilon_k}f_3\ .            
\label{ba10} \end{eqnarray}                                                      
                                                                                
\chapter{Two-body kinematical relations}\label{kr}                              
In the calculation of the electromagnetic amplitude we have to express the
scalar products of the four vectors $\omega$, $k_1$,  $k_2$, $k'_1$, $k'_2$,
$p$, $p'$,  $q$ with each other in terms of the integration variables and the
four-momentum transfer squared.    The variables and, hence, the kinematical
relations for the one loop  diagram of fig. \ref{f1-ks92-b} (impulse
approximation) and for the two loop diagram of fig. \ref{ct} are different. 

\section{One loop diagram}
The one loop diagram is shown in fig. \ref{f1-ks92-b}.
All four-momenta are on the corresponding mass shells:
$$k_1^2=m_1^2,\, k_2^2=k_2'^2=m_2^2,\, p^2=M^2,\, p'^2={\cal M}^2,\,
q^2=-\vec{\Delta}^2=-Q^2=t,\, \omega^2=0.$$
We consider the general case of different masses, M being the mass of 
the initial state, while $\cal M$ is the mass of the final state.

We start from the four-vector $R_1$ already defined in
eq.(\ref{sc8}):            
 $$R_1=k_1-xp,\quad x=\omega\cd k_1/\omega\cd p.$$

Since $R_1\cd \, \omega=0$, it can be represented as $R_1=(R_0,
\vec{R}_{\perp}, \vec{R}_{\|})$ with  
$\vec{R}_{\perp}\cd\, \vec{\omega}=0$ and $\vec{R}_{\|}$ is parallel to
$\vec{\omega}$.  
Similarly we represent $q=(q_0,\vec{\Delta},\vec{q}_{\|})$ with 
$\vec{\Delta}\cd\, \vec{\omega}=0$.  With these definitions, we 
immediately get (with $\omega\cd q=0$):  
\begin{equation}\label{ap0} 
\omega\cd k_1 =x\omega\cd p,\,  \omega\cd k_2 =(1-x)\omega\cd p,\, 
\omega\cd k'_2=(1-x)\omega\cd p,\, \omega\cd p=\omega\cd p'\ . 
  \end{equation} 

We get for the various scalar products the following relations.

\noindent
Scalar products with $k_1$:
\begin{eqnarray}\label{ap1}
k_1\cd k_2&=& (s-m_1^2-m_2^2)/2,
\nonumber\\
k_1\cd k_2'&=& (s'-m_1^2-m_2^2)/2,
\nonumber\\                                                                  
k_1\cd p&=& xM^2/2+(1-x)s/2+(m_1^2-m_2^2)/2,
\nonumber\\                                                                  
k_1\cd p'&=&k_1\cd p+k_1\cd q,
\nonumber\\                
k_1\cd q&=&-\vec{R}_{\perp}\cd\vec{\Delta}+xp\cd q\ ,
\end{eqnarray}
with $s$ and $s'$ given by:
\begin{eqnarray}\label{ap6}
s&\equiv& (k_1+k_2)^2=\frac{\vec{R}_{\perp}^2+m_1^2}{x} 
+\frac{\vec{R}_{\perp}^2+m_2^2}{1-x}\ , \nonumber \\
s'&\equiv& (k_1'+k_2')^2=\frac{\vec{R}_{\perp}'^2+m_1^2}{x}
+\frac{\vec{R}_{\perp}'^2+m_2^2}{1-x} \ .
\end{eqnarray}
where $\vec{R}'_{\perp}=\vec{R}_{\perp}-x\vec{\Delta}$.

Scalar products with $k_2$:
\begin{eqnarray}\label{ap2}
k_2\cd k_2'&=&m_2^2+k_2\cd q +(1-x)(s'-{\cal M}^2-s+M^2)/2,
\nonumber\\                
k_2\cd p&=&(1-x)M^2/2+xs/2+(m_2^2-m_1^2)/2,
\nonumber\\                
k_2\cd p'&=&k_2\cd p+k_2\cd q,
\nonumber\\                
k_2\cd q&=&\vec{R}_{\perp}\cd\vec{\Delta}+(1-x)p\cd q.
\end{eqnarray}
Scalar products with $k_2'$:
\begin{eqnarray}\label{ap3}
k_2'\cd p&=&k_2'\cd p'-k_2'\cd q,
\nonumber\\                
k_2'\cd p'&=&(1-x){\cal M}^2/2+xs'/2+(m_2^2-m_1^2)/2,
\nonumber\\                
k_2'\cd q&=&k_2\cd q-\vec{\Delta}^2.
\end{eqnarray}
Scalar products with $p$:
\begin{eqnarray}\label{ap4}
p\cd p'&=&(\vec{\Delta}^2+M^2+{\cal M}^2)/2,
\nonumber\\                
p\cd q&=&(\vec{\Delta}^2-M^2+{\cal M}^2)/2.
\end{eqnarray}
Scalar product with $p'$:
\begin{equation}\label{ap5}
p'\cd q=({\cal M}^2-M^2-\vec{\Delta}^2)/2.
\end{equation}

\section{Two-loop diagram}\label{twoloop}
The two-loop diagrams indicated on fig. \ref{ct4} corresponds to the
contribution of the contact term. One of them is reproduced in
fig.\ref{ct}, where all the momenta are explicitly indicated.

\begin{figure}[htbp]
\centerline{\epsfbox{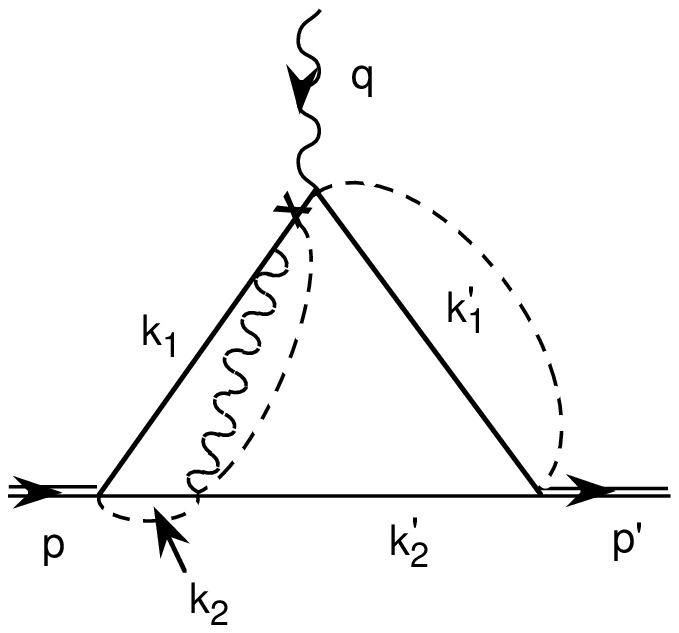}}
\figcap{Contact term of fig. \protect{\ref{ct4}}(a) contributing to the
deuteron electromagnetic form factors.} 
\label{ct}
\end{figure}

Some of the relations coincide with the ones given above.  We 
repeat them here for completeness.  For the scalar products of $\omega$ 
we immediately get:
\begin{equation}\label{kin0}                                                    
\omega\cd k_1=x\omega\cd p,\, \omega\cd k_2 =(1-x)\omega\cd p,\,                        
\omega\cd k'_1=x'\omega\cd p,\, \omega\cd k'_2=(1-x')\omega\cd p,\,                     
\omega\cd p=\omega\cd p',\, \omega\cd q=0.                                               
\end{equation}                                                                  
                                                                                
All the following invariants can be transformed to the variables                   
$\vec{k},\vec{k}\,'$ and $\vec{n}$ by the substitution:                            
\begin{eqnarray}\label{kin10}                                                   
\vec{R}_{\perp}&=&\vec{k}-(\vec{n}\cd\vec{k})\vec{n},\;                              
x=\frac{\varepsilon_1}{\varepsilon_1+\varepsilon_2}(1-\frac{\vec{n}\cd\vec{k}}{\
varepsilon_1});\\               
\vec{R}\,'_{\perp}&=&\vec{k}\,' -(\vec{n}\cd\vec{k}\,')\vec{n},\;                    
x'=\frac{\varepsilon'_1}{\varepsilon'_1+\varepsilon'_2}(1- 
\frac{\vec{n}\cd\vec{k}\,'}{\varepsilon'_1}),                   
\end{eqnarray}                                                                  
taking into account that
$\vec{\Delta}\cd\vec{n}=0$.                                                                                                          
We denote in this equation $\varepsilon_i=\sqrt{k^2+m_i^2}$, and similarly for 
quantities with prime. We thus get  the following scalar
products.                                                                                                                    

Scalar products with $k_1$:                                                 
 \begin{eqnarray}\label{kin2}                                                   
k_1\cd k_2&=&(s-m_1^2-m_2^2)/2, \nonumber \\ 
k_1\cd k'_1&=&-\vec{R}_{\perp}\cd\vec{R}'_{\perp} 
+x\vec{\Delta}\vec{R}'_{\perp} 
-x'\vec{\Delta}\cd\vec{R}_{\perp} + x(1-x')s'/2 + x'(1-x)s/2 +                     
x x'\vec{\Delta}\,^2/2, \nonumber \\                                            
k_1\cd k'_2&=&k_1\cd p'+x(s'-{\cal M}^2)/2-k_1\cd k'_1,\quad k_1\cd 
p=(1-x)s/2+xM^2/2,\nonumber\\ 
k_1\cd p'&=&k_1\cd q+k_1\cd p,\quad k_1\cd 
q=-\vec{R}_{\perp}\cd\vec{\Delta}+xp\cd q.  
\end{eqnarray} 
Scalar products with $k_2$:  
\begin{eqnarray}\label{kin3} 
k_2\cd k'_1&=&k'_1\cd 
p+x'(s-M^2)/2-k_1\cd k'_1, \nonumber \\
k_2\cd k'_2&=&k_2\cd p'-k_2\cd 
k'_1+(1-x)(s'-{\cal M}^2)/2, \nonumber\\ 
k_2\cd p&=&xs/2+(1-x)M^2/2, \nonumber \\
k_2\cd p'&=&k_2'\cd q+k_2\cd p, \nonumber \\   
k_2\cd q&=&\vec{R}_{\perp}\cd\vec{\Delta}+(1-x)p\cd q\
.                                     
\end{eqnarray}                                                                  
Scalar products with $k'_1$:                                                
\begin{eqnarray}\label{kin4}                                                    
k'_1\cd
k'_2&=&(s'-m_1^2-m_2^2)/2,
\nonumber \\  
 k'_1\cd p&=&k'_1\cd p'-k'_1\cd q, 
 \nonumber \\     
k'_1\cd p'&=&(1-x')s'/2+x'{\cal M}^2/2, \nonumber \\ 
k'_1\cd q&=&-\vec{R}'_{\perp}\cd\vec{\Delta}+x'p'\cd q\ .
\end{eqnarray}                                                                  
Scalar products with $k'_2$:                                                
\begin{eqnarray}\label{kin5}                                                    
k'_2\cd p&=&k'_2\cd p'-k'_2\cd q, \nonumber \\  
k'_2\cd p'&=&x's'/2+(1-x'){\cal M}^2/2, \nonumber \\  
k'_2\cd q&=&\vec{R}'_{\perp}\cd\vec{\Delta}+(1-x')p'\cd q.
\end{eqnarray}                                                                  
The scalar products with $p$ and $p'$ are the same as (\ref{ap4}), (\ref{ap5}).

The kinematics for the final plane wave contribution (neglecting the
final state energy) is obtained from the above formulas
by  the substitution ${\cal M}=2m$,
$\vec{R}'_{\perp}=0$, $x'=1/2$ (or $\vec{k}\,'=0$).                                                                
The kinematics for the elastic deuteron  form factors is obtained 
by the replacement ${\cal M} \rightarrow M$.  

\chapter{Three-body kinematical relations}\label{e}

The scalar products appearing in the calculation of the amplitude indicated in 
fig.
\ref{nem} are  expressed through the integration variables
$\vec{R}_{i\perp},x_i$ and  $\vec{\Delta}$.  We give them below in the general
case of different  quark masses.  For $p\cd p'$ we have:
\begin{equation}\label{a1p}
p\cd p'=M^2+\vec{\Delta}^2/2\ .
\end{equation}
In order to find $p\cd k_i$ $(i=1,2,3)$, we calculate $(k_i-x_ip)^2= 
m_i^2-2x_ip\cd k_i+x_i^2M^2 =R_i^2=-\vec{R}_{i\perp}^2$. From here we 
find:
\begin{equation}\label{a1}
p\cd k_i=\frac{\vec{R}_{i\perp}^2+m_i^2}{2x_i}+\frac{1}{2}M^2x_i \ .
\end{equation}
Other scalar products are found analogously. We have:
\begin{equation}\label{a2}
p'\cd k_i=p\cd k_i-\vec{R}_{i\perp}\cd\vec{\Delta}+x_i\vec{\Delta}^2/2\ ,
\end{equation}
\begin{equation}\label{a3}
p'\cd k_i'=\frac{\vec{R}'^2_{i\perp}+m_i^2}{2x_i}+\frac{1}{2}M^2x_i \ ,
\end{equation}
\begin{equation}\label{a4}
p\cd k_i'=p'\cd k_i'+\vec{R}_{i\perp}'\cd\vec{\Delta}+x_i\vec{\Delta}^2/2\ ,
\end{equation}
\begin{equation}\label{a5}
k_i\cd k_j'=-\vec{R}_i\cd\vec{R}_j'+x_ip\cd k_j'+x_jp'\cd k_i-x_ix_jp\cd p'
\qquad\mbox{for}\,\,\,i,j=1,2,3.
\end{equation}
The scalar product $k_i\cd k_j$ is obtained from eq.(\ref{a5}) by 
deleting primes at all  variables, the scalar product $k_i'\cd k_j'$ 
is obtained by the following replacement in eq.(\ref{a5}):
$\vec{R}_i\rightarrow \vec{R}'_i$, $k_i\rightarrow k'_i$, 
$p\rightarrow p'$. 
The variables $x_i$ are the same for initial and final states.

Let us now express $\vec{R}_{i\perp}'$ through $\vec{\Delta}$ and the 
integration variables $\vec{R}_{i\perp}$ and $x_i$. The expressions are 
not symmetrical relative to the quark 1, which interacts with the photon,
 and to 
the spectator quarks 2 and 3.  We have $R_2'=k_2'-x_2p'$, 
$R_3'=k_3'-x_3p'$. With $k_2'=k_2$, $k_3'=k_3$ we find $R_2'=R_2-x_2q$, 
$R_3'=R_3-x_3q$. With
$R_1'=k_1'-x_1p'=R_1+(1-x_1)q+ \omega(\tau'-\tau)$,  we get:
\begin{equation}\label{a6}
\vec{R}_{1\perp}'=\vec{R}_{1\perp}+(1-x_1)\vec{\Delta}\ ,\quad
\vec{R}_{2,3\perp}'=\vec{R}_{2,3\perp}-x_{2,3}\vec{\Delta}\ .
\end{equation}
Note that $\vec{R}_{1\perp}'$ can be also found from the relation
$\vec{R}_{1\perp}'=-\vec{R}_{2\perp}'-\vec{R}_{3\perp}'$ .
The arguments $\vec{q}\,^2_i$ $(i=1,2,3)$ of the initial wave function 
are given by
$$
\vec{q}\,^2_i=\varepsilon_{q_i}^2-m^2_i\ ,
$$
where $\varepsilon_{q_i}$ is the energy of the particle in the
reference system where $\vec{{\cal P}}=\vec{k}_1+\vec{k}_2+\vec{k}_3=0$.
It can be represented in the invariant form as 
$\varepsilon_{q_i}={\cal P}\cd k_i/{\cal M}$ with 
${\cal P}=p+\omega\tau$, $\tau=({\cal M}^2-M^2)/(2\omega\cd p)$,
that gives:
\begin{equation}\label{a8}
\varepsilon_{q_i}=[p\cd k_i+\frac{1}{2}x_i({\cal M}^2-M^2)]/{\cal M}\ ,
\end{equation}
where ${\cal M}^2$ is the invariant mass squared:
\begin{equation}\label{a7}
{\cal M}^2=(k_1+k_2+k_3)^2=\sum_{i=1}^3 
\frac{\vec{R}_{i\perp}^2+m_i^2}{x_i}\ .
\end{equation}
The arguments $\vec{q}\,'^2_{i}$ of the final
wave function are constructed similarly from the final momenta.

\bibliography{database}

\end{document}